\documentclass[12pt]{report}
\usepackage{dbl12,rac,epsfig,amsmath,graphicx,feynmf,float,amssymb,arydshln,psfrag,verbatim,mathrsfs}

\def\tr{\mathrm{Tr}}

\newcommand{\nc}{\newcommand}
\nc{\beaa}{\begin{eqnarray*}} \nc{\eeaa}{\end{eqnarray*}} \nc{\beq}{\begin{equation}}
\nc{\eeq}{\end{equation}} \nc{\bea}{\begin{eqnarray}}   \nc{\eea}{\end{eqnarray}}
\nc{\baa}{\begin{array}}      \nc{\eaa}{\end{array}} \nc{\bit}{\begin{itemize}}
\nc{\eit}{\end{itemize}} \nc{\ben}{\begin{enumerate}}  \nc{\een}{\end{enumerate}}
\nc{\bce}{\begin{center}}     \nc{\ece}{\end{center}} \nc{\non}{\nonumber}

\newcommand{\nnmb}{\nonumber}

\newcommand{\cvev}{\left<\chi\right>}

\def\bed{\begin{description}}
\def\eed{\end{description}}

\def\non{\nonumber}

\def\k1slash{k_1\hspace{-10.5pt}/\ \ }

\def\phiphi{\phi_{{}_\Phi}}
\def\psiphi{\psi_{{}_\Phi}}
\def\phiphibar{\phi_{{}_{\overline{\Phi}}}}

\def\phichi{\phi_{{}_\chi}}
\def\psichi{\psi_{{}_\chi}}

\def\simge{\mathrel{%
   \rlap{\raise .57ex \hbox{$>$}}{\lower .57ex \hbox{$\sim$}}}}
\def\simle{\mathrel{
   \rlap{\raise 0.512ex \hbox{$<$}}{\lower 0.512ex \hbox{$\sim$}}}}

\def\subCDM{\mathit{CDM}}
\def\DM{\mathit{DM}}
\def\Lvis{L_{\mathit{vis}}}
\def\Lhid{L_{\mathit{hid}}}
\def\RHSn{\tilde{\nu}_R}
\def\im{\mathrm{Im}}
\begin{document}
\titlepage{Theory and Phenomenology of Dirac           
  Leptogeneis      }                  
    {Brooks D. Thomas   }                                 
    {Doctor of Philosophy}                          
    {Physics}                
    {2007}                                          
    {Associate Professor James D.\ Wells,     Chair \\       
     Professor Fred C.\ Adams    \\                       
     Professor Gordon L.\ Kane  \\                        
     Associate Professor Timothy A.\ McKay \\
     Assistant Professor Aaron T.\ Pierce}
\unnumberedpage                             
\copyrightpage{Brooks D. Thomas}            
\initializefrontsections                    

\dedicationpage{
\begin{center}
\begin{verse}
From the digital fountains\\
To the analog mountains\\
Let the mirror express the room.\\
\end{verse}
\vspace{.5cm}~\hspace{1.5cm}~---~David Berman\end{center}}           
\startacknowledgementspage                  
\indent I would like to thank my advisor, James D. Wells, for all the help and guidance
he's offered me over the past five years, and the members of my committee for all their
assistance.  I would also like to think all the colleagues in the MCTP with whom I've
worked and interacted during my tenure as a graduate student, and especially my research
collaborators David Morrissey and Manuel Toharia.


\tableofcontents                     
\listoffigures                       
\listoftables                        
\listofappendices                    

\pagestyle{empty} \startabstractpage{Theory and Phenomenology of Dirac Leptogenesis}
{Brooks D. Thomas}{Chair: James D. Wells}   
Dirac leptogenesis, in which neutrinos are purely Dirac and develop small but nonzero
effective masses without the aid of the see-saw mechanism, provides an interesting
alternative to the standard leptogenesis picture.  Here we review the theory and
phenomenology of Dirac leptogenesis and show that it is a viable theory capable of
simultaneously satisfying all relevant bounds from cosmology, neutrino physics, and
flavor violation.  In addition, we also explore several potential extensions of the
model, such as the possibility of right-handed sneutrino dark matter and the potential
for relating the leptogenesis mechanism to the origin of the \(\mu\)-term.  Theories with
a heavy gravitino and gaugino masses generated by anomaly mediation emerge as one natural
context for Dirac leptogenesis.  In such models the lightest neutralino is often expected
to be predominately wino or Higgsino, and is a viable dark matter candidate.  We conclude
with an examination of the prospects for detecting the effectively monoenergetic photon
signal that results from the annihilation of such a dark matter particle in the galactic
halo.

\startthechapters                    
\chapter{BARYOGENESIS\ AND\ LEPTOGENESIS\label{ch:introduction}}

\section{The Problem of the Baryon Asymmetry\label{sec:sakharov}}

\indent

The universe we live in is manifestly asymmetric between baryons and antibaryons.  This
statement can be quantified by introducing a parameter \(\eta\), defined as
\(\eta=n_{B}/n_{\gamma}\).  Here \(n_{B}\equiv n_{b}-n_{\overline{b}}\), where \(n_{b}\)
and \(n_{\overline{b}}\) are the baryon density and antibaryon density of our universe,
respectively, and \(n_{\gamma}\) is the present number density of photons.  The value of
\(\eta\) has recently been measured with great precision by WMAP~\cite{Bennett:2003bz} to
be within the range
\begin{equation}
  \eta=\left(6.1 \pm 0.3\right)\times 10^{-10},
  \label{eq:WMAPeta}
\end{equation}
which implies that the relic density of baryonic matter in our universe is somewhere in
the range
\begin{equation}
  0.021 < \Omega_{b}h^2< 0.025.\label{eq:OmegaBaryon}
\end{equation}
Since the standard cosmology is manifestly symmetric with respect to matter and
antimatter, it must be supplemented with some additional physics which can account for
the presence of a baryon asymmetry on the observed scale.  The generic name for such
scenarios is baryogenesis.

\indent Any successful baryogenesis model must satisfy a set of three requirements which
were originally pointed out by Sakharov~\cite{Sakharov:1967dj}.  The first and most
self-evident of these conditions is that there must be baryon number violation;
otherwise, a universe with an initial value of \(\eta=0\) could not evolve to one where
\(\eta\neq 0\).  The second is that the charge conjugation symmetry \(C\) and its
composition with parity \(CP\) must also be violated.  If this were not the case, baryon
number violation and antibaryon number violation would occur at the same rate, and hence
no net baryon asymmetry would be created.  Finally, one must have either a departure from
thermal equilibrium or else posit that \(CPT\), the composition of \(CP\) with the time
evolution operator, is somehow violated during the early universe.  This condition can be
obtained by calculating the thermal average \(\langle B\rangle\) of the baryon number, a
dimensionless quantity defined by the \(B\equiv (n_{b}-n_{\overline{b}})/s\) in terms of
the entropy density
\begin{equation}
  s=\frac{2\pi^2}{45}g_{\ast}T^3,
\end{equation}
where \(T\) is the temperature of the thermal bath and \(g_{\ast}\) is the number of
interacting degrees of freedom (in the minimal supersymmetric model (MSSM) during the
baryogenesis epoch, \(g_{\ast}\approx205\)).  The result is~\cite{Riotto:1999yt}:
\begin{equation}
  \langle B\rangle =\tr[e^{-\beta H} B]=\tr[(CPT)(CPT)^{-1}e^{-\beta H} B].
\end{equation}
If \(CPT\) commutes with the Hamiltonian, this becomes
\begin{equation}
  \langle B\rangle =\tr[e^{-\beta H} B]=\tr[e^{-\beta H}(CPT)^{-1} B(CPT)]
  =-\tr[e^{-\beta H} B],
\end{equation}
which implies \(B=0\) (conversely, if thermal equilibrium is not maintained, all particle
species will not have a common temperature, and trace cyclicity can no longer be
invoked). If any of these three conditions is not met, \(\eta\) will not deviate from
zero. In general, since Lorentz invariance is believed to be necessary for the
formulation of a consistent quantum field theory, this condition has been taken to imply
a departure from thermal equilibrium rather than the presence of \(CPT\)-violation. The
alternative, in which Lorentz invariance is temporarily, dynamically violated in the
early universe, was originally explored in the spontaneous baryogenesis scenario
of~\cite{Cohen:1987vi} and has been elaborated on in models like the radion-induced
baryogenesis of~\cite{Alberghi:2003ws} and the gravitational baryogenesis
of~\cite{Davoudiasl:2004gf}, in which a net baryon number is produced via a
\(CP\)-violating effective interaction between the baryon number current and the
derivative of the Ricci Scalar.  Indeed Lorentz invariance is violated whenever a tensor
field receives a nonzero vacuum expectation value (VEV), and indeed may arise
spontaneously in string theories, brane-world scenarios, and modified gravity
models~\cite{Carroll:2005dj}.  Still, no experimental signal of \(CPT\) violation has
ever been detected~\cite{Carosi:1990ms,Schwingenheuer:1995uf}, and most theories of
baryogenesis (including the one with which this work is principally concerned) assume
that \(CPT\) is a good symmetry of the Hamiltonian and that the generation of a baryon
number for the universe involves out-of-equilibrium dynamics rather than Lorentz
violation.

\section{Baryogenesis via Leptogenesis}

\subsection{Leptogenesis and the See-Saw Mechanism}

\indent

A variety of viable baryogenesis models exist, including electroweak baryogenesis, in
which $CP$-violation occurs at a bubble wall, or phase boundary, and Affleck-Dine
baryogenesis~\cite{Affleck:1984fy}, in which the baryon asymmetry is generated by moduli
fields charged under \(B-L\), where \(L=(n_{\ell}-n_{\overline{\ell}})/s\) denotes the
lepton number of the universe. In this paper, we will focus on models which achieve
baryogenesis through a framework known as
leptogenesis~\cite{Fukugita:1986hr,Luty:1992un}, where decays of heavy particles in the
early universe which violate both \(CP\) and lepton number \(L\) produce an initial
lepton asymmetry, which is then converted to a nonzero baryon asymmetry by sphaleron
processes associated with the \(SU(2)\) electroweak anomaly~\cite{'tHooft:1976fv}.
Leptogenesis is a particularly attractive model because in addition to its ability to
yield a realistic value for \(\eta\)~\cite{Buchmuller:2002xm}, it can also explain why
the standard model neutrinos have small but nonzero masses.  In its most common form,
which we will call Majorana leptogenesis, the role of the heavy, decaying particles is
played by the right-handed neutrinos $\nu_{R}$ required to fill out the multiplets
housing the Standard Model fields in many grand unified groups. As they are gauge
singlets, nothing prevents them from obtaining large (GUT-scale) Majorana masses
\(M_{\nu_{R}}\), and such masses are inherently lepton-number-violating.

\indent In the following brief review of Majorana leptogenesis, and for the remainder of
this work, we will work under the assumption that the universe we live in is inherently
supersymmetric, but that supersymmetry is broken at some high scale
\(M_{\mathit{SUSY}}\).  Although leptogenesis certainly does not require supersymmetry,
we choose to work in a supersymmetric framework for a variety of reasons. Supersymmetry
naturally explains the stability of the weak scale against quadratic
divergences~\cite{Wess:1974tw,Veltman:1980mj} and improves the outlook for the
unification of the Standard Model gauge couplings~\cite{Langacker:1991an,Amaldi:1991cn}.
Furthermore, the assumption of conserved (or nearly conserved) \(R\)-parity makes the
lightest supersymmetric particle (LSP) a convenient dark matter
candidate~\cite{Ellis:1983ew}.

\indent In supersymmetric Majorana leptogenesis, the most general realizable
superpotential one can write for the relevant fields is
\begin{equation}
  \mathcal{W}\ni y_{\alpha\beta}L_{\alpha} N_{\beta} H_{u} + M_{\nu_{R}\alpha}N_{\alpha} N_{\alpha},
  \label{eq:MajLepSuperpot}
\end{equation}
where \(L_{\alpha}\) and \(N_{\alpha}\) represent the left- and right-handed neutrino
superfields respectively, \(H_{u}\) is the up-type Higgs superfield, and
\(y_{\alpha\beta}\) is a dimensionless trilinear coupling.\footnote{We have chosen to
work in a basis where the \(M_{\nu_{R}\alpha}\) are diagonal.}  This superpotential
contains both Majorana and Dirac masses for neutrinos (once the Higgs field gets a VEV)
and yields a mass matrix
\begin{equation}
  \overline{\nu}\mathbf{M}\nu=(\overline{\nu}_{L}~,~\overline{\nu}_{R})
  \left(\begin{array}{cc} 0 & m_{D} \\ m_{D}^{T} & M_{\nu_{R}}\end{array}\right)
  \left(\begin{array}{c} \nu_{L}\\ \nu_{R}\end{array}\right),
\end{equation}
where flavor indices have been suppressed to simplify notation.  Since \(m_{D}=y v
\sin\beta\), where \(\tan\beta=v_{u}/v_{d}\) denotes the ratio of the up- and down-type
Higgs VEVs, the off-diagonal terms mixing \(\nu_{L}\) and \(\nu_{R}\) will be small
compared to \(M_{\nu_{R}\alpha}\).  As a result, when this matrix is diagonalized, the
mass spectrum of the theory contains three light neutrinos with masses
\begin{equation}
  m_{\nu}=
  m_{D}\frac{1}{M_{\nu_{R}}}m^{\dagger}_{D},
\end{equation}
which are identified with the Standard Model neutrinos, as well as three heavy neutrinos
with masses \(\mathcal{O}(M_{\nu_{R}})\).  This method for obtaining small but nonzero
neutrino masses is known as the see-saw
mechanism~\cite{Yanagida:1979as,Schechter:1980gr}, and the liaison it forges between the
origin of the observed baryon asymmetry and the lightness of the Standard Model neutrinos
is one of the most compelling aspects of leptogenesis scenarios.

\begin{figure}[t!]
\begin{center}
\begin{fmffile}{MajCPtree4}
  \fmfframe(20,20)(20,20){\begin{fmfchar*}(80,50)
    \fmfleft{i1}
    \fmfright{o1,o2}
    \fmflabel{$N_{R_1}$}{i1}
    \fmflabel{$H_{d}^c$}{o1}
    \fmflabel{$\ell_{\alpha}$}{o2}
    \fmf{vanilla,tension=.5}{i1,v1,o2}
    \fmf{dashes,tension=.5}{v1,o1}
  \end{fmfchar*}}
\end{fmffile}~~~~~
\begin{fmffile}{MajCPloop4}
  \fmfframe(20,20)(20,20){\begin{fmfchar*}(80,50)
    \fmfleft{i1}
    \fmfright{o1,o2}
    \fmflabel{$N_{R_1}$}{i1}
    \fmflabel{$H_{d}^c$}{o1}
    \fmflabel{$\ell_{\alpha}$}{o2}
    \fmf{vanilla,tension=.5}{i1,v1}
    \fmf{dashes,tension=.3,left=.6,label=$H_{d}$}{v1,v2}
    \fmf{vanilla,tension=.1,left=.5,label=$N_{R_i}$}{v2,v3}
    \fmf{dashes,tension=.5}{v3,o1}
    \fmf{vanilla,tension=.5}{v2,o2}
    \fmf{vanilla,tension=.3,right=.6,label=$\ell_{\beta}$}{v1,v3}
  \end{fmfchar*}}
\end{fmffile}~~~~~
\begin{fmffile}{MajLoopleg3}
  \fmfframe(20,20)(20,20){\begin{fmfchar*}(80,50)
    \fmfleft{i1}
    \fmfright{o1,o2}
    \fmflabel{$N_{R_1}$}{i1}
    \fmflabel{$H_{d}^c$}{o1}
    \fmflabel{$\ell_{\alpha}$}{o2}
    \fmf{vanilla,tension=.5}{i1,v1}
    \fmf{dashes,tension=.3,left,label=$\ell_{\beta}$}{v1,v2}
    \fmf{vanilla,tension=.1,left,label=$H_{d}$}{v2,v1}
    \fmf{dashes,tension=.5}{v3,o1}
    \fmf{vanilla,tension=.5}{v3,o2}
    \fmf{vanilla,tension=.3,label=$N_{R_i}$}{v2,v3}
  \end{fmfchar*}}
\end{fmffile}
\end{center}
  \caption{The diagrams whose interference yields the leading contribution to
  $CP$-violation from $N_{R_1}$ decays in the
  traditional, Majorana version of leptogenesis.  In a supersymmetric theory, the
  supersymmetrized versions of these diagrams must also be included.
  \label{fig:MajCPviolDiagram}}
\end{figure}
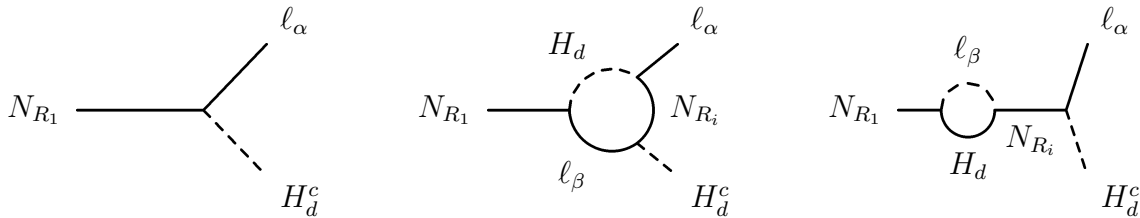

\begin{figure}
\begin{center}
\begin{displaymath}
\begin{fmffile}{NT43}
  \fmfframe(20,20)(20,20){\begin{fmfchar*}(150,100)
    \fmfleft{i1,i2}
    \fmfright{o1,o2}
    \fmftop{t1}
    \fmfbottom{b1}
    \fmflabel{$n$}{i1}
    \fmflabel{$n$}{i2}
    \fmflabel{$p$}{t1}
    \fmflabel{$p$}{b1}
    \fmflabel{$e^{-}$}{o2}
    \fmflabel{$e^{-}$}{o1}
    \fmf{fermion}{i1,v1}
    \fmf{fermion,tension=.5}{v1,b1}
    \fmf{photon,label=$W$,tension=1.5}{v1,v2}
    \fmf{fermion}{v2,o1}
    \fmfcmd{%
      vardef cross_bar (expr p, len, ang) =
        ((-len/2,0)--(len/2,0))
        rotated (ang + angle direction length(p)/2 of p)
        shifted point length(p)/2 of p
      enddef;
        style_def majorana expr p=
        cdraw p;
        cfill (tarrow (reverse p, .70));
        cfill (tarrow (p, .70));
        ccutdraw cross_bar (p, 5mm, 45);
        ccutdraw cross_bar (p, 5mm, -45)
    enddef;}
    \fmf{majorana,tension=2.0,label=$\nu$}{v2,v3}
    \fmf{fermion}{v3,o2}
    \fmf{photon,label=$W$,tension=1.5}{v3,v4}
    \fmf{fermion}{i2,v4}
    \fmf{fermion,tension=.5}{v4,t1}
  \end{fmfchar*}}
\end{fmffile}
\end{displaymath}
\end{center}
\caption{A diagrammatical representation of neutrinoless double-$\beta$
decay.\label{fig:NlessDubBetaDec}}
\end{figure}
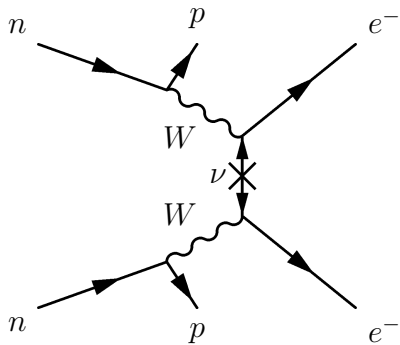

\indent One of the useful consequences of Majorana component to the light neutrino masses
is that such a source of lepton number violation would be observable in neutrinoless
double-$\beta$ decay experiments~\cite{Schechter:1981bd}.  The contribution to this
process from a Majorana neutrino mass is shown diagrammatically in
figure~\ref{fig:NlessDubBetaDec}.  It should be noted that a Majorana neutrino mass is
not required for it to occur (certain other other beyond-the-Standard-Model sources of
lepton number violation can also contribute), but if neutrinoless double-$\beta$ decay
were ever observed, it would serve as compelling evidence in favor of the see-saw
mechanism.

\indent The production of lepton number in Majorana leptogenesis occurs when the heavy
right-handed neutrinos decay out of equilibrium.  The leading $CP$-violating contribution
comes from the interference between the diagrams shown in
figure~\ref{fig:MajCPviolDiagram} (in supersymmetric models, the supersymmetrized
versions of these diagrams must also be included), and results in the generation of a
nonzero lepton number $L$ for the universe in the traditional, Majorana version of
leptogenesis.  In the limit where there is a large splitting between the right-handed
sneutrino masses ($M_{\nu_R 1}\ll M_{\nu_R 2},~M_{\nu_R 3}$) The contribution from a
single \(N_{R_1}\) decay is commonly called the decay asymmetry and is given
by~\cite{Buchmuller:2005eh}
\begin{equation}
  \epsilon\approx\frac{3}{16\pi}\,\frac{\im\left[
  y_{1\alpha}^{\ast}y_{i\beta}y_{1\beta}^{\ast}y_{i\alpha}
  \right]}{(y_{1\beta}y_{1\beta}^{\ast})}\,
  \left(\frac{M_{\nu_R 1}}{M_{\nu_R i}}\right)
\end{equation}
where a sum over repeated indices is assumed.  This translates into a net lepton
number~\cite{Fukugita:1986hr}
\begin{equation}
  L=\kappa_{W}\frac{\epsilon}{g_{\ast}}
\end{equation}
for the universe, where \(\kappa_{W}\) is a dimensionless, model-dependent prefactor
included to account for the washout of lepton number by \(2\leftrightarrow 2\)
lepton-number-violating processes, etc. that deplete the lepton number genrated during
\(N_{R_1}\) decay.  Whether Majorana leptogenesis is capable of reproducing the observed
value of \(\eta\) depends both on \(\epsilon\) and on \(\kappa_{W}\).

\subsection{Effects of the Electroweak Anomaly\label{sec:sphalerons}}

\indent

So far we have seen that the leptogenesis mechanism is capable of generating a net lepton
number \(L\neq 0\) for the universe.  A mechanism by which this lepton number can be
transformed into a net baryon number is built into the Standard Model
(SM)~\cite{'tHooft:1976up}. The baryon and lepton number currents
\begin{eqnarray}
  J_{\mu}^{B}&=&\frac{1}{3}\sum_{i}\left(
  \overline{q}_{L_i}\gamma_{\mu}q_{L_i}+
  \overline{u}_{R\alpha}\gamma_{\mu}u_{R\alpha}+
  \overline{d}_{R\alpha}\gamma_{\mu}d_{R\alpha}
  \right)\\
  J_{\mu}^{L}&=&\frac{1}{3}\sum_{_i}\left(
  \overline{\ell}_{L_i}\gamma_{\mu}\ell_{L_i}+
  +\overline{e}_{R_i}\gamma_{\mu}e_{R_i}
  \right),
\end{eqnarray}
where the sum is over generations, are anomalous due to the triangle anomaly, with
divergence
\begin{equation}
  \partial^{\mu}J_{\mu}^{B}=\partial^{\mu}J_{\mu}^{L}=
  \frac{N_f}{32\pi^2}\left(g_2^2W_{\mu\nu}^a\tilde{W}^{a\mu\nu}+
  g_Y^2F_{\mu\nu}\tilde{F}^{\mu\nu}
   \right).
\end{equation}
Here, \(W_{\mu\nu}^{a}\) and \(F_{\mu\nu}\) are the respective field-strength tensors of
the \(SU(2)\) and \(U(1)_{Y}\) gauge fields, \(g_{2}\) and \(g_{Y}\) are the respective
gauge coupling constants, \(N_f\) denotes the number of fermion generations, and
\begin{equation}
  \tilde{F}^{\mu\nu}=\frac{1}{2}\epsilon^{\mu\nu\rho\sigma}F_{\rho\sigma}
\end{equation}
is the dual of $F^{\mu\nu}$ (the expression for $\tilde{W}^{a\mu\nu}$ is analogous).  The
change in baryon number over some duration \(t\) is then given in terms of the
Chern-Simons number
\begin{equation}
  N_{\mathit{CS}}(t)=\frac{g_2^3}{96\pi^2}\int d^3x\epsilon^{ijk}\epsilon_{abc}
  W_{i}^{a}W_{j}^{b}W_{k}^{c}(t)
\end{equation}
by
\begin{eqnarray}
  B(t)-B(0)&=&\int_{0}^{t}\int d^3x\partial^{\mu}J_{\mu}^{B}\nonumber\\
  &=& N_{f}\left[N_{\mathit{CS}}(t)-N_{\mathit{CS}}(0)\right].\label{eq:SphalNCSdef}
\end{eqnarray}
An analogous equation exists for lepton number.

\indent This theory has an infinite number of quasi-degenerate vacua, in each of which
the \(W^{a}_{i}\) are pure gauge and consequently \(N_{\mathit{CS}}\) becomes an integer.
Equation~(\ref{eq:SphalNCSdef}) tells us that vacuum-to-vacuum transitions involve
\(\Delta B=\Delta L=N_f\Delta N_{\mathit{CS}}\), so the minimum change for such a
transition in the SM, where \(N_f=3\), is \(\Delta B=\Delta L=3\).  This corresponds to
an effective 12-fermion interaction\footnote{The physical scale \(\Lambda\) associated
with this effective operator is the inverse of the magnetic screening length
$R_{\mathit{Sph}}\sim1/(\alpha_{2}T)$.}
\begin{equation}
  \mathcal{O}_{B+L}=\prod_{i}\left(q_{L_{i}}q_{L_i}q_{L_i}\ell_{i}\right),
  \label{eq:SphalOpDef}
\end{equation}
where we have included the \(B+L\) subscript to draw attention to another important
consequence of equation~(\ref{eq:SphalNCSdef}): since the same equation describes the
evolution of both baryon and lepton number, the combination \(B-L\) is conserved in any
process of the form~(\ref{eq:SphalOpDef}), while \(B+L\) is violated by at least 6 units.
These effective interactions serve as the primary means of conversion between \(B\) and
\(L\) in leptogenesis.

\indent In calculating the transition rate for these interactions, we will make use of
the fact that the potential energy of the electroweak theory between any pair of adjacent
($\Delta N_{\mathit{CS}}=1$) vacua contains a saddle point, and that transitions through
that point in field space should dominate over all others.  This field configuration is
known as the electroweak sphaleron, and the height of the potential barrier separating
any two adjacent vacua is known as the sphaleron energy, usually denoted by
\(E_{\mathit{sph}}\). At zero temperature, the rate for such transitions is determined by
the instanton action and turns out to be completely negligible~\cite{'tHooft:1976up}:
\begin{equation}
  \Gamma_{\mathit{inst}}\sim e^{-S_{\mathit{inst}}}=e^{-1/g_{2}^{2}}\sim
  10^{-165}.
\end{equation}
At finite temperatures, however, the situation is modified by the possibility of thermal
excitations over the barrier.  For low temperatures \(T<E_{\mathit{sph}}\), the rate
calculation is reasonably straightforward and the result~\cite{Arnold:1987mh} is
\begin{equation}
  \Gamma_{\mathit{sph}} = c \left[\frac{E_{\mathit{sph}}^3 M^4_{W}(T)}{T^6}\right]
  e^{-E_{\mathit{sph}}/T},
  \label{eq:LowTempSphRate}
\end{equation}
where \(c\) is an \(\mathit{O}(1)\) constant and \(M_{W}(T)\) is the mass of the \(W\)
boson as a function of temperature.

\indent At higher temperatures \(T>E_{\mathit{sph}}\), the calculation of the sphaleron
interaction rate is somewhat difficult, but it can be estimated on dimensional grounds by
examining the scales of the processes involved~\cite{Arnold:1996dy}. Non-perturbative
fluctuations of the gauge field that mediate the sphaleron transition are associated with
magnetic fluctuations with characteristic distance scale
$R_{\mathit{Sph}}\sim1/(\alpha_{2}T)$ on the order of the magnetic screening length. If
one naively assumes that the time scale for the process is of the same order, one obtains
the result \(\Gamma_{\mathit{Sph}}\sim\alpha_2^4 T\). However, when one takes into
account damping effects in the plasma~\cite{Arnold:1998cy,Huet:1996sh} one finds that the
time scale is slowed to \(t_{\mathit{Sph}}\sim 1/(\alpha_2^2 T)\), which leads to a
rate\footnote{It has been argued~\cite{Bodeker:1998hm} that the characteristic time scale
of these processes may also have a logarithmic suppression, so that
\(\Gamma_{\mathit{sph}}\sim [\alpha_2^5 T\ln(1/\alpha)]\). Since what is of interest to
us is the numerical result for the rate of baryon-number-changing transitions, we can
remain agnostic on this issue.}
\begin{equation}
  \Gamma_{\mathit{Sph}}=(25.4\pm2.0)\alpha_{2}^{5}T,
  \label{eq:HighTempSphRate}
\end{equation}
where \(\alpha_{2}\equiv g_{2}^{2}/4\pi\approx1/30\) and the value given for the
proportionality constant is the result of numerical
calculations~\cite{Moore:1997sn,Bodeker:1999gx}.

\indent Sphaleron interactions will be in equilibrium whenever the sphaleron transition
rate \(\Gamma_{\mathit{sph}}\) exceeds the expansion rate of the universe, expressed by
the Hubble parameter \(H\).  When the universe is radiation-dominated, the expansion rate
is given by
\begin{equation}
  H=1.66 g_{\ast}^{1/2} T^{2}/M_{P}.\label{eq:HubbleBubble}
\end{equation}
From equations~(\ref{eq:LowTempSphRate}) and~(\ref{eq:HighTempSphRate}) we find that
sphaleron interactions are in equilibrium when
\begin{equation}
  100~\mathrm{GeV}\,< T <10^{13}~\mathrm{GeV},
\end{equation}
which means that the interconversion of \(B\) and \(L\) can be considered rapid perhaps
during and certainly soon after the lepton number asymmetry \(L\) is built up by the
decays of the heavy right-handed neutrino fields and will remain rapid down to around the
weak scale.

\indent In order to relate these baryon and lepton number asymmetries to each other in a
quantitative way, we can take advantage of the set of conditions among the chemical
potentials \(\mu_{i}\) implied by equilibrium conditions among the species \(i\) present
in the thermal bath.  If chemical equilibrium is established between particle species
\(a_1,a_2\ldots,a_m,b_1,b_2,\ldots b_n\) by sufficiently rapid interactions of the form
\(a_1,a_2\ldots a_f\leftrightarrow b_1b_2\ldots b_f\), their chemical potentials obey the
relation
\begin{equation}
  \sum_{i=1}^{m}\mu_{a_i}=\sum_{i=1}^{n}\mu_{b_i}.
\end{equation}
For example, when sphaleron interactions are in equilibrium,
equation~(\ref{eq:SphalOpDef}) implies that
\begin{equation}
  \sum_{i}(3\mu_{q_{i}}+\mu_{\ell_{i}})=0.
\end{equation}
Similarly, \(SU(3)\) QCD instanton processes and global hypercharge conservation
respectively require that
\begin{equation}
  \sum_{i}(2\mu_{q_{i}}-\mu_{u_i}-\mu_{d_i})=0
\end{equation}
\begin{equation}
  \sum_{i}\left(\mu_{q_{i}}+2\mu_{u_i}-\mu_{d_i}-
  \mu_{\ell_i}-\mu_{e_i}+\frac{2}{N_{f}}\mu_{H}\right)=0,
\end{equation}
and from the fermion Yukawa interactions
\begin{eqnarray}
    \sum_{i}(2\mu_{q_{i}}-\mu_{H}-\mu_{d_i})=0\\
    \sum_{i}(2\mu_{q_{i}}+\mu_{H}-\mu_{u})=0\\
    \sum_{i}(2\mu_{\ell_{L i}}-\mu_{H}-\mu_{e_i})=0.
\end{eqnarray}
Since particles of different generations will also be in equilibrium with one another at
high temperatures, we can take \(\mu_{q_{i}}=\mu_{q_{L}}\), \(\mu_{u_{i}}=\mu_{u}\),
\(\mu_{d_{i}}=\mu_{d}\), \(\mu_{\ell_{i}}=\mu_{\ell}\), \(\mu_{e_{i}}=\mu_{e}\)  and
solve this system of equations for one of the \(\mu_{i}\) (we choose \(\mu_{\ell}\)). The
resulting chemical potentials are
\begin{eqnarray}
  \mu_{e}=\frac{2N_f+3}{6N_f+3}\mu_{\ell}& \mu_{d}=-\frac{6N_f+1}{6N_f+3}\mu_{\ell}
  & \mu_{u}=\frac{2N_f-1}{6N_f+3}\mu_{\ell}\nonumber
  \\ \mu_{q}=\frac{1}{3}\mu_{\ell} & \mu_{H}=\frac{4N_f}{6N_f+3}\mu_{\ell},
\end{eqnarray}
and since the values of \(B\) and \(L\) are related to these chemical potentials (see
equation~(\ref{eq:MuAsymmetryRel}) in appendix~\ref{app:BoltzDiracLep}) by
\begin{eqnarray}
  B=\frac{1}{g_{\ast s}T}\sum_{i=\mathit{baryon}}g_{i}\mu_{i}=
  \frac{N_f}{g_{\ast s}T}(2\mu_{q}+\mu_u+\mu_d)\\
  L=\frac{1}{g_{\ast s}T}\sum_{i=\mathit{baryon}}g_{i}\mu_{i}=
  \frac{N_f}{g_{\ast s}T}(2\mu_{\ell}+\mu_e),
\end{eqnarray}
one finds that
\begin{equation}
  B=\frac{8N_f+4}{22N_f+13}(B-L)~~~\hspace{1cm}~~~
  L=-\frac{14N_f+9}{22N_f+13}(B-L)
\end{equation}
in the Standard Model.  From this, we can state the relationship between baryon and
lepton number when sphaleron interactions are in equilibrium:
\begin{equation}
  B=\frac{28}{51}L.\label{eq:BLrelNotSUSY}
\end{equation}
In the MSSM, things are modified by the presence of a second Higgs doublet, and the
result becomes result is
\begin{equation}
  B=\frac{8}{23}L.\label{eq:BLrelSUSY}
\end{equation}
In Majorana leptogenesis, we know that since \(B-L\) is conserved by sphalerons and
violated only by the lepton-number-producing decays of \(N_{R_1}\), the initial value
\((B-L)_{\mathit{init}}=-L_{\mathit{init}}\) generated during the leptogenesis epoch and
the present value will be equal; hence (in the MSSM)
\begin{equation}
  B_{\mathit{today}}=-\frac{8}{23}L_{\mathit{init}}=
  -\frac{8}{23}\,\kappa_{W}\,\frac{\epsilon}{g_{\ast}}
\end{equation}
and the universe receives a nonzero baryon number.

\subsection{An Alternative Leptogenesis Scenario}

\indent

Although it is not our aim to discuss the details of Majorana leptogenesis models, we
note that they have been shown to be able to yield a realistic value of \(\eta\),
reproduce the observed light neutrino spectrum, and evade problems associated with
relevant astrophysical constraints (for recent reviews,
see~\cite{Buchmuller:2002xm,Buchmuller:2005eh,Cline:2006ts,Chen:2007fv}).  In this form,
leptogenesis emerges as a viable phenomenological theory.  There may be other forms of
leptogenesis that are just as successful, however.  The aim of this work is to conduct a
thorough examination of one such alternative, supersymmetric Dirac leptogenesis, and to
show that it is also a phenomenologically viable model.  We will begin by discussing the
symmetries and field content which determine the form of superpotential and discuss the
consequences of that superpotential for baryogenesis and neutrino masses in
chapter~\ref{ch:DiracLep}.  In chapter~\ref{ch:constraints}, we enumerate the
phenomenological constraints on Dirac leptogenesis and construct a simple,
theoretically-motivated model. In chapter~\ref{ch:Boltzmann}, we solve the Boltzmann
equations for the evolution of baryon and lepton number in the early universe numerically
and discuss the consequences of these results on the model parameters.  We investigate
some potentially promising extensions of the model in chapter~\ref{ch:extensions}, and in
chapter~\ref{ch:Conclusions}.

\chapter{DIRAC\ LEPTOGENESIS\label{ch:DiracLep}}

\section{Superpotential and Fields\label{sec:WandFields}}

\indent

While Majorana leptogenesis is a certainly successful model, it is not the only way in
which leptogenesis can be realized.  There are several reasons why exploring potential
alternatives is a worthwhile endeavor, and in particular why it is advantageous to have a
viable leptogenesis mechanism in a model without Majorana neutrinos.  One is that the
non-observation of neutrinoless double-$\beta$ decay at future
experiments~\cite{Schechter:1981bd,Vogel:2006sq} could significantly constrain the
parameter space of Majorana leptogenesis to the point where severe model tensions might
arise between such constraints and others from astrophysics, flavor-physics, etc. Another
is that it is not at all obvious that massive right-handed neutrinos emerge naturally
from a string-theory context.  While landscape surveys, even for particular classes of
models and specified orbifold compactifications, are limited by computational complexity,
preliminary results~\cite{Giedt:2005vx} performed for heterotic \(\mathrm{BSL}_{A}\)
models on the \(Z_{3}\) orbifold suggest that a see-saw mechanism in the traditional
sense is not a generic feature of otherwise phenomenologically promising string models.

\begin{figure}[t!]
\hspace{1.5cm}
\includegraphics[width=14cm]{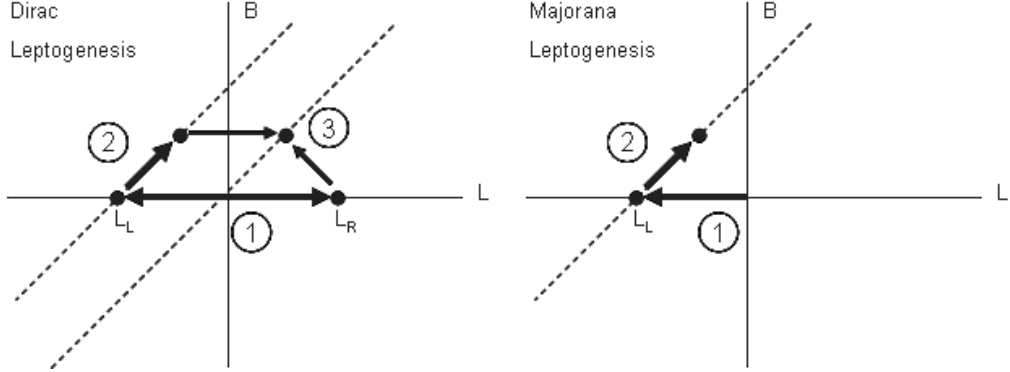}
  \caption{A schematic representation, after~\cite{Dick:1999je,Thomas:2005rs}, of the
  evolution of baryon number $B$ (vertical axis) and lepton number
  $L$ (horizontal axis) in Dirac and Majorana leptogenesis.  In
  Dirac leptogenesis (left panel), the evolution of $B$ and $L_{\mathit{tot}}$ proceeds in three steps: first,
  two stores of lepton number $L_{\ell}$ (stored in left-handed neutrinos) and
  $L_{\nu_{R}}$
  (stored in right-handed neutrinos) are produced during heavy particle decays; second, sphaleron processes
  (which act along lines of constant $B-L$) mix $L_{\ell}$ and $B$ while leaving $L_{\nu_{R}}$ alone; third, after
  sphaleron interactions have effectively shut off, equilibration between $L_{\ell}$ and $L_{\nu_{R}}$ results in a net
  positive $B$ and $L$ for the universe.  This is qualitatively quite different from the situation in
  Majorana leptogenesis (right-hand panel).  Here, lepton-number-violating heavy particle decays first build up a single,
  nonzero store or lepton number $L_{\ell}=L_{\mathit{tot}}$; second, sphalerons transmute this lepton
  number into a nonzero baryon number for the universe.  Only one store of
  lepton number is created, and the result is a universe with negative $L_{\mathit{tot}}$ and positive $B$.
  \label{fig:BLSpaceDiagram1}}
\end{figure}

\indent For all these reasons, it would be useful if one could find a way to link the
smallness of the physically-observed neutrino masses to a successful baryogenesis
mechanism without having to introduce singlet neutrinos with Majorana masses.  It is
indeed possible to do this in the context of a scenario that has come to be known as
Dirac leptogenesis or Dirac neutrinogenesis~\cite{Dick:1999je,Murayama:2002je}. In this
scenario, an additional symmetry (the precise form of which is not terribly important or
stringently constrained by the model framework) is introduced, and charges are assigned
under this new symmetry in a manner which forbids, at tree level, both the Majorana and
Dirac neutrino mass terms appearing in~(\ref{eq:MajLepSuperpot}).  A set of heavy,
vector-like pairs of fields introduced whose couplings to the standard model fields
contain nontrivial, \(CP\)-violating phases are also introduced.  These fields will play
the role that heavy right-handed neutrinos play in Majorana leptogenesis, and their
decays during the early universe will lead to the buildup of equal and opposite lepton
asymmetries \(L_{\ell}\) and \(L_{\nu_{R}}\) in the left-handed lepton and right-handed
neutrino sectors, while conserving the overall lepton number for the universe
\(L_{\mathit{tot}}=L_{\ell}+L_{\nu_{R}}=0\).  The equilibration rate between these stores
is suppressed by the smallness of the effective neutrino Dirac mass term; as a result,
the electroweak sphaleron processes which convert \(L_{\ell}\) into a baryon asymmetry
\(B\) effectively shut off before \(L_{\ell}\) and \(L_{\nu_{R}}\) have a chance to
equilibrate.  Consequently, unlike in Majorana leptogenesis, the universe ends up with a
net positive lepton number as well as a net positive baryon number---a result which is
depicted schematically in figure~\ref{fig:BLSpaceDiagram1}.

\indent We will begin by writing down the superpotential for Dirac
leptogenesis,\footnote{Once again, we remark that Dirac leptogenesis does not require
supersymmetry and functions perfectly well without it~\cite{Dick:1999je}.} which is
modified from that of the MSSM only in the lepton sector. The field content and charge
assignments of the model are essentially the same as those presented
in~\cite{Murayama:2002je}.  Of the usual quark and lepton supermultiplets of the MSSM,
the left-handed lepton multiplets \(L_{\alpha}\) (\(\alpha\) is a family index) and the
Higgs multiplets \(H_{u}\) and \(H_{d}\) will be pertinent to leptogenesis.  We also
include a right-handed neutrino superfield \(N_{\alpha}\) for each family, an exotic
chiral multiplet \(\chi\), and a number \(N_{\Phi}\) of vector-like pairs of chiral
multiplets \(\Phi_{i}\) and \(\overline{\Phi}_{i}\).  While at least two such pairs are
required for leptogenesis, any \(N_{\Phi}\geq2\) is in principle allowed from a
baryogenesis standpoint.  We also need to introduce an additional symmetry to forbid
Majorana masses for the \(N_{\alpha}\), which we choose to be an additional global
\(U(1)\) we will call \(U(1)_{N}\), due to the fact that the right-handed neutrinos are
charged under it. The charge configurations of these fields under this additional
\(U(1)\), as well as the rest of the relevant symmetries of the theory, are shown in
table~\ref{tab:U1Charges}.\footnote{With the field content given in
table~\ref{tab:U1Charges}, the global $U(1)_{N}$ is anomalous.  To deal with this issue,
one can modify the theory at high energies by introducing additional heavy fields,
appealing to the Green-Schwarz mechanism~\cite{Green:1984sg}, etc. Alternatively, one may
modify the field content of the low-energy effective theory and introduce a set of fields
with the appropriate charges to cancel the $U(1)_{N}$ anomalies.  We will explore one
such choice in section~\ref{sec:sneutrinoCDM}.} The most general superpotential that can
be constructed out of this set of fields is
\begin{equation}
  \mathcal{W}\ni\lambda_{i\alpha}N_{\alpha}\Phi_{i}H_{u}+
  h_{i\alpha}L_{\alpha}\overline{\Phi}_{i}\chi+
  M_{ij}\Phi_{i}\overline{\Phi}_{j}+\mu H_{u}H_{d},
  \label{eq:DiracLepSuperpotential}
\end{equation}
where \(\alpha\) is a family index, \(\lambda_{i\alpha}\) and \(h_{i\alpha}\) are Yukawa
couplings, \(M_{ij}\) is a matrix of supersymmetry-respecting mass terms coupling the
\(\Phi_{i}\) and \(\overline{\Phi}_{i}\) fields, and \(\mu\) is the usual Higgs mass
parameter.  We note that we can always choose to work in a basis where the the mass are
diagonal and real, and hence from this point onward, without loss of generality, we will
adopt the convention that \(M_{ij}\Phi_{i}\Phi_{j}=M_{\Phi_{i}}\Phi_{i}\Phi_{i}\) and
absorb all phases into the coupling matrices \(\lambda_{i\alpha}\) and \(h_{i\alpha}\).

\begin{table}[t!]
 \begin{center}
  \begin{tabular}{|cccccc|}
     \hline Field & \(U(1)_{L}\) & \(U(1)_{N}\) & \(SU(2)\) & \(U(1)_{Y}\) & $P_M$ \\ \hline
     \(N\) & -1 & +1 & \(\mathbf{1}\) & 0 & -1 \\
     \(L\) & +1 & 0 & \(\mathbf{2}\) & \(-\frac{1}{2}\) & -1 \\
     \(H_{u}\) & 0 & 0 & \(\mathbf{2}\) & \(\frac{1}{2}\) & +1 \\
     \(H_{d}\) & 0 & 0 & \(\mathbf{2}\) & \(-\frac{1}{2}\) & +1 \\
     \(\phi\) & +1 & -1 & \(\mathbf{2}\) & \(-\frac{1}{2}\) & -1 \\
     \(\overline{\phi}\) & -1 & +1 & \(\mathbf{2}\) & \(\frac{1}{2}\) & -1 \\
     \(\chi\) & 0 & -1 & \(\mathbf{1}\) & 0 & +1 \\ \hline
   \end{tabular}
 \end{center}
 \caption{One possible set of charge assignments, taken from~\cite{Murayama:2002je},
 that leads to the Dirac leptogenesis superpotential given
 in~(\ref{eq:DiracLepSuperpotential}).  Here the additional symmetry
 employed is a $U(1)$ (which may in principle be either global or
 local).  Only the charges of the fields relevant to leptogenesis, which include the Higgs doublets
 $H_{u}$ and $H_{d}$, the left-handed lepton superfield $L$,
 the right-handed neutrino superfield $N$, the heavy fields
 $\Phi$ and $\overline{\Phi}$, and the additional field
 $\chi$, have been included.  Here,  $U(1)_L$, $SU(2)$, and $U(1)_{Y}$ respectively
 denote lepton number, $SU(2)$, and $U(1)$ hypercharge quantum numbers, $P_M$ denotes
 matter parity, and $U(1)_N$
 denotes the charge under the additional $U(1)$.\label{tab:U1Charges}}
\end{table}

\section{Dirac Leptogenesis, Baryogenesis, and Neutrino Masses\label{sec:MneutfromDlep}}

\indent

In order for successful baryogenesis to occur, we must satisfy the Sakharov criteria
discussed in section~\ref{sec:sakharov}.  The out-of-equilibrium condition is satisfied
during the decays of the component fields in \(\Phi\) and \(\overline{\Phi}\) (both the
scalar and fermionic components of the \(\Phi\) and \(\overline{\Phi}\) supermultiplets,
which we denote by \(\phi\), \(\overline{\phi}\), \(\psi_{\Phi}\), and
\(\psi_{\overline{\Phi}}\) will play the role that the \(\nu_{R}\) play in Majorana
leptogenesis).  The mass of the lightest \(\Phi-\overline{\Phi}\) pair, which we will
denote \(M_{\Phi_1}\), defines the leptogenesis scale.  Since the parameters
\(\lambda_{i\alpha}\), and \(h_{i\alpha}\) may in general be complex (as may the elements
of (\(M_{ij}\)) in a general basis), they will in general contain nontrivial
\(CP\)-violating phases that cannot be rotated away.  This allows us to satisfy the
\(CP\)-violation criterion.  As in Majorana leptogenesis, a source of baryon number
violation is provided by electroweak sphaleron processes.  The primary difference between
the two models is that explicit lepton-number-violating terms are present in
equation~(\ref{eq:MajLepSuperpot}) but absent in
equation~(\ref{eq:DiracLepSuperpotential}). Unlike in Majorana leptogenesis, where \(L\)
is explicitly violated by the decays of heavy fields, here the overall values of \(B\)
and the total lepton number \(L_{\mathit{tot}}\) are altered only by sphaleron processes,
and \(B-L\) is never violated.

\begin{figure}[t]
  \begin{center}
\includegraphics[width=7cm]{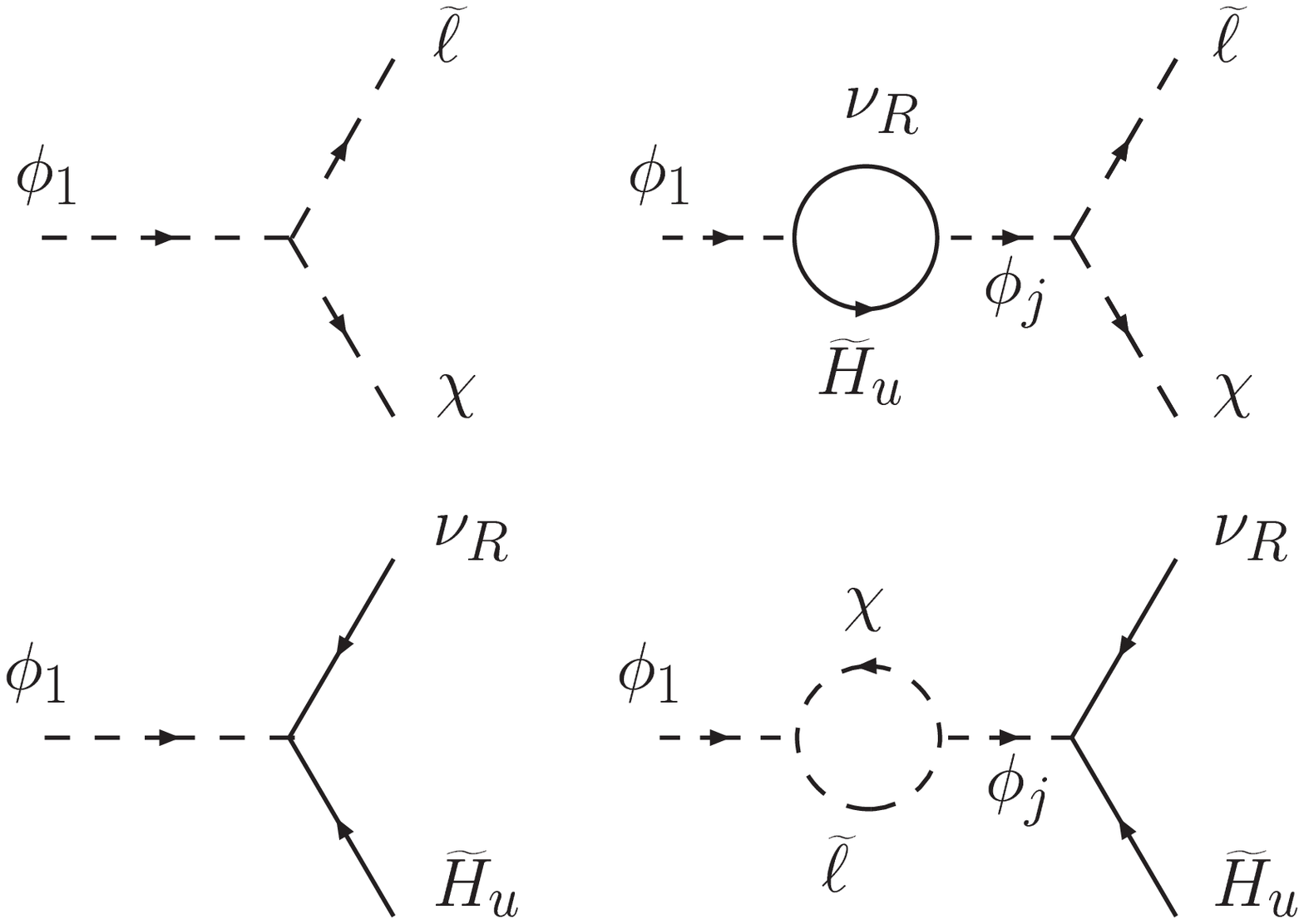}\hspace{.5cm}
\includegraphics[width=7cm]{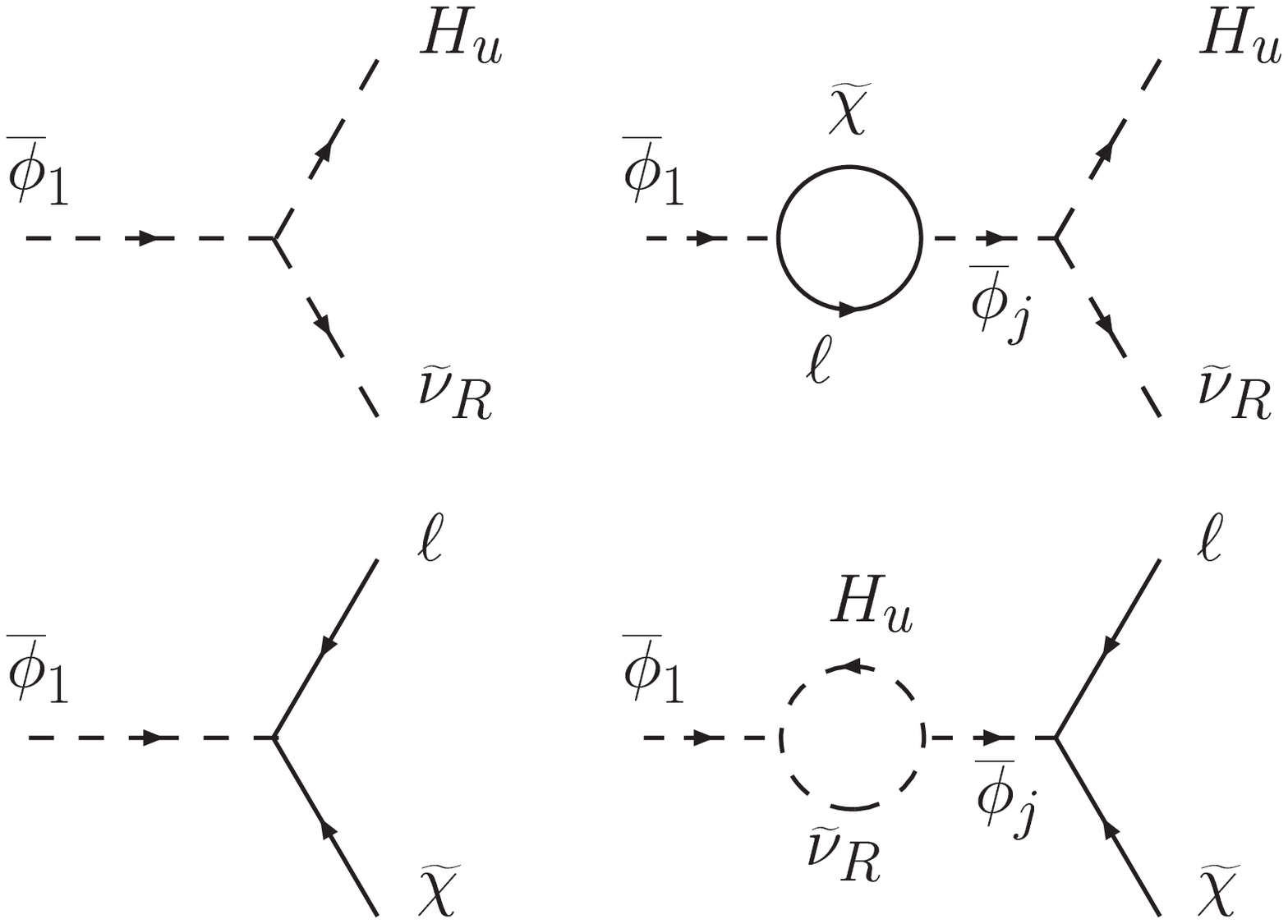}
\end{center}
\caption{Diagrams that give the leading contribution to the $CP$ asymmetry
    from decays of the scalar fields $\phi_{1}$ and $\overline{\phi}_{1}$.
    Similar $CP$ asymmetries are generated during the decay of the fermionic
    fields $\psi_{{}_{\Phi_{1}}}$ and $\psi_{{}_{\overline{\Phi}_{1}}}$.\label{fig:DecayDiagrams}}
\end{figure}

\indent When the temperature of the thermal bath drops below the leptogenesis scale
\(M_{\Phi_1}\), both the scalar and fermionic components of the \(\Phi_{1}\) and
\(\overline{\Phi}_{1}\) superfields will decay, generating a net \(CP\) asymmetry and
building up stores of lepton number in the lepton fields \(\nu_{R}\) and \(\ell\), and in
the associated slepton fields \(\tilde{\nu}_{R}\), \(\tilde{\ell}\).  The leading
contribution to \(CP\)-violation arises due to the interference of tree-level and
one-loop-level diagrams.  Those relevant to \(\phi\) and \(\overline{\phi}\) decay are
shown in figure~\ref{fig:DecayDiagrams}; the fermion fields \(\psi_{\Phi_{1}}\) and
\(\psi_{\overline{\Phi}_{1}}\) undergo similar decays, but in the approximation of
unbroken supersymmetry
the amplitudes (and
resultant \(CP\)-asymmetries) in the fermion case will be the same as those for the
scalar case, so the rates need not be separately evaluated.  For completeness, we have
included contributions involving the scalar component of \(H_{d}\), which start to become
important when the supersymmetric Higgs mass parameter \(\mu\) is of the same order as
the \(M_{\Phi_{i}}\), but in what follows we will assume that \(\mu\ll M_{1}\) and hence
the contribution from these diagrams is negligibly small.

\indent For purposes of illustration, let us begin by examining a simple toy model.  We
will consider the case where there are only two sets of \(\Phi\) and \(\overline{\Phi}\),
the minimum number required for \(CP\)-violation, in which case~\cite{Flanz:1996fb}, one
may parameterize the associated lepton number violation by defining a single decay
asymmetry \(\epsilon\), which represents the amount of lepton number generated in any
particular lepton-number-carrying species by the decay of a single heavy particle.  This
implies the relations
\begin{eqnarray}
  \Gamma(\Phi_{1}\longrightarrow N_{\alpha}^{c}H^{c}_{u})-\Gamma(\Phi_{1}^{c}\longrightarrow N_{\alpha}H_{u}) & \equiv & \epsilon\Gamma_{D} \label{eq:1steps} \\
  \Gamma(\Phi_{1}\longrightarrow L_{\alpha}\chi)-\Gamma(\Phi_{1}\longrightarrow L_{\alpha}^{c}\chi^{c}) & \equiv & -\epsilon\Gamma_{D} \label{eq:2ndeps} \\
  \Gamma(\overline{\Phi}_{1}\longrightarrow L_{\alpha}^{c}\chi^{c})-\Gamma(\overline{\Phi}_{1}^{c}\longrightarrow L_{\alpha}\chi) & \equiv & \epsilon\Gamma_{D} \label{eq:3rdeps} \\
  \Gamma(\overline{\Phi}_{1}\longrightarrow N_{\alpha}H_{u})-\Gamma(\overline{\Phi}_{1}^{c}\longrightarrow N_{\alpha}^{c}H_{u}^{c}) & \equiv & -\epsilon\Gamma_{D}
  \label{eq:4theps}
\end{eqnarray}
among the rates for the processes depicted in figure~\ref{fig:DecayDiagrams},
where \(\Gamma_{D}\) is the total decay width of any of the heavy fields in the
\(\Phi_{1}\) or \(\overline{\Phi}_{1}\) supermultiplets, and we have used the superfield
notation for \(\Phi_{1}\), \(N\), etc.\ because in the assumption of unbroken
supersymmetry, the supersymmetrized versions of the diagrams appearing in
fig.~\ref{fig:DecayDiagrams} yield the same result as the unsupersymmetrized ones.
Explicit calculation of \(\Gamma_{D}\) and \(\epsilon\) yields
\begin{equation}
  \Gamma_{D}=\frac{1}{16\pi}M_{\Phi_{1}}
  \sum_{\alpha}\left(|\lambda_{1\alpha}|^{2}+|h_{1\alpha}|^{2}\right).
  \label{eq:GammaD}
\end{equation}
and
\begin{equation}
  \epsilon=\frac{\mathrm{Im}(\lambda^{\ast}_{1\alpha}\lambda_{2\alpha}h^{\ast}_{1\beta}h_{2\beta}M_{\Phi_{1}}M^{\ast}_{\Phi_{2}})}{4\pi(|M_{\Phi_{2}}|^{2}-|M_{\Phi_{1}}|^{2})(|\lambda_{1\gamma}|^{2}+|h_{1\gamma}|^{2})},
  \label{eq:epsilon}
\end{equation}
where in both equations, a sum over the repeated indices \(\alpha\), \(\beta\) and
\(\gamma\) is implied.  It will be convenient to define a parameter
\(\delta\equiv|M_{\Phi_{1}}|/|M_{\Phi_{2}}|\), and in terms of \(\delta\),
\begin{equation}
  \epsilon=\frac{\mathrm{Im}(\lambda^{\ast}_{1\alpha}\lambda_{2\alpha}h^{\ast}_{1\beta}h_{2\beta}e^{i\psi})}{4\pi(|\lambda_{1\gamma}|^{2}+|h_{1\gamma}|^{2})}\left(\frac{\delta}{1-\delta^{2}}\right),\label{eq:epsilond}
\end{equation}
where \(\psi\) is the relative phase between \(M_{\Phi_{1}}\) and \(M_{\Phi_{2}}\).  This
tells us that for small values of \(\delta\), the final baryon-to-photon ratio will be
approximately proportional to \(\delta\).  Since \(\ell_{\alpha}\) and \(n_{R\alpha}\)
have equal and opposite charges under the global \(U(1)_{L}\) symmetry, the individual
\(L_{\ell}\) and \(L_{\nu_{R}}\) lepton numbers respectively stored in left-handed
leptons and right-handed neutrinos will likewise be equal and opposite.  Supersymmetry
enforces a similar condition $L_{\tilde{\ell}}=-L_{\tilde{\nu}_R}$ in the sneutrino
sector, and consequently no net lepton number is produced by the decays of \(\Phi_1\) and
\(\overline{\Phi_1}\).

\indent In order to discuss the subsequent evolution of these stores of lepton number, we
must now take a moment to examine how things look at temperatures far below the
leptogenesis scale \(M_{1}\). Here, the theory can be described by an effective
superpotential \(\mathcal{W}_{\mathit{eff}}\) in which the heavy \(\Phi_{i}\) and
\(\overline{\Phi}_{i}\) have been integrated out:
\begin{equation}
  \mathcal{W}_{\mathit{eff}}\ni
  \frac{\lambda_{i\alpha} h_{i\beta}^{\ast}}{M_{\Phi_{i}}}\chi L_{\beta} H_{u}
  N_{\alpha}+\mu H_{u} H_{d}.
  \label{eq:EffDiracLepSuperpotential}
\end{equation}
If we arrange for the scalar component of the \(\chi\) superfield to acquire a VEV
\(\langle\chi\rangle\), an effective Yukawa matrix for neutrinos, proportional to the
ratio \(\langle\chi\rangle/M_{\Phi_1}\) will result; then, when \(H_{u}\) acquires its
VEV during electroweak symmetry breaking, this will translate into a neutrino mass matrix
with entries given by
\begin{equation}
  m_{\nu\alpha\beta}= \langle\chi\rangle v \sin\beta\   \sum_{i}
  \frac{\lambda_{i\alpha} h_{i\beta}^{\ast}}{M_{\Phi_{i}}}.
  \label{eq:NeutMassMatrixGaugeBasis}
\end{equation}
If \(\langle\chi\rangle\ll M_{1}\), this setup ostensibly yields small but nonzero masses
for neutrinos, even when the elements in \(\lambda_{i\alpha}\) and \(h_{i\alpha}\) are
\(\mathcal{O}(1)\), and thus stands as an alternative to the traditional see-saw
mechanism. Additionally, since the mass matrix in~(\ref{eq:NeutMassMatrixGaugeBasis}) has
a reasonably simple structure, it can yield interesting predictions about the mass
hierarchy among the standard model neutrinos, especially when certain additional,
well-motivated constraints are applied, as we shall see in
section~\ref{sec:neutrinophys}.

\indent We have so far said nothing about how \(\chi\) receives its requisite VEV, but
there are a variety of ways of engineering such a thing.  One workable example is the
O'Raifeartaigh-type model employed in~\cite{Borzumati:2000mc}, in which the $F$-term of
\(\chi\) acquires a large VEV \(\langle F\rangle\simeq m_{3/2}M_{P}\), and supergravity
effects give rise to a nonzero VEV \(\langle \chi\rangle\simeq16\pi m_{3/2}\kappa^{-3}\)
for the scalar component of \(\chi\), where \(\kappa\) is an undetermined dimensionless
coupling constant.  Another, in which the VEV is induced by introducing new
superpotential couplings involving \(\chi\) with additional exotic superfields charged
under \(U(1)_{N}\) and no $F$-term VEV develops for \(\chi\), is presented in
chapter~\ref{ch:extensions}.  For the moment, we will not concern ourselves with the
precise manner in which a \(\chi\) VEV comes about, but it will be important to
distinguish theories based on the presence or absence of a large $F$-term VEV \(\langle
F_{\chi}\rangle\).  The reason for this is that when \(\langle F_{\chi}\rangle\) is
nonzero, the interaction lagrangian resulting from~(\ref{eq:EffDiracLepSuperpotential})
includes an effective $A$-term
\begin{equation}
  \mathcal{L}_{\mathit{eff}}\ni
  \frac{\langle F_{\chi}\rangle}{M_{\Phi_{i}}}\lambda{i\alpha}h_{i\beta}H_{u}\tilde{\ell}_{\beta}
  \tilde{\nu}_{\alpha}+c.c.\label{eq:EffectiveAtermL}
\end{equation}
through which left- and right-handed sneutrinos can equilibrate.  When \(\langle
F_{\chi}\rangle\) is large, this interaction will result in the lepton number asymmetries
\(L_{\tilde{\ell}}\) and \(L_{\tilde{\nu}_{R}}\) stored in the slepton sector being
rapidly equilibrated away; when it is small or vanishing, these two stores of lepton
number do not equilibrate until late times.  This effective $A$-term can induce
potentially large flavor-violating effects, as will be addressed in
chapter~\ref{ch:constraints}.

\indent As discussed in chapter one, \(B\) and \(L\) are not separately conserved in the
early universe due to the electroweak anomaly, and as we saw in
section~\ref{sec:sphalerons}, sphaleron processes associated with that anomaly will
intermix the two, conserving \(B-L\) and violating \(B+L\).  Since the net lepton number
for the universe is zero in Dirac leptogenesis, we would normally expect these processes
to wash out \(B\), but this conclusion can be avoided if the individual stores of lepton
number we have generated in charged (s)leptons and (s)neutrinos do not have a chance to
equilibrate among themselves until well below the scale of electroweak phase transition
\(T_{c}\), at which point sphalerons have effectively shut off.  For simplicity's sake,
let us begin by assuming that any lepton number stored in sleptons and sneutrinos is
rapidly annihilated through an effective $A$-term of the sort described above.  In this
case, there are only two stores of lepton number \(L_{\ell}\) and \(L_{\nu_R}\) present.
Since the \(\ell\) fields are charged under \(SU(2)\times U(1)_{Y}\) whereas \(\nu_{R}\)
are not, sphaleron effects will act on \(L_{\ell}\) only, transmuting this store of
lepton number into a nonzero baryon number along lines of constant \(B-L_{\ell}\) in the
manner described in section~\ref{sec:sphalerons}, while leaving \(L_{\nu_{R}}\)
untouched.  As discussed above, when \(L_{\ell}\) and \(L_{\nu_{R}}\) finally do
equilibrate, a net baryon number for the universe will already have frozen in, and the
universe will end up with net positive \(B\) and \(L\), as shown in
figure~\ref{fig:BLSpaceDiagram1}.

\begin{figure}[t!]
\begin{center}
\begin{fmffile}{LREquilProcDID11}
  \fmfframe(20,20)(20,20){\begin{fmfchar*}(80,50)
    \fmfleft{i1,i2}
    \fmfright{o1}
    \fmflabel{\footnotesize{$\nu_{R}$}}{i1}
    \fmflabel{\footnotesize{$\ell$}}{i2}
    \fmflabel{\footnotesize{$h^{0},H^{0},H^{\pm}$}}{o1}
    \fmf{vanilla,tension=.5}{i1,v1,i2}
    \fmf{dashes,tension=.5}{v1,o1}
  \end{fmfchar*}}
\end{fmffile}~~\hspace{.5cm}~~
\begin{fmffile}{LREquilProcScatb10}
  \fmfframe(20,20)(20,20){\begin{fmfchar*}(100,50)
    \fmfleft{i1,i2}
    \fmfright{o1,o2}
    \fmflabel{\footnotesize{$\nu_{R}$}}{i1}
    \fmflabel{\footnotesize{$\ell$}}{i2}
    \fmflabel{\footnotesize{$\overline{b},W^{\pm},Z,\mathit{etc.}$}}{o1}
    \fmflabel{\footnotesize{$b,W^{\pm},Z,\mathit{etc.}$}}{o2}
    \fmf{vanilla,tension=1}{i1,v1,i2}
    \fmf{vanilla,tension=1}{o1,v2,o2}
    \fmf{dashes,tension=.5,label=\footnotesize{$h^{0},,H^{0},,H^{\pm}$}}{v1,v2}
  \end{fmfchar*}}
\end{fmffile}\\
\begin{fmffile}{LREquilProcScath10}
  \fmfframe(20,20)(20,20){\begin{fmfchar*}(100,50)
    \fmfleft{i1,i2}
    \fmfright{o1,o2}
    \fmflabel{\footnotesize{$\nu_{R}$}}{i1}
    \fmflabel{\footnotesize{$\ell$}}{i2}
    \fmflabel{\footnotesize{$Z,W^{\pm}$}}{o2}
    \fmflabel{\footnotesize{$h^{0},H^{0},H^{\pm}$}}{o1}
    \fmf{vanilla,tension=1}{i1,v1,i2}
    \fmf{photon,tension=1}{v2,o2}
    \fmf{dashes,tension=.5,label=\footnotesize{$h^{0},,H^{0},,H^{\pm}$}}{v1,v2}
    \fmf{dashes,tension=1}{v2,o1}
  \end{fmfchar*}}
\end{fmffile}
\end{center}
\caption{Diagrams corresponding to the leading order contribution to the equilibration of
left-handed leptons with right-handed neutrinos. These include, from left to right, Higgs
decays and inverse decays, scatterings off Standard Model fermions and electroweak gauge
bosons through a virtual Higgs, and scattering off a Higgs boson in conjunction with the
emission of an electroweak gauge boson.  The right two diagrams also have $t$-channel
equivalents.  The amplitude associated with each diagram is proportional to the (small)
effective neutrino Yukawa given in
equation~(\ref{eq:EffDiracLepSuperpotential}).\label{fig:LRneutEquilib}}
\end{figure}
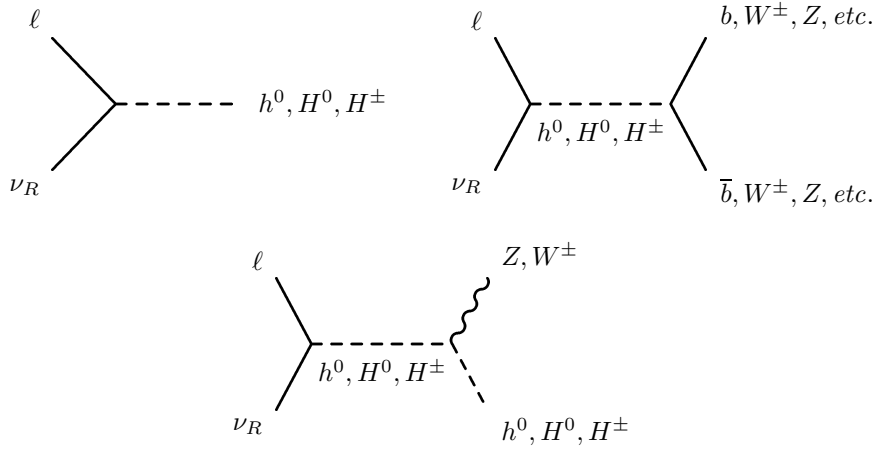

\indent The leading contributions to the equilibration rate between left- and
right-handed neutrinos are shown in figure~\ref{fig:LRneutEquilib}.  The amplitude for
each of these processes is proportional to the effective Higgs coupling term in
equation~(\ref{eq:EffDiracLepSuperpotential}), and hence suppressed by
\(\langle\chi\rangle/M_{\Phi_1}\). In order for Dirac leptogenesis to work, we must
ensure that the net equilibration rate does not become significant compared to the
expansion rate of the universe until well after the electroweak phase transition. The
equilibration rate may be estimated on dimensional grounds to be
\begin{equation}
  \Gamma_{\mathit{eq}}\sim
  \frac{|\lambda|^{2}|h|^{2}\langle\chi\rangle^{2}}{M_{\Phi_{1}}^{2}}g_{\mathit{sc}}^{2}T,
\end{equation}
where \(g_{\mathit{sc}}\) is an \(\mathcal{O}(1)\) gauge or top Yukawa coupling (see
figure~\ref{fig:LRneutEquilib}) and \(T\) is temperature.  Requiring that
\(\Gamma_{\mathit{eq}}<H\) for \(T>T_{c}\) leads to the condition
\begin{equation}
  \frac{|\lambda||h|\langle\chi\rangle}{M_{\Phi_{1}}}\leq10^{-8},
  \label{eq:EquilibrationBound}
\end{equation}
which is easily satisfied, as it is essentially a recapitulation of the statement that
neutrino masses must be small.  This limit can be translated into a bound on the neutrino
mass using equation~(\ref{eq:NeutMassMatrixGaugeBasis})
\begin{equation}
m_\nu\leq 1.74\,(\sin\beta)~{\rm keV}
\end{equation}
and thus the stipulation that left-right equilibration happen only well below \(T_{c}\)
is satisfied automatically when neutrinos are given realistic masses.

\indent In addition to this constraint, there are several other consistency checks which
Dirac leptogenesis must pass in order to be considered a legitimate baryogenesis model.
One of these is that the Sakharov criterion that the abundances of \(\phi_{1}\) and
\(\overline{\phi}_{1}\) depart from their equilibrium values must be satisfied.  In order
for this to occur, the decay rate \(\Gamma_{D}\)~(\ref{eq:GammaD}) must be slower than
the rate of the expansion of the universe $H(T)$~(\ref{eq:HubbleBubble}) evaluated at the
temperature \(T\sim M_{\Phi_{1}}\) when \(\phi_{1}\) can no longer be treated as
effectively massless and its abundance begins to fall off, or in other words
\begin{equation}
  \frac{\Gamma_{D}}{H(M_{\Phi{1}})}=
  9.97\cdot 10^{-2}g_{\ast}^{-1/2}\frac{M_{P}}{M_{\Phi_{1}}}
  \sum_{\alpha}(|\lambda_{1\alpha}|^{2}+|h_{1\alpha}|^{2})\lesssim
  1.
  \label{eq:GammaOverH}
\end{equation}
This constraint also favors small couplings and large \(M_{\Phi_{1}}\).  However, in
addition to these two requirements, we must ensure that the present value of \(\eta\)
satisfies the bounds in equation~(\ref{eq:WMAPeta}).   To do this properly, one must
solve the full system of Boltzmann equations, which we do in chapter~\ref{ch:Boltzmann},
though a rough estimate can be made using the ``drift-and-decay''
approximation~\cite{Kolb:1990vq}, in which we assume that the heavy particle decays occur
well out of equilibrium and that the effects of inverse decays and \(2\leftrightarrow2\)
processes where $\Delta L_{\ell}\neq0$ are negligible. Including contributions from both
scalar and fermion decays, this gives the result
\begin{equation}
  L_{\ell}=
  \frac{2\epsilon n^{\mathit{init}}_{\phi_{1}}}{s}=
  \frac{90\epsilon}{\pi^{4}g_{\ast}}K_{2}(1)=
  7.32\times10^{-3}\epsilon,
  \label{eq:DriftAndDecay}
\end{equation}
where \(n_{\phi_{1}}^{\mathit{init}}\) is the initial number density of \(\phi\),
\(K_{2}(x)\) denotes the modified Bessel function of the second kind, evaluated at $x$,
and \(\epsilon\) is the decay asymmetry given in~(\ref{eq:epsilon}). Since the final
baryon-to-entropy ratio \(B\) (related to \(\eta\) by \(B=\eta/7.04\)) generated by
sphaleron processes will be on the same order ($B\simeq 0.35 L_{\ell}$), this can serve
as a rough estimate for \(B\). Thus even if equation~(\ref{eq:GammaOverH}) is satisfied,
the final baryon-to-entropy ratio of the universe will be proportional to \(\epsilon\);
and from equation~(\ref{eq:epsilon}), we see that for \(\epsilon\) to be large, either
the couplings must be large or the splitting between \(M_{\Phi_{1}}\) and
\(M_{\Phi_{2}}\) must be small.

\indent The upshot of all this is that while there are tensions among the model
parameters in Dirac leptogenesis, they are not difficult to reconcile---in part because
relevant physical scales in the theory are determined by the interplay of a large number
of model parameters: (\(M_{\Phi_{1}}\), \(\delta\), \(\langle\chi\rangle\)), and the
elements of the complex coupling matrices \(\lambda_{i\alpha}\) and \(h_{i\alpha}\).
Unlike in Majorana leptogenesis, where neutrino masses are determined solely from the
neutrino Yukawa \(y_{\nu}\) matrix and the masses of the heavy fields according to the
see-saw mechanism, neutrino masses in Dirac
leptogenesis~(\ref{eq:NeutMassMatrixGaugeBasis}) depend not only on the superpotential
couplings \(\lambda\) and \(h\) and the masses \(M_{\Phi_i}\), but also on
\(\langle\chi\rangle\). Consequently, the model parameters of Dirac leptogenesis are far
less constrained.  Of course this versatility comes at the price of introducing an
additional intermediate scale, though we will show that it is possible to relate
\(\langle\chi\rangle\) to other physical scales, for example the Higgs \(\mu\)-term, in
chapter~\ref{ch:extensions}.

\indent So far the constraints we have discussed have been limited to those which
function as consistency checks on the model.  The real tensions among \(M_{\Phi_{1}}\),
\(\delta\), \(\lambda_{i\alpha}\), \(h_{i\alpha}\), and \(\langle\chi\rangle\) are not
those inherent in the Dirac leptogenesis framework, however, but those that arise when we
demand that the model respect the full battery of additional constraints from neutrino
physics, flavor physics, and cosmology.  We now turn to address these constraints and
their implications for Dirac leptogenesis.


\chapter{CONSTRAINTS\ ON\ THE\ MODEL\label{ch:constraints}}

\section{Astrophysical Constraints\label{sec:AstroConstraints}}

\subsection{Baryogenesis and the Gravitino Problem}

\indent

So far, we have seen that a simple toy model of Dirac leptogenesis is capable of yielding
a nonzero baryon-to-photon ratio for the universe and explaining the observed scale of
neutrino masses, but we have not yet shown that this scenario is a viable
phenomenological model. To do this satisfactorily, we must first ensure that none of the
modifications we have made disrupt the standard cosmology, i.e.\ that it is compatible
with Big Bang nucleosynthesis (BBN), cosmic inflation, etc.~\cite{Thomas:2005rs}.
Furthermore, we must investigate whether the theory is simultaneously capable of yielding
a realistic neutrino spectrum and avoiding all current bounds on flavor-violation in the
lepton sector~\cite{Thomas:2006gr}.  At the same time, we must show that the theory is
capable of reproducing the value of \(\eta\) observed by WMAP~(\ref{eq:WMAPeta}) once all
these constraints are applied.

\indent There are two primary ways supersymmetric Dirac leptogenesis could potentially
disrupt BBN.  First, constraints on the light element abundances place a bound on the
number of additional light neutrino species~\cite{Olive:1999ij}
\begin{equation}
\Delta N_{\nu}\leq 0.3\label{eq:Nnumberbound}
\end{equation}
might cause one to worry that the presence of three additional light, sterile neutrino
fields \(\nu_{R\alpha}\) in the theory would violate this bound.  However, the
\(\nu_{R\alpha}\) are not in thermal equilibrium with the bath during the BBN epoch;
hence their contribution to \(\Delta N\) is suppressed by and entropy factor
\begin{equation}
  \Delta N=3\left(\frac{T_{\nu_{R}}}{T_{\mathit{bath}}}\right)^{4}=
  3\left(\frac{g_{\ast}(\mathrm{1~MeV})}{g_{\ast}(\mathrm{MSSM}+N_{i})}\right)=0.02,
\end{equation}
and thus the bound in~(\ref{eq:Nnumberbound}) is respected.  The second way in which
supersymmetric Dirac leptogenesis can impact BBN is that, as in any supersymmetric
theory, energy and entropy released during late decays of heavy sparticles (gravitinos
will be of particular concern) could potentially distort the light element abundances
from their observed values.  This merits an in-depth discussion, as gravitino physics can
place quite stringent constraints on the leptogenesis scale \(M_{\Phi_1}\).

\indent The connection between leptogenesis and gravitino physics, which might not at
first seem intimately interrelated, occurs through the reheating temperature \(T_{R}\)
associated with cosmic inflation.  Since leptogenesis requires that a thermal population
of \(\Phi_{1}\) and \(\overline{\Phi}_{1}\) be generated from the thermal bath during
reheating, we require that \(M_{\Phi_{1}}\lesssim T_{R}\); hence an upper bound on
\(T_{R}\) translates into an upper bound on \(M_{\Phi_{1}}\).  On the other hand, if
\(m_{3/2}\lesssim T_{R}\) heavy gravitinos will also be generated from the thermal bath,
and one must take care that they do not cause problems for the standard cosmology.  There
are two distinct varieties of gravitino problem that must be addressed. First, as alluded
to above, late gravitino decays can disrupt BBN by releasing energy in the form of
photons and other energetic particles into the system; second, as the assumption of
$R$-parity conservation implies that at least one LSP will be produced at the end of the
decay chain resulting from each late gravitino decay, the potentially large, non-thermal
population of stable particles produced in this manner could overclose the universe---or
if the right amount is produced, could make up the majority of cold dark matter (CDM).
Both of these issues are contingent on the gravitino lifetime \(\tau_{3/2}\), and for
cases where \(m_{3/2}\gg m_{s}\) (where \(m_{s}\) is the rough scale of the MSSM
sparticle masses), this lifetime is approximated by~\cite{Moroi:1995fs}
\begin{equation}
  \tau_{3/2}=4.0\times 10^{8} \left(\frac{m_{3/2}}{100\mathrm{GeV}}\right)^{-3}\mathrm{s}.
\end{equation}
Careful analysis of the BBN constraints (see, for example,~\cite{Kawasaki:2004qu} and
references therein) reveals that unless \(m_{3/2}\gtrsim10^{5}\)~GeV, \(T_{R}\) cannot be
greater than around \(10^{8}\)~GeV, and for \(m_{3/2}\lesssim5\times10^{3}\)~GeV, cannot
exceed \(10^{6}\)~GeV. In models where \(m_{3/2}\) is at the PeV scale, however,
\(\tau_{3/2}\sim 10^{-4}\) s, which implies that gravitinos produced in the thermal bath
decay long before the BBN epoch (at \(t_{\mathit{universe}}\sim 1\) s), and thus there is
no gravitino problem of the former type for models with \(m_{3/2}\) at or above these
scales. However, since a weakly interacting LSP decouples on a timescale \(t_{f}\sim
10^{-11}\) s, it will have long frozen out by the time gravitino decay occurs unless
\(m_{3/2}\) is larger than around \(10^{8}\)~GeV; hence in theories with smaller
gravitino masses (including simple PeV-scale supersymmetry with anomaly-mediated gaugino
masses), LSPs produced by gravitino decay will be unable to thermalize and the latter
type of gravitino issue cannot be ignored.

\indent In order to avoid any complications from late gravitino decay, we require not
only that the LSP not overclose the universe, but that its surviving relic density
\(\Omega_{\mathit{LSP}}\) must be less than (or ideally, if the LSP is to constitute the
majority of cold dark matter, equal to) the relic density of CDM as measured by
WMAP~\cite{Spergel:2003cb},
\begin{equation}
  \Omega_{\mathrm{CDM}}h^{2}=0.11\pm0.01 \mbox{ (WMAP 68\% C.L.)}.
  \label{eq:WMAPBounds}
\end{equation}
In general, \(\Omega_{\mathit{LSP}}\) will have both a thermal and a non-thermal
component, so that \(\Omega_{\mathit{LSP}}=\Omega^{\mathit{Th}}_{\mathit{LSP}}+
\Omega^{\mathit{NT}}_{\mathit{LSP}}\).  The thermal component
\(\Omega^{\mathit{Th}}_{\mathit{LSP}}\) may be ascertained by solving the relevant set of
Boltzmann equations for the LSP abundance at freeze-out.  The results, for the case where
the LSP is essentially either a pure Wino or Higgsino, are~\cite{Giudice:2004tc}
\begin{eqnarray}
  \Omega^{\mathit{Th}}_{\mathit{LSP}}h^{2}=0.02\left(\frac{|M_{2}|}{\mathrm{TeV}}\right)
    & \mbox{ for Wino LSP}\label{eq:OmegaThermWino} \\
  \Omega^{\mathit{Th}}_{\mathit{LSP}}h^{2}=0.09\left(\frac{|\mu|}{\mathrm{TeV}}\right)
    & \mbox{ for Higgsino LSP}.\label{eq:OmegaThermHiggsino}
\end{eqnarray}

\indent Of course it is also possible that the gravitino itself is the LSP, in which case
the primary concerns are that next-to-lightest supersymmetric particle (NLSP) decays to
gravitino do not disrupt BBN and that the gravitino abundance does not exceed the WMAP
bound~(\ref{eq:WMAPBounds}).  In Gauge-Mediated supersymmetry breaking and other theories
where the gravitino is exceedingly light (with a mass on the order of a few keV or less),
the gravitino contribution to the energy density of the universe during the BBN
epoch~\cite{Giudice:1998bp} is also a concern.  It has been
shown~\cite{Gherghetta:1998tq} that these constrains can be translated to a reheating
temperature limit $T_{R}\lesssim 10^{7}$~GeV, which, as we shall see in
section~\ref{sec:numerres} turns out to be incompatible with Dirac leptogenesis.  When
the gravitino is reasonably heavy, with a mass of \(\mathcal{O}(100~\mbox{GeV})\), the
gluino mass is around 500~GeV, and the NLSP is Higgsino-like (the best-case scenario) the
constraints from NLSP decay and $\Omega_{\mathrm{CDM}}$ conspire to produce a reheating
temperature bound $T_{R}\lesssim 10^{9}$~GeV~\cite{Bolz:1998ek}, which will also turn out
to be problematic for the theory.  It thus appears that a gravitino LSP is essentially
incompatible with Dirac leptogenesis; thus we shall henceforth focus our efforts on
models in which the gravitino is not the LSP.

\indent Now we turn to evaluating \(\Omega^{\mathit{NT}}_{\mathit{LSP}}\).  We begin by
addressing the regime in which there is no significant reduction in
\(\Omega_{\mathit{LSP}}^{\mathit{NT}}\) from LSP annihilations. Assuming for the moment
that the dominant contribution to the non-thermal relic abundance comes from late
gravitino decays and that all the LSPs produced from such decays survive until present
day, \(\Omega^{\mathit{NT}}_{\mathit{LSP}}\) is given by
\begin{equation}
  \Omega^{\mathit{NT}}_{\mathit{LSP}}=
  \frac{m_{LSP}\zeta(3)T_{0}^{3}}{\pi^{2}\rho_{\mathit{crit}}}
  Y_{3/2}(T_{3/2}),
  \label{eq:OmegaNTForm}
\end{equation}
where \(\rho_{\mathit{crit}}\) is the critical density of the universe, \(T_{0}\) is the
present temperature of the universe, and \(Y_{3/2}(T_{3/2})\) is the number of gravitinos
per co-moving volume at the characteristic temperature \(T_{3/2}\) at which the gravitino
decays, which is given by~\cite{Kawasaki:1994af}
\begin{equation}
  Y_{3/2}(T_{3/2})=0.856\times 10^{-11}
  \left(\frac{T_{R}}{10^{10}\mathrm{GeV}}\right)
  \left(1-0.0232\ln\left(\frac{T_{R}}{10^{10}\mathrm{GeV}}\right)\right).
\end{equation}
Substituting this into equation (\ref{eq:OmegaNTForm}) yields
\begin{equation}
  \Omega^{\mathit{NT}}_{\mathit{LSP}}h^{2}=2.96\times 10^{-4}
  \left(\frac{m_{\mathit{LSP}}}{\mathrm{GeV}}\right)
  \left(\frac{T_{R}}{10^{10}\mathrm{GeV}}\right)
  \left(1-0.0232\ln\left(\frac{T_{R}}{10^{10}\mathrm{GeV}}\right)\right).
  \label{eq:OmegaNTNot}
\end{equation}

\begin{figure}[ht!]
  \begin{center}
    \includegraphics{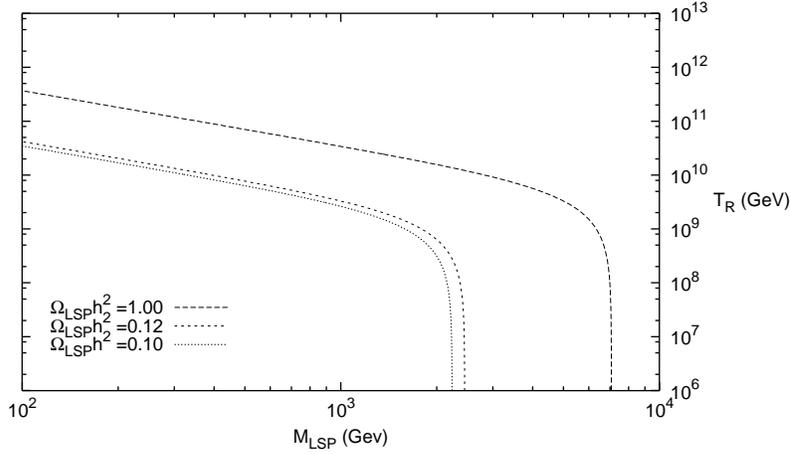}
  \end{center}
  \caption{Contours of $\Omega_{LSP}h^{2}$~\cite{Thomas:2005rs} corresponding to the upper and lower bounds from WMAP
(\ref{eq:WMAPBounds}), as well as a simple overclosure bound, as a function of the LSP
mass $ m_{\mathit{LSP}} $ and the reheating temperature $T_{R}$ associated with cosmic
inflation for a Wino LSP.  The requirement that $\Omega_{\mathit{LSP}}$ not overclose the
universe severely constrains $T_{R}$, and hence the temperature scale of thermal
leptogenesis, for a theory like PeV-scale loop-split supersymmetry, in which the LSP is
particularly heavy.  For a Higgsino LSP, the results are similar, but the contours move
slightly to the left.\label{fig:OmegaTot}}
\end{figure}

\indent In figure~\ref{fig:OmegaTot}, we plot the contours corresponding to the WMAP
upper and lower bounds from equation (\ref{eq:WMAPBounds}) on the total relic abundance
of a Wino LSP, as well as the simple overclosure bound \(\Omega_{\mathit{LSP}}h^{2}=1\),
taking into account both thermal and non-thermal contributions, as a function of
\(m_{\mathit{LSP}}\) and \(T_{R}\).  The gently-sloping portion of each contour
corresponds to the nonthermal abundance in equation~\ref{eq:OmegaNTNot}, which becomes
significant when \(T_{R}\) is large; the nearly vertical portion on the right side of the
graph corresponds to the thermal abundance given in~\ref{eq:OmegaThermWino}.  In the
region above and to the right of the WMAP upper bound given in~(\ref{eq:WMAPBounds}),
dark matter is overproduced and hence excluded.  In the narrow strip between the upper
and lower WMAP bounds, the thermal LSP abundance and the nonthermal LSP abundance
resulting from late gravitino decay conspire to reproduce the observed value for
\(\Omega_{\mathit{CDM}}\).  In the region below and to the left of the lower WMAP
contour, the experimental upper bound is not violated and hence this region of parameter
space is phenomenologically allowed, provided there are other sources of dark matter to
make up the deficit between \(\Omega_{\mathit{LSP}}\) and \(\Omega_{\mathit{CDM}}\).
These sources could include other non-thermally-generated contributions to
\(\Omega_{\mathit{LSP}}\) or contributions, contributions from other exotic particles, or
some combination of both.  It can be seen from figure~\ref{fig:OmegaTot} that when
\(m_{\mathit{LSP}}\) is large enough that \(\Omega^{\mathit{Th}}_{\mathit{LSP}}\approx
\Omega_{\mathit{CDM}}\) and CDM is essentially thermal in origin, the ceiling on
\(T_{R}\) is quite low--around \(10^{9}\)~GeV.  In regions of parameter space where
\(m_{\mathit{LSP}}\) is smaller and the majority of CDM is generated non-thermally,
\(T_{R}\) may be raised a bit, but is still constrained to be below \(\sim 5\times
10^{10}\)~GeV. As we shall see in chapter~\ref{ch:Boltzmann}, \(T_{R}\gtrsim10^{9}\) GeV
turns out to be problematic for Dirac leptogenesis (in terms of the final
baryon-to-photon ratio generated), largely due to the out-of equilibrium condition
in~(\ref{eq:GammaOverH}).  This implies that if we want to raise \(T_{R}\) above
\(10^{9}\)~GeV and still have the LSP relic density dominate \(\Omega_{\mathit{CDM}}\),
the majority of the dark matter abundance (be it from late gravitino decays or something
else) must be essentially non-thermal in origin.

\indent We now turn to address the regime where LSP annihilations do play a role in
reducing \(\Omega_{\mathit{LSP}}^{\mathit{NT}}\), and thus the upper bound on \(T_{R}\)
may be raised.  This effect becomes important when \(m_{3/2}\gg m_{\mathit{LSP}}\).  When
it is taken into account~\cite{Moroi:1999zb}, the non-thermal LSP relic density is
modified to
\begin{equation}
  \Omega^{\mathit{NT}}_{\mathit{LSP}}=\min
    \left(\Omega^{\mathit{NT} (0)}_{\mathit{LSP}},\Omega^{\mathit{NT} \mathit{(ann)}}_{\mathit{LSP}}\right),
\end{equation}
where \(\Omega^{\mathit{NT} (0)}_{\mathit{LSP}}\) is the relic density given in equation
(\ref{eq:OmegaNTNot}), and \(\Omega^{\mathit{NT} \mathit{(ann)}}_{\mathit{LSP}}\) is the
relic density obtained by solving the full system of Boltzmann equations for the LSP.
For a Wino LSP, \(\Omega^{\mathit{NT} \mathit{(ann)}}_{\mathit{LSP}}\) is given by
\begin{eqnarray}
  \Omega^{\mathit{NT} \mathit{(ann)}}_{\mathit{LSP}} & = &
  2.41\times10^{-2}
  \frac{(2-x_{W})^{2}}{(1+x_{W})^{3/2}}
  \left(\frac{m_{\mathit{LSP}}}{100\mathrm{GeV}}\right)^{3}
  \left(\frac{m_{3/2}}{100\mathrm{TeV}}\right)^{-3/2} \nonumber \\ & \times &
  \left(1-\left(\frac{m_{\mathit{LSP}}}{m_{3/2}}\right)\right)^{3}
  \left(1+\frac{1}{3}\left(\frac{m_{\mathit{LSP}}}{m_{3/2}}\right)\right),
  \label{eq:OmegaNTAnn}
\end{eqnarray}
where \(x_{W}\equiv m_{W}/m_{\mathit{LSP}}\); for a Higgsino LSP, which annihilates far
less efficiently, \(\Omega^{\mathit{NT} (\mathit{ann})}_{\mathit{LSP}}\) will be even
higher.

\begin{figure}[ht!]
\hspace{1.5cm}
  \includegraphics{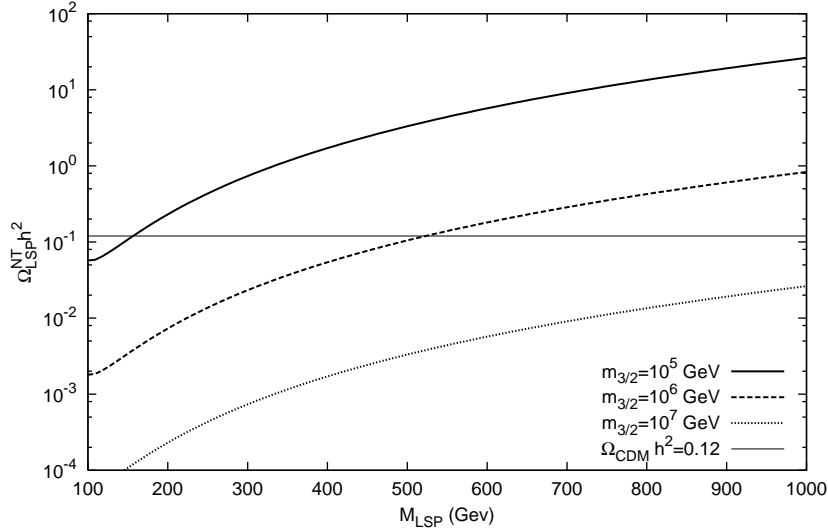}
  \caption{Here, we show the variation of $\Omega^{\mathit{NT} \mathit{(ann)}}_{\mathit{LSP}}$ (the improved expression
    for the LSP relic density that accounts for the effect of LSP annihilations) with LSP mass for
    several different values of $m_{3/2}$~\cite{Thomas:2005rs}.  The horizontal line corresponds to the WMAP upper
    bound on $\Omega_{\mathit{CDM}}$.  For a given choice of $m_{3/2}$, portions of the
    contour that fall below the horizontal line respect the WMAP constraint and are
    phenomenologically allowed, and the reheating temperature $T_{R}$ be increased beyond
    the naive upper bound from figure~\ref{fig:OmegaTot}.  Portions that lie above it overproduce the LSP and are
    excluded (for example, when $m_{3/2}=10^6$~GeV, an LSP mass $m_{LSP}\geq500$~GeV is
    excluded).\label{fig:OmegaAnn}}
\end{figure}

\indent In figure~\ref{fig:OmegaAnn}, we show the relationship between
\(\Omega^{\mathit{NT} (\mathit{ann})}_{\mathit{LSP}}\) and \(m_{\mathit{LSP}}\) for
several values of \(m_{3/2}\).  The horizontal line corresponds to the WMAP upper bound
on \(\Omega^{\mathit{CDM}}\). From this plot it is evident that annihilations are only
effective in reducing the LSP relic abundance below this bound when \(m_{3/2}\) is much
larger than \(m_{\mathit{LSP}}\).  However, when \(m_{3/2}\) is increased beyond around
\(10^{8}\)~GeV, \(\tau_{3/2}\) becomes short enough that gravitino decay occurs before
LSP freeze-out, and \(\Omega_{\mathit{LSP}}^{\mathit{NT}}\) drops to zero regardless of
what the ratio of \(m_{\mathit{LSP}}\) to \(m_{3/2}\) is, and nonthermal LSP
overproduction from late gravitino decays no longer remains a concern.

\indent While the problems that can arise for small gravitino masses have now been
thoroughly addressed, the caveats associated with extremely large \(m_{3/2}\) should also
be mentioned. As has been shown in~\cite{Arvanitaki:2005fa}, split supersymmetry models
with a large hierarchy between the gravitino and gaugino masses can suffer from
phenomenological problems associated with the overproduction of gluinos, including the
distortion of both the CMB and the light element abundances through their late decays.
Since gluinos will be bound into \(R\)-hadrons at temperatures below the scale
\(\Lambda_{\mathit{QCD}}\) associated with the QCD phase transition, a precise analysis
of their decay rate at late times has not yet been performed.  Still, while a precise
ceiling for \(m_{3/2}\) must wait until the decay of \(R\)-hadrons is better understood,
it is known that this ceiling falls somewhere in the \(m_{3/2}\simeq10^{10} - 10^{12}\)
GeV range.  For this reason, one should be wary about making the gravitino mass
arbitrarily large. There are also caveats associated with the gravitino-producing decays
of scalar sparticles in models where one or more scalars has a mass larger than
\(m_{3/2}\)~\cite{Allahverdi:2005rh}.

\begin{figure}[ht!]
\hspace{1.5cm}
  \includegraphics{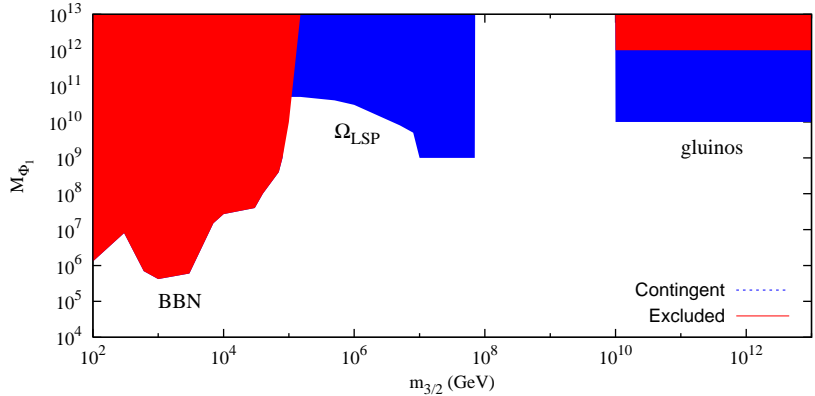}
  \caption{A schematic representing the bounds on the leptogenesis scale $M_{\Phi_{1}}$
  from gravitino physics as a function of the gravitino mass $m_{3/2}$.  When
  $m_{3/2}\lesssim10^{5}$~GeV, constraints from BBN are quite severe.  In the range
  $10^{5}$~GeV~$\lesssim m_{3/2}\lesssim10^{8}$~GeV, $M$ is still limited by
  nonthermal decays to the LSP.  For $m_{3/2}<10^{10}$~GeV, decays to gluinos are a
  concern.  This leaves a window $10^{5}$~GeV~$\lesssim m_{3/2}\lesssim10^{10}$~GeV within
  which, depending on the mass of the LSP, Dirac leptogenesis can be successful.
  \label{fig:TRboundsSchematic}}
\end{figure}

\indent As discussed above, bounds on the reheating temperature in supersymmetric
theories can be viewed as bounds on the leptogenesis scale \(M_{\Phi_{1}}\).  This is not
in any way peculiar to Dirac leptogenesis either: in a Majorana leptogenesis model, the
constraints still apply with the lightest right-handed neutrino mass \(M_{\nu_{R}}\) in
place of \(M_{\Phi_{1}}\).  In figure~\ref{fig:TRboundsSchematic}, we display these
constraints graphically.  For light gravitinos (\(m_{3/2}\lesssim10^{5}\)~GeV), BBN
limits severely constrain \(T_{R}\), and hence \(M_{\Phi_{1}}\).  For slightly heavier
gravitinos (\(10^{5}\)~GeV~\(\lesssim m_{3/2}\lesssim10^{8}\)~GeV), there are still
constraints on \(T_{R}\) from nonthermal decays which are only alleviated (via LSP
annihilations) when \(m_{\mathit{LSP}}\ll m_{3/2}\). For extremely heavy gravitinos, with
mass \(m_{3/2}\gg 10^{10}\)~GeV, decays to gluinos become worrisome, and when
\(m_{3/2}\gg 10^{12}\)~GeV they will almost certainly become problematic.

\indent While we have not said much about other constraints on \(M_{\Phi_{1}}\), but it
is not difficult to see that substantial model tension arises when the reheating
temperature is constrained to be below around \(10^{10}\)~GeV.  The out-of-equilibrium
condition~(\ref{eq:GammaOverH}) demands that \(\lambda_{1\alpha}\) and \(h_{1\alpha}\) be
at most \(\mathcal{O}(10^{-4})\) for \(M_{\Phi_{1}}\sim10^{10}\) GeV; on the other hand,
(\ref{eq:epsilon}) and~(\ref{eq:DriftAndDecay}) imply that unless there is a large
hierarchy among the Yukawa couplings to different sets of \(\Phi\) and
\(\overline{\Phi}\) (so that \(\lambda_{1\alpha}\ll\lambda_{2\alpha}\) for all $\alpha$),
decreasing the \(\lambda_{i\alpha}\) and \(h_{i\alpha}\) below \(\mathcal{O}(10^{-4})\)
will yield insufficient baryon number---and we have not yet taken into account the
effects of \(2\leftrightarrow2\) processes and inverse decays.\footnote{One can also get
around this by making \(M_{\Phi_{1}}\) and \(M_{\Phi_{2}}\) essentially degenerate.  We
will elaborate on this possibility in section~\ref{sec:Resonating}.}  Furthermore,
engineering \(\lambda_{i\alpha}\) and \(h_{i\alpha}\) to be extremely small while the
rest of the trilinear couplings in the superpotential are \(\mathcal{O}(1)\) seems to
defeat the purpose of leptogenesis, the advantage of which was its ability to explain the
smallness of neutrino masses without resorting to arbitrary fine-tuning. While any
precise statement about the value of \(M_{\Phi_{1}}\) must wait until after we solve the
Boltzmann equations governing the development of \(B\) and \(L\) during the leptogenesis
epoch, leptogenesis seems to become difficult or contrived when
\(M_{\Phi_{1}}\lesssim10^{10}\)~GeV.  This means that when \(m_{3/2}\lesssim10^8\)~GeV
and LSP annihilations are ineffective, substantial tensions arise among the
out-of-equilibrium decay criterion, overclosure bounds related to the reheating
temperature \(T_{R}\), the equation that determines the decay asymmetry \(\epsilon\),
etc., and problems are likely to arise.  We shall confirm these suspicions in
chapter~\ref{ch:Boltzmann}.

\indent The indication that \(m_{3/2}\) must be quite large in order for Dirac
leptogenesis to work suggests that the model could be quite successful in the context of
split supersymmetry~\cite{Arkani-Hamed:2004fb,Giudice:2004tc}.  In this scenario, a
hierarchy generated between the gaugino masses and the masses of the scalar sparticles in
which the latter are elevated to a high scale in order to evade unwanted flavor violation
effects, while the former are kept at the TeV scale or below to constitute dark matter.
The gravitino mass also tends to be quite large in split supersymmetry, hence it would
provide a solution to the model tensions that gravitino cosmology presents Dirac
leptogenesis.  It will soon be made clear that Dirac leptogenesis and split supersymmetry
are indeed compatible models.

\indent As a case in point, a particularly simple and phenomenologically interesting
split supersymmetry scenario in which it is difficult (though not impossible) to get
Dirac leptogenesis to work is the PeV-scale supersymmetry of~\cite{Wells:2003tf},
sometimes also referred to as loop-split supersymmetry, in which anomaly mediation is
invoked in the gaugino sector, but not in the scalar sector.  One assumes that the
messenger fields \(X\) responsible for transmitting the effects of supersymmetry-breaking
to the visible scale are charged under some symmetry, and consequently, while the scalar
masses are still given by
\begin{equation}
  m_s^2=c\frac{F^{\dagger}_X F_X}{M_P^2},\label{eq:AMSBScalarMass}
\end{equation}
where \(c\) is an \(\mathcal{O}(1)\) constant, the term which normally yields dominant
contribution to the gaugino masses
\begin{equation}
  \int d^2 \theta \frac{X}{M_{P}} W^{a}W^{a}\to \frac{F_X}{M_{P}} \lambda^{a}\lambda^{a}
\end{equation}
is not longer gauge-invariant.  The leading contribution arises at one loop, via the
anomaly-mediated expression
\begin{equation}
  M_{\lambda}=\frac{\beta_{g_{\lambda}}}{g_{_{\lambda}}}
  \left(\frac{\langle F_{X}^{\dagger}F_{X}\rangle}{M_{P}^{2}}\right)^{1/2}.
  \label{eq:AMSBMass}
\end{equation}
If we assume that supersymmetry is broken at an intermediate scale, around
\(\sim10^{5}-10^{7}\) GeV, then all scalars in the theory, (with the exception of one
light Higgs particle) receive masses around the PeV scale while the gauginos acquire
masses at the TeV scale~\cite{Randall:1998uk}.  This model is attractive in its
simplicity, and furthermore it is connected to the rich phenomenology associated with
anomaly-mediated models which includes possibilities for the detection of dark matter via
observations at the next generation of \(\gamma\)-ray
telescopes~\cite{Masiero:2004ft,Arvanitaki:2004df,Thomas:2005te} and characteristic
gluino decay signatures that could be observed at the
LHC~\cite{Toharia:2005gm,Gambino:2005eh}.

\indent The problem with getting Dirac leptogenesis to work in loop-split supersymmetry
is that not only does the theory require \(m_{3/2}\) to be around the PeV scale (which is
in itself not particularly worrisome), but also mandates a particular relationship
between \(m_{3/2}\) and the LSP mass through equation~(\ref{eq:AMSBMass}) and the AMSB
gravitino mass relation~\cite{Randall:1998uk,Giudice:1998xp}
\begin{equation}
  m_{3/2}^2=\frac{F_{X}^{\dagger}F_{X}}{M_P^2}.\label{eq:AMSBGravitinoMass}
\end{equation}
Figure~\ref{fig:OmegaAnn} indicates that Dirac leptogenesis prefers a splitting
\(m_{3/2}/m_{\mathit{LSP}}\) substantially larger than that dictated by
equation~\ref{eq:AMSBMass}, which means that \(\Omega_{\mathit{LSP}}^{\mathit{NT}}\) will
generally exceed the WMAP bound.  Loop-split supersymmetry can still be made to
work---for example in a situation where lepton number production is amplified by
resonance effects---but tuning the \(M_{\Phi_{i}}\) to this degree begs some sort of
motivation or additional theoretical machinery. It should be emphasized, however, that
the splitting between \(m_{LSP}\) and \(m_{3/2}\) can be much
larger~\cite{Arkani-Hamed:2004fb,Ibe:2004tg} in more general split supersymmetry
scenarios, making Dirac leptogenesis far easier to realize.

\subsection{Goldstone Bosons and Symmetry Breaking\label{sec:Goldstones}}

\indent

Finally, in addition to the battery of constraints outlined above, Dirac leptogenesis
must respect cosmological bounds associated with the production of light scalars.  As was
mentioned in chapter~\ref{ch:DiracLep}, some new symmetry must be posited in order to
construct the Dirac leptogenesis superpotential and forbid Majorana masses for the
right-handed neutrinos. In Dirac leptogenesis, neutrino masses are the result of the
scalar component of the \(\chi\) superfield acquiring a VEV which breaks this new
symmetry, producing a Goldstone boson or pseudo-Goldstone boson, depending on the way the
symmetry is broken (which we have not yet specified).  Let us first examine the case
where the breaking has an explicit component and the relevant scalar is a
pseudo-Goldstone boson. Constraints on such particles arise from both BBN and cosmic
microwave background (CMB) considerations~\cite{Hall:2004yg} as well as from the
detection of abnormalities in the neutrino flux associated with supernova
events~\cite{Goldberg:2005yw}, and they can become problematic (depending on the mass of
the Goldstone boson) when the symmetry-breaking VEV is less than around 1~GeV.

\indent In the case where symmetry breaking is completely spontaneous, a true Goldstone
boson will result, which these constraints seem to rule out.  Of course this assumes that
the Goldstone boson is a physical state: in the case where the additional symmetry which
forbids neutrino Majorana masses is a gauge symmetry, the Goldstone boson is ``eaten'' by
the gauge field and the relevant constraints become those associated with extensions of
the Standard Model gauge structure.  If it is an Abelian gauge symmetry, for example, one
must take care that bounds associated with \(U(1)\) mixing are not violated and that the
gauge theory is free of anomalies (which would not be the case given the field content in
table~\ref{tab:U1Charges}). We will discuss these requirements and investigate their
implications further in section~\ref{sec:sneutrinoCDM}.

\section{Neutrino Physics\label{sec:neutrinophys}}


\subsection{Experimental Constraints}

\indent

In addition to respecting constraints arising from cosmological considerations, in order
to be phenomenologically viable, a given Dirac leptogenesis model must yield a neutrino
spectrum that accords with current experimental constraints.  The most stringent such
constraints come from solar and atmospheric neutrino oscillation
experiments~\cite{Fogli:2001vr,Bahcall:2001zu}, and place limits both on the mass
splittings
\begin{equation}
  \Delta m_{ab}^{2}\equiv m_{\nu_{a}}^{2}-m_{\nu_{b}}^{2},
\end{equation}
where the indices \(a\) and \(b\) label the different neutrino mass eigenstates, and on
the mixing angles \(\theta_{ab}\) between these eigenstates.  The primary connection
between the latter and observable physics occurs through the leptonic mixing matrix
\begin{equation}
  U_{\mathit{MNS}}=U^{(\nu)}_{}U^{(e)}_{},
  \label{eq:UMNSDef}
\end{equation}
where \(U^{(\nu)}_{}\) is the neutrino mixing matrix and \(U^{(e)}_{}\) is the charged
lepton mixing matrix. In the basis where the charged lepton mass matrix is diagonal,
\(U_{\mathit{MNS}}\) is the neutrino mixing matrix and may be expressed in terms of the
neutrino mixing angles \(\theta_{ab}\) as
\begin{equation}
  U_{\mathit{MNS}}=
  \left(\begin{array}{ccc} c_{12}c_{13} & s_{12}c_{13} & s_{13}e^{i\delta_{\mathit{CP}}} \\ -s_{12}c_{23}-c_{12}s_{23}s_{13}e^{i\delta_{\mathit{CP}}} & c_{12}c_{23}-s_{12}s_{23}s_{13}e^{-i\delta_{\mathit{CP}}} & s_{23}c_{13} \\ s_{12}s_{23}-c_{12}c_{23}s_{13}e^{-i\delta_{\mathit{CP}}} & -c_{12}s_{23}-s_{12}c_{23}s_{13}e^{-i\delta_{\mathit{CP}}} & c_{23}c_{13} \end{array}\right),
  \label{eq:UMNS}
\end{equation}
where \(c_{ab}=\cos\theta_{ab}\), \(s_{ab}=\sin\theta_{ab}\), and
\(\delta_{\mathit{CP}}\) is a \(CP\)-violating phase.  The present limits\footnote{We do
not take into account the LSND result which would require an extra neutrino mass
eigenstate. In the case that forthcoming data from experiments such as MiniBooNE
corroborate the LSND signal, it will be necessary to extend the neutrino content of our
model.} on the \(\Delta m_{ab}^{2}\) and \(\theta_{ab}\) are~\cite{Hagedorn:2005kz}
\begin{eqnarray}
  \begin{array}{ccc}
    \sin^{2}\theta_{12}=0.30^{+0.04}_{-0.05}, & \sin^{2}\theta_{23}=0.50^{+0.14}_{-0.12}, &
    \sin^{2}\theta_{13}\leq0.031,
  \end{array} \nonumber \\
  \begin{array}{cc}
    \Delta m_{21}^{2}=\left(7.9^{+0.6}_{-0.6}\right)\times 10^{-5} \mathrm{eV}^{2}, &
    |\Delta m_{31}^{2}|=\left(2.2^{+0.7}_{-0.5}\right)\times 10^{-3} \mathrm{eV}^{2}.
  \end{array}
  \label{eq:NeutLimits}
\end{eqnarray}
The smaller of the two mass splittings, \(\Delta m_{21}^{2}\), is to be identified with
the \(\Delta m_{\odot}^{2}\) obtained from solar neutrino data (the MSW-LMA solution);
the larger, \(\Delta m_{31}^{2}\), with the \(\Delta m_{A}^{2}\) from atmospheric
neutrino data.  When we take these constraints and substitute them into the
\(U_{\mathit{MNS}}\) matrix, we arrive a set of bounds
\begin{equation}
  |U_{\mathit{MNS}}|=\left(\begin{array}{ccc} .79-.86 & .49-.58 & 0-.18 \\ .30-.58 & .40-.68 & .61-.80 \\ .19-.46 & .50-.77 & .59-.79 \end{array}\right),\label{eq:MNSdata}
\end{equation}
However, the constraints in~(\ref{eq:NeutLimits}) say nothing about the sign of the
largest mass squared difference $\Delta m_{31}^{2}$.  As as result, we are left with two
possibilities: the physical neutrino $\nu_3$ can be either the heaviest of the three mass
eigenstates, i.e. $m_1<m_2\ll m_3$ (a situation dubbed the ``normal hierarchy'') or the
lightest, i.e. $m_3\ll m_1<m_2$ (the ``inverted hierarchy'').

\indent  The reason $U_{\mathit{MNS}}$ is of particular importance is that in the basis
where the charged lepton mass matrix is diagonal, it becomes the unitary matrix
responsible for diagonalizing the squared neutrino mass matrix:
\begin{equation}
  \left(m_\nu^2\right)_{\mathit{diag}}=\ U_{\mathit{MNS}}^\dagger\ m_\nu m_\nu^\dagger\
  U_{\mathit{MNS}}.
  \end{equation}
We can therefore estimate \cite{Hagedorn:2005kz} the generic form of the neutrino mass
matrix squared, since it has to be diagonalized by  $U_{\mathit{MNS}}$.  In the normal
hierarchy scenario, we find that
\begin{equation}
  (m^{2}_{\nu})_{\mathit{norm}}\sim \Delta m_{31}^{2}
  \left(\begin{array}{ccc}\xi&\xi&\xi\\ \xi&1&1\\ \xi&1&1\end{array}\right),
    \label{eq:NeutMassNormalGBForm}
\end{equation}
where the $\xi$ are small compared to the $\mathcal{O}(1)$ entries and not necessarily
the equal to one another.  Similarly, in the inverted hierarchy scenario, we find
\begin{equation}
  (m^{2}_{\nu})_{\mathit{inv}}\sim \Delta m_{31}^{2}
  \left(\begin{array}{ccc}1&\xi&\xi\\ \xi&1&1\\ \xi&1&1\end{array}\right).
    \label{eq:NeutMassInvertedGBForm}
\end{equation}
In either case, the ratio of the \(\mathcal{O}(1)\) entries to the small $\xi$ is
constrained to be at least of order \(\rho_{32}\equiv \Delta m_{31}^{2}/\Delta
m_{21}^{2}\), which, according to~(\ref{eq:NeutLimits}), must respect the bounds
\begin{equation}
  20.0 < \rho_{23} < 39.7\ .
\end{equation}
In order for Dirac leptogenesis to be phenomenologically viable leptogenesis model we
need simultaneously to be able to satisfy the above constraints and reproduce the form of
the neutrino mass matrix given in~(\ref{eq:NeutMassNormalGBForm})
or~(\ref{eq:NeutMassInvertedGBForm}). Ideally, we should like to find a simple way of
arriving at one of these matrix structures that is motivated by theoretical
considerations as well as data pressures.  Of course this solution must also be
compatible with successful baryon number generation, the astrophysical constraints
discussed in section~\ref{sec:AstroConstraints}, etc.

\subsection{The Flavor Structure of the Trilinear Couplings}

\indent

As we saw in chapter~\ref{ch:DiracLep}, the neutrino mass-squared matrix in Dirac
leptogenesis is given by
\begin{equation}
  |m_{\nu}|^{2}_{\alpha\beta}= \Big(v \langle\chi\rangle \sin\beta\Big)^2\
  \sum_{i,j=1}^{2\ (or\ 3)} \sum_{\gamma=1}^3\lambda^{\ast}_{i\gamma}\lambda_{j\gamma}
  h^{\ast}_{i\alpha } h_{j\beta} \frac{1}{M^\ast_{\Phi_i}M_{\Phi_j}}.
  \label{eq:NeutMassSpelledOut}
\end{equation}
As long as \(\lambda\) and \(h\) are completely generic and there are at least three sets
of \(\Phi\) and \(\overline{\Phi}\), it is apparent that a matrix of this form can yield
an arbitrary neutrino mass spectrum.  On the one hand this is good, for it means that the
theory is perfectly viable, in the sense that there exists some set of parameters that
will satisfy the battery of constraints given in equation~(\ref{eq:NeutLimits}); on the
other hand, this arbitrariness comes at the price of introducing many additional free
parameters, whose relative values must be determined by some additional underlying
physics.  As a first step toward understanding what that additional physics ought to
involve, let us examine the spectrum of a simplified model containing only a single pair
of heavy fields \(\Phi_1\) and \(\overline{\Phi}_1\) of mass \(M_{\Phi_1}\), or
alternatively, a theory in which \(M_{\Phi_{1}}\ll M_{\Phi_{i}}\) for all \(i>1\).  In
such cases the mass-squared matrix is proportional to an outer product of the
family-space vectors \(\lambda_{1\alpha}\) and thus its eigenvalues are
\begin{eqnarray}
  m_{\nu_{1}}=0 & m_{\nu_{2}}=0 &
  m_{\nu_{3}}=\sum_{\alpha}^{3}\lambda_{1\alpha}h_{1\alpha}.
\end{eqnarray}
Here, two of the physical neutrinos are massless.  For each additional set of \(\Phi\)
and \(\overline{\Phi}\) with a mass similar to \(M_{\Phi_{1}}\), an additional neutrino
acquires a nonzero mass.  Thus in ``short-suited'' models where \(M_{\Phi_1} \ll
M_{\Phi_2}\) and \(M_{\Phi_2}\ll M_{\Phi_i}\) for all \(i> 2\) (i.e. where there are
effectively only two sets of \(\phi\) and \(\overline{\phi}\) involved in determining the
neutrino spectrum), one neutrino mass eigenstate is massless and a hierarchy will exist
between the other two, determined by \(\delta=M_{\Phi_1}/M_{\Phi_2}\). One can then
arrange for the other two neutrino masses to take the experimentally observed values
\(m^{2}_{\nu_{2}}\approx\Delta m^{2}_{21}\) and \(m^{2}_{\nu_{3}}\approx\Delta
m^{2}_{31}\) by tinkering with \(\delta\) and the structures of \(\lambda\) and \(h\).
Conveniently, leptogenesis in such models is well-approximated by the toy model
considered in section~\ref{ch:DiracLep}, in which there were only two sets of \(\Phi\)
and \(\overline{\Phi}\).

\indent Let us now examine the effect of \(\lambda\) and \(h\) on the neutrino spectrum
in models with this sort of hierarchy among the heavy particle masses.  It is apparent
from equation~(\ref{eq:NeutMassSpelledOut}) that the matrix structure of $m_{\nu}^2$ is
primarily determined by $h$ rather than \(\lambda\).  In the limit where \(\delta\ll 1\),
this matrix becomes
\begin{eqnarray}
|m_\nu|^2=\Big(v \langle\chi\rangle \sin\beta\Big)^2 \sum_\gamma |\lambda_{1\gamma }|^2
\frac{1}{M^2_{\Phi_1}}\
\left(\begin{array}{ccc}|h_{11}|^2&h^{\ast}_{11}h_{12}&h_{11}^{\ast}h_{13}\\
h^{\ast}_{12}h_{11}&|h_{12}|^2&h_{12}^{\ast}h_{13}\\
h_{13}^{\ast}h_{11}&h_{13}^{\ast}h_{12}&|h_{13}|^2\end{array}\right) + {\cal
O}(\delta),\label{eq:hierarA}.
\end{eqnarray}
If for some reason the off-diagonal elements of \(h\) are much smaller than the diagonal
elements (i.e.\ $h_{12},\,h_{13}\gg h_{11}$), this matrix will take
form~(\ref{eq:NeutMassNormalGBForm}) that reproduces the normal hierarchy solution.  If
\(h_{12}\) and \(h_{13}\) are roughly of the same order, so that \(h_{12}\approx\
h_{13}\equiv \tilde{h}\), equation~(\ref{eq:hierarA}) becomes
\begin{eqnarray}
|m_\nu|^2\approx\left(v \langle\chi\rangle \sin\beta\right)^2 \sum_\gamma
|\lambda_{1\gamma}|^2 \frac{|\tilde{h}|^2}{
M^2_{\Phi_1}}\ \left(\begin{array}{ccc}\xi^2&\xi&\xi\\ \xi&1&1\\
\xi&1&1 \end{array}\right)+ {\cal O}(\delta),
\end{eqnarray}
where \(\xi=h_{11}/\tilde{h}\).  Here we have implicitly assumed that contributions of
$\mathcal{O}(h_{11}/\tilde{h})$ dominate over the \(\mathcal{O}(\delta)\) contributions.
If this is not the case, and in particular if \(h_{11}=0\), the matrix structure becomes
\begin{equation}
  (m^{2}_{\nu})\propto
  \left(\begin{array}{ccc}\delta^2&\delta&\delta\\ \delta&1&1\\ \delta&1&1\end{array}\right).
  \label{eq:DeltasInFirstRow}
\end{equation}
Again the generic structure yielding the normal hierarchy scenario is obtained, but in
this case the value of $\delta$ fixes the ratio $\rho_{23}$.  In any case, what is
significant here is that imposing the two simple hierarchical requirements,
$M_{\Phi_1}\ll M_{\Phi_{2,3}}$ and $h_{11}\ll h_{12}\sim h_{13}$ generically gives rise
to the correct neutrino phenomenology.

\indent These considerations suggest that we ought to look for some setup that will
ensure that the diagonal element $h_{11}$ is small compared to the off diagonal terms
$h_{12}$ and $h_{13}$.  One particularly simple way to do this is to posit an
antisymmetry condition on the matrix $h$, in which case the neutrino mass matrix will
take the form given in~(\ref{eq:DeltasInFirstRow}), with small entries of
\(\mathcal{O}(\delta)\). This option is of particular interest because there is also a
theoretical motivation for it: antisymmetric Yukawa matrices emerge quite naturally in
certain grand unified theories and in models with non-Abelian flavor
symmetries~\cite{Davidson:1980sx,King:2004tx,Froggatt:1978nt,Barbieri:1996ww}.  We shall
put off discussion of such matters until section~\ref{sec:FlavonExtension}, where we
provide a theoretical motivation for the hierarchies among the \(M_{\Phi_i}\) and
elements of \(\lambda\) and \(h\).

\subsection{Constrained Hierarchical Dirac Leptogenesis}


\indent

Motivated by the preceding remarks, we now define a particular model that ought to be
able to yield a phenomenologically acceptable neutrino spectrum.  We define {\bf
constrained hierarchical Dirac leptogenesis (CHDL)}~\cite{Thomas:2005rs} as the setup in
which:
\begin{enumerate}
  \item The mass matrix \((M_{\Phi})\) is real and diagonal.
  \item The coupling matrices $\lambda$ and $h$ are antisymmetric
  \item The large mixing angles in the neutrino sector are result from the smallness of
  \(\delta=M_{\Phi_{1}}/M_{\Phi_{2}}\).
\end{enumerate}
In this model, the antisymmetry of \(\lambda\) and \(h\) allows us to parameterize them
in the manner
\begin{eqnarray}
  \lambda=f
  \left(\begin{array}{ccc} 0 & 1 & a_{2}\\-1 & 0 & a_{3} \\ -a_{2} & -a_{3} & 0\end{array}\right) &
  \hspace{.5cm} &
  h=f
  \left(\begin{array}{ccc} 0 & b_{1} & b_{2}\\ -b_{1} & 0 & b_{3} \\ -b_{2} & -b_{3} & 0\end{array}\right),
  \label{eq:YukawaParametrization}
\end{eqnarray}
which is convenient when the \(a_i\) and \(b_i\) are all roughly \(\mathcal{O}(1)\).
Since the assumption of a hierarchy among the \(M_{\Phi_{i}}\) leads to the neutrino
mass-squared matrix of the form~(\ref{eq:NeutMassNormalGBForm}), we expect that \(a_{3}\)
and \(b_{3}\), the effects of which show up only at the \(\mathcal{O}(\delta)\) level,
will be less tightly constrained than the rest of the \(a_{i}\) and \(b_{i}\), which
contribute to the leading term.  This is in fact the case: if \(a_{1}\), \(a_{2}\),
\(b_{1}\), or \(b_{2}\) deviates significantly from one, the neutrino spectrum cannot
satisfy the constraints in~(\ref{eq:NeutLimits}).  It is therefore appropriate, since the
value of \(f\) is unimportant as far as this set of constrains are concerned (any
rescaling of \(f\) can be compensated for by a similar rescaling of
\(\langle\chi\rangle\)), to analyze constrained hierarchical models as functions of
$a_{3}$ and $b_{3}$ alone.

\indent

\begin{figure}[ht!]
  \begin{center}
  \includegraphics{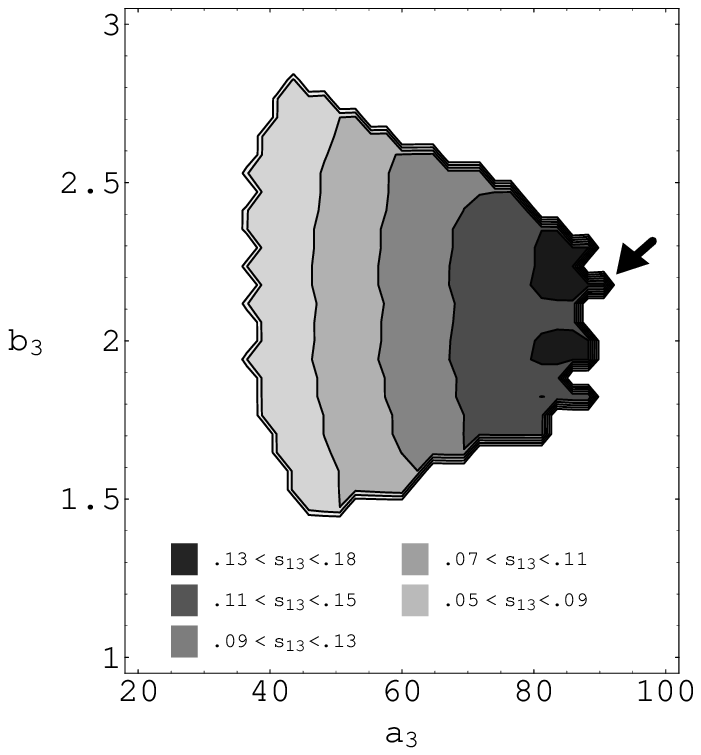}
  \includegraphics{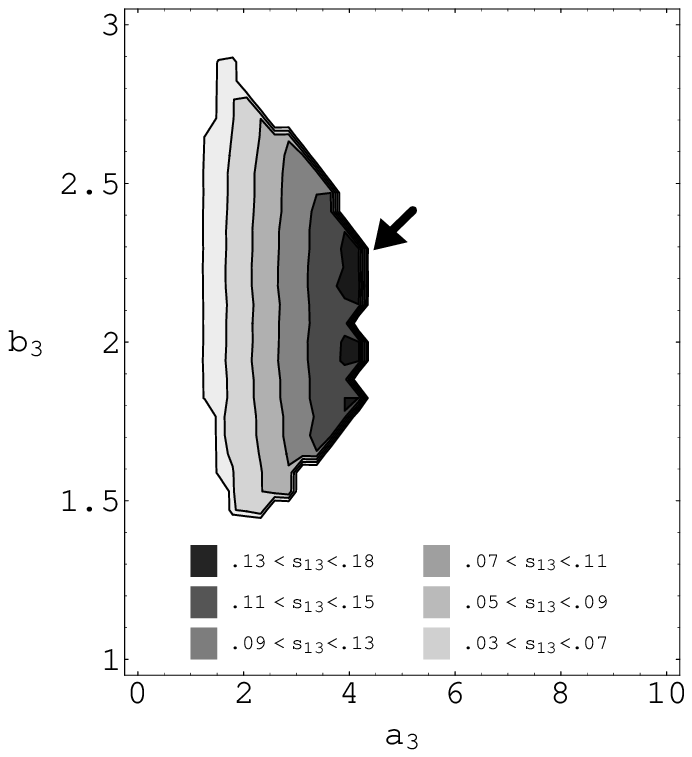}
  \end{center}
  \caption{Here, the regions of $a_{3}-b_{3}$ space (see
    equation~(\ref{eq:YukawaParametrization}) for a description of the
    Yukawa-matrix parametrization used) for which all
    constraints on neutrino masses and mixings~(\ref{eq:NeutLimits}) are simultaneously satisfied for some
    combination of the remaining $a_{i}$ and $b_{i}$ are shown for two different values
    of $\delta$~\cite{Thomas:2005rs}.  Additionally, contours
    depicting the ranges for $s_{13}=\sin\theta_{13}$ (which depends primarily on
    $a_{3}$, but varies slightly with the remaining
    $a_{i}$ and $b_{i}$) are shown.  In the right panel $\delta=10^{-1}$;
    in the left panel, $\delta=m_{e}/m_{\mu}= 4.83\times 10^{-3}$
    The plots reveal that while
    $b_{3}$ is constrained, along with most of the other $a_{i}$ and $b_{i}$, to lie
    reasonably near 1, $a_{3}^{\mathit{max}}(\delta)$ can be quite large and increases with
    decreasing $\delta=10^{-1}$, making it easier to obtain a realistic baryon-to-photon ratio $\eta$.
    For $\delta=10^{-1}$, $a_{3}^{\mathit{max}}\simeq4.5$; for $\delta= 4.83\times 10^{-3}$,
    $a_{3}^{\mathit{max}}\simeq 95$.
    In each panel, the configuration that yields the
    greatest decay asymmetry $\epsilon$ is marked with an arrow.
    \label{fig:NeutrinoMassPlots}}
\end{figure}

In figure~\ref{fig:NeutrinoMassPlots}, we show the region of viability in \(a_{3}-b_{3}\)
space for two different values of \(\delta\): in the left-hand panel, we set
\(\delta=m_{e}/m_{\mu}\) (and \(M_{\Phi_{2}}/M_{\Phi_{3}}=m_{\mu}/m_{\tau}\)) as required
in the minimal version of CHDL discussed above; in the right-hand panel, we set
\(\delta=10^{-1}\). We consider a given combination of \(a_{3}\) and \(b_{3}\) to be
phenomenologically viable if there is any combination of the remaining \(a_{i}\) and
\(b_{i}\) for which the combination simultaneously obeys all the neutrino oscillation
constraints in~(\ref{eq:NeutLimits}).  The entirety of the shaded region shown in each
panel, including all differently-shaded bands, is permitted by these constraints: the
bands represent of contours $\sin\theta_{13}$ the value of which will be measured or
constrained in future neutrino experiments.  This plot demonstrates two important
features of the Yukawa matrices in CHDL: first, it is indeed possible to satisfy the
neutrino oscillation constraints for \(\delta=m_{e}/m_{\mu}\); second, while \(b_{3}\),
like most of the other \(a_{i}\) and \(b_{i}\), is constrained to lie fairly close to 1,
\(a_{3}\) is permitted to be quite large when \(\delta\) is small.  In the same figure,
we also show contours for the value of $\sin\theta_{13}$, the value of which will be
measured or constrained in future neutrino experiments. It is seen that the value of
$\sin\theta_{13}$ increases with increased $a_3$ until reaching its maximum experimental
bound.

\indent Let us now take a moment to address how these results affect leptogenesis. Since
we are assuming that \(M_{\Phi_{3}}\gg M_{\Phi_{1}},M_{\Phi_{2}}\), the
formula~(\ref{eq:epsilond}) for the decay asymmetry \(\epsilon\) tells us that
\(\mathrm{Im}(\lambda^{\ast}_{1\alpha}\lambda_{2\alpha}h^{\ast}_{1\beta}h_{2\beta}M_{\Phi_{1}}
M^{\ast}_{\Phi_{2}})\) will vanish (when it involves diagonal elements of \(\lambda\) or
\(h\)) unless \(\beta=\alpha=3\).  This means that
\begin{equation}
  \epsilon\propto\mathrm{Im}
  \left(a^{\ast}_{2}a_{3}b^{\ast}_{2}b_{3}\right){\delta\over 1-\delta}
\end{equation}
in CHDL; the two panels in figure~\ref{fig:NeutrinoMassPlots} then show that for a given
\(\delta\), the largest amount of left-handed lepton number \(L_{L}\) (and therefore the
largest baryon asymmetry) will be obtained when $a_{3}=a_{3}^{\mathit{max}}(\delta)$,
where \(a_{3}^{\mathit{max}}(\delta)\) is the maximum possible value of \(a_{3}\) for a
given $\delta$ consistent with neutrino masses and mixings. It is interesting to note
that since the maximum experimental value of $\sin\theta_{13}$ sets the value of
$a_{3}^{\mathit{max}}$, then the maximum baryon asymmetry will be obtained when
$\sin\theta_{13}$ acquires its maximal experimental value, which we take to be
$\sin_{max}^2\theta_{13}=0.031$.  For \(\delta=10^{-1}\), the result is
\(a_{3}^{\mathit{max}}(1/10)\approx4.5\), as indicated in the right panel of
figure~~\ref{fig:NeutrinoMassPlots}.  It is also of note that
\(a_{3}^{\mathit{max}}(\delta)\) rises sharply as \(\delta\) is decreased, as indicated
by the result for $\delta= 4.83\times 10^{-3}$ (a value which, as we shall see in
section~\ref{sec:FlavonExtension}, is predicted by certain theoretically-motivated
extensions of CHDL) shown in the left panel, in which case \(a_{3}^{\mathit{max}}\simeq
95\).

\indent One of the assumptions we have been implicitly making here is that we have
maximal CP violation in the decays of the fields $\Phi$ and $\overline{\Phi}$ (i.e. the
overall phase in the product $ a^{\ast}_{2}a_{3}b^{\ast}_{2}b_{3}$ must be $\pi/2$). In
this case one can obtain a value for the effective CP violation in the lepton sector.
This can be defined in a phase invariant way in terms of the quantity $J={\rm
Im}(U_{12}U^*_{22}U_{23}U^*_{13})$ \cite{Jarlskog:1985ht,Dunietz:1985uy}, where $U_{ij}
\equiv (U_{\mathit{MNS}})_{ij}$. Taking for example the two points that will generate the
most baryon number in the left and right handed panels of
figure~\ref{fig:NeutrinoMassPlots} (marked with an arrow) we find $J\simeq0.034$ for the
point in the left panel and $J\simeq0.030$ for the point in the right panel (since the
maximal value for \(J\) is \(|U_{12}U^*_{22}U_{23}U^*_{13}|\), which for these points
yields a value of \(J\simeq.038\), the \(CP\)-violation here is close to maximal). It
could be interesting to do a more detailed study of the issue of linking more generally
the effective CP violation in the lepton sector to the CP violation in the interactions
of the heavy fields $\Phi$ and $\overline{\Phi}$. Nevertheless, we will not pursue this
issue further in the present work.

\indent We have shown here that it is possible to construct Dirac leptogenesis scenarios
that are capable of satisfying current constraints on the light neutrino spectrum.  In
CHDL, the set of parameters relevant to neutrino physics and to leptogenesis is quite
small, given the antisymmetry condition on the two coupling matrices $\lambda$ and $h$:
the leptogenesis scale \(M_{\Phi_{1}}\), the intermediate scale \(\langle\chi\rangle\),
the trilinear matrix entries \(a_{3}\) and \(b_{3}\), the mass ratio \(\delta\), and the
overall coupling strength \(f\) are essentially the only free parameters in the model. In
fact, even this represents an overcounting: equations~(\ref{eq:NeutMassSpelledOut})
and~(\ref{eq:NeutLimits}) imply an additional constraint on these parameters.  For
example, take the case where \(\delta=10^{-1}\) and \(a_3\) and \(b_3\) have been set to
the values most advantageous for baryogenesis, \(a_{3}=4.5\) and \(b_{3}=2.2\) (see
figure~\ref{fig:NeutrinoMassPlots}).  In this case, the constraint becomes
\begin{equation}
 \frac{f^{2}\langle\chi\rangle}{M_{\Phi_{1}}}\sin\beta= 1.009\times10^{-13},
 \label{eq:FchiMconstraint}
\end{equation}
and the model contains only five free parameters (and that's including \(b_{3}\), which
is rather stringently constrained).  We will generally use this constraint to eliminate
\(\langle \chi\rangle \) or \(f\), depending on which is most convenient for the purpose
at hand.

\indent While we again emphasize that it is certainly not the only scenario for obtaining
a realistic neutrino spectrum in Dirac leptogenesis, CHDL is a particularly simple and
predictive model---and it will be the one on which we will focus our attentions from this
point forward.  It still remains to be seen whether CHDL (or indeed any Dirac
leptogenesis model) can also simultaneously satisfy the constraints from gravitino
physics and yield a realistic baryon number for the universe, however, though as we shall
see in chapter~\ref{ch:Boltzmann}, thermal Dirac leptogenesis will indeed turn out to be
workable in a variety of models.

\section{Soft Masses and Flavor Violation\label{sec:FlavorConstraints}}


\subsection{Flavor Violation in Supersymmetric Models}

\indent

Another requirement which Dirac leptogenesis must satisfy in order to be considered
viable is that the theory must be compatible with present constraints on flavor violation
in the lepton sector.  The most stringent bounds come from measurements of the branching
ratios for flavor-violating decays and conversions of heavy leptons, such as
\(\mu\rightarrow e\gamma\),  \(\tau\rightarrow \mu\gamma\), $\mu\to eee$ and  $\mu A\to
eA$.  The current experimental limits on the 2-body decay processes
are~\cite{Eidelman:2004wy}
\begin{eqnarray}
\mathit{BR}(\mu\rightarrow e\gamma)&<&1.2\times 10^{-11},\label{eq:CurrentExpConstraints}\\
\mathit{BR}(\tau\rightarrow \mu\gamma)&<&1.1\times 10^{-6}.
\end{eqnarray}
The MEG experiment~\cite{MEGHome:2006mh} is expected to improve on bound on
\(\mu\rightarrow e\gamma\) by several orders of magnitude in the near future, bringing it
to \(\mathcal{O}(10^{-13}-10^{-14})\) or lower. Other related projects, such as
PRIME~\cite{Kuno:2005mm} (sensitive to \(\mu A\rightarrow e A\) conversion), are expected
to go online over the next few years, and projects have also been
proposed~\cite{Calibbi:2006nq} that would lower the bound on \(BR(\tau\rightarrow
\mu\gamma)\) to \(\mathcal{O}(10^{-9})\).

\indent Supersymmetric theories with generic soft parameters tend to result in
unacceptable levels of flavor violation due to flavor misalignment between quark and
lepton mass eigenstates on the one hand, and squark and slepton eigenstates on the other.
There are a variety of ways to address this problem, the most common one being the
assumption of soft mass universality.  Here, one assumes that whatever mechanism gives
rise to supersymmetry breaking at some high scale $M$ (the Planck scale, the GUT scale,
etc.) results in a set of squark and slepton soft mass-squared matrices that are
flavor-blind and diagonal and $A$-terms proportional to the Standard Model Yukawa
couplings, i.e.\ that
\begin{eqnarray}
  \mathbf{m}_{Q}^2=m_{Q}^2\mathbf{1},~~~~
  \mathbf{m}_{L}^2=m_{L}^2\mathbf{1},~~~~
  \mathbf{m}_{\overline{u}}^2=m_{\overline{u}}^2\mathbf{1},\hspace{.5cm}\nonumber\\
  \mathbf{m}_{\overline{d}}^2=m_{\overline{d}}^2\mathbf{1},~~~~
  \mathbf{m}_{\overline{e}}^2=m_{\overline{e}}^2\mathbf{1},~~~~
  \mathbf{m}_{\overline{\nu}}^2=m_{\overline{\nu}}^2\mathbf{1},\hspace{.5cm}\nonumber\\
  \mathbf{a}_{u}=A_{u}\mathbf{y}_{u},~~~~
  \mathbf{a}_{u}=A_{u}\mathbf{y}_{d},~~~~
  \mathbf{a}_{u}=A_{u}\mathbf{y}_{e},~~~~
  \mathbf{a}_{u}=A_{u}\mathbf{y}_{\nu},~~~~
\end{eqnarray}
then all squarks and sleptons become degenerate in mass and can be freely rotated into
one another up to $A$-term-induced mixings, which will only be large for sfermions of the
third generation.  Flavor-violating effects would then be expected to be very small.

\indent In Dirac leptogenesis, even if soft masses are universal at scale $M$, as they
are run down from $M$ to the leptogenesis scale $M_{\Phi_1}$ flavor-off-diagonal terms
will be generated by quantum corrections, due to the nontrivial matrix structure of the
trilinear coupling matrices $\lambda$ and $h$. It can be shown that, under assumption of
universal supersymmetry breaking, this is not a damning problem for the MSSM (for a brief
review, see~\cite{Martin:1997ns}). However, since the lepton sector of the Dirac
leptogenesis superpotential~(\ref{eq:DiracLepSuperpotential}) is modified from that of
the MSSM and includes new coupling matrices with nontrivial flavor structure, we must
ensure that any off-diagonal contributions to the slepton mass matrices \(m^{2}_{LL}\),
\(m^{2}_{RR}\) and \(m^{2}_{LR}\) generated by these modifications do not result in an
unacceptable level of flavor violation.  The renormalization group evolution (RGE)
equations for parameters appearing in a general superpotential soft supersymmetry
breaking Lagrangian are well known~\cite{Martin:1993zk}, and the RGE equation for a soft
mass is given by
\begin{eqnarray}
  \beta_{(m^2)^i_j}&=&\frac{1}{16\pi^2}\left[
  \frac{1}{2}y^{\ast}_{ipq}y^{pqn}(m^{2})^j_n+
  \frac{1}{2}y^{jpq}y^{\ast}_{pqn}(m^{2})^n_i+
  2y_{ipq}^{\ast}y^{jpr}(m^2)^q_r\right.\nonumber\\ &&\left.
  +a^{\ast}_{ipq}a^{jpq}-8g_{a}^2C_{a}(i)|M_{a}|^2\delta_{i}^{j}+
  2g_{a}^2(T^{a})_{i}^{j}\tr[T^a m^2]\right],
\end{eqnarray}
where the \(a_{ijk}\) are soft $A$-terms, \(M_{a}\) are gaugino masses,  \(y_{ijk}\) are
trilinear couplings appearing in the superpotential, and a sum over gauge groups is
implied.  The \(C_{a}(i)\) are quadratic Casimir group invariants, defined in terms of
the generators \(T^a\) by the relation
\begin{equation}
  C_{a}(i)\delta^j_i=(T^a T^a)^j_i.
\end{equation}
Assuming the Universality condition at the high scale and that all soft $A$-terms are
equal to the relevant Yukawa coupling multiplied by the universal soft supersymmetric
mass $m_s$, one can simply estimate the flavor violating corrections to the mass matrix
by integrating the RGE equations iteratively~\cite{Hisano:1995cp,Petcov:2005jh}, and
using this method, one obtains off-diagonal contributions to the slepton masses $\delta
m_{LL}^{2}$ and $\delta m_{RR}^{2}$:
\begin{eqnarray}
  \delta m_{LL}^{2}&\approx &
  -\frac{1}{2\pi^2}\ln\left(\frac{M}{M_{\Phi_{1}}}\right)h_{i\alpha}^{\ast}h_{i\beta}m_{s}^{2}\label{eq:MLL}\\
  \delta m_{RR}^{2}&\approx &
  -\frac{1}{2\pi^2}\ln\left(\frac{M}{M_{\Phi_{1}}}\right)\lambda_{i\alpha}^{\ast}\lambda_{i\beta}m_{s}^{2}\label{eq:MRR}.
\end{eqnarray}
Another off-diagonal scalar mass term arises from the effective $A$-term
in~(\ref{eq:EffectiveAtermL}) once electroweak symmetry breaking occurs in models where
the $F$-term of \(\chi\) acquires a VEV \(\langle F_{\chi}\rangle\neq 0\) (a corollary in
many mechanisms where its scalar component obtains its VEV).  After electroweak symmetry
breaking, this term results in a contribution
\begin{equation}
  \delta m_{LR}^{2}=h^{\dagger}_{i\alpha}\lambda_{i\beta}\frac{\langle
  F_{\chi}\rangle}{M_{\Phi_{1}}}v\sin\beta\label{eq:MLR}
\end{equation}
to the sneutrino mass matrix which mixes left-handed and right-handed
sneutrinos\footnote{The CP-violating phases in \(\lambda\) and \(h\) required for
leptogenesis can induce new phases in the slepton mass matrices during the RGE running or
through the mixing term of Eq.~(\ref{eq:MLR}). We have checked that the effect of these
phases is too small to reach the experimental bounds on lepton EDM's, within the region
of parameter space considered here in which leptogenesis is successful. As for the rest
of phases of the MSSM, they are assumed to be small enough to avoid violating these same
experimental bounds.}.

\indent In order to examine the effect of these mixings, the full mass matrices for both
the charged sleptons and sneutrinos must be taken into account. For simplicity, we will
continue to assume that the leading soft breaking sector is flavor diagonal and universal
with a common scalar mass $m_s$. The resulting additional contributions to the slepton
mass squared matrices, given by equations~(\ref{eq:MLL}),~(\ref{eq:MRR}),
and~(\ref{eq:MLR}), can thus be expressed in terms of the \(3\times3\) submatrices
\(\delta m_{LL}^{2}\), \(\delta m_{RR}^{2}\), and \(\delta m_{LR}^{2}\) as
\begin{equation}
  \delta m_{\tilde{\ell}^{\pm}}^{2}=
  {\small
  \left(\begin{array}{c:c}
  \delta m^2_{LL} & ~~0~~ \\[.25cm] \hdashline \\[-.25cm] ~~0~~ & ~~0~~
  \end{array}\right)}~~~~~~
  \delta m_{\tilde{\nu}}^{2}={\small
  \left(\begin{array}{c:c}
  \delta m^2_{LL} & \delta m^2_{LR} \\[.25cm] \hdashline \\[-.25cm] (\delta m^{2}_{LR})^\dagger & \delta m^2_{RR}
  \end{array}\right)}. \label{eq:MassMatricesDiracLep}
\end{equation}
The only contribution to the charged slepton mass squared matrix comes from \(\delta
m_{LL}^{2}\) (no off-diagonal terms will be generated radiatively in the RR part of
charged slepton matrix $m_{\tilde{\ell}^{\pm}}^{2}$ since the Yukawa coupling matrix
$Y_e$ can always be chosen diagonal at the high scale), while the sneutrino mass squared
matrix receives not only additional flavor mixings among left-handed and among
right-handed sneutrinos, but also an effective $A$-term from \(\delta m_{LR}^{2}\) which
intermixes left-handed and right-handed sneutrinos.

\indent In CHDL, where we have a specific flavor structure for the matrices \(\lambda\),
\(h\) and \(M_{\Phi}\), one has a specific prediction for flavor mixing among sleptons
once the electroweak and hidden symmetries are broken.  Using the neutrino mass
constraint~(\ref{eq:FchiMconstraint}), we can express the overall dependence of the
slepton mass terms in~(\ref{eq:MLL}), (\ref{eq:MRR}), and~(\ref{eq:MLR}) on the relevant
mass scales in the theory:
\begin{equation}
  \begin{array}{lcccc}
  \delta m_{LL}^{2}&\propto& f^2&\propto&
  \frac{M_{\Phi_{1}}}{\langle\chi\rangle}\\
  \delta m_{RR}^{2}&\propto& f^2& \propto&
  \frac{M_{\Phi_{1}}}{\langle\chi\rangle}\\
  \delta m_{LR}^{2}&\propto& \frac{f^2}{M_{\Phi_{1}}}&\propto&
  \frac {1}{\langle\chi\rangle}.
  \end{array}\label{eq:DeltaMassProps}
\end{equation}
It should be noted here that the proportionality constants for the bottom two equations
are not dimensionless: the ones associated with $\delta m_{LL}^{2}$ and $\delta
m_{RR}^{2}$ each contain a factor of \(m_s^{2}\) and have mass dimension $[m]^{2}$, while
the one associated with $\delta m_{LR}^{2}$ contains a factor of \(\langle
F_{\chi}\rangle v\) and has mass dimension $[m]^{3}$.

\subsection{Lepton-Sector Flavor Violation in Dirac Leptogenesis\label{sec:FlavorViolation}}


\indent

\begin{figure}[ht!]
\begin{center}
\includegraphics[width=10cm]{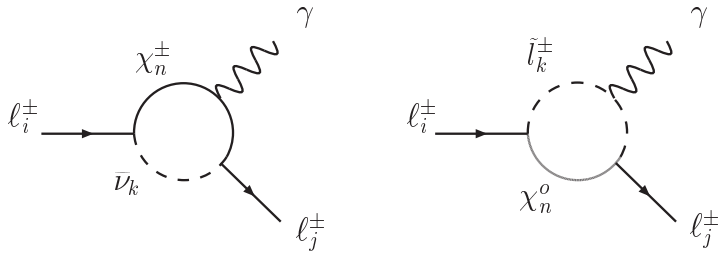}
\end{center} \caption{Feynman diagrams giving the two leading order
  contributions to the flavor-changing process
  $\ell_{i}^{-}\rightarrow\ell_{j}^{-}\gamma$ due to sneutrino (left
  diagram) and charged slepton (right diagram) mass
  mixings.\label{fig:FeynMixingsFull}}
\end{figure}

\indent The effective interaction leading to lepton flavor
  violating decays of the form
\(\ell_{i}\rightarrow\ell_{j}\gamma\), where $\ell_{i}$ and $\ell_{j}$ are charged
leptons, can be written as
\begin{equation}
{\cal I}=i\hspace{.125em}e\hspace{.125em} m_{\ell_j}\hspace{.125em}
\bar{u}_i(q-p)\hspace{.125em} \sigma_{\alpha\beta}\hspace{.125em}q^\beta\left(A^L P_L+A^R
P_R\right)u_j(p)\hspace{.125em} \epsilon^*(q),
\end{equation}
where $q$ and $p$ are the momenta of the photon and the outgoing lepton $\ell_j$
respectively, and $m_{\ell_j}$ is the outgoing lepton mass.  The resulting decay rate is
\begin{eqnarray}
\Gamma(l_j^- \rightarrow l_i^-~\gamma) = \frac{e^2}{16 \pi} m_{l_j}^5 (|A^L|^2+|A^R|^2).
\label{eq:ViolationRate}
\end{eqnarray}
The leading contributions to the amplitudes $A^L$ and $A^R$ appear at one loop level and
are shown in figure~\ref{fig:FeynMixingsFull}. They involve both a sneutrino (and
chargino) mass eigenstate and charged slepton (and neutralino) mass eigenstate running in
the loop. These amplitudes were computed in~\cite{Hisano:1995cp} for a general MSSM
scenario and we have included the modified expressions which account for the presence of
light right-handed neutrino and sneutrino fields in appendix~\ref{app:Looplitudes}.

\begin{figure}[ht!]
\vspace{1cm}
\begin{center}
\includegraphics[width=14cm]{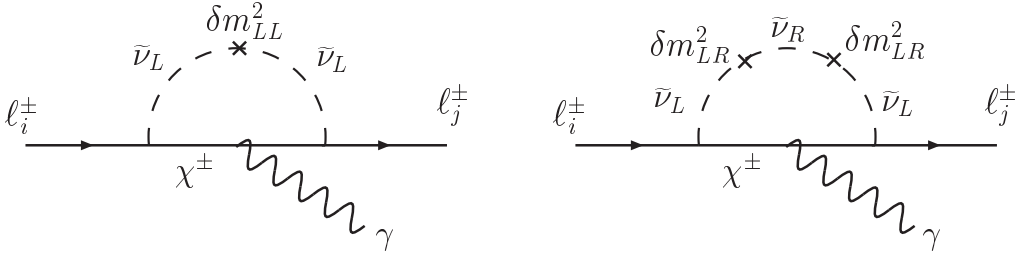}
\end{center} \caption{Feynman diagrams for the leading-order process involving
  $\delta m^{2}_{LL}$
  (left diagram), and for the leading process involving $\delta
  m^{2}_{LR}$ (right diagram),
  with sneutrinos running in the loop in the mass-insertion
  approximation.  Note that any process involving $\delta
  m^{2}_{LR}$ necessarily involves two mass insertions, and any one
  involving $\delta m^{2}_{RR}$ (given by the diagram on the right
  with an additional $\delta m^{2}_{RR}$ insertion)
  necessarily involves three.
  \label{fig:FeynMixingsMassInsert}}
\end{figure}

\indent Since the relationship between the Dirac leptogenesis model parameters \(f\),
\(\langle\chi\rangle\), and \(M_{\Phi_1}\), which enter into
equation~(\ref{eq:ViolationRate}) via the masses and mixings defined in
appendix~\ref{app:Looplitudes}, and the flavor-violation rates~(\ref{eq:ViolationRate})
is somewhat obscure, let us take a moment to make it a bit more transparent.  Consider an
exclusion contour in parameter space corresponding to the \(\mu\rightarrow e\gamma\)
bound (which will turn out to be the more stringent of the two) in
equation~(\ref{eq:CurrentExpConstraints}). Along this bound, the branching ratio for
\(BR(\mu\rightarrow e\gamma)\) is by definition quite low and flavor-violating effects
will be small.  It is therefore valid to use the mass-insertion approximation there and
treat \(\delta m^{2}_{LL}\), \(\delta m^{2}_{LR}\), and \(\delta m^{2}_{RR}\) as small
corrections to the slepton propagators. Let us focus on corrections to the sneutrino
propagator, which enters in the left diagram in figure~\ref{fig:FeynMixingsFull}, since
it can receive corrections from all three of these.  The leading contributions involving
each of \(\delta m^{2}_{LL}\) (left panel) and \(\delta m^{2}_{LR}\) (right panel) are
shown in figure~\ref{fig:FeynMixingsMassInsert} (the leading contribution to \(\delta
m^{2}_{RR}\), which would require three mass insertions, is the least important of the
three, and is not shown). Since there is no coupling between leptons and right-handed
sneutrinos, corrections from \(\delta m^{2}_{LR}\) and \(\delta m^{2}_{RR}\) only appear
at second and third order in the mass insertion expansion, respectively. Therefore, if
there is no substantial hierarchy among these three sets of mixing terms, mixings between
left-handed sleptons provide the primary source of flavor violation. In the approximation
that all slepton, chargino, and neutralino masses are roughly on the same scale
$m_{\mathit{soft}}$, the branching ratio may be estimated as
\begin{equation}
  \mathit{BR}(\mu\rightarrow e\gamma)\propto
  \frac{\alpha^{3}}{G_{F}^{2}} \frac{(\delta
  m_{LL}^{2})^{2}}{m_{s}^{8}}.
  \label{eq:BRwithLefts}
\end{equation}
In interpreting this result, let us treat \(M_{\Phi_1}\) and \(\langle\chi\rangle\) as
free parameters, treat \(\delta\), $a_3$ and \(b_{3}\) as fixed, and eliminate \(f\)
using the constraint~(\ref{eq:FchiMconstraint}). Equation~(\ref{eq:BRwithLefts}) tells us
that contours of branching ratio in the $M_{\Phi_1}-\langle\chi\rangle$ plane correspond
to contours of $\delta m_{LL}^{2}$.  According to equation~(\ref{eq:DeltaMassProps}),
$\delta m_{LL}^{2}=c_{1}M_{\Phi_1}/\langle\chi\rangle$, where $c_{1}$ is a dimensionless
proportionality constant with dimension \([m]^2\), so the exclusion contour associated
with left-left mixing takes the form
\begin{equation}
\ln M_{\Phi}=\ln\langle\chi\rangle+C_{LL}, \label{eq:ContourEquationLL}
\end{equation}
where \(C_{LL}=-\ln(\delta m_{LL}^{2}/c_{1})\) is an as yet undetermined constant.

\indent In the absence of any large hierarchy between \(\delta m^{2}_{LL}\), \(\delta
m^{2}_{LR}\), and \(\delta m^{2}_{RR}\) the oblique contour defined by
equation~(\ref{eq:ContourEquationLL}) is the only important one.  However, there is no a
priori reason why such a hierarchy should not exist.  The \(\delta m^{2}_{LR}\)
contribution~(\ref{eq:MLR}) is proportional to \(\langle F_{\chi}\rangle\), which has
little relevance to baryogenesis other than that it serves to equilibrate the left- and
right-handed sneutrino fields and is essentially unconstrained.  As was pointed out in
section~\ref{sec:MneutfromDlep}, \(\sqrt{\langle F\rangle}\) can potentially be quite
large (\(10^{6}~\mathrm{GeV}\) or higher), and if this is the case, contribution from
\(\delta m^{2}_{LR}\) could be as important as those from \(\delta m^{2}_{LL}\).  Let us
assume for a moment that this is the case and examine the constraints related to \(\delta
m^{2}_{LR}\) and \(\delta m^{2}_{LR}\) together. In regions of parameter space where
\(\langle\chi\rangle\) is small, we now have
\begin{equation}
BR(\mu\rightarrow e\gamma)\propto \frac{\alpha^{3}}{G_{F}^{2}m_{S}^{8}} \frac{(\delta
m^{2}_{LR})^{4}}{m_{\tilde{\nu}_{R}}^{4}}=\mathrm{constant} \label{eq:BRwithRights}
\end{equation}
along any exclusion contour.  Equation~(\ref{eq:DeltaMassProps}) tells us that,
\begin{equation}
  \delta m_{LR}^2=c_{2}\frac{1}{\chi},
\end{equation}
where $c_{2}$ has mass dimension $[m]^{3}$. The associated contour is therefore given by
\begin{equation}
\ln\langle\chi\rangle=C_{LR} \label{eq:ContourEquationLR},
\end{equation}
where \(C_{LR}=\ln(c_{2}/\delta m_{LR}^{2})\).  As for \(\delta m^{2}_{RR}\),
equation~(\ref{eq:DeltaMassProps}) implies that it becomes important (and in fact the
dominant contribution to the right-handed sneutrino mass) in regions of parameter space
where \(\langle\chi\rangle\) is small and \(M_{\Phi_{1}}\) is large.  In this regime the
mass insertion approximation can no longer be used, but we can use the approximation
\(m_{\tilde{\nu}_{R}}^2\simeq\delta m^{2}_{RR}\) in the sneutrino propagator in
equation~(\ref{eq:BRwithRights}).  The amplitude for the process in the left panel
of~\ref{fig:FeynMixingsFull} then implies
\begin{equation}
\frac{(\delta m^{2}_{LR})^{2}}{\delta m^{2}_{RR}}=
c_{3}\frac{1}{M_{\Phi_{1}}\langle\chi\rangle},
\end{equation}
where $c_{3}$ is a proportionality constant with mass dimension \([m]^{4}\).  This
implies yet another oblique exclusion contour corresponding to the line
\begin{equation}
\ln M_{\Phi_1}=-\ln\langle\chi\rangle+C_{RR} \label{eq:ContourEquationRR},
\end{equation}
where \begin{equation}
  C_{RR}= \ln\left(c_{3}\ \frac{\delta
m^{2}_{RR}}{(\delta m^{2}_{LR})^{2}}\right).
\end{equation}
This line runs parallel to the one from left-left mixing~(\ref{eq:ContourEquationLL}). As
discussed above, it will generally be the case that the \(\delta_{LL}^2\) contour
provides a more stringent bound than \(\delta m^{2}_{RR}\), and hence this contour can
generally be ignored.

\indent Taken together, the contours determined by equations~(\ref{eq:ContourEquationLL})
and~(\ref{eq:ContourEquationLR}) suggest that unacceptable amounts of flavor violation
will occur in regions with large \(M_{\Phi_1}\) and small \(\chi\). It is not yet obvious
exactly how large \(M_{\Phi_1}\) and how small \(\chi\) must be before problems arise (or
to put it another way, what the precise values of \(C_{LL}\) and $C_{LR}$ are) in a given
Dirac leptogenesis model, however, nor is it obvious that these constraints are
compatible with those from gravitino cosmology, neutrino physics, etc.  These issues will
be addressed in the next chapter through a careful numerical calculation which takes into
account the full formulae for the masses, mixings, et al.\ given in
appendix~\ref{app:Looplitudes}.

\chapter{EVOLUTION\ OF\ THE\ BARYON\ ASYMMETRY\label{ch:Boltzmann}}

\section{Boltzmann Equations\label{sec:BoltzEqIntro}}

\indent

Let us now turn to the numerical calculation of \(\eta\) and the solution of the full
Boltzmann equations.  This is slightly more complicated for Dirac than for Majorana
leptogenesis because in the latter, it is only necessary to keep track of the overall
lepton number.  In contrast, the former involves the creation of several distinct stores
of lepton number with different properties.  The heavy fields aside, in Dirac
leptogenesis there are six particle species charged under lepton number (\(\nu_{R}\),
\(\tilde{\nu}_{R}\), \(\ell\), \(\tilde{\ell}\), and the right-handed charged lepton and
slepton fields \(e_{R}\) and \(\tilde{e}_{R}\)), and thus six individual stores of lepton
number to keep track of: \(L_{\ell}\), \(L_{\nu_{R}}\), \(L_{\tilde{L}}\),
\(L_{\tilde{\nu}_{R}}\), \(L_{e_{R}}\), and \(L_{\tilde{e}_{R}}\).  Some of these stores
are positive and some negative, some of the particle species are involved in sphaleron
interactions while others are not, and so forth, so it might appear necessary that we
keep track of each field and each store individually.  Including an equation for the
overall baryon number of the universe \(B\) and accounting for the dynamics of the heavy
fields in the \(\Phi_{1}\) and \(\overline{\Phi}_1\) supermultiplets raises the total
number of equations in the Boltzmann system to twenty-one.  The situation can be greatly
simplified, however, by noting that the fields \(\ell\), \(\tilde{\ell}\), \(e_{R}\), and
\(\tilde{e}_{R}\) participate in \(SU(2)\) and/or \(U(1)_{Y}\) gauge interactions, which
should be sufficiently rapid (compared to other processes relevant to leptogenesis) that
these species will always be in chemical equilibrium with one another. As a result, any
lepton number stored in any one of them will be rapidly distributed among \(L_{\ell}\),
\(L_{\tilde{\ell}}\), \(L_{e_{R}}\), and \(L_{\tilde{e}_{R}}\) in proportion to the
relative number of degrees of freedom of each respective field.  These fields then
compose a distinct, ``visible'' sector of the theory with an aggregate lepton number
\(\Lvis\).

\indent As for the remaining two species charged under lepton number, there are two
possible choices.  While the \(\nu_{R}\), by construction, have no interactions with the
visible sector and thus can be seen as forming a ``hidden'' sector which only
equilibrates with the visible sector fields at late times through the effective neutrino
Dirac Yukawa, no such requirement exists for the right-handed sneutrinos
\(\tilde{\nu}_{R}\).  When \(\langle F_{\chi}\rangle\) is large, the \(\tilde{\nu}_{R}\)
fields have additional, rapid interactions with the visible sector fields (and in
particular \(\tilde{\ell}\)) through effective $A$-terms~(\ref{eq:EffectiveAtermL}) and
should thus be considered part of the visible sector.  As shown in
section~\ref{sec:FlavorViolation}, an $F$-term VEV of order \(\langle F_{\chi}\rangle\sim
10^8\)~GeV is still permitted by flavor violation constraints in theories with weak-scale
slepton squared masses, and in split supersymmetry scenarios with much heavier sfermions,
even this bound no longer applies, so rapid sneutrino equilibration is far from excluded.
On the other hand, when \(\langle F_{\chi}\rangle\) is small, the \(\tilde{\nu}_{R}\)
fields decouple and become part of the hidden sector.  In either case, since none the
hidden sector fields interacts via electroweak sphalerons, an aggregate lepton \(\Lhid\)
can be defined for the hidden sector, which is just the sum of the lepton numbers
individually stored in each of its constituent fields.  We will henceforth refer to these
two situations as the large-\(\langle F_{\chi}\rangle\) and small-\(\langle
F_{\chi}\rangle\) scenarios.

\indent The definitions of \(\Lvis\) and \(\Lhid\) simplify our task considerably; in the
limit of rapid equilibration within the visible sector, only three Boltzmann equations
are required to describe the evolution of lepton and baryon number asymmetries of the
light fields in Dirac leptogenesis: one for \(\Lvis\), one for \(\Lhid\), and one for the
overall baryon number \(B\).  In addition to these, it turns out (see
appendix~\ref{app:BoltzDiracLep}) that only three additional equations are needed to
describe the dynamics of the heavy fields: two to describe the evolution of the
individual lepton numbers \(L_{\phiphi}\) and \(L_{\phiphibar}\) stored respectively in
\(\phi_1\) and \(\overline{\phi}_1\), and one to track the abundance of one of the heavy
fields (we choose \(Y_{\phiphi}^c\)).  Our Boltzmann system has now been reduced from
twenty-one equations down to a far more manageable six.

\indent In addition to the rapid gauge interactions that equilibrate the fields in
\(\Lvis\), there are a variety of additional processes which we must take into account.
First, we must include the decays and inverse decays of the fields in \(\Phi_{1}\) and
\(\overline{\Phi}_1\).  This introduces a pair of rates \(\Gamma_{L}=\Gamma(\phi\to
\widetilde{\ell}+\chi)\) and \(\Gamma_{R}=\Gamma(\phi\to \nu^c_R+\tilde{H}^c_u)\), which
are related to the overall \(\phi\) decay rate \(\Gamma_{D}\) given in~(\ref{eq:GammaD})
by the defining relation
\begin{equation}
  \Gamma_{D}=\Gamma_{L}+\Gamma_{R}.
\end{equation}
Second of all, we must include interactions involving virtual \(\phi_{i}\) and
\(\overline{\phi}_{i}\) fields (and their fermionic superpartners) that can transfer
lepton number between \(\Lvis\) and \(\Lhid\). The dominant contribution to the transfer
rate comes from \(2\leftrightarrow2\) processes, and since this rate, which we will call
\(\Gamma_{2\leftrightarrow2}\), will be suppressed by inverse factors of the leptogenesis
scale for \(T\ll M_{\Phi_{1}}\), these processes cannot be considered rapid compared to
the gauge interactions. Third, since annihilation processes second-order in the heavy
fields can serve to reduce the abundances of these particles during the leptogenesis
epoch~\cite{Hambye:2005tk,Chun:2005ms}, we should include them as well, with rate
\(\Gamma_{A}\). Fourth and finally, we must include sphaleron processes with rate
\(\Gamma_{\mathit{sph}}\), which will interconvert \(B\) and \(\Lvis\).

\indent We are now ready to write the full set of Boltzmann equations governing the
evolution of baryon number \(B\) in the early universe, a full derivation for which is
provided in appendix~\ref{app:BoltzDiracLep}, along with explicit definitions for all
quantities used therein.  Since these equations will differ slightly between the
large-\(\langle F_{\chi}\rangle\) and small-\(\langle F_{\chi}\rangle\) scenarios, due to
the differing constitution of \(\Lvis\) and \(\Lhid\), they need to be written down
separately for each case.  We will concentrate on the large-\(\langle F_{\chi}\rangle\)
scenario, in which they are
\begin{eqnarray}
 {dB \over dz} & = & \frac{z}{H(M_{\Phi_{1}})}\left[ -\langle\Gamma_{\mathit{sph}}\rangle
 (B+\frac{8}{15}L_{\mathit{vis}})\right] \label{eq:BoltzB}\\
 {d L_{\mathit{vis}}\over dz}&=& 
\frac{z}{H(M_{\Phi_{1}})}\left[\rule[0pt]{0pt}{16pt}-2\epsilon\langle
\Gamma_{D}\rangle(Y_{\phiphi}^c-Y_{\phiphi}^{{eq}})+\langle
\Gamma_L\rangle (L_{\phiphi} + L_{\phiphibar})\right.\non\\
&& \hspace{1.5cm} \left.+\langle\Gamma_R\rangle L_{\phiphibar}- 2
L_{\mathit{vis}}\Big(\langle \Gamma_{D}\rangle_{{}_{ID}}+\langle
\Gamma_L\rangle_{{}_{ID}}\Big)\right.\non\\
&& \hspace{1.5cm} \left. + (\Lhid -{1\over7}L_{\mathit{vis}}) \langle
\Gamma_{2\leftrightarrow2}\rangle- \langle\Gamma_{\mathit{sph}}\rangle
 (B+\frac{8}{15}L_{\mathit{vis}})\right] \label{eq:BoltzLnet}\\
{d\Lhid\over dz}&=& \frac{z}{H(M_{\Phi_{1}})}\left[2\epsilon
\langle\Gamma_{D}\rangle(Y_{\phiphi}^c-Y_{\phiphi}^{eq}) + L_{\phiphi}
\langle\Gamma_R\rangle - 2 \Lhid \langle \Gamma_R\rangle_{{}_{ID}}\right.\non \\ & &
\hspace{1.5cm} \left.-(\Lhid
-{1\over7}L_{\mathit{vis}}) \langle \Gamma_{2\leftrightarrow2}\rangle\right] \label{eq:BoltzR}\\
{dY_{\phiphi}^c \over dz} &=& \frac{z}{H(M_{\Phi_{1}})}
\left[\rule{0pt}{18pt}-\langle\Gamma_{D}\rangle(Y_{\phiphi}^c - Y_{\phiphi}^{eq})
+{1\over2} L_{\mathit{vis}}
\langle\Gamma_L\rangle_{{}_{ID}}+{1\over2} \Lhid\ \langle\Gamma_R\rangle_{{}_{ID}}\right.\non\\
& &\left. -\langle\Gamma_A\rangle \left(\left(Y_{\phiphi^c}/Y_{\phiphi}^{eq}\right)^2 - 1
\right)\rule{0pt}{18pt}\right] \label{eq:BoltzPhiconj}\\ 
{dL_{\phiphi} \over dz} &=& \frac{z}{H(M_{\Phi_{1}})}\left[-\langle \Gamma_{D}\rangle
L_{\phiphi} +2 L_{\mathit{vis}} \langle \Gamma_L\rangle_{{}_{ID}}+2 \Lhid
\langle \Gamma_R\rangle_{{}_{ID}}\right]\label{eq:BoltzLPhi}\\
{dL_{\phiphibar} \over dz} &=& \frac{z}{H(M_{\Phi_{1}})}\left[ -\langle \Gamma_{D}\rangle
 L_{\phiphibar} +2 L_{\mathit{vis}}\langle
 \Gamma_{D}\rangle_{{}_{ID}}\label{eq:BoltzLPhibar}\right]
\end{eqnarray}
in terms of the variable $z\equiv M_{\Phi_{1}}/T$.  Here, the inverse decay rates
\(\langle\Gamma_D\rangle_{{}_{ID}}\), \(\langle\Gamma_L\rangle_{{}_{ID}}\), and
\(\langle\Gamma_R\rangle_{{}_{ID}}\) are defined by
\begin{eqnarray}
\langle\Gamma_D\rangle_{{}_{ID}}={1\over7}{n_{\phiphi}^{eq}\over
n_\gamma}\left({K_1(z)\over K_2(z)}\right)\Gamma_D,\\
\langle\Gamma_L\rangle_{{}_{ID}}={1\over7}{n_{\phiphi}^{eq}\over
n_\gamma}\left({K_1(z)\over K_2(z)}\right)\Gamma_L, & \mathrm{and} & \\
\langle\Gamma_R\rangle_{{}_{ID}}={n_{\phiphi}^{eq}\over n_\gamma}\left({K_1(z)\over
K_2(z)}\right)\Gamma_R,
\end{eqnarray}
where \(\Gamma_{D}\) is the total decay width of \(\phi_{1}\) given in
equation~(\ref{eq:GammaD}), \(n_{\phi}^{\mathit{eq}}\) is the equilibrium number density
of \(\phi_{1}\) (and \(\overline{\phi}_{1}\), etc.), and the quantities \(\Gamma_{L}\)
and \(\Gamma_{R}\) represent the partial decay widths for \(\phi\to\nu_{R}\tilde{H}\) (or
\(\overline{\phi}\to\tilde{\nu}_{R}H\)) and \(\overline{\phi}\to\ell\tilde{\chi}\) (or
\(\phi\to\tilde{\ell}\chi\)).  The ratio of modified Bessel functions \(K_{1}(z)\) and
\(K_{2}(z)\) appearing in these rates is a result of averaging over time-dilation
factors: see~(\ref{eq:timedilation}) in Appendix~\ref{app:BoltzDiracLep}.
We will examine this scenario from this point forward.

%

\section{Numerical Analysis of the Boltzmann System\label{sec:NumSolBigF}}


\subsection{Calculation of Rates\label{sec:rates}}

\indent

From this point forward we will focus our attention on the large-$\langle
F_{\chi}\rangle$ case and proceed to solve the Boltzmann
equations~(\ref{eq:BoltzB}\,-\,\ref{eq:BoltzLPhibar}) numerically.  In order to do this
we must first calculate the relevant rates appearing in these equations in terms of the
model parameters \(M_{\Phi_{1}}\), \(\delta\), \(a_3\), $b_3$, and \(f\) (or
alternatively \(\langle\chi\rangle\)). We already have expressions for
\(\Gamma_{\mathit{sph}}\)~(\ref{eq:HighTempSphRate}) and
\(\Gamma_{D}\)~(\ref{eq:GammaD}), from the latter of which \(\Gamma_{L}\) and
\(\Gamma_{R}\) can be obtained trivially.  This leaves only the rate
\(\Gamma_{2\leftrightarrow 2}\) for processes that shuffle lepton number between
\(L_{\mathit{vis}}\) and \(L_{\mathit{hid}}\) and the heavy field annihilation rate
\(\Gamma_{A}\).  We now turn to address each of these in turn.

\begin{figure}[ht!]
\begin{center}
\includegraphics[width=8.5cm]{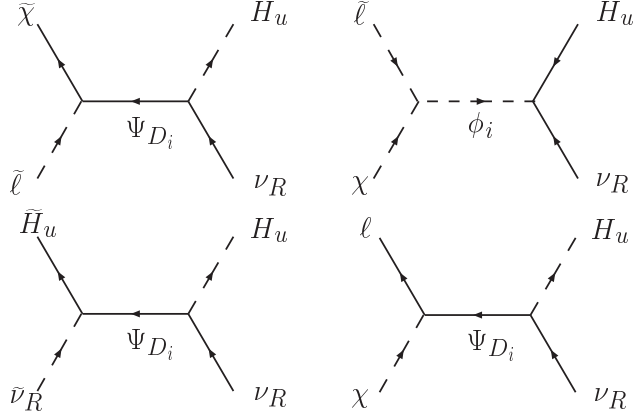}
\end{center}
  \caption{Diagrams for $2\leftrightarrow2$ $s$-channel processes which transfer lepton
    number between $L_{\mathit{aggVis}}$ (the aggregate lepton number in the sector comprising the fields
    $\ell$, $\tilde{\ell}$, $\tilde{\nu}_{R}$, $e_{R}$, and
    $\tilde{e}_{R}$, which are assumed to be in chemical equilibrium with one another due to rapid gauge and
    $\langle F_{\chi}\rangle$-term equilibration interactions);
     and $\Lhid$ (the lepton number stored in right handed neutrinos).  The
     two $t$-channel
    interactions associated with each diagram appearing above must also be included in
    calculating the full thermally averaged cross-section.
    \label{fig:2to2SleptonProc}}
\end{figure}

\indent  While there are a large number of \(2\leftrightarrow2\) processes which shuffle
lepton number between different particle species, only a few will transfer it between
\(\Lvis\) and \(\Lhid\).  The rest, which collectively serve to assist the rapid
\(SU(2)\times U(1)_{Y}\) gauge interactions in equilibrating lepton number among the
fields in the \(L_{\mathit{vis}}\) sector, can be ignored.  The relevant $s$-channel
diagrams are pictured in figure~\ref{fig:2to2SleptonProc}.  In addition to these, there
are contributions from the $t$-channel transforms (two per diagram) of these diagrams. In
order to evaluate diagrams containing virtual heavy fermions, we define the Dirac spinor
\begin{equation}
    \Psi_{Di}=\left(\begin{array}{c} (\psi_{\Phi_{i}})_{\alpha} \\ (\psi_{\overline{\Phi}_{i}})^{\dagger\dot{\alpha}}
    \end{array}\right),
\end{equation}
where \((\psi_{\Phi_{i}})_{\alpha}\) and \((\psi_{\overline{\Phi}_{i}})_{\alpha}\) are
the Weyl spinor components of the \(\Phi_{i}\) and \(\overline{\Phi}_{i}\) superfields.
Numerical calculation of the thermally averaged cross-sections for the diagram pictured
on the left in the top row of figure~\ref{fig:2to2SleptonProc}, which involves two
Yukawa-type couplings of a scalar to two fermions, yields \(\langle\sigma
v\rangle_{i\alpha\beta}\simeq\sigma_{i}^{\mathit{(2Y)}}|\lambda_{i\alpha}|^{2}|\lambda_{i\beta}|^{2}\),
where~\cite{Thomas:2005rs}
\begin{equation}
  \sigma_{i}^{\mathit{(2Y)}}\equiv10^{-2}\frac{T^{2}}{(M_{\Phi_{i}}^{2}+T^{2})^{2}}.
\end{equation}
For the diagram on the right in the first row of figure~\ref{fig:2to2SleptonProc}, which
includes one Yukawa-type coupling and one trilinear scalar coupling proportional to
\(M_{\Phi_{i}}\), the result is very nearly temperature independent and well approximated
by \(\langle\sigma
v\rangle_{i\alpha\beta}\simeq\sigma_{i}^{\mathit{(1Y1S)}}|\lambda_{i\alpha}|^{2}|h_{i\beta}|^{2}\),
where
\begin{equation}
  \sigma_{i}^{\mathit{(1Y1S)}}\equiv0.5\times\frac{1}{M^{2}_{\Phi_{i}}}.
  \label{eq:1Y1S}
\end{equation}
The contribution from each of the two diagrams in the second row of
figure~\ref{fig:2to2SleptonProc}, which involve two Yukawa couplings and a mass insertion
from the heavy fermions, is equal to that in~(\ref{eq:1Y1S}).  These interactions
dominate among \(2\leftrightarrow2\) processes. The diagram in the bottom row of
figure~\ref{fig:2to2SleptonProc}, which involves a trilinear scalar coupling to the
down-type Higgs, may be approximated by \(\langle\sigma
v\rangle_{i\alpha\beta}\simeq\sigma_{i}^{\mathit{(Hd)}}|\lambda_{i\alpha}|^{2}|\lambda_{i\beta}|^{2}\),
where
\begin{equation}
  \sigma_{i}^{\mathit{(Hd)}}\propto\frac{\mu}{M^{3}_{\Phi_{i}}}.
\end{equation}
The constant of proportionality in this equation is \(\mathcal{O}(1)\), and is thus
suppressed relative to the rate given in~(\ref{eq:1Y1S}) by \(\mu/M_{\Phi_{i}}\).  Here,
we will assume that \(\mu\) is several orders of magnitude smaller than all the
\(M_{\Phi_{i}}\), and therefore the effect of these processes can be neglected.

Taking into account contributions involving virtual fields in the \(\Phi_{2}\) and
\(\overline{\Phi}_{2}\) supermultiplets, as well as those in \(\Phi_{1}\) and
\(\overline{\Phi}_{1}\), we find the total interconversion rate between \(\Lvis\) and
\(\Lhid\) in the large-\(\langle F_{\chi}\rangle\) scenario to be
\begin{eqnarray}
  \Gamma_{2\leftrightarrow2} & \simeq & 3 n_{\gamma}\sum_{\alpha}\sum_{\beta}
    \left(\sigma_{1}^{\mathit{(2Y)}}|\lambda_{1\alpha}|^{2}|\lambda_{1\beta}|^{2}+\sigma_{2}^{\mathit{(2Y)}}|\lambda_{2\alpha}|^{2}|\lambda_{2\beta}|^{2}\right)\label{eq:2to2R} \non\\  & + &
    9n_{\gamma}\sum_{\alpha}\sum_{\beta}
    \left(\sigma_{1}^{\mathit{(1Y1S)}}|\lambda_{1\alpha}|^{2}|h_{1\beta}|^{2}+\sigma_{2}^{\mathit{(1Y1S)}}|\lambda_{2\alpha}|^{2}|h_{2\beta}|^{2}\right)\nonumber \\ & + &
    3n_{\gamma}\sum_{\alpha}\sum_{\beta}
    \left(\sigma_{1}^{\mathit{(Hd)}}|\lambda_{1\alpha}|^{2}|\lambda_{1\beta}|^{2}+\sigma_{2}^{\mathit{(Hd)}}|\lambda_{2\alpha}|^{2}|\lambda_{2\beta}|^{2}\right),
\end{eqnarray}
where we have assumed an equilibrium number density for all non-leptonic light species
(e.g.\ \(H_{u}\), \(\chi\)) involved.

\indent The annihilation rate for the heavy fields \(\Gamma_{A}\), which is associated
with second order processes of the form \(\phi_{1}\phi_{1}^c\rightarrow ij\) and
\(\phi_{1}\overline{\phi}_{1}^c\rightarrow ij\) (and miscellaneous supersymmetrizations
thereof), is most readily expressed in terms of the reaction density, given by the
general expression
\begin{equation}
  \gamma\equiv\frac{T}{64\pi^{4}}\int^{\infty}_{s_{\mathit{min}}}s^{1/2}
  K_{1}\left(\frac{\sqrt{s}}{T}\right)\hat{\sigma}(s),
  \label{eq:ReacDensDef}
\end{equation}
where $T$ is temperature, $s$ is the usual Mandelstam variable, and \(\hat{\sigma}(s)\)
is the total reduced cross section for annihilations of \(\phi_{1}\phi_{1}^c\),
\(\phi_{1}\overline{\phi}_{1}^c\), etc.\ into light fields.  This is defined by the
formula
\begin{equation}
  \hat{\sigma}(s)=\frac{1}{8\pi s}\int^{t_{+}}_{t_{-}}
  \sum_{i}|\mathcal{M}_{i}(t)|^{2}dt,
  \label{eq:SigmaWithHatOnIt}
\end{equation}
where both $t$ and $s$ denote the Mandelstam variables. The limits of integration are
given by $t_\pm=M_{\Phi_1}^2-{s}(1\mp r)/2$, with $r$ defined below.  The relationship
between the reaction density for the annihilation of two particles \(i\) and \(j\) with
number densities \(n_i\) and $n_j$, the thermally-averaged cross-section
$\langle\sigma|v|\rangle$ for annihilation, and the annihilation rate \(\Gamma\) via
\begin{equation}
  \gamma=n_{i}n_{j}\langle\sigma|v|\rangle=
  \frac{n_{i}n_{j}}{n_{\gamma}}\Gamma.\label{eq:RelBetweenGammasandSigmas}
\end{equation}
Hence, in the case under consideration here, the annihilation reaction
density\(\gamma_{A}\) between $\phi_{1}$ and \(\phi^c_1\) is given by
\begin{equation}
  \gamma_A\simeq\frac{s^2}{n_\gamma}Y_{\phiphi^c}^2\Gamma_{A}
\end{equation}
in the (very good) approximation than \(L_{\phiphi}\ll Y_{\phiphi^c}\).

\begin{figure}[ht!]
\begin{center}
\includegraphics[width=7cm,height=9.5cm]{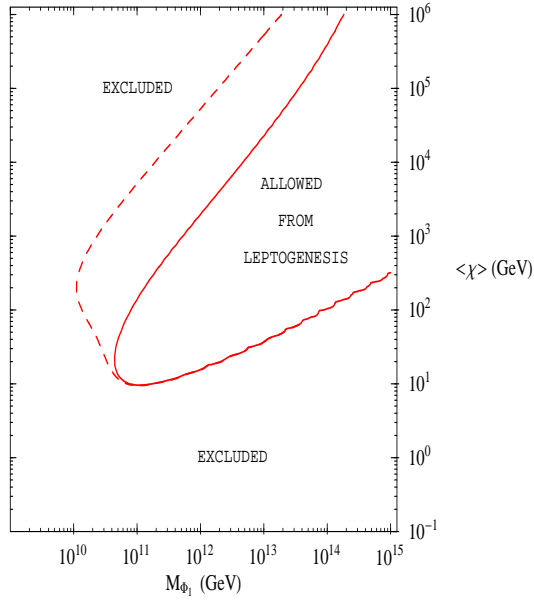}
\vspace{-1cm}
\end{center} \caption{This figure, taken from~\cite{Thomas:2006gr}, illustrates the effect of
  second-order processes of the form $\phi_{1}\phi_{1}\rightarrow
  ij$ and $\tilde{\phi}_{1}{\phi}_{1}\rightarrow ij$ on the
  exclusion contours from leptogenesis. Contours are displayed for the
  case without annihilation (dashed line) and with annihilation (solid
  line). \label{fig:SecondOrderFX}} \vspace{.5cm}
\end{figure}

\indent In supersymmetric Dirac leptogenesis, the total reduced cross-section
\(\gamma_{A}\), including all relevant decay processes, is~\cite{Thomas:2006gr}
\begin{eqnarray}
  \hat{\sigma}_{\mathit{SUSY}}^{\mathit{tot}}&=&
  \frac{1}{16\pi}\left[
  6g_{Y}^{2}g_{2}^{2}\left(\left(-7+\frac{4}{x}\right)r
    +\left(\frac{8}{x^2}-\frac{4}{x} +
    9\right)\ln\left(\frac{1+r}{1-r}\right)\right)\right.\nonumber\\&&+
  g_{2}^{4}\left(\left(32+\frac{66}{x}\right)r+
    3\left(-\frac{16}{x^{2}}-\frac{16}{x}+9\right)
    \ln\left(\frac{1+r}{1-r}\right)\right)\nonumber\\&&+\left.
  g_{Y}^{4}\left(\left(19-\frac{36}{x}\right)r+
    \left(\frac{16}{x^2}-\frac{8}{x}+17\right)
    \ln\left(\frac{1+r}{1-r}\right)\right)\right],
\end{eqnarray}
where $x\equiv s/M_{\Phi_{1}}^2$, $r=\sqrt{1-4/x}$, and \(g_{2}\) and \(g_{Y}\) are the
\(SU(2)\) and \(U(1)_{Y}\) coupling constants.
The effect of such second order annihilation processes on the parameter space of Dirac
leptogenesis is shown in figure~\ref{fig:SecondOrderFX}, where, for comparison, we show
two sets of leptogenesis exclusion contours: one representing no second-order processes
and one representing annihilation in a supersymmetric model.  It is evident from this
graph that second order processes do indeed lower the upper exclusion contour, though the
effect is not a dramatic one.

\subsection{Numerical Results\label{sec:numerres}}

\indent

\begin{figure}[t!]
  \begin{center}
\hspace{.25cm}
  \includegraphics{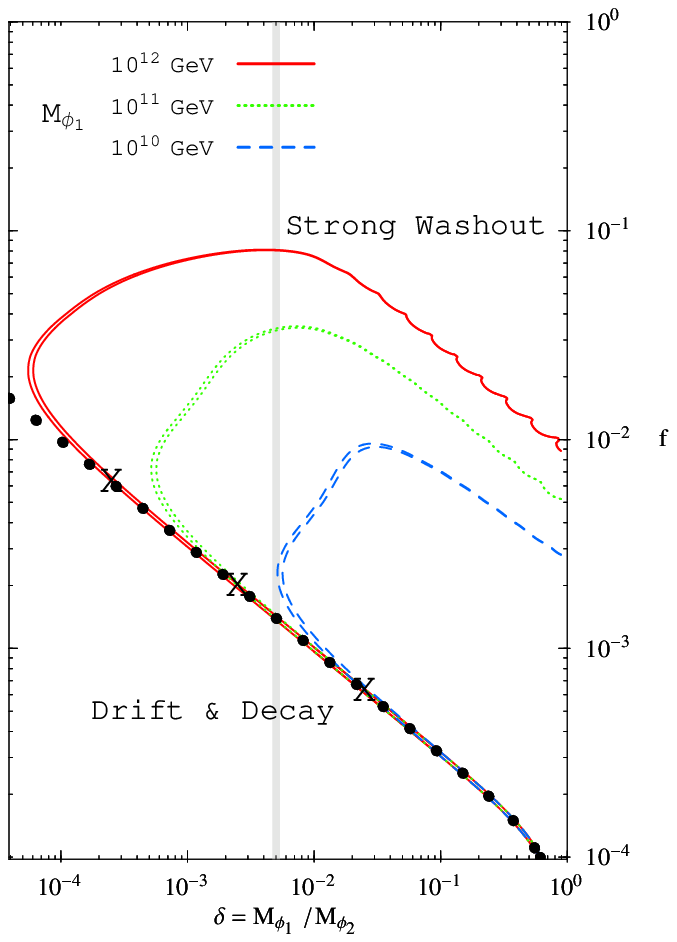}
\hspace{.5cm}
  \includegraphics{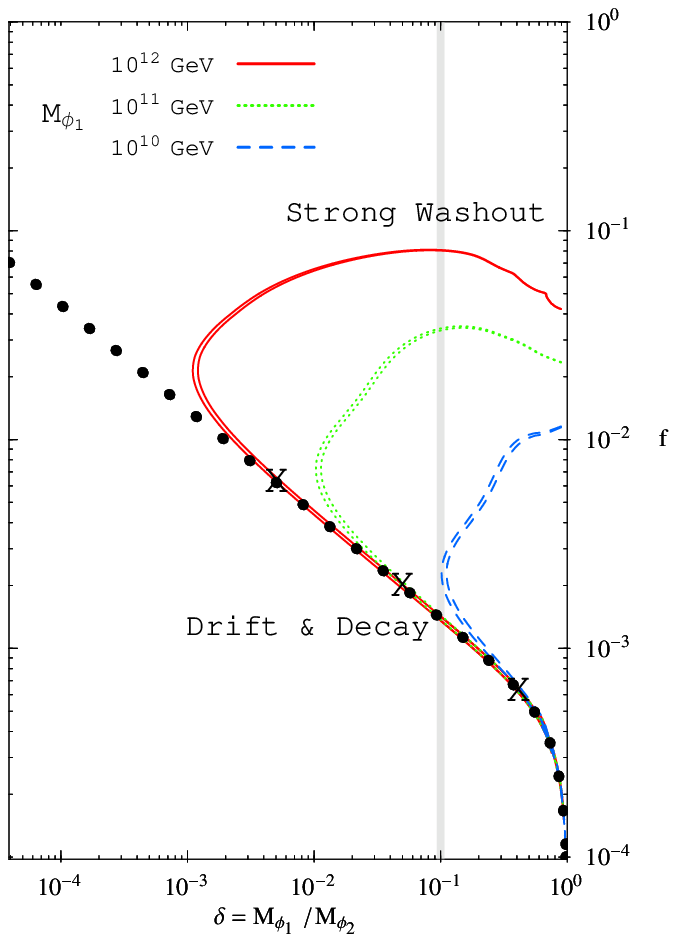}
  \end{center}
\vspace{-1cm}
  \caption{Bands in $f-\delta$ parameter space for which the final baryon
  number $B_{f}$ 
falls within the range permitted by WMAP, for different
  choices of $M_{\Phi_{1}}$~\cite{Thomas:2005rs}. Here $f$ 
parameterizes the couplings of the $\Phi_i$ fields (see
eq.~(\ref{eq:YukawaParametrization})).
The configuration of the left (right) panel is the one marked with a dot in the left
(right) panel of figure~\ref{fig:NeutrinoMassPlots}, in which $a_{3}=95$ ($a_{3}=4.5$).
The shaded vertical lines show the constraint on the value of $\delta$ coming from
neutrino mixings and masses ($\delta=m_{e}/m_{\mu}\approx4.83\times10^{-3}$ in the left
and $\delta=10^{-1}$ in the right).
  Note that there are two points that yield a realistic value of $B_{f}$ for a given $\delta$ (the points at which the grey vertical line intersects a given ribbon).
  When $f$ is small enough, the initial baryon
  number generated is just enough to be consistent with WMAP, while the washout effect
  of inverse decays and $2\leftrightarrow2$ processes is negligible. In
  this situation the baryon number generated is independent
  of $M_\Phi$ and its the final value is proportional to the CP
violating parameter $\epsilon$ (see eq.~(\ref{eq:DriftAndDecay})). The dark dotted curve
  shows the band of consistent baryon number calculated in this ``drift
  and decay'' limit. Each ``X'' marks the point in which
  ${\Gamma_D/ H}=1$ for each different $M_{\Phi_1}$.
At these points, the ``strong washout'' regime starts. The now active
  washout processes reduce an initial surfeit of baryon number (due to
  a larger $f$) down to an acceptable level (see figure~\ref{fig:SpeciesEvolution}).}
  \label{fig:EtaContoursCCwithMRatios}
\end{figure}

Now that we have calculated the requisite rates, we proceed to a numerical evaluation of
the Boltzmann system.  As discussed above, CHDL contains five free parameters, so in
presenting our results, we will begin by selecting a particular value of \(\delta\) small
enough to be consistent with the neutrino physics constraints in~(\ref{eq:NeutLimits})
(we choose $\delta=10^{-1}$) and tune \(a_{3}\) and \(b_{3}\) to the values most
advantageous for leptogenesis (in this case \(a_{3}=4.5\) and \(b_{3}=2.2\)).  Treating
the reparameterized coupling strength \(f\), the overall scaling factor in both
\(\lambda\) and \(h\) in the parametrization defined in
equation~(\ref{eq:YukawaParametrization}), as a free parameter, we display the results
corresponding to several different choices of \(M_{\Phi_1}\) in \(f-\delta\) parameter
space in the right panel of figure~\ref{fig:EtaContoursCCwithMRatios}.  Here, the regions
of \(f-\delta\) parameter space in which the final value of \(\eta\) generated falls
within the WMAP-allowed range given in~(\ref{eq:WMAPeta}), appear as thin `ribbons'
corresponding to each value of \(M_{\Phi_{1}}\).  We should reemphasize that \(\delta\)
specifies the values of \(a_{3}\) and \(b_{3}\) and hence the results displayed are
precisely valid only for points lying along the grey, vertical line corresponding to the
specific value of \(\delta\) chosen.  To illustrate the effect of changing \(\delta\) on
the shape of the ribbons, we have included a second plot for which
\(\delta=4.83\times10^{-3}\) in the left panel of
figure~\ref{fig:EtaContoursCCwithMRatios}.  The motivation for selecting this particular
value of \(\delta\) for contrast will be made apparent in
section~\ref{sec:FlavonExtension}.

\begin{figure}[ht!]
    \includegraphics{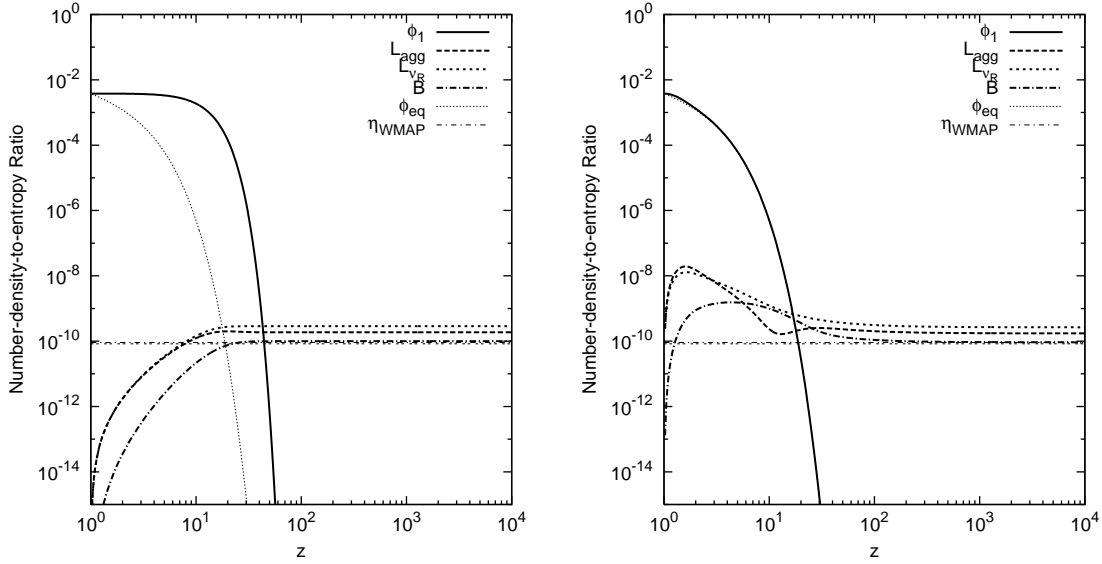}
  \caption{These two plots, originally appearing in~\cite{Thomas:2005rs},
  show the evolution of baryon number $B$
  for $M_{\Phi_{1}}=10^{12}$~GeV and
  $\delta=m_{e}/m_{\mu}=4.83\times10^{-3}$, in the two different
  regimes that produce a realistic value for the final baryon number of the universe, $B_{F}$.  For
  the rescaled coupling strength ${f}=\sqrt{\lambda_{23}h_{23}}=1.5\times10^{-3}$, as
  shown in the left panel, the effects of $2\leftrightarrow2$
  lepton-number-changing processes are negligible and the final
  baryon-to-entropy ratio is the same as that initially produced
  by $\phi$ and $\overline{\phi}$ decays.  For stronger coupling
  ${f}=3.8\times10^{-2}$, as shown in the right panel,
  baryon number is initially overproduced, but $2\leftrightarrow2$
  processes, which are stronger for stronger coupling, reduce $B$
  to an acceptable level by the time they freeze out.  The left and
  right panels correspond respectively to the lower and upper parts of the
  `ribbon' in figure~\ref{fig:EtaContoursCCwithMRatios}.
  \label{fig:SpeciesEvolution}}
\end{figure}

In figure~\ref{fig:EtaContoursCCwithMRatios}, the effects of the
processes detailed in section~\ref{sec:rates} are apparent, and certainly nontrivial.
Physically, these effects can be interpreted as follows: increasing the strength of the
neutrino-sector couplings (here parameterized by \({f}\)) increases \(\Gamma_{D}\), which
in turn increases the initial value of \(B\); however, from equation~(\ref{eq:2to2R}),
increasing \({f}\) also increases the rates for the \(2\leftrightarrow2\) processes which
shuffle lepton number back and forth between \(L_{\mathit{vis}}\) and \(\Lhid\).
Furthermore, it increases the rate for inverse decays.  This allows two possibilities for
generating a realistic final value for \(B\).  In the first case, where \({f}\) is small,
the initial baryon number produced by \(\phi\) and \(\overline{\phi}\) decays is
approximately within the range allowed by WMAP, and \(2\leftrightarrow2\) and inverse
decay processes are so slow as to be negligible; this is the ``drift-and-decay limit" of
equation~(\ref{eq:DriftAndDecay}). In the second case, where \({f}\) is large, a surfeit
of baryon number will initially be produced, but these processes, which occur more
rapidly for larger \({f}\), subsequently reduce \(B\) to a phenomenologically acceptable
level; this we refer to as the ``strong-washout regime''.  These two possibilities are
shown in the two panels of figure~\ref{fig:SpeciesEvolution}, in which the dynamical
evolution of \(B\), \(L_{\mathit{vis}}\) and other relevant quantities has been plotted,
for \(\delta=4.83\times10^{-3}\) and \(M_{\Phi}=10^{12}\) GeV.
In figure~\ref{fig:EtaContoursCCwithMRatios}, the two regimes are represented
respectively by the lower and upper portions of each ribbon---or more properly, by the
two points at which these two portions of the ribbon intersect the grey line
corresponding to the chosen value of \(\delta\) in each graph.  In the \(\delta=10^{-1}\)
case, the strong washout corresponds to \(f=8.3\times10^{-2}\) and the drift-and-decay
limit corresponds to \(f=1.5\times10^{-3}\).  Alternatively, one can use
equation~(\ref{eq:FchiMconstraint}) to express things in terms of \(\langle \chi\rangle\)
once \(M_{\Phi_1}\) is specified: for \(M_{\Phi_{1}}=10^{12}\)~GeV,
\(\langle\chi\rangle\sim5\)~GeV in the strong washout case and
\(\langle\chi\rangle\sim50\)~TeV in the drift-and-decay limit; for
\(M_{\Phi_{1}}=10^{10}\)~GeV the two cases converge and \(\langle\chi\rangle\sim1\)~TeV.

\indent  Having discussed the general effects of the washout processes described above as
a group, it is also important to address their characteristics relative to one another.
Inverse decays dominate over \(2\leftrightarrow2\) processes only for a brief period,
where \(1\lesssim z \lesssim 50\), but during this period they are extremely effective in
reducing lepton number, and in fact are the primary factor in determining the final value
of \(\eta\).  For larger \(z\), until they freeze out, the \(2\leftrightarrow2\)
interactions dominate and further reduce \(L_{\mathit{vis}}\) and \(\Lhid\) (and
consequently \(B\)). It should be noted, however, that the total \(B-L_{\mathit{tot}}\)
number of the universe is manifestly conserved by the Boltzmann
equations~(\ref{eq:BoltzB}) - (\ref{eq:BoltzLPhibar}) (the sum of the rates for the
various lepton numbers involved is zero), and since we began with
\(B-L_{\mathit{tot}}=0\), we end up with \(B-L_{\mathit{tot}}=0\) in any case, as
expected.

\indent The upshot of figure~\ref{fig:EtaContoursCCwithMRatios} is that Dirac
leptogenesis can work even when the constraints from chapter~\ref{ch:constraints} are
taken into consideration, but that these constraints have important implications for the
theory.  Combined constraints from baryogenesis and neutrino physics make it extremely
difficult simultaneously to obtain a realistic neutrino spectrum and obtain the correct
baryon number when \(M_{\Phi_{1}}<10^{10}\)~GeV, which in turn requires a reheating
temperature \(T_{R}\gtrsim10^{10}\)~GeV.  This means that the constraints from gravitino
cosmology discussed in section~\ref{sec:AstroConstraints}, and especially the
particularly stringent BBN constraints that arise when \(m_{3/2}\lesssim10^{5}\)~GeV, are
of genuine concern.  Even for heavier heavy gravitino masses \(10^{5}\)~GeV~\(\lesssim
m_{3/2} \lesssim 10^{8}\)~GeV, we must require either that \(m_{\mathit{LSP}}\) is light
enough that the naive reheating temperature bound permits a reheating temperature above
\(10^{10}\)~GeV (see figure~\ref{fig:OmegaTot}), or else that the ratio
\(m_{3/2}/m_{\mathit{LSP}}\) be large enough that LSP annihilations are effective
(figure~\ref{fig:OmegaAnn}).  Barring some auxiliary mechanism (such as resonant
leptogenesis) for increasing baryon number generation, then Dirac leptogenesis indeed
appears to require heavy gravitinos.
However, when \(m_{3/2}\) is sufficiently large that \(M_{\Phi_{1}}>10^{10}\)~GeV is
permitted, CHDL succeeds in providing an explanation for the origin of the observed
baryon asymmetry and neutrino mixings.

\section{Satisfying the Flavor Constraints\label{sec:FlavorNumerics}}


\indent

The results in the previous section are enough to support the claim that Dirac
leptogenesis works in theories with a heavy gravitino and sfermion masses heavy enough to
evade the flavor violation constraints in~\ref{eq:CurrentExpConstraints}.  We now expand
our analysis to the case where the sparticle spectrum is light.  Doing so greatly
increases the number of parameters in our theory, however: the flavor violation rates
in~(\ref{eq:ViolationRate}) depend not only on the CHDL model parameters but also on the
gaugino masses $M_1$ and $M_2$, the Higgs mass parameter \(\mu\), the ratio of Higgs VEVs
\(\tan\beta\), and the soft masses for the sleptons in the manner discussed in
appendix~\ref{app:Looplitudes}. In order to obtain precise predictions for the rates,
these need to be specified. In our analysis, we choose the values \(M_{1}=160\)~GeV,
\(M_{2}=220\)~GeV, \(\mu=260\)~GeV, and \(\tan\beta=10\). As for the slepton soft masses
we will assume a common scale $m_{s}=200$~GeV for them.  We will also examine what effect
varying \(m_{s}\) and \(\tan\beta\) from these chosen values has on
\(\mathit{BR}(\mu\rightarrow e\gamma)\) and \(\mathit{BR}(\tau\rightarrow \mu\gamma)\).
We will assume that the high scale at which soft masses are universal is
\(M=2\times10^{16}\)~GeV, though the results are not particularly sensitive to this
choice.  We will continue to work in the large-$\langle F_{\chi}\rangle$ regime, and
since the flavor violation rate is sensitive to $\langle F_{\chi}\rangle$, through the
left-right slepton mixing term~(\ref{eq:MLR}) (whereas baryon number is not), we need to
specify it explicitly; we choose $\sqrt{\langle F_\chi\rangle}=10^7$~GeV.

\begin{figure}[ht!]
\begin{center}
  \includegraphics[height=10.26cm,width=7.0cm,trim=0cm .65cm 0cm 0cm]{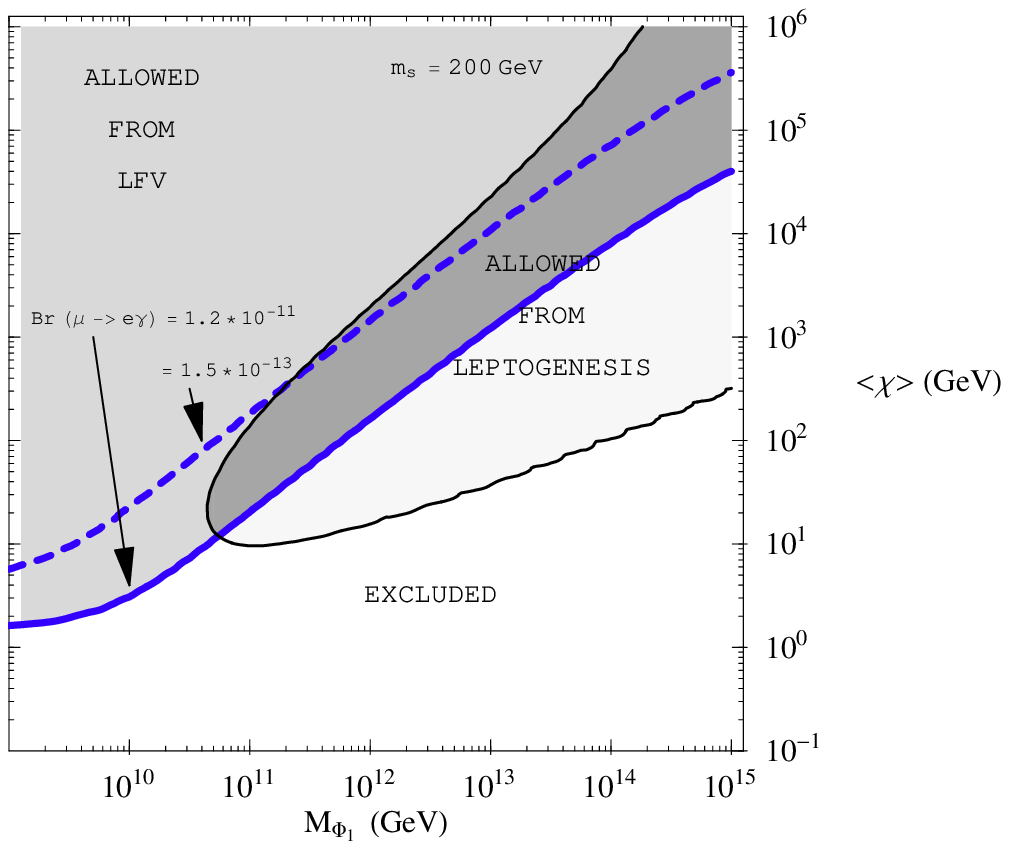}
\hspace{.3cm}
\includegraphics[height=10.07cm,width=7.0cm,trim=0cm 0cm 0cm 0cm]{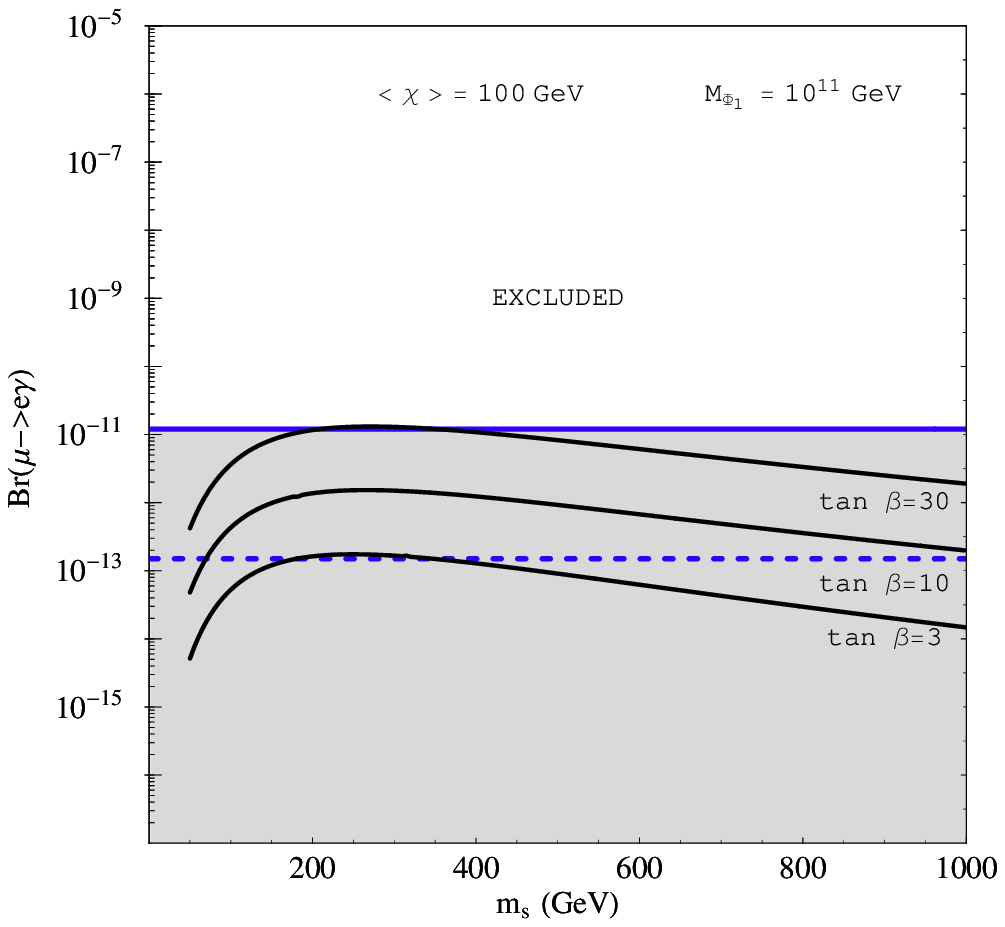}
\end{center}
\vspace{-1cm} \caption{Exclusion plots~\cite{Thomas:2006gr} combining constraints from
  both leptogenesis and flavor violation in the process
  $\mu\rightarrow e\gamma$.  The left-hand panel shows exclusion
  contours in $M_{\Phi_{1}}$-$\langle\chi\rangle$ space for a
  universal scalar soft mass $m_{s}=200$~GeV, with $\tan\beta=10$; the right hand panel
  shows the variation of the branching ratio
  $\mathit{BR}(\mu\rightarrow e\gamma)$ with respect to $m_{s}$ using
  $\tan\beta=3, 10$ and $30$.
  In both plots, we have
  taken $M_{1}=160$~GeV, $M_{2}=220$~GeV, and
  $\mu=260$~GeV.  We have also assigned the $\chi$ superfield an
  $F$-term VEV $\sqrt{\langle
    F_\chi\rangle}=10^7$ GeV.  Such a large VEV results in large trilinear couplings
  between Higgs fields and sneutrinos and therefore induces
  potentially sizeable mixings between left-handed and right-handed
  sneutrinos after electroweak symmetry breaking.  In each plot, the
  thick solid contours represent the current experimental bound on the
  branching fraction~(\ref{eq:CurrentExpConstraints}); the dashed
  lines represent the expected future experimental bound of
  $1.5\times10^{-13}$ from MEG.
  The thin solid contour in the left-hand panel delimits the region
  allowed by leptogenesis constraints.\label{fig:ExclusionPlot}}
\end{figure}

\indent The results of our calculation are displayed in figure~\ref{fig:ExclusionPlot}.
In the left panel, we show exclusion contours in \(M_{\Phi_{1}}\)-\(\langle\chi\rangle\)
space for $m_{s}=200$~GeV. The areas below and to the right of the lower contour (the
white region) are excluded by the experimental bound from \(\mu\rightarrow e\gamma\). The
contours associated with \(\tau\rightarrow\mu\gamma\) are far weaker and do not eliminate
any otherwise acceptable region of parameter space, and as such we do not display them
here.  We also include contours demarcating the region wherein baryogenesis can succeed.
The exclusion contours take the form anticipated by
equations~(\ref{eq:ContourEquationLL}) and~(\ref{eq:ContourEquationLR}), and things
become problematic when \(M_{\Phi_1}\) is large and when \(\langle\chi\rangle\) is small,
as predicted.  As discussed in section~\ref{sec:FlavorViolation}, the lower contour
represents the result of left-right sneutrino mixing contributions \(\delta m_{LR}^2\)
and is primarily controlled by $\langle F_{\chi}\rangle$.  When $\langle F_{\chi}\rangle$
is increased, the lower contour in figure~\ref{fig:ExclusionPlot} is raised, and we find
that when \(\sqrt{\langle F_{\chi}\rangle}\gtrsim 10^{9}\)~GeV, the entirety of parameter
space is excluded.

\indent In the right panel, we also show how varying the universal scalar mass affects
the branching ratio for \(\mu\rightarrow e\gamma\), which reaches a maximum when
\(m_{s}\) is around the weak scale.  This is to be expected: when \(m_{s}\) is much
larger than the weak scale both the slepton mass-squared eigenvalues and the
flavor-violating terms scale like $m_{s}^{2}$ and the sneutrino and charged slepton
mixing matrices asymptote to a constant value, while the branching ratio is still
suppressed by the masses running in the loop; as $m_{s}$ decreases below the weak scale,
\(\delta m_{LL}^{2}\) and \(\delta m_{LL}^{2}\) go to zero and the slepton masses are
dominated by flavor diagonal electroweak contributions. We also observe that, as in the
supersymmetric see-saw case~\cite{Hisano:1995cp,Petcov:2005jh}, the flavor violation rate
is quite sensitive to \(\tan\beta\).  The lepton-flavor-violating processes depicted in
figure~\ref{fig:FeynMixingsFull} involve a \(\tan\beta\)-dependent chirality flip, and
hence are augmented when \(\tan\beta\) is large.

\indent The important result we learn from figure~\ref{fig:ExclusionPlot} is that Dirac
leptogenesis and the satisfaction of flavor violation constraints associated with
weak-scale soft masses are indeed compatible---a conclusion that would not have been
obvious a priori.  There is substantial overlap between the region of parameter space
within which CHDL is capable of reproducing the observed baryon asymmetry of the universe
and the region in which lepton-sector flavor violation accords with experimental bounds.
The improvement of these bounds from the next generation of lepton flavor-violation
experiments will either confirm that some beyond-the-standard-model source of flavor
violation exists or else rule out a substantial amount of the available parameter space.
Thus these experiments will serve as an important check on Dirac leptogenesis models with
slepton masses near the weak scale. We again emphasize that these tensions can be relaxed
by elevating the universal sfermion mass parameter $m_{s}$.  In theories like split
supersymmetry where $m_{s}$ is extremely large, flavor violation constraints can be
ignored entirely.

\section{The Resonant Escape\label{sec:Resonating}}

\indent It is clear at this point that the primary model tension in Dirac leptogenesis is
a conflict between the ceiling gravitino cosmology places on the reheating temperature
after inflation and the condition that baryogenesis reproduce the WMAP \(\eta\).  In
CHDL, this tension is overcome by arranging a large hierarchy among the Yukawa couplings
to different sets of \(\Phi\) and \(\overline{\Phi}\).  Couplings to the second lightest
set, which are crucial to baryogenesis but otherwise have little physical effect, are
large enough to produce substantial baryon number while couplings to the first remain
small so as to avoid violating the out-of-equilibrium
condition~(\ref{eq:EquilibrationBound}), which forces \(\lambda_{1\alpha}\) and
\(h_{1\alpha}\) down when \(M_{\Phi_1}\) is itself required to be small.  There is,
however, a second possibility for avoiding this bound.  It can be seen from the
definition of the decay asymmetry \(\epsilon\) in equation~(\ref{eq:epsilon}) that if
\(M_{\Phi_{1}}\approx M_{\Phi_{2}}\) (in other words, if \(\delta\approx1\)), a resonance
condition in is obtained.  The perturbation theory we have used in calculating equation
(\ref{eq:epsilon}) is good as long as the separation of \(M_{\Phi_{1}}\) and
\(M_{\Phi_{2}}\) is substantially greater than the value of the off-diagonal elements
\(M_{ij}\) in the \(\Phi-\overline{\Phi}\) mass mixing matrix induced at the one-loop
level:
\begin{equation}
  M_{ij}\sim g^{\ast} g' (M_{\Phi_{1}}+M_{\Phi_{1}})I_{ij},
\end{equation}
where \(g_{i}\) and \(g'_{j}\) represent the appropriate \(\lambda\) and \(h\), summed
over the fermion family index, and \(I_{ij}\) is a numerical factor on the order of
\(1/16\pi^{2}\) from the loop integral. Thus for small \(\lambda\) and \(h\), \(\delta\)
can be set very close to one and \(M_{\phi_{1}}\) and \(M_{\phi_{2}}\) may be very nearly
degenerate.  This allows for the possibility of resonant leptogenesis, as has been done
in the Majorana leptogenesis
case~\cite{Flanz:1996fb,Covi:1996fm,Pilaftsis:1997dr,Pilaftsis:1997jf}, which can be
invoked to generate a large baryon number in cases where the bounds on \(T_{R}\) are more
severe.  This possibility could potentially allow Dirac leptogenesis to work in scenarios
such as PeV-scale supersymmetry, where is is otherwise all but ruled out.

\chapter{EXTENSIONS\ OF\ THE\ MODEL\label{ch:extensions}}

\indent Thus far, we have shown that Dirac leptogenesis is a self-consistent model and
that there exists a simple version of that model, CHDL, which can simultaneously
reproduce both the observed baryon asymmetry of the universe and the observed neutrino
spectrum without violating any additional constraints from astrophysical observations,
flavor-violation experiments, etc.  At this point, we turn to outline some of the
theoretical motivations for CHDL and to examine some interesting possible extensions of
the model, including a potential connection between the Dirac leptogenesis superpotential
and the origin of the Higgs \(\mu\)-term, a method for canceling \(U(1)_{N}\) anomalies,
and the possibility for right-handed sneutrino dark matter.

\section{Origin of the $\mu$-term\label{sec:muterm}}

\indent

One of the advantages of Dirac leptogenesis is that unlike in Majorana leptogenesis,
where they are essentially determined by the leptogenesis scale, neutrino masses are set
not only by the masses $M_{\Phi_{i}}$ of the heavy fields, but also by the scalar VEV
$\langle\chi\rangle$.  This leaves us more freedom to choose the mass scale of the heavy
fields, but can also be thought of as a drawback since we have lost some predictiveness.
For this reason, it would be interesting if some of these additional intermediate scales
could be tied to other intermediate scales that arise in supersymmetric models.  One
interesting possibility involves making use of the \(U(1)_{N}\) symmetry to link the
dynamics of the $\chi$ field to the origin of the supersymmetric Higgs mass parameter
$\mu$.  A simple way of doing this is to posit an interaction coupling the up- and
down-type Higgs fields with \(\chi\) and modifying
equation~(\ref{eq:EffDiracLepSuperpotential}) to
\begin{equation}
  \mathcal{W}_{\mathit{eff}}\ni
  \frac{\lambda_{i\alpha} h_{i\beta}^{\ast}}{M_{\Phi_{i}}}\chi L_{\beta} H_{u}
  N_{\alpha}
  +Y_\chi\ \chi H_{u} H_{d},
\end{equation}
where $Y_\chi$ is an \(\mathcal{O}(1)\) coupling constant.  The Higgs superfields are now
necessarily charged under the new hidden sector symmetry, and therefore the charge
assignments given in table~\ref{tab:U1Charges} must be revised.  Furthermore, the rest of
the standard model quark and lepton fields which couple to the Higgs fields must also be
assigned nontrivial charges under this symmetry.  This makes the task of arranging
anomaly cancelation, arranging for gauge couplings to unify at some high scale, etc.\
vastly more difficult.  In this situation, we have $\mu=Y_\chi \langle\chi\rangle$ and
any observation or limits on Higgsino dark matter would directly constrain the VEV
$\langle \chi\rangle$. (Split) supersymmetry might also give us some ideas as to how to
relate the heavy fields $\Phi$ to the SUSY breaking scale.

\section{Theoretical Motivations for CHDL\label{sec:FlavonExtension}}

\indent

While the combination of small \(\delta\) and antisymmetric \(h\) and \(\lambda\) which
constitutes CHDL may yield the correct neutrino phenomenology, we have not offered any
explanation of how such a situation might arise.  Ideally, we would like to have a
theoretical framework that would provide both the form of the superpotential required for
Dirac leptogenesis (\ref{eq:DiracLepSuperpotential}) and the necessary flavor structure.
We would like to take a moment and indicate one interesting possibility in which simple
assumptions enable us to reproduce the needed conditions, while at the same time reducing
the number of new free parameters that must be introduced.

In many grand unified theories and models with non-Abelian flavor symmetries, the Yukawa
matrices (and in general the flavor interactions) can be symmetric, antisymmetric, or
both (see for example~\cite{Davidson:1980sx,King:2004tx,Froggatt:1978nt} and references
therein).  Operating in this paradigm, let us assume that the SM left-handed leptons have
the same flavor charge as the heavy fields $\bar{\Phi}$, and the SM right-handed charged
leptons have same flavor charge as the fields $\Phi$. Upon breaking of the flavor
symmetry, let us assume that the charged lepton Yukawa matrix is symmetric due to a
symmetric flavon VEV configuration $\langle S_{\alpha\beta}\rangle$. The corresponding
effective superpotential is
\begin{equation}
  W\supset Y_l\ {\langle S_{\alpha
  \beta}\rangle \over M_F} H_d L_\alpha e_\beta + Y_{\Phi}\ \langle S_{\alpha \beta}\rangle
  \bar{\Phi}_\alpha\Phi_\beta {\rm +h.c.}
\end{equation}
where $Y_l$ and $Y_{\Phi}$ are dimensionless couplings, and $\alpha$ and $\beta$ are
flavor indices. The charged lepton Yukawa matrix $(y_{{}_l})_{\alpha\beta}$ and the mass
matrix $(M_{\Phi})_{\alpha\beta}$ of the heavy fields $\Phi$ would be both symmetric and
proportional
\begin{equation}
  (M_{\Phi})_{\alpha\beta} = M_F {Y_{\Phi}\over Y_l}\
  (y_{{}_l})_{\alpha\beta}\label{eq:MandYProp}
\end{equation}
This specific structure predicts exactly the mass spectrum for the $\phi$ fields in terms
of the flavor scale $M_F$, which is at the origin of the intermediate scale required for
successful thermal leptogenesis.
\begin{eqnarray}
  M_{\Phi_1}=m_e {M_F\over v},
  \hspace{.5cm} M_{\Phi_2}=m_\mu {M_F\over v} \hspace{.2cm}{\rm and}\hspace{.2cm}
  M_{\Phi_3}=m_\tau {M_F\over v} \hspace{.5cm}
\end{eqnarray}
If the flavor scale is of the order of some GUT scale $M_F\sim 10^{16}$~GeV, we would
then expect $M_{\Phi_1}\sim 10^{11}$~GeV, with $\delta=M_{\Phi_1}/M_{\Phi_2}=m_e/m_\mu =
4.83\times 10^{-3}$ being a small parameter.  Because this ratio is quite small and
\(\epsilon\) is approximately proportional to \(\delta\) for small \(\delta\), one might
worry that this might doom baryogenesis. However, as indicated in the left panel of
figure~\ref{fig:NeutrinoMassPlots}, \(a_{3}^{\mathit{max}}(m_{e}/m_{\mu})\approx 95\), so
that a hierarchy between couplings to different sets of \(\Phi\) and \(\overline{\Phi}\)
is permitted, and the result is that \(\epsilon\) is only suppressed by an
\(\mathcal{O}(1)\) numerical factor.  This is corroborated by the results of our
numerical analysis, displayed in the left panel of
figure~\ref{fig:EtaContoursCCwithMRatios} and explains why we chose to examine the
\(\delta=4.83\times 10^{-3}\) case.

\indent Another related possibility that can help reduce the proliferation of scales
(\(M_{\Phi_1}\), $M_{F}$, etc.) in the model is to link the supersymmetry-breaking scale
$M_{\mathit{SUSY}}$ to other physical scales in the theory by coupling the spurion field
\(X\) responsible for supersymmetry breaking to the \(\Phi_{i}\) and
\(\overline{\Phi}_{i}\) fields.  If the spurion field is charged under the same symmetry
(e.g.\ the $U(1)_N$ introduced in section~\ref{sec:WandFields}), responsible for
dictating the form of the superpotential in~(\ref{eq:DiracLepSuperpotential}), one could
arrange that
\begin{equation}
  W\supset {S_{\alpha \beta}\over M_F}\left(Y_l\ H_d
  L_\alpha e_\beta + Y_{\Phi}\ X\ \bar{\Phi}_\alpha\Phi_\beta\right) {\rm +h.c.}
\end{equation}
where again $Y_l$ and $Y_{\Phi}$ are dimensionless couplings. Supersymmetry breaking
effects can provide the field $X$ with a VEV $\langle A_{X}\rangle\sim \left(m_{3/2}
M_{Pl}\right)^{1/2}$, and upon flavor symmetry breaking we could get the effective
superpotential
\begin{equation}
  W_{\mathit{eff}}\supset (y_{{}_l})_{\alpha\beta}\
  \left(H_d L_\alpha e_\beta +\sqrt{m_{3/2} M_{Pl}}\ \bar{\Phi}_\alpha\Phi_\beta\right)
  {\rm +h.c.}
\end{equation}
where we have assumed that the original constants $Y_l\sim Y_\Phi$. In this situation,
for example with $m_{3/2}\sim 10^{10}$~GeV (a value sufficiently high as to avoid any
model tensions associated with gravitino cosmology), the mass of the lightest $\Phi$
field would be $M_{\Phi_1}\sim10^{11}$ GeV.

\indent Now, let us also assume that the flavon VEV $ \langle S_{\alpha\beta}\rangle$ is
symmetric and that the coupling $\bar{\Phi} L \chi$ becomes antisymmetric upon flavor
breaking, i.e. the superpotential can be written as
\begin{equation}
  W\supset
  (y_{{}_l})^{sym}_{\alpha\beta} H_d L_\alpha e_\beta\ + (M_\Phi)^{sym}_{\alpha \beta}
  \bar{\Phi}_\alpha\Phi_\beta\ +\lambda_{\alpha\beta}N_\alpha \Phi_\beta H_u +
  h_{\alpha\beta}^{antisym} \bar{\Phi}_\alpha L_\beta \chi
\end{equation}
where we have $\lambda_{\alpha\beta}\equiv \langle A^N_{\alpha\beta}\rangle$ and
$h_{\alpha\beta}^{anti}\equiv \langle A^\chi_{\alpha\beta}\rangle$, with the flavon $A^N$
acquiring an antisymmetric VEV configuration and with the VEV configuration of $A^\chi$
being arbitrary in flavor space.  As alluded to above, this assumption is well-motivated
from GUT considerations.  While the flavor structure of $A^\chi$ is not important for our
present purposes, in a specific model of flavor it would likely end up being some linear
combination of symmetric and antisymmetric VEV configurations.

\indent Let us see what the implications of these ingredients are for our model. In
general, the charged lepton Yukawa matrix \((y_{{}_l})_{\alpha\beta}\) can be
diagonalized by a biunitary transformation of the type
\begin{equation}
  U^{\dagger}_{}y_{{}_l} V_{}=y_{{}_l}^{\mathit{(diag)}},
\end{equation}
but when \(y_{l}\) is symmetric, this biunitary transformation takes the simpler form
\begin{equation}
    U^{T}_{}y_{{}_l} U_{}=y_{{}_l}^{\mathit{(diag)}}.
\end{equation}
If \((M_{\Phi})\propto  (y_{{}_l})\), as in the setup described above (see
equation~(\ref{eq:MandYProp})), then \((M_{\Phi})\) will be diagonalized by the same
transformation. When the mass matrices for the charged leptons and the
\(\phi-\overline{\phi}\) system are simultaneously diagonalized (that is, when we go to
the charged lepton basis), \(\lambda\) and \(h\) transform as
\begin{eqnarray}
  \lambda'_{}=U^{T}_{}\lambda_{}U_{} &\hspace{.5cm} &
  {h^{'\ }}^{anti}_{}=U^{T}_{}h^{anti}_{}U_{}.
  \label{eq:UTtransforming}
\end{eqnarray}
Transformations of this type preserve the antisymmetry of \(h\) (and $\lambda$ if also
antisymmetric); thus in the charged lepton basis the matrix $h$ remains antisymmetric and
$(M_{\Phi})$ is real and diagonal; this reproduces the neutrino mass matrix structure
given in~(\ref{eq:DeltasInFirstRow}), in which the diagonal elements of $h$ are zero and
it is the smallness of $\delta$ that is responsible for the large mixing angles observed
in the lepton mixing matrix $U_{\mathit{MNS}}$.

\indent From a purely structural point of view, it will be noted that any matrix of the
form
\begin{equation}
  (M_{\Phi})=Ay_{l}+BI_{3\times3},
\end{equation}
where $A$ and $B$ are arbitrary constants and \(I_{3\times3}\) it the \(3\times3\)
identity matrix, can be diagonalized along with $y_{l}$.  Thus \((M_{\Phi})\propto
y_{l}\) is not required for the transformation rules in~(\ref{eq:UTtransforming}) to
hold.  It follows from this that \(M_{\Phi}\) may receive arbitrary diagonal
contributions and \(\delta=M_{\Phi_{1}}/M_{\Phi_{2}}=m_{e}/m_{\mu}\) is not required:
\(\delta\) can in principle take any value (as long as it is consistent with the
observational bounds on the neutrino spectrum).  This gives us a great deal more freedom
to adjust model parameters without sacrificing the theoretical motivations we have
developed here.

\section{Sneutrino Dark Matter\label{sec:sneutrinoCDM}}


\subsection{Dark Matter in Dirac Leptogenesis}

\indent

In order to be a viable dark matter candidate, the LSP must be stable (which follows
automatically from \(R\)-parity conservation) and interact only weakly with the standard
model fields.  The most obvious choice is the lightest neutralino, \(\tilde{N}_{1}\), and
this choice is perfectly compatible with Dirac leptogenesis.  The choice of model
parameters outlined in section~\ref{sec:FlavorNumerics}, for example, yields a stable,
predominately bino LSP with a mass of around 160~GeV.  In the context of PeV-scale
supersymmetry or split supersymmetry, little changes except for that the LSP mass may be
quite large, potentially facilitating the detection of energetic photons from cold dark
matter annihilation at the galactic center by the next generation of Cherenkhov
telescopes~\cite{Ullio:2001qk,Masiero:2004ft,Arvanitaki:2004df,Thomas:2005te}.  We
explore this possibility further in chapter~\ref{ch:PhotonsFromCDM}.

\indent Dirac leptogenesis does, however, offer a new potential candidate for LSP: the
lightest right-handed sneutrino \(\tilde{\nu_{R_{1}}}\).  Unlike in Majorana leptogenesis
scenarios, where right-handed sneutrino masses receive a supersymmetry-respecting
contribution on the order of the leptogenesis scale, here they acquire mass only through
soft terms, and since the \(\tilde{\nu_{R}}\) are still singlets under the Standard Model
\(SU(3)\times SU(2)\times U(1)_{Y}\) gauge group, it is certainly possible that their
soft masses are significantly smaller than---and perhaps of a different origin entirely
from---those of the SM squarks and sleptons.  Obtaining the observed value of
\(\Omega_{\mathit{CDM}}\)~(\ref{eq:WMAPBounds}) with a pure right-handed sneutrino LSP is
somewhat difficult, however, owing to the fact that the particle's interactions are so
weak that annihilations are insufficient in reducing its relic density to the appropriate
level.  One solution to this problem is to make the LSP a mixture of left- and
right-handed sneutrinos~\cite{Hall:1997ah,Arkani-Hamed:2000bq} with mixing angle
\(\theta\). The \(SU(2)\) gauge interactions of the \(\tilde{\nu}_{L}\) then permit the
particles to annihilate efficiently enough to reduce \(\Omega_{\mathit{LSP}}\) to an
acceptable level, but (since the annihilation rate is suppressed by \(\sin^4\theta\)) not
too efficiently. Here we will take another approach: we will gauge the global
\(U(1)_{N}\) symmetry we introduced in order to forbid Majorana masses for neutrinos and
allow \(\tilde{\nu_{R_{1}}}\) to annihilate down to an acceptable level via their
interactions with \(U(1)_{N}\) gauge bosons and gauginos.

\indent Since neutralino dark matter still works in the context of Dirac leptogenesis,
investigating the viability of another, more exotic CDM candidate might appear a somewhat
frivolous endeavor.  However, there is a compelling reason why sneutrino dark matter is
of particular interest.  It is a puzzling but well-documented fact about our universe
that the dark matter abundance \(\Omega_{\mathit{CDM}}\)~(\ref{eq:WMAPBounds}) and the
abundance of baryonic matter \(\Omega_{b}\)~(\ref{eq:OmegaBaryon}) are roughly the same
order, and in the standard picture of neutralino dark matter there is no known reason why
this should be the case. However, in situations where an asymmetry develops in some
globally conserved quantum number under which the LSP is charged~\cite{Hooper:2004dc}, it
is possible to link these two quantities.  Consider for a moment an effective theory
containing both visible and hidden sectors, coupled only by dynamics at some high scale
whose precise nature is unimportant.  As the universe evolves down from the scale of the
unifying dynamics, an asymmetry in this conserved quantum number develops between the two
sectors.  Suppose the hidden sector contains a number of light fields charged under the
relevant global symmetry, each with charge \(Q_{i}\) and number density \(n_{i}\), and
their antiparticles, with charge \(-Q_{i}\) and number density \(\overline{n}_{i}\). The
respective particle asymmetries \(A_{\mathit{vis}}\) and \(A_{\mathit{hid}}\) in the
visible and hidden sectors, normalized with respect to the entropy density of the
universe \(s\), are then
\begin{equation}
  A_{\mathit{hid}}=\sum_{i}\frac{Q_{i}(n_{i}-\overline{n}_{i})}{s}=-A_{\mathit{vis}}.
\end{equation}

\indent If the hidden sector contains a massive, stable (or at least extraordinarily
long-lived) particle \(a\) with mass \(m_a\), this particle becomes a potential dark
matter candidate. Let us assume that it also carries a sizeable fraction of this
asymmetry \(A_{a}=\alpha A_{\mathit{hid}}\) at the time of the freeze-out epoch and
consider the limit where annihilation processes are efficient and conserve \(A_{a}\).  In
this case, \(a\) and its conjugate \(a^{\ast}\) will annihilate rapidly with each other
until only one or the other remains (even if the annihilation rate is quite large, \(a\)
and \(a^{\ast}\) cannot annihilate further via such processes once either \(a\) or
\(a^{\ast}\) is entirely depleted). The particles left over, having no conjugates with
which to annihilate, become the dark matter, with a relic abundance
\begin{equation}
  \Omega_{\subCDM}=\frac{m_{a}|A_{a}|s_{0}}{\rho_{\mathit{crit,0}}}=
  2.236\times10^{10}m_{a}|A_{a}|\label{eq:OmegaFromAsymmetry}
\end{equation}
proportional to the asymmetry.  An asymmetry on the order
\(|A_{a}|\sim10^{-12}-10^{-10}\) (depending on \(m_{a}\)) will thus replicate adequately
the observed dark matter abundance.  Since Majorana neutralinos are their own
antiparticles, they must have vanishing \(U(1)\) charge, and thus cannot be used to link
\(\Omega_{b}\) and \(\Omega_{\mathit{CDM}}\) in this manner.  In contrast, when effective
$A$-terms mixing left- and right-handed sneutrinos (such as those engendered by \(\langle
F_{\chi}\rangle\)) are small, a right-handed sneutrino LSP in Dirac leptogenesis fits the
bill perfectly: it is stable; it resides in a hidden sector essentially decoupled from
the visible sector fields until well after the freeze-out epoch; and it is charged under
the globally conserved quantum number \(B-L\), under which the visible sector fields are
also charged.  The \(B-L\) asymmetry between the hidden and visible sectors created
during the decay of the heavy \(\Phi_{1}\) and \(\overline{\Phi}_{1}\) superfields will
also determine the relic density of \(\tilde{\nu}_{R_{1}}\), and consequently the
mechanism responsible for baryogenesis will also be responsible for generating
\(\Omega_{\mathit{CDM}}\).

\indent The simplified scenario resulting in~(\ref{eq:OmegaFromAsymmetry}) is somewhat
difficult to realize in practice, however, due the fact that it simultaneously requires
the net rate \(\Gamma(a^{\ast}a\to X)\) (where \(X\) represents some unspecified final
state comprising light, hidden-sector particles) for $A_{a}$-conserving annihilation
processes to be efficient in reducing the dark matter abundance to a cosmologically
acceptable level and the rate \(\Gamma(aa\to X)\) for $A_{a}$-violating
processes\footnote{We assume that CP-violation can be neglected and that the rates for
the conjugate processes \(aa\rightarrow X\) and \(a^{\ast}a^{\ast}\rightarrow X^{\ast}\)
are equivalent, hence \(\Gamma_{aa}=\Gamma_{a^{\ast}a^{\ast}}\).} to be negligibly small.
The correct hierarchy of rates can be achieved in certain circumstances---in the presence
of large \(s\)-channel contributions to \(\Gamma(a^{\ast}a\to X)\) that have no
\(t\)-channel equivalent, for example---and we will discuss these possibilities when we
solve the Boltzmann equations for a right-handed sneutrino LSP.  Before we do this,
however, we turn briefly to address the modifications necessary for gauging the
\(U(1)_N\) symmetry, as this is the simplest and most convenient method of making
\(L\)-conserving sneutrino annihilations efficient.

\subsection{Extending the Hidden Sector}

\indent

In order for our theory to be consistent, we must require that our new \(U(1)_{N}\)
symmetry is non-anomalous.  With the field content listed in table~\ref{tab:U1Charges},
we find that there are no mixed $SU(2)_L^2\,U(1)_N$, $U(1)_Y^2U(1)_N$ or
$U(1)_Y\,U(1)_N^2$ anomalies, but that
\begin{eqnarray}
(\mathit{grav.})^2U(1)_N:~~3(+1) + 3(-1) + 3(+1)+ (-1) = 2\\
U(1)_N^3:~~3(-1)^3 + 3(-1)^3+3(+1)^3 + (-1)^3 = 2
\end{eqnarray}
are non-vanishing.  It has been shown~\cite{Morrissey:2005uz} that in the case where the
only non-vanishing anomalies for a hidden sector \(U(1)\) are \(U(1)^3\) and
\((\mathit{grav.})^2U(1)\) (i.e.\ mixed anomalies vanish), it is possible to cancel those
anomalies by introducing new fields charged only under that \(U(1)\).  We therefore look
for an appropriate expansion of the field content in the hidden sector that will cancel
the above anomalies.  The simplest way to cancel these anomalies would be to introduce
two additional fields with the same charges as \(\chi\), but mixing between the fermionic
components of these fields will ensure that at least one linear combination of them will
be massless.  Since these fields are necessarily odd under \(R\)-parity, this choice is
problematic for dark matter and therefore must be excluded.  The next simplest
option\footnote{Here we have assumed that all new superfields $\Xi_{i}$ have \(U(1)_{N}\)
charges (not necessarily integral) \(Q_{i}<20\).} requires four new fields $\xi_a$
($a=1,2$), $\varphi$, and $\rho$, the charge assignments for which are recorded in
table~\ref{tab:newcharges} (the reason for the \(R\)-parity assignments therein will
become evident in a moment).  Adding these fields permits us to write down two new
superpotential terms
\begin{equation}
  W\supset \tilde{\mu}\,\chi\,\rho + \zeta_a\, \chi\,\xi_a\,\varphi,
  \label{eq:wnew}
\end{equation}
where \(\tilde{\mu}\) is a supersymmetric mass term and \(\zeta_a\) ($a=1,2$) are
dimensionless couplings.

\begin{table}[t!]
 \begin{center}
  \begin{tabular}{|cccccc|}
     \hline Field & \(U(1)_{L}\) & \(U(1)_{N}\) & \(SU(2)\) & \(U(1)_{Y}\) & $P_M$ \\ \hline
     \(\xi_a\) & 0 & -4 & \(\mathbf{1}\) & 0 & +1 \\
     \(\varphi\) & 0 & 5 & \(\mathbf{1}\) & 0 & +1 \\
     \(\rho\) & 0 & 1 & \(\mathbf{1}\) & 0 & +1 \\ \hline
   \end{tabular}
 \end{center}
 \caption{This table shows the charge assignments for minimal set of additional superfields
 necessary to cancel the $U(1)_{N}^3$ and $(\mathit{grav.})^2U(1)_N$ anomalies of the
 fields in table~\ref{tab:U1Charges}.\label{tab:newcharges}}
\end{table}

\indent Since our aim is to arrange for \(\tilde{\nu}_{R_1}\) to be the LSP, we must make
sure that none of the new \(R\)-parity-odd fields in the hidden sector ends up massless
and that the scalar component of \(\chi\) can still acquire a VEV on the appropriate
order to yield a realistic set of neutrino masses.

The scalar potential in the hidden sector, which contains both an \(F\)-term contribution
from~(\ref{eq:wnew})
\begin{eqnarray}
  V_F &=& |\tilde{\mu}\rho+\zeta_a\xi_a\varphi|^2 +
  |\tilde{\mu}\chi|^2 + |\zeta_a\chi\xi_a|^2
  +(|\zeta_1|^2+|\zeta_2|^2)|\chi\varphi|^2 \phantom{\frac{X}{Y}}
  \nnmb\\
  &=& |\tilde{\mu}|^2|\rho|^2 + |\tilde{\mu}|^2|\chi|^2
  +(\tilde{\mu}^*\zeta_a\xi_a\varphi\rho^* + h.c.)\nnmb\\
  &&~~~~~+ |\zeta_a|^2|\xi_a\varphi|^2 + |\zeta_a|^2|\xi_a\chi|^2 +
  (|\zeta_1|^2+|\zeta_2|^2)|\chi\varphi|^2
\end{eqnarray}
and a $D$-term contribution from the new $U(1)_{N}$ gauge interaction given by
\begin{equation}
  V_D = \frac{g^2}{2}\left(|\tilde{n}|^2-|\tilde{\chi}|^2 +|\rho|^2 -
  4|\xi_1|^2 - 4|\xi_2|^2 + 5|\varphi|^2\right)^2,
\end{equation}
both of which serve to stabilize the fields.  In addition, a contribution \(V_{S}\) will
arise from soft supersymmetry-breaking Lagrangian.  As for the fermions (which, as they
are singlets under the Standard Model \(SU(3)\times SU(2)\times U(1)_Y\), will be
referred to as ``neutralinos'' from this point forward), their mass matrix, in the
$\{\tilde{Z}',\tilde{\chi},\tilde{\rho}, \tilde{\xi}_1,\tilde{\xi_2},\tilde{\varphi}\}$
basis, is
\begin{equation} \mathcal{M} = \left({\small
\begin{array}{cccccc}
M'&-\sqrt{2}g\,\cvev&\sqrt{2}g\,\left<\rho\right>& -4\sqrt{2}g\,\left<\xi_1\right>&
-4\sqrt{2}g\,\left<\xi_2\right>&5\sqrt{2}g\,\left<\varphi\right>\\
\cdot&0&\tilde{\mu}&\zeta_1\left<\varphi\right>& \zeta_2\left<\varphi\right>&
\zeta_1\left<\xi_1\right>+\zeta_2\left<\xi_2\right>\\
\cdot&\cdot&0&0&0&0\\
\cdot&\cdot&\cdot&0&0&
\zeta_1\left<\chi\right>\\
\cdot&\cdot&\cdot&\cdot&0&
\zeta_2\left<\chi\right>\\
\cdot&\cdot&\cdot&\cdot&\cdot&0
\end{array}}
\right).\label{eq:HSneutralinoMassMat}
\end{equation}
As long as the scalar components of $\chi$ and $\xi_1$ receive nonzero VEVs, all the
physical mass eigenstates resulting from this matrix will be massive, as desired.  This
can be engineered by introducing tachyonic soft squared-masses for these fields in
\(V_{S}\).  Assuming the other scalars \(\xi_2\), \(\varphi\), and \(\rho\) receive
large, positive squared-masses the effective (tree-level) potential for the remaining
fields becomes
\begin{equation}
  V = (|\tilde{\mu}|^2+m_{\chi}^2)|\chi|^2 +
  (|\tilde{\mu}|^2+m_{\xi_1}^2)|\xi_1|^2 + |\zeta_1|^2|\xi_1|^2|\chi|^2 +
  \frac{g^2}{2}\left(|\chi|^2+4|\xi_1|^2\right)^2.\label{eq:lowenergyHSpot}
\end{equation}
It should be noted that since the \(U(1)_{N}\) charges of $\chi$ and $\xi_1$ are of the
same sign, there are no problems with $D$-flat directions.

\indent Minimizing the potential in equation~(\ref{eq:lowenergyHSpot}), we find that in
terms of the quantities
\begin{equation}
\tilde{m}_{\chi}^2 \equiv -(|\tilde{\mu}|^2+m_{\chi}^2),~~~~~ \tilde{m}^2_{\xi_1} \equiv
-(|\tilde{\mu}|^2+m_{\xi_1}^2),\label{eq:mchisqdmx1sqd}
\end{equation}
the expectation values for \(\xi_1\) and \(\chi\) are given by
\begin{eqnarray}
\left<|\xi_1|^2\right> &=&
\frac{(4\,\tilde{m}_{\chi}^2-\tilde{m}_{\xi_1}^2)}{3|\zeta_1|^2}\nonumber\\
\left<|\chi|^2\right> &=& \frac{1}{g^2}\left[ \tilde{m}_{\chi}^2
-\frac{(g^2+|\zeta_1|^2)}{3|\zeta_1|^2}(4\,\tilde{m}_{\chi}^2-\tilde{m}_{\xi_1}^2)
\right].
\end{eqnarray}
Equation~\ref{eq:mchisqdmx1sqd} implies that in order to avoid having to introduce
fine-tunings among the hidden sector mass parameters \(\tilde{\mu}\), $m_{\chi}$,
and~\(m_{\xi_1}\), these parameters ought to be of roughly the same magnitude as the
physical scales $\langle\chi\rangle$ and $\langle\xi_1\rangle$ (if \(\tilde{\mu}\) were
small or vanishing, some of the hidden-sector neutralinos would be too light). In order
for both VEVs to be positive, we must have
\begin{equation}
\left(\frac{4\alpha-1}{\alpha}\right) ~~<~~
\frac{\tilde{m}_{\xi_1}^2}{\tilde{m}_{\chi}^2} ~~<~~ 4,\label{eq:alphainequality}
\end{equation}
where $\alpha = (g^2+|\zeta_1|^2)/3|\zeta_1|^2 \geq 1/3$.  Both of these inequalities
must be satisfied in order for $\xi_1$ and $\chi$ to receive
VEVs.\footnote{Equation~(\ref{eq:alphainequality}) also indicates why the matter parity
assignments given in table~\ref{tab:newcharges} were chosen as they were.  If $\xi_1$ and
\(\varphi\) were taken to be odd under matter parity, the coupling constant \(\zeta_{1}\)
in~(\ref{eq:wnew}) would vanish, sending \(\alpha\to\infty\).}  As for $Z'$ vector boson
associated with the gauged \(U(1)_{N}\) symmetry, when the symmetry is broken by the VEVs
of \(\xi_1\) and \(\chi\), it acquires a mass
\begin{equation}
M_{Z'}^2 = 2g^2\left(\cvev^2 + 16\left<\xi_1\right>^2\right).\label{eq:Zprimemass}
\end{equation}
Thus all fields in the hidden sector acquire masses on the order of the scales
$\langle\chi\rangle$ and $\langle\xi_a\rangle$ or larger, and the lightest right-handed
sneutrino remains a viable LSP candidate. In fact, it should be pointed out that gauging
\(U(1)_{N}\) introduces new $D$-term contributions to the physical mass of the
right-handed sneutrino which serve to drive down $m_{\tilde{\nu}_{R}}^2$. The total
result, including soft terms, but assuming that left-right mixing through effective
$A$-terms is negligible, is
\begin{equation}
  m_{\tilde{\nu}_{R}}^2 = m_{\tilde{\nu}_{R},\mathit{soft}}^2 - g^2\left( \left<|\chi|^2\right> +
  4\,\left<|\xi_1|^2\right> + 4\,\left<|\xi_2|^2\right> - \left<|\rho|^2\right> -
  5\,\left<|\varphi|^2\right>\right),
\end{equation}
and since \(m_{\tilde{\nu}_{R},\mathit{soft}}^2\) is essentially a free parameter, it can
be tuned freely so that the mass of the lightest right-handed sneutrino is small, as
desired.  It should also be pointed out that this setup results in a VEV for the scalar
field \(\chi\) without necessarily producing an $F$-term VEV \(\langle \chi\rangle\), and
hence the right-handed fields are indeed part of the hidden sector, as is required for
relating \(\Omega_{\mathit{CDM}}\) and \(\Omega_{b}\) in the manner discussed above.

\subsection{Evolution of the Dark Matter Abundance}


\indent

In order to ascertain the dark matter abundance in any nontrivial model, it is necessary
to solve the coupled system of Boltzmann equations that govern the evolution of
\(\tilde{\nu}_{R_1}\) and \(\tilde{\nu}_{R_1}^{\ast}\) during the freeze-out epoch.  The
two-to-two processes which yield the leading contribution to dark matter annihilation
include both processes of the form \(\tilde{\nu}_{R_1}\tilde{\nu}_{R_1}^{\ast}\rightarrow
X\), which contribute to washout of the dark matter density but do not alter
\(L_{\tilde{\nu}_R}\) (which is equivalent to $(B-L)_{\tilde{\nu}_R}$, since electroweak
sphalerons do not affect the fields in the hidden sector and hence \(B\) and \(L\) are
separately conserved there), and those of the form
\(\tilde{\nu}_{R_1}\tilde{\nu}_{R_1}\rightarrow X\), which contribute to the depletion of
both \(\Omega_{\mathit{CDM}}\) and \((B-L)_{\tilde{\nu}_R}\). Let us call the net rate
for the former processes \(\Gamma_{\tilde{\nu}\tilde{\nu}^{\ast}}\) and the rate for the
latter \(\Gamma_{\tilde{\nu}\tilde{\nu}}\).  The Boltzmann equations can be written
either in terms of \(\tilde{\nu}_{R_1}\) and \(\tilde{\nu}_{R_1}^{\ast}\) or,
equivalently, for the dark matter abundance
\(Y_{\DM}=Y_{\tilde{\nu}_R}+Y_{\tilde{\nu}_R^{\ast}}\) and the lepton number asymmetry
\(L_{\tilde{\nu}_R}\). We find these equations to be
\begin{eqnarray}
{dL_{\tilde{\nu}_R} \over dz} &=&-{2z\over H(m_{\tilde{\nu}_R})s } \Gamma_{aa}\
\left(L_{\tilde{\nu}_R} Y_{DM} -
  \frac{g_{*s}}{2}(\Lhid - L_{\tilde{\nu}_R}) {Y^{eq^2}_{DM}} \right)\label{eq:LeqSnoot}\\
{dY_{DM} \over dz} &=&{z\over2 H(m_{\tilde{\nu}_R})s} \left[\Gamma^{(+)}\
\left({Y^{eq}_{DM}}^2-{Y^2_{DM}}\right) +\Gamma^{(-)}\ {L_{\tilde{\nu}_R}^2} \right]
\label{eq:YeqSnoot},
\end{eqnarray}
where \(z\equiv m_{\tilde{\nu}_R}/T\), and \(H(m_{\nu_R})\) is the Hubble parameter at
scale \(T=m_{\tilde{\nu}_R}\).  The effective rates
\begin{equation}
  \Gamma_{\tilde{\nu}\tilde{\nu}}=\gamma_{\tilde{\nu}\tilde{\nu}}/
  Y_{DM}^{eq^2}~\hspace{.5cm}~\mathrm{and}~~\hspace{.5cm}~
  \Gamma^{\pm}=\Gamma_{\tilde{\nu}\tilde{\nu}^{\ast}}\pm
  2\Gamma_{\tilde{\nu}\tilde{\nu}}=4(\gamma_{\tilde{\nu}\tilde{\nu}^{\ast}}\pm
  \gamma_{\tilde{\nu}\tilde{\nu}})/Y_{DM}^{eq^2}
\end{equation}
are determined from the reaction densities \(\gamma_{\tilde{\nu}\tilde{\nu}}\) and
\(\gamma_{\tilde{\nu}\tilde{\nu}^{\ast}}\), defined as in~(\ref{eq:ReacDensDef}), which
are in turn determined from the reduced cross sections
\(\hat{\sigma}(\tilde{\nu}\tilde{\nu})\) and
\(\hat{\sigma}(\tilde{\nu}\tilde{\nu}^{\ast})\), a calculation of which is included in
appendix~\ref{app:SneutInts}. Given in terms of $r=\sqrt{1-4m_{\tilde{\nu}_R}^2/s}$,
$x_{\chi_1}=s/m^2_{\chi_1}$, $x_n=s/m_{\tilde{\nu}_{R}}^2$, and
$A=(2x_n^{-1}-2x_\chi^{-1}-1)$, the results are
\begin{eqnarray}
  \hat{\sigma}(\tilde{\nu}\tilde{\nu})&=&{g^4\over \pi}{1\over x_\chi}\left({1\over
  A}\ln{\left(A+r\over A-r\right)^2}-4{r\over r^2-A^2}\right)\label{eq:SigmaHatsofSneutConserve}\\
  \hat{\sigma}(\tilde{\nu}\tilde{\nu}^{\ast})&=&
  {g^4\over 12\pi}\left(\frac{s}{s-M_Z^2}\right)^2 r^3+
  {g^4\over4\pi}\left(-4r+A \ln{\left(A+r\over
  A-r\right)^2}\right)\nonumber\\&&+
  {g^4\over8\pi}\left(\frac{s}{s-M_Z^2}\right)\left(2Ar+
  {(r^2-A^2)\over2} \ln{\left(A+r\over A-r\right)^2}\right). \label{eq:SigmaHatsofSneutViolate}
\end{eqnarray}

\indent We note that the equations~(\ref{eq:LeqSnoot}\,-\,\ref{eq:YeqSnoot}) are actually
quite general and can be applied in a variety of scenarios in which an asymmetry in some
globally conserved quantum number between visible and hidden sectors (or between a pair
of distinct hidden sectors) develops during the evolution of the universe and the dark
matter particle is located in the hidden sector and charged under that quantum
number.\footnote{In special cases, such as that of a nearly-degenerate LSP-NLSP pair in
supersymmetric models, the Boltzmann equations must be modified~\cite{Griest:1990kh}, but
we do not consider such exceptions here.}  Any model-dependent aspects of the Boltzmann
evolution enter only in the relevant annihilation rates, in the mass of the dark matter
particle, and in the input value for \(\Lhid\).

\indent Before we move on to address the Boltzmann
system~(\ref{eq:LeqSnoot}\,-\,\ref{eq:YeqSnoot}) numerically, we note that in certain
limiting cases it reduces to a particularly simple form.  One of these cases is the
trivial case, where $L_{\RHSn}=0$ and we are left with the single equation
\begin{equation}
  {dY_{DM} \over dt} ={1\over2 s} \Gamma^{(+)}\ \left({Y^{eq}_{DM}}^2-{Y^2_{DM}}\right)
  \label{eq:BoringCDMCase}
\end{equation}
whose solution is the standard one for a dark matter abundance.  Perhaps more interesting
is the case where $\Gamma_{RR}\simeq 0$ and $L_{\RHSn}\neq 0$.  In this case, $L_{\RHSn}$
is fixed and the equation for the dark matter abundance becomes
\begin{equation}
  {dY_{DM} \over dt} ={1\over2 s} \Gamma_{RR^c}
  \left[\left({Y^{eq}_{DM}}^2-{Y^2_{DM}}\right) + {L_{\RHSn}^2} \right].
\end{equation}
This represents the case described in~\cite{Hooper:2004dc}.  As mentioned above, this
situation is somewhat difficult to engineer for a right-handed sneutrino in Dirac
leptogenesis, but in the next section we will provide one or two possible schemes in
which it can be realized.

\subsection{Numerical Results and Discussion}


\indent

We are now ready to solve the Boltzmann system given
in~(\ref{eq:LeqSnoot})\,-\,(\ref{eq:YeqSnoot}) for right-handed sneutrino dark matter in
supersymmetric Dirac leptogenesis.  Obtaining the correct dark matter abundance in this
scenario is not difficult, provided the \(U(1)_{N}\) gauge coupling constant is on the
appropriate order, but generating \(\Omega_{\mathit{CDM}}\) via the mechanism described
in equation~(\ref{eq:OmegaFromAsymmetry}) and thereby forging a link between this
quantity the observed baryon asymmetry of the universe is not.  The reason for this is
that the \(L_{\tilde{\nu}}\)--conserving and \(L_{\tilde{\nu}}\)--violating rates
determined respectively by~(\ref{eq:SigmaHatsofSneutConserve})
and~(\ref{eq:SigmaHatsofSneutViolate}) are naturally on the same order unless a hierarchy
exists among the masses of the virtual particles exchanged.  The $Z'$ mass given
in~(\ref{eq:Zprimemass}) depends primarily on the hidden-sector gauge coupling $g$, the
soft squared masses $m_{\chi}^2$ and $m_{\xi_1}^2$, and the supersymmetric mass parameter
\(\tilde{\mu}\) (which ought to be of roughly the same order as the aforementioned soft
masses if we wish to avoid fine-tuning); while the neutralino mass and \(U(1)_{N}\)
gaugino content depend primarily on $M'$.  Thus unless a hierarchy between scalar and
gaugino masses emerges naturally from supersymmetry breaking in which gaugino masses are
far heavier than scalar masses, $\hat{\sigma}(\tilde{\nu}\tilde{\nu})$ and
$\hat{\sigma}(\tilde{\nu}\tilde{\nu}^{\ast})$ will be of roughly the same order.

\begin{figure}
\begin{center}
 \includegraphics{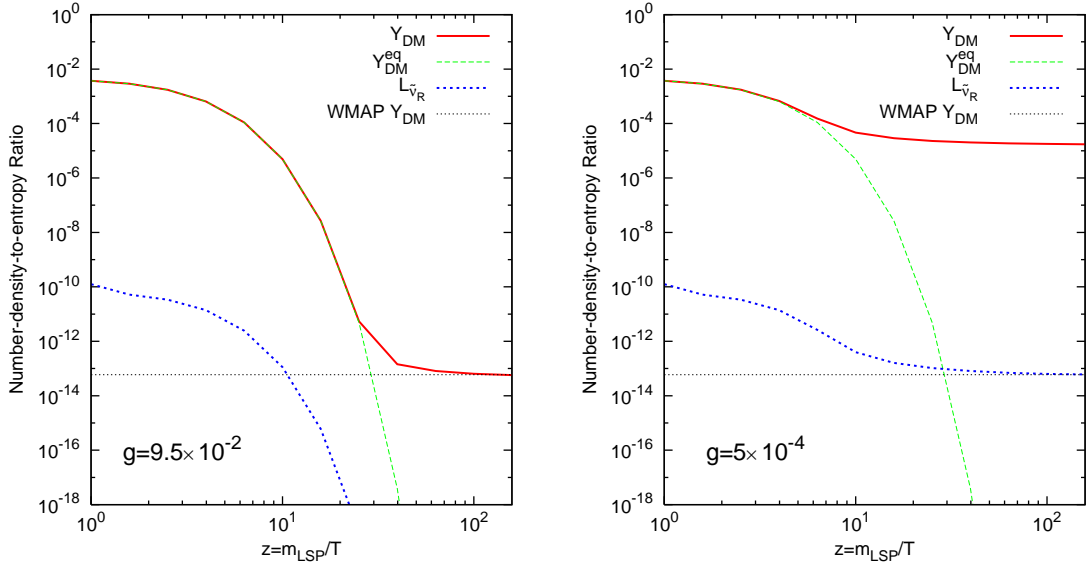}
 \end{center}
\caption{Here, we illustrate the two representative cases for sneutrino dark matter in
Dirac neutrinogenesis.  In each case, we have set $M'=200$~GeV and
$m_{\tilde{\nu}_{R}}=150$~GeV and maintained the relations $g\langle\chi\rangle=200$~GeV,
$g\langle\xi_1\rangle=100$~GeV, $\tilde{\mu}=500$~GeV, and $\zeta_{1}/g=\zeta_{2}/g=1$.
The left panel depicts the situation where the $U(1)_{N}$ gauge coupling is reasonably
strong ($g=9.5\times 10^-2$).  As the universe evolves, $T$ decreases and $z\to \infty$.
Here, the number-density-to-entropy-density ratio $Y_{DM}$ of a right-handed sneutrino
LSP reproduces the WMAP prediction (the horizontal dashed line), but the lepton number
$L_{\tilde{\nu}_R}$ stored in right-handed sneutrinos is washed out almost entirely and
all connection between the dark matter abundance and the observed baryon number of the
universe is lost.  The right panel shows the situation where the gauge coupling is small
$g=5\times 10^{-4}$. Here, washout is insignificant and $L_{\tilde{\nu}_R}$ persists
essentially unchanged from its initial value until present time, but sneutrino
annihilations are inefficient in reducing the dark matter abundance to an acceptable
level.  The equilibrium right-handed sneutrino abundance $Y^{eq}_{DM}$ is also shown in
each panel for reference.  It should be emphasized that in the left panel, the correct
dark matter relic density is obtained, and that while $\Omega_{\mathit{CDM}}$ and
$\Omega_{b}$ are not tied to one another, there is no problem for cosmology, whereas the
situation depicted in the right panel is cosmologically unacceptable, since dark matter
is overproduced.\label{fig:SneutCDMEvolPlots}}
\end{figure}

\indent This leads us to choose between the two less-than-ideal outcomes depicted in
figure~\ref{fig:SneutCDMEvolPlots}.  Here, we show the evolution of \(Y_{DM}\) and
\(L_{\tilde{\nu}_{R}}\) in two representative regimes: one in which the \(U(1)_{N}\)
gauge coupling \(g\) is comparatively strong ($g=9.5\times 10^{-2}$), and one in which it
is weaker ($g=5\times10^{-4}$).  In each case, we have maintained the relations
$g\langle\chi\rangle=200$~GeV, $g\langle\xi_1\rangle=100$~GeV, $\tilde{\mu}=500$~GeV, and
$\zeta_{1}/g=\zeta_{2}/g=1$ and set $M'=200$~GeV.  By examining
figure~\ref{fig:ExclusionPlot}, one can verify that these parameter choices are both
compatible with successful leptogenesis, given an appropriate choice of $M_{\Phi_1}$. The
mass of the lightest right-handed sneutrino is taken to be $m_{\tilde{\nu}_{R}}=150$~GeV,
which makes it the LSP.  In the strong coupling case presented in the left panel,
$\Gamma(\tilde{\nu}\tilde{\nu})$ and $\Gamma(\tilde{\nu}\tilde{\nu}^{\ast})$ are large
enough to reduce the sneutrino abundance to a level in accord with the WMAP bound, but
$\Gamma(\tilde{\nu}\tilde{\nu})$ is so large that \(L_{\nu_R}\) is washed out rapidly and
essentially completely.  Here, the model is successful in yielding the correct dark
matter abundance, but all connection between \(L_{\nu_{R}}\) (and therefore \(B\)) and
\(\Omega_{\mathit{CDM}}\) is lost, and the evolution of \(Y_{DM}\) reduces to the
standard situation described in equation~(\ref{eq:BoringCDMCase}).  In the right panel,
for weak gauge coupling, things go even more awry: $\Gamma(\tilde{\nu}\tilde{\nu})$ is
small enough to prevent the washout of \(L_{\nu_R}\), but since
$\Gamma(\tilde{\nu}\tilde{\nu}^{\ast})$ is likewise small, sneutrino annihilations are
inefficient in reducing the right-handed sneutrino abundance to an acceptable level---in
essence, we recover (literally, in the \(g\to0\) limit) the situation we had before
gauging the \(U(1)_{N}\) symmetry.  The message here is that the right-handed sneutrino
in Dirac leptogenesis is a viable dark matter candidate, but that a relationship between
\(\Omega_{\mathit{CDM}}\) and \(\Omega_{b}\) does not emerge naturally from the Dirac
leptogenesis framework, as least in the simple model presented here.

\indent Arranging a hierarchy between $\Gamma_{\tilde{\nu}\tilde{\nu}^{\ast}}$ and
$\Gamma_{\tilde{\nu}\tilde{\nu}}$, though difficult, is not altogether impossible,
however.  As mentioned previously, one way this could occur is if a hierarchy between
gaugino and scalar masses, in which the former were much higher than the latter, emerged
naturally from the supersymmetry-breaking mechanism, but this is not easy to achieve. The
most promising model would appear to be gaugino
mediation~\cite{Mirabelli:1997aj,Kaplan:1999ac,Chacko:1999mi,Schmaltz:2000ei}, in which
the supersymmetry-braking sector is confined to a ``source'' brane parallel to our own,
visible brane in a five-dimensional bulk.  The matter fields of the theory (the MSSM
fields, right-handed neutrino superfields, etc.) are confined to the visible brane, while
gauge supermultiplets are allowed to propagate in the bulk.  As a result, one can arrange
for a loop-splitting between gaugino and scalar soft masses, but it is difficult to
arrange for this splitting to be as pronounced as we require.  We thus turn to
alternative methods for engineering a scale-separation between
$\Gamma_{\tilde{\nu}\tilde{\nu}^c}$ and $\Gamma_{\tilde{\nu}\tilde{\nu}}$.

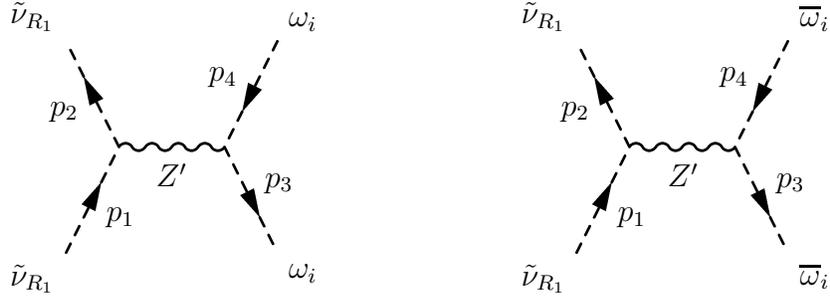
\begin{figure}
\begin{center}
\begin{fmffile}{LightScal4}
  \fmfframe(20,20)(20,20){\begin{fmfchar*}(100,80)
    \fmfleft{i1,i2}
    \fmfright{o1,o2}
    \fmflabel{$\tilde{\nu}_{R_1}$}{i1}
    \fmflabel{$\tilde{\nu}_{R_1}$}{i2}
    \fmflabel{$\omega_i$}{o2}
    \fmflabel{$\omega_i$}{o1}
    \fmf{scalar,label=$p_1$}{i1,v1}
    \fmf{scalar,label=$p_2$}{v1,i2}
    \fmf{scalar,label=$p_4$}{o2,v2}
    \fmf{scalar,label=$p_3$}{v2,o1}
    \fmf{photon,label=$Z'$}{v1,v2}
  \end{fmfchar*}}
\end{fmffile}~~~\hspace{.5cm}~~~
\begin{fmffile}{LightScalbar4}
  \fmfframe(20,20)(20,20){\begin{fmfchar*}(100,80)
    \fmfleft{i1,i2}
    \fmfright{o1,o2}
    \fmflabel{$\tilde{\nu}_{R_1}$}{i1}
    \fmflabel{$\tilde{\nu}_{R_1}$}{i2}
    \fmflabel{$\overline{\omega}_i$}{o2}
    \fmflabel{$\overline{\omega}_i$}{o1}
    \fmf{scalar,label=$p_1$}{i1,v1}
    \fmf{scalar,label=$p_2$}{v1,i2}
    \fmf{scalar,label=$p_4$}{o2,v2}
    \fmf{scalar,label=$p_3$}{v2,o1}
    \fmf{photon,label=$Z'$}{v1,v2}
  \end{fmfchar*}}
\end{fmffile}
\end{center}
\caption{The additional contribution to $\hat{\sigma}(\tilde{\nu}\tilde{\nu}^{\ast})$
that arises in the right-handed sneutrino dark matter scenario in the presence of
additional light scalars $\omega_i$ and $\overline{\omega}_{j}$ charged under the
$U(1)_{N}$ gauge group.\label{fig:LightScalarDiagram}}
\end{figure}

\indent A second way of arranging a hierarchy between these rates is to further expand
the field content of the theory and introduce additional light scalars into which the
\(\tilde{\nu}_{R_1}\) can decay. This can be arranged by introducing pairs of vector-like
superfields \(\omega_i\) and \(\overline{\omega}_i\) that are even under matter parity
(so \(\tilde{\nu}_{R_1}\) is still the LSP) and charged only under $U(1)_{N}$. Charges
$Q_i$ are assigned them so that the form of the Dirac leptogenesis
superpotential~(\ref{eq:DiracLepSuperpotential}) and its hidden sector
extension~(\ref{eq:wnew}) are unchanged,\footnote{This is trivial and can be easily
accomplished, for example, by giving assigning fractional charges to all the \(\omega_i\)
and \(\overline{\omega}_i\).} save for the addition of mass terms
\begin{equation}
  W\supset M_{\omega_i}\omega_{i}\overline{\omega}_i.
\end{equation}
Since the new superfields come in vector-like pairs, they will not spoil anomaly
cancelation.  The soft Lagrangian will also contain new soft mass terms for the scalar
fields in \(\omega_i\) and \(\overline{\omega}_i\), and a partial cancelation between a
tachyonic soft squared mass and the contribution \(M_{\omega_i}^2\) from the
superpotential can arrange for the scalar components of \(\omega_i\) and
\(\overline{\omega}_i\) (also denoted \(\omega_i\) and \(\overline{\omega}_i\)) to remain
light compared to $\tilde{\nu}_{R_1}$, while their fermionic superpartner
\(\tilde{\omega}_i\) and \(\tilde{\overline{\omega}}_i\) acquire masses
\(M_{\omega_i}\geq m_{\nu_{R}}\).  In this case, an additional \(s\)-channel contribution
to the right-handed sneutrino annihilation cross-section via the diagrams in
figure~\ref{fig:LightScalarDiagram} will arise, which has no \(t\)-channel equivalent.
The contribution to $\hat{\sigma}(\tilde{\nu}\tilde{\nu}^{\ast})$ from each pair of
\(\omega_i\) and \(\overline{\omega}_i\) is
\begin{equation}
  \Delta \hat{\sigma}(\tilde{\nu}\tilde{\nu}^{\ast})= \frac{g^4Q_i^2}{8\pi}
  \left(\frac{s}{s-M_{Z'}^2}\right)^2 r,\label{eq:AugmentwitheScalarsRedXsec}
\end{equation}
where $r$ is defined as above equation~(\ref{eq:SigmaHatsofSneutConserve}), and if the
number of additional scalars is large (or the charges assigned them are large), the
\(L_{\nu_{R}}\)--conserving rate can be increased without simultaneously increasing the
\(L_{\nu_{R}}\)--violating one.

\begin{figure}
\begin{center}
\includegraphics{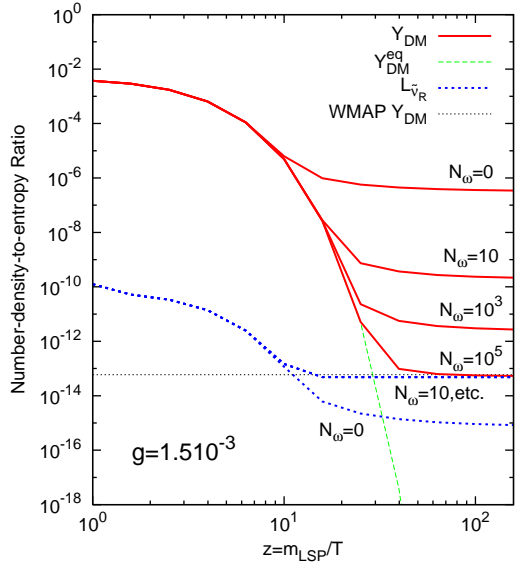}
\end{center}
\caption{The effect on the dark matter number-density-to-entropy-density ratio $Y_{DM}$
of adding a number $N_{\omega}$ of vector-like pairs of light scalars to the theory.
Here, all such scalars have $U(1)_{N}$ have charge magnitude $|Q|=20$, and results are
shown for $N_{\omega}=10,~10^3$, and~$10^5$, as well as for the unmodified case
$N_{\omega}=0$.  As in the right panel of figure~\ref{fig:SneutCDMEvolPlots},
$M'=200$~GeV and $m_{\tilde{\nu}_{R}}=150$~GeV and maintained the relations
$g\langle\chi\rangle=200$~GeV, $g\langle\xi_1\rangle=100$~GeV, $\tilde{\mu}=500$~GeV,
$\zeta_{1}/g=\zeta_{2}/g=1$, and $g=1.5\times 10^{-3}$.  The addition of light scalars
does not affect the lepton number $L_{\tilde{\nu}_{R}}$ stored in right-handed sneutrinos
appreciably for $N_{\omega}\geq 10$. Because $g$ is small, an inordinately large number
of light fields are required in order to achieve a situation in which
$L_{\tilde{\nu}_{R}}$ dictates the size of $Y_{DM}$. When $N_{\omega}$ is large enough,
$L_{\nu_R}$ creates an effective ``floor'' for the dark matter abundance (corresponding
to a dark matter abundance consisting only of $\tilde{\nu_R}$ and none of its
antiparticle $\tilde{\nu_R}^{\ast}$).\label{fig:BoltzWithScalarsAdded}}
\end{figure}

\indent While this idea seems promising, it turns out to be difficult to realize in
practice.  The reason is that the reduced
cross-section~(\ref{eq:AugmentwitheScalarsRedXsec}) for sneutrino annihilations, like the
usual expressions~(\ref{eq:SigmaHatsofSneutConserve})
and~(\ref{eq:SigmaHatsofSneutViolate}), is proportional to $g^4$, and \(g\) must be kept
small in order to prevent $L_{\tilde{\nu}_{R}}$ from being washed out.  This
substantially hinders the effectiveness of this new contribution at reducing \(Y_{DM}\):
one must either add a very large number \(N_{\omega}\) of vector-like superfield pairs to
the theory---or else make the charges \(Q_{i}\) of the new superfields inordinately
large---in order to achieve the desired hierarchy between
\(\Gamma_{\tilde{\nu}\tilde{\nu}}\) and \(\Gamma_{\tilde{\nu}\tilde{\nu}^{\ast}}\).  In
figure~\ref{fig:BoltzWithScalarsAdded}, we show the effect on $Y_{DM}$ of adding light
scalars to the theory.  Here, we have chosen the (quite large) value $|Q|=20$ for the
magnitude of the charge assigned to all \(\omega_{i}\,-\,\overline{\omega}_i\) pairs and
set $g=1.5\times10^{-3}$ (the input values for all other relevant model parameters are
the same as in figure~\ref{fig:SneutCDMEvolPlots}).  To compensate for the smallness of
the $g^4\sim10^{-12}$ prefactor in~(\ref{eq:AugmentwitheScalarsRedXsec}), an additional
\(N_{\omega}\sim 10^5\) field pairs are needed.  This is highly unmotivated, to say
nothing of the havoc such fields would wreak on the RGE running of the \(U(1)_{N}\)
coupling in terms of the prospects for gauge coupling unification, etc., and therefore
not a particularly compelling solution to the rate correlation problem.

\indent The possibility that Dirac leptogenesis presents for right-handed sneutrino dark
matter is indeed an interesting one, and as the left panel of
figure~\ref{fig:SneutCDMEvolPlots}, one that can reproduce the observed value of
\(\Omega_{\mathit{CDM}}\).  The disappointment is that it it a challenge to tie the size
of the observed dark matter matter abundance to the baryon number of the universe through
their mutual connection to \(L_{\nu_R}\) in a way where the former is driven
automatically driven by the latter to the correct value.  It should again be emphasized
that the link between $L_{\nu_R}$ the baryon number $B$ of the universe is forged at the
leptogenesis scale \(M_{\Phi_1}\gtrsim 10^{10}\)~GeV.  The two asymmetries do not
communicate with one another during the CDM freeze-out epoch, and the the rate
\(\Gamma_{\tilde{\nu}\tilde{\nu}}\) does not alter the total lepton number $\Lhid$ stored
in the hidden sector, but rather transfers the portion of that asymmetry \(L_{\nu_{R}}\)
stored in right-handed sneutrinos to right-handed neutrinos (or perhaps light scalars).
Thus the dynamics associated with \(\Gamma_{\tilde{\nu}\tilde{\nu}}\) have no effect on
$B$, and baryogenesis proceeds unhindered no matter to what degree \(L_{\nu_{R}}\) is
washed out.  Thus sneutrino dark matter is perfectly compatible with baryogenesis in
Dirac leptogenesis scenarios.  However, it is generally not feasible to forge a
connection between \(\Omega_{\mathit{CDM}}\) and the relic abundance \(\Omega_{b}\) of
baryonic matter through a mutual connection to $L_{\tilde{\nu}}$: a correlation between
the the rates \(\Gamma_{\tilde{\nu}\tilde{\nu}}^{\ast}\) and
\(\Gamma_{\tilde{\nu}\tilde{\nu}}\) for $L_{\tilde{\nu}}$--conserving and
$L_{\tilde{\nu}}$--violating processes makes it difficult simultaneously to preserve
$L_{\tilde{\nu}}$ while reducing $Y_{\tilde{\nu}}$ to an acceptable level.


\chapter{INDIRECT\ DARK\ MATTER\ DETECTION\label{ch:PhotonsFromCDM}}

\section{Detecting Heavy Neutralino Annihilations\label{sec:RatesandProfiles}}

\indent

In section~\ref{sec:sneutrinoCDM}, it was mentioned that Dirac leptogenesis is compatible
with neutralino dark matter.  Furthermore, it strongly prefers a heavy gravitino and
hence is quite naturally compatible with split supersymmetry, which is associated with a
particular, characteristic type of neutralino dark matter.  In theories where anomaly
mediation is responsible for the generation of gaugino masses \(M_{1}\), \(M_2\), and
\(M_{3}\) (as is the case in split supersymmetry), the neutralino spectrum is dictated by
the $\beta$-functions of the Standard Model gauge groups by equation~(\ref{eq:AMSBMass}).
In the MSSM, to lowest order, the relationship is
\begin{equation}
  \label{eq:M123}
  M_{3} \simeq 3 M_{1} \simeq 9 M_{2},
\end{equation}
which implies that if a neutralino is in fact the LSP,
it will be predominately either Wino or (depending on the value of
\(\mu\)), Higgsino.  We now turn to investigate the possibility for detecting such an LSP
experimentally.

\indent In section~\ref{sec:AstroConstraints}, we discussed the relic abundance of such
an LSP, which had both a thermal component $\Omega^{\mathit{Th}}_{\mathit{LSP}}$ given by
equation~(\ref{eq:OmegaThermWino}) or~(\ref{eq:OmegaThermHiggsino}) for a majority Wino
or Higgsino LSP, respectively, and in theories with a heavy gravitino, an additional,
nonthermal component $\Omega^{\mathit{NT}}_{\mathit{LSP}}$, given by
equations~(\ref{eq:OmegaNTNot})\,-\,(\ref{eq:OmegaNTAnn}), resulting from gravitino
decay. As is evident from figure~\ref{fig:OmegaTot}, the thermal component places an
upper bound on \(m_{LSP}\), which turns out in each case to be~\cite{Thomas:2005te}
\begin{eqnarray}
  m_{\mathit{LSP}}\leq 2.5\mbox{ TeV~~~(for Wino)} \label{eq:mwUPb} \\
  m_{\mathit{LSP}}\leq 1.2\mbox{ TeV~~~(for Higgsino)}.\label{eq:mhUPb}
\end{eqnarray}

\indent The direct detection of a heavy Wino or Higgsino LSP with a mass \(m_{LSP}\gg
M_{W}\)~GeV is all but precluded at present and planned facilities~\cite{Masiero:2004ft}
for all but the most unnaturally peaked halo models, and the cross-section for the
annihilation of such an LSP into positrons is too small to yield a detectable signal. The
only promising detection method available is to search for high-energy photons resulting
from LSP annihilation processes in the galactic halo, and in particular
\(\tilde{N}_{1}\tilde{N}_{1}\rightarrow \gamma\gamma\) and
\(\tilde{N}_{1}\tilde{N}_{1}\rightarrow \gamma Z\). Expressions for the cross-sections
for these processes have been computed~\cite{Bergstrom:1997fh}, and in the limit where
\(m_{\mathit{LSP}}\) is much larger than the weak scale, $\sigma
v(\tilde{N}_{1}\tilde{N}_{1}\rightarrow\gamma\gamma)$ and $\sigma
v(\tilde{N}_{1}\tilde{N}_{1}\rightarrow Z\gamma)$ take the asymptotic forms
\begin{eqnarray}
  \sigma v(\tilde{W}\tilde{W}\rightarrow\gamma\gamma) \simeq 4.0 \times 10^{-27} \mbox{ }
    \mathrm{cm}^{3}\mathrm{s^{-1}}\label{eq:sigWWannDM1eq}   \\
  \sigma v(\tilde{W}\tilde{W}\rightarrow Z\gamma) \simeq 9.0 \times 10^{-27} \mbox{ }
  \mathrm{cm}^{3}\mathrm{s^{-1}},\label{eq:sigWWannDM2eq}
\end{eqnarray}
for a Wino LSP, and
\begin{eqnarray}
  \sigma v(\tilde{H}\tilde{H}\rightarrow\gamma\gamma) \simeq 9.0 \times 10^{-29} \mbox{ }
    \mathrm{cm}^{3} \mathrm{s^{-1}} \label{eq:sigHHannDM1eq}  \\
  \sigma v(\tilde{H}\tilde{H}\rightarrow Z\gamma) \simeq 2.0 \times 10^{-29} \mbox{ }
    \mathrm{cm}^{3}\mathrm{s^{-1}}.\label{eq:sigHHannDM2eq}
\end{eqnarray}
for a Higgsino LSP.  The reason these processes are of particular interest is that the
photons produced thereby are effectively monoenergetic, with respective energies
\begin{eqnarray}
  E_{\gamma}=m_{LSP} & \hspace{1cm} \mathrm{and} \hspace{1cm} &
  E_{\gamma}=m_{LSP}\left(1-\frac{m_{Z}^{2}}{4m_{\mathit{LSP}}^{2}}\right).\label{eq:TwoLines}
\end{eqnarray}
The detection of either signal would be compelling evidence for heavy Wino or Higgsino
dark matter, since effectively monoenergetic signals at TeV-scale energies tend to
originate in particle physics processes (rather than astrophysical ones); taken together,
they would serve as a distinctive signal---were it possible to resolve them.  Our aim
here is to analyze the prospects for the detection of this signal at present and future
$\gamma$-ray telescopes.

\indent Calculation of the observed integral flux $\Phi$ of $\gamma$-rays from the
galactic center is complicated somewhat by our ignorance of the precise density
distribution profile \(\rho(\psi,s)\) of dark matter in the galactic halo (expressed here
as a function of line-of-sight distance $s$ and the angle \(\psi\) away from the galactic
center).  The density profile enters into the integral flux (usually expressed in
\(\mbox{cm}^{-2}s^{-1}\)) through the expression
\begin{equation}
  \Phi=[\sigma v (\tilde{N}_1\tilde{N}_1\rightarrow X)]\times N_{\gamma}(X)\frac{1}{4\pi m_{\mathit{LSP}}^2}
  \int_{L}\rho^{2}(\psi,s)\mathit{ds}.
\end{equation}
where \(N_{\gamma}(X)\) is the number of photons in the final state $X$, and the integral
is evaluated along the line of sight.  It should be noted that \(\rho(\psi,s)\) is
proportional to the LSP mass and the asymptotic expressions for $\sigma v
(\tilde{N}_1\tilde{N}_1\rightarrow X)$ given
in~(\ref{eq:sigWWannDM1eq}\,-\,\ref{eq:sigHHannDM2eq}) are independent of
\(m_{\mathit{LSP}}\), so $\Phi$ is actually insensitive to \(m_{\mathit{LSP}}\) when
\(m_{\mathit{LSP}}\gg M_{W}\). Since the density integral is essentially independent of
the particle physics, it is common practice to abstract it by defining the quantity
\begin{equation}
  J(\psi)\equiv\frac{1}{8.5\mbox{ kpc}}
  \left(\frac{1}{0.3\mbox{ GeV}}\right)^{2}\int_{L}\rho^{2}(\psi,s)ds.\label{eq:defofJ}
\end{equation}
While for an ideal detector, one would be interested only in \(J(0)\), the value for the
line passing directly through the galactic center, a real detector will receive incoming
photons originating over some finite slice of solid angle.  The relevant quantity is
angular acceptance \(\Delta\Omega\), the solid angle from which the detector actively
receives light, which for a given detector may be adjusted as desired over a window
ranging from a detector's angular resolution to its field of view.  Thus instead of
\(J(0)\), the quantity of interest is $\langle J(\psi)\rangle_{\Delta\Omega}$, the
average of \(J(\psi)\) over \(\Delta\Omega\).

\begin{table}[t!]
  \begin{center}
    \begin{tabular}{|l|cccc|} \hline
    & $\alpha$ & $\beta$ & $\gamma$ & $R$ \\ \hline
    Isothermal profiles & 2.0 & 2.0 & 0 & 3.5 \\
    NFW & 1.0 & 3.0 & 1.0 & 20.0 \\
    Moore et al. & 1.5 & 3.5 & 1.5 & 28.0 \\ \hline
    \end{tabular}
    \end{center}
  \caption{The defining parameters $\alpha$, $\beta$, $\gamma$ and $R$
    (see equation \ref{eq:albega}) for the halo models we examine.  $R$ is given in
    kpc.\label{tab:abgvals}}
\end{table}

\begin{figure}[ht!]
  \begin{center}
    \includegraphics[width=13cm]{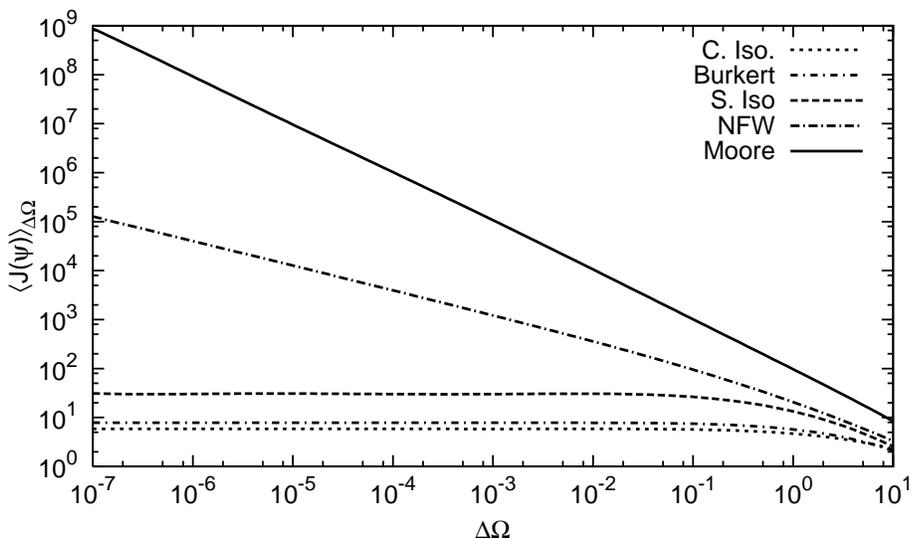}
  \end{center}
  \caption{$\langle J(\psi)\rangle_{\Delta\Omega}$ (the line-of-sight integral
    through the halo density squared, averaged over the angular acceptance
    $\Delta\Omega$) vs. $\Delta\Omega$ for
    several halo profiles: smooth isothermal, clumpy isothermal, Burkert,
    NFW, and Moore et al.  Taken from~\cite{Thomas:2005te}.\label{fig:jpsis}}
\end{figure}

\indent We have yet to discuss the dark matter density profile itself, and since its
precise shape is not well known, we will survey a representative sample of profiles
rather than focussing on any single model. In general, these profiles are derived from
numerical simulations~\cite{Moore:1999gc}, and most take the form
\begin{equation}
  \label{eq:albega}
  \rho(r)=
  \frac{\rho_{0}}{(r/R)^{\gamma}(1+(r/R)^{\alpha})^{(\beta-\gamma)/\alpha}},
\end{equation}
where the density function \(\rho\) has been expressed in terms of the radius $r$ away
from the galactic center and spherical symmetry is assumed. The three power-law indices
\(\alpha\), \(\beta\), and \(\gamma\), along with the characteristic radius $R$, serve to
define a given model.  The models we examine here include the relatively cuspy Moore et
al. profile~\cite{Moore:1997sg}, the widely used Navarro-Frenk-White
profile~\cite{Navarro:1995iw}, and a pair of isothermal models~\cite{Casertano:1991sh},
one with a smooth density distribution, the other with a greater level of dark matter
clumping.\footnote{The dark matter density given by the Moore et al. profile is divergent
for \(r\ll R\) and is assumed to be truncated at some small radius \(r_{t}\).  Here, we
choose \(r_t=10^{-5}\)~kpc.} We list the \(\alpha\), \(\beta\), \(\gamma\), and \(R\)
assignments which define each of these models in table~\ref{tab:abgvals}.  We also
include in our survey one model in which the dark matter profile is not defined by
equation~(\ref{eq:albega}), but by the relation
\begin{equation}
  \rho(r)=
  \frac{\rho_{0}r_{0}^{3}}{(r+r_{0})(r^{2}+r_{0}^{2})},
\end{equation}
where \(r_{0}\) and $\rho_{0}$ are fiducial distance and density parameters, respecively
(for our galaxy, \(r_{0}\approx 11.7\)~kpc and $\rho_0\approx 0.34~\mathrm{GeV}~\mathrm{
cm}^{-3}$~\cite{Edsjo:2004pf}). This model was originally proposed by Burkert et
al.~\cite{Burkert:1995yz} and is also in use. Figure~\ref{fig:jpsis} shows the
relationship between \(\langle J(\psi)\rangle_{\Delta\Omega}\) and \(\Delta\Omega\) that
results from plugging each of the distributions we consider into
equation~(\ref{eq:defofJ}).  As one would expect, the more sharply peaked distributions
(the Moore and Burkert profiles), in which the dark matter is concentrated near the
galactic center, yield a higher value of \(\langle J(\psi)\rangle_{\Delta\Omega}\) for a
given $\Delta\Omega$, and therefore to a higher \(\gamma\)-ray flux.\footnote{It is worth
noting that the models surveyed here assume a spherically symmetric halo, and it has been
observed~\cite{Jing:2002np} that triaxial models provide a better fit to the cosmological
density profiles obtained in cosmological simulations than do spherical ones.  The
corrections introduced by taking such considerations into account would amount to an
\(\mathcal{O}(1)\) factor in the overall result, and since it will soon become evident
that detection prospects will be an issue of orders of magnitude rather than numerical
prefactors, such prefactors do not significantly compromise our results.} It should be
noted that different profiles can lead to drastically different values for $\Phi$. The
situation is further complicated by the possibility that the presence of a massive black
hole at the center of the galaxy could significantly modify the dark matter abundance
near the galactic center and lead to a pronounced density spike~\cite{Gondolo:1999ef},
though there is some debate over the precise effect such a black hole would have, and for
this reason we will not consider such modifications here.

\section{Outlook at Present and Planned Facilities\label{sec:CDMFacilitiesOutlook}}


\indent

Now that we have calculated $\Phi$ (up to considerations involving the shape of the dark
matter profile), let us turn to a discussion of the performance of present and future
$\gamma$-ray telescopes. These can be divided into two main types: satellite facilities,
which detect incoming photons directly, and atmospheric Cherenkhov detectors (ACTs),
which operate by observing showers of Cherenkhov light that result when high-energy
\(\gamma\)-rays enter the Earth's upper atmosphere. In assessing performance and the
ability to register a discovery at the \(5\sigma\) level, the relevant detector
attributes to consider are angular acceptance $\Delta\Omega$, energy resolution \(\Delta
E/ E\), and effective collection area \(A_{\mathit{eff}}\). In table~\ref{tab:detectors},
we provide a list of these performance specifics for a variety of present and planned
experiments, including both satellite facilities and ACTs, as well as those for two
hypothetical facilities (one a satellite detector, one an ACT) slightly more advanced
than any currently planned facility of its kind, which we will use when examining the
potential for observation of heavy Wino or Higgsino dark matter annihilation at the next
generation of $\gamma$-ray experiments.

\begin{table}[t!]
  \begin{center}
    \begin{tabular}{|l|ccccc|} \hline \rule{0pt}{14pt} Facility
      & $A_{\mathrm{eff}}$ $(\mbox{cm}^2)$ & $\Delta E/E$ &
      $\Delta\Omega_{\mathrm{min}}$ $(\mbox{sr})$
      & $\epsilon_{\mathrm{had}}$ & \\ \hline \rule{0pt}{14pt}
      WHIPPLE (Arizona) & $3.5 \times 10^{8}$ & 30\% & $1.88 \times 10^{-5}$ & 1.0 & \\
      GRANITE II (Arizona)  & $5\times 10^{8}$ & 20\% & $9.56 \times 10^{-6}$ & 1.0 & \\
      HESS (Namibia) & $7 \times 10^{8}$ & 15\% & $9.56 \times 10^{-6}$ & 0.25 & \\
      VERITAS (Arizona) & $1 \times 10^{9}$ & 15\% & $3.83 \times 10^{-7}$ & 0.25 & \\
      EGRET (Satellite) & $1 \times 10^{4}$ & 15\% & $3.22 \times 10^{-2}$ & - & \\
      GLAST (Satellite) & $1.5 \times 10^{4}$ & 4\% & $9.56 \times 10^{-6}$ & - & \\ \hline
      \rule{0pt}{14pt}Next generation ACT & $1.5 \times 10^{9}$ & 10\% & $1.00 \times 10^{-7}$ &
      0.25 & \\
      Next generation PPT & $2 \times 10^{4}$ & 1\% & $1.00 \times 10^{-7}$ & - & \\
      \hline
    \end{tabular}
  \end{center}
  \caption{The performance parameters~\cite{Weekes:2003,Bertone:2004pz} for current and planned $\gamma$-ray
    telescopes, including both ACTs (WHIPPLE, GRANITE II, HESS~\cite{Hinton:2004eu}, and VERITAS~\cite{Veritas:2006ve})
    and space telescopes (EGRET and GLAST~\cite{Glast:2006gl}).  Also included are the parameters
    corresponding to the hypothetical ``next generation'' atmospheric Cerenkhov telescope (ACT)
    and space-based pair production telescope (PPT) we have used in our analysis.\label{tab:detectors}}
\end{table}

\begin{figure}[ht!]
  \begin{center}
    \includegraphics[width=13cm]{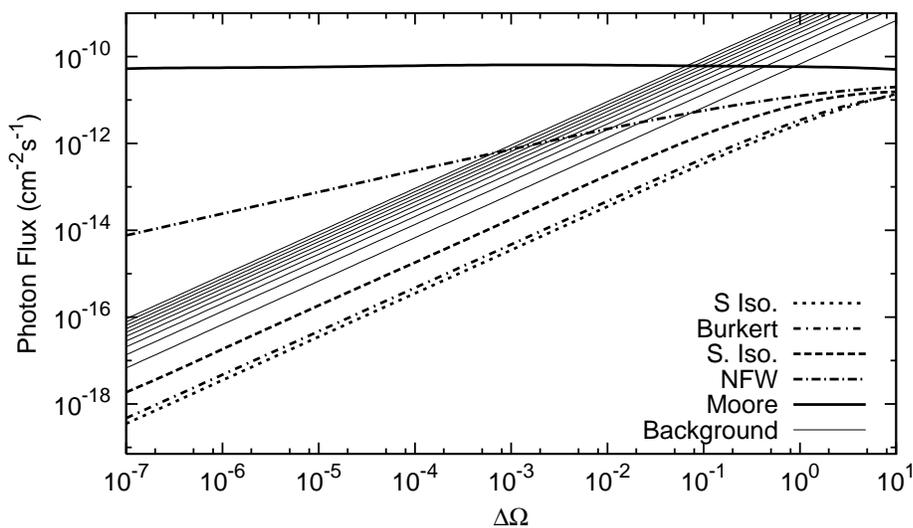}
  \end{center}
  \caption{The expected flux from the annihilation of a heavy Wino LSP, as a
    function of $\Delta\Omega$, that would be detected by a
    satellite detector aimed at the galactic center~\cite{Thomas:2005te}.  Also included
    is the
    anticipated background flux for a 2.3~TeV Wino
    at such a detector for different values of detector
    energy resolution, ranging from $\Delta E/E = 50\%$ (top line) to
    $\Delta E/E = 5\%$ (bottom line).  It
    should be noted that the spread in the signal is
    smaller than the width of the energy bin.  From this, it is apparent that
    for the NFW and Moore
    et al.\ profiles, the prospects for detection increase with better angular
    resolution (decreasing
    $\Delta\Omega$).  For a Higgsino LSP, the resulting curves are
    similar, but the signal is two orders of magnitude lower.  The
    background flux for a 1.1~TeV Higgsino also is increased by a factor of
    $\sim10$ over the 2.3~TeV Wino
    case, owing to the lower energy of the signal photons.\label{fig:fluSp}}
\end{figure}

\indent The advantage of satellite facilities, examples of which include EGRET and the
soon-to-be-launched pair-production telescope GLAST~\cite{Glast:2006gl}, is that they can
have excellent angular resolution and energy resolution.  The field of view for GLAST,
for example, will be on the order of a steradian, its angular resolution in the TeV range
will be \(\sim0.1^{\circ}\), and its energy resolution will be on the order of 4\%.  The
background seen by such facilities is---as one might expect---the actual diffuse
gamma-ray background, which is not currently well known for energies in the TeV range.
The best that can currently be done is to make the assumption that the power law spectrum
from EGRET data (good up to \(\sim100 \mbox{ GeV}\)) can be extrapolated to higher
energies~\cite{Hunger:1997we,Sreekumar:1997un,Strong:2004de}. In doing this, one obtains
a power-law of the form
\begin{equation}
  \label{eq:alphaSpace}
  \frac{dn_{\mathrm{BG}}}{d\Omega dE}=N_{0}
  \left(\frac{E}{1\mbox{ GeV}}\right)^{-\alpha}\mathrm{cm}^{-2}\mathrm{s}^{-1}
  \mathrm{GeV}^{-1}\mathrm{sr}^{-1},
\end{equation}
with a numerical prefactor \(N_{0}\) on the order of \(10^{-6} \mbox{ cm}^{2}\) and an
exponent \(\alpha\) somewhere between 2.0 and 2.5 (following~\cite{Sreekumar:1997un}, we
take \(\alpha\) to be \(2.1\) and \(N_{0}\) to be \(7.32\times10^{6}\)).  Since GLAST
will provide a great deal of information about the diffuse \(\gamma\)-ray background at
high energies, we can expect the uncertainties in \(\alpha\) and \(N_{0}\) to be
dramatically reduced once it begins taking data.  In figure~\ref{fig:fluSp}, we plot the
$\gamma$-ray flux that a satellite detector would observe as a function of angular
acceptance along with the $\gamma$-ray background flux (for several different values of
\(\Delta E/E\)) determined from equation~(\ref{fig:fluSp}).  It should be noted that
while the signal contours are largely independent of \(m_{\mathit{LSP}}\), the background
contours are not: the power-law relation~(\ref{eq:alphaSpace}) indicates that with
positive \(\alpha\), background flux will decrease with increasing \(m_{\mathit{LSP}}\).
Thus the results given for \(m_{\mathit{LSP}}\) near the
bounds~(\ref{eq:mwUPb})\,-\,(\ref{eq:mhUPb}) represent the best-case scenario.  One can
see that for the sharply peaked Moore et al.\ and NFW profiles, the signal exceeds the
background significantly when \(\Delta\Omega\) is small.

\begin{figure}[ht!]
  \begin{center}
    \includegraphics[width=15cm]{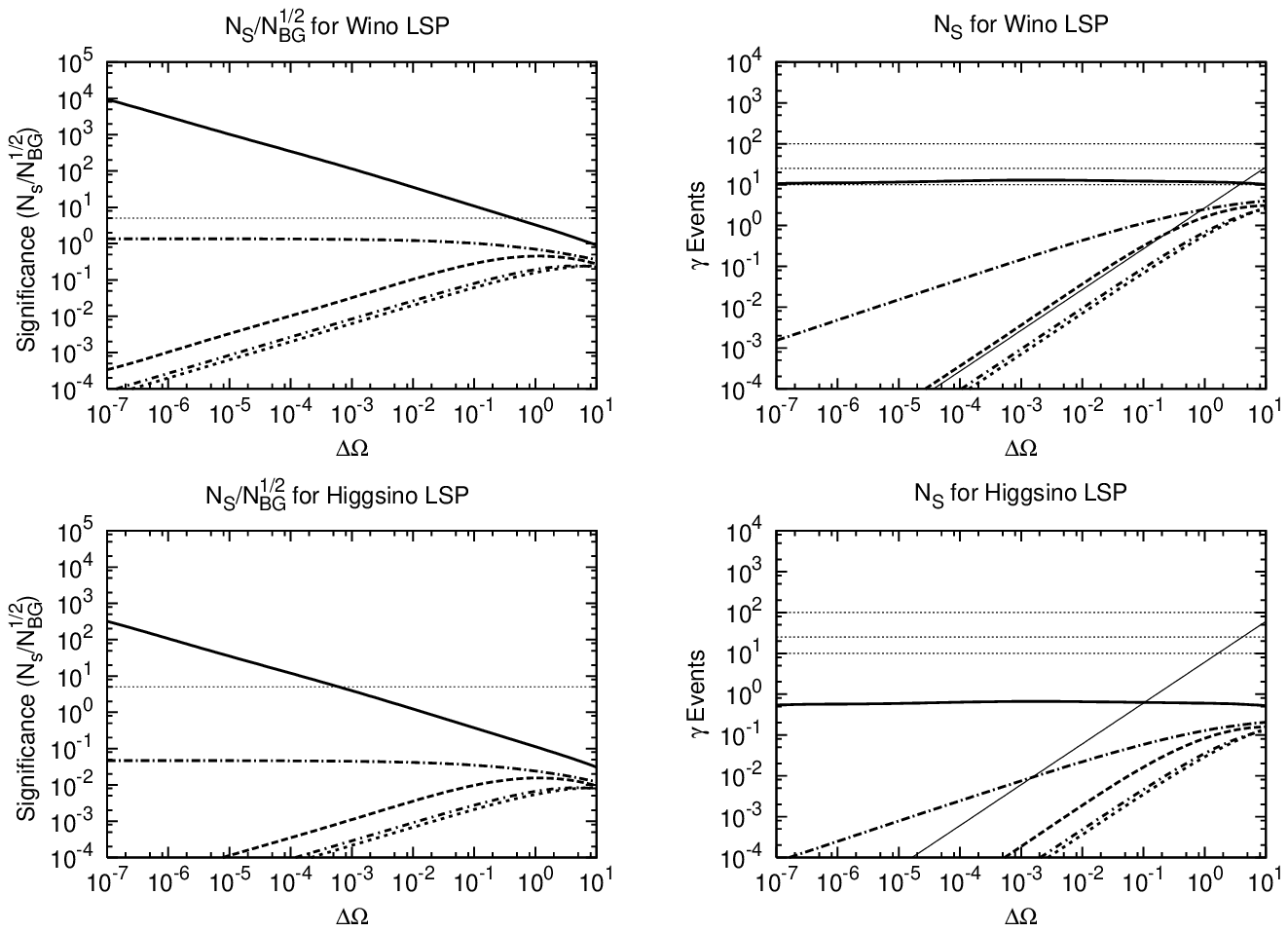}
  \end{center}
  \caption{The ratio of $N_{\mathrm{signal}}/\sqrt{N_{\mathrm{background}}}$
    (left panels) and total number of photons (right panels) collected by a generic
    space telescope with an effective area of
    $2 \times 10^{4}\mbox{ cm}^{2}$
    and an energy resolution of 1\%, over a range of $\Delta\Omega$, after
    $10^{7}\mbox{ s}$ (about 1/3 of an active year) of viewing time, and for both
    Wino (top panels)
    and Higgsino LSP (bottom panels)~\cite{Thomas:2005te}.  See figure~\ref{fig:fluCh}
    caption for the halo model key.  The
    threshold for $5\sigma$ discovery has been
    included for reference in the significance graphs, and contours corresponding
    to 10, 25, and 100 events have been included in the event
    count graphs.  It can be seen here that for such a space telescope, no halo
    profile is capable of producing the
    25 events necessary for detection, and that only for the Moore et al.\ profile is
    the significance criterion even achieved.\label{fig:EventsSigSp}}
\end{figure}

\indent Satellite detectors do have a major drawback, however, which is that they are
limited by collection area constraints.  In order for the signal registered at any
detector to be interpreted as a discovery, not only must the significance level (the
ratio of \(N_{S}\), the total number of signal photons registered, to
\(\sqrt{N_{\mathrm{BG}}}\), where \(N_{\mathrm{BG}}\) is the total number of background
photons registered) exceed \(5\sigma\), but the total number of detected photons must
exceed 25, the threshold below which Poisson statistics give an equivalent confidence
limit\footnote{While the likelihood of random statistical fluctuations at the \(5\sigma\)
level increases with improved energy resolution, these can be differentiated from a true
signal by requiring the signal to be consistent over multiple trials.}.  These
requirements, when written explicitly in terms of \(A_{\mathrm{eff}}\), \(\Delta E/E\),
\(\Delta\Omega\), and observation time, are
\begin{equation}
  \label{eq:fivesigma}
  (.68)^{2}\left(\frac{\Phi(\Delta\Omega)\sqrt{A_{\mathrm{eff}}t}}{\sqrt{\Phi_{\mathrm{BG}}(\Delta\Omega,\Delta E/E, \epsilon_{\mathrm{had}})}}\right)\geq 5
\end{equation}
\begin{equation}
  \label{eq:eventcount}
  \Phi(\Delta\Omega)A_{\mathrm{eff}}t\geq 25.
\end{equation}
Since the effective collection area is constrained by the size of the telescope itself,
present satellite facilities tend to have an \(A_{\mathrm{eff}}\) on the order of
\(10^{4}\mbox{ }\mathrm{cm}^{2}\), and dramatic (i.e.\ order of magnitude) improvements
on this in the future are unlikely.  The consequences of this are shown in
figure~\ref{fig:EventsSigSp}.  Here, contours of total event count at an advanced,
hypothetical space telescope (the ``Next generation PPT'' from table~\ref{tab:detectors})
are presented alongside significance plots corresponding to each of the halo models
discussed above for both a Wino and Higgsino LSP, assuming an exposure time of \(10^{7}\)
s.  While obtaining the necessary \(N_{S}/\sqrt{N_{\mathrm{BG}}}\) to claim a discovery
is difficult in itself (and only achieved for the Moore et al.\ profile), the most
troublesome issue is that none of the halo profiles surveyed yield enough events to
register a \(5\sigma\) discovery.  Space telescopes are thus ``event-count-limited,'' and
as a result, the prospects for detecting monoenergetic photons from heavy LSP decay at
such facilities (including GLAST) are quite dim.

\indent For ACTs, the situation is quite different.  This class of detector (examples of
which include the HESS~\cite{Hinton:2004eu} and VERITAS~\cite{Veritas:2006ve} arrays)
operates by observing showers of Cherenkhov light produced by the incidence of
high-energy \(\gamma\)-rays on the Earth's upper atmosphere.  The general scale of the
effective area associated with an ACT is determined by the area of the Cherenkhov light
pool on the ground, which is \(\sim5\times10^{8}\mbox{ }\mathrm{cm}^{2}\).  Compared to
the aforementioned \(A_{\mathrm{eff}}\) scale associated with satellite facilities, this
is enormous. The trade-off is that uncertainties in reconstructing the energy of the
primary photon from the properties of the radiation shower place limits on the energy
resolution, which is generally quite poor: a single imaging detector can achieve \(\Delta
E/E\simeq 30 -40\%\); an array of parallel detectors, \(10-15\%\)).  Furthermore, the
background ``seen'' by an ACT is not simply the diffuse $\gamma$-ray background.  Any
event that precipitates a similar Cherenkhov cascade is a source of background, and the
primary background at such detectors actually comes from cosmic-ray protons, electrons,
etc., which dominate over the diffuse gamma-ray background by an order or two of
magnitude.  It is possible to discriminate between hadronic showers and those initiated
by $\gamma$-rays to a degree due to the shape of the cascade and to the time spread of
the light pulse, but showers initiated by leptons (predominately electrons) are
indistinguishable from $\gamma$-ray cascades.  These backgrounds are higher than those
seen by satellite detectors, though their spectra are reliably known up to 5~TeV .  The
power-law behavior~\cite{Bergstrom:1997fj} for hadronic and leptonic background events is
given by
\begin{eqnarray}
  \frac{dN_{\mathrm{had}}}{dEd\Omega}=1.0\cdot 10^{-2}\epsilon_{\mathrm{had}}
    \left(\frac{E_{0}}{1\mbox{ GeV}}\right)^{-2.7}\mbox{ }
    \mathrm{cm}^{-2}\mbox{s}^{-1}\mbox{ GeV}^{-1}\mathrm{sr}^{-1}\label{eq:alphaACThad} \\
  \frac{dN_{\mathrm{e^{-}}}}{dEd\Omega}=6.9\cdot 10^{-2}
    \left(\frac{E_{0}}{1\mbox{ GeV}}\right)^{-3.3}\mbox{ }
    \mathrm{cm}^{-2}\mbox{s}^{-1}\mbox{ GeV}^{-1}\mathrm{sr}^{-1},\label{eq:alphaACTlep}
\end{eqnarray}
where we have replaced \(N_{0}\) and the power-law index \(\alpha\) with their explicit
numerical values.  Here, \(\epsilon_{\mathrm{had}}\) is a dimensionless coefficient which
represents the detector's ability to reject hadronic events based on the criteria
discussed above, normalized to the performance of the WHIPPLE telescope (for which
\(\epsilon_{\mathrm{had}}=1\)).\footnote{In other words, $\epsilon_{\mathrm{had}}=0.5$
indicates a hadronic rejection twice as good as WHIPPLE's and $\epsilon_{\mathrm{had}}=0$
indicates perfect hadronic rejection and a background consisting purely of leptonic and
photonic events.} The total observed background is the sum of the hadronic and leptonic
backgrounds given in equations~(\ref{eq:alphaACThad}) and~(\ref{eq:alphaACTlep}).
Improvements in hadronic rejection techniques have since lowered
\(\epsilon_{\mathrm{had}}\) to around \(0.25\) at instruments such as HESS and VERITAS.
Once again, both hardonic and leptonic contributions to the overall Cherenkhov background
scale with photon energy by a power law with a negative index; hence the best-case
scenario will involve an LSP mass near the upper bounds given in
equations~(\ref{eq:mwUPb})\,-\,(\ref{eq:mhUPb}).

\begin{figure}[ht!]
  \begin{center}
    \includegraphics[width=13cm]{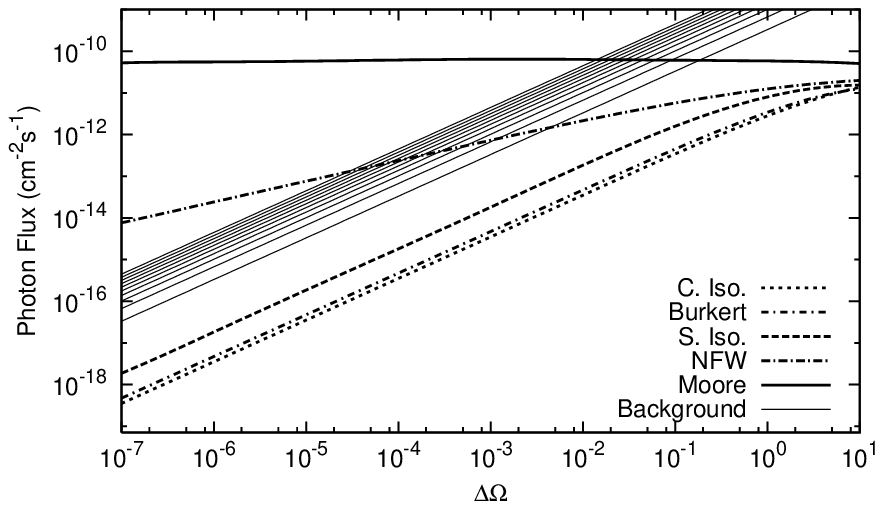}
  \end{center}
  \caption{The expected flux from the annihilation of a heavy Wino LSP, as a
    function of $\Delta\Omega$, that would be detected by a
    Cherenkhov detector aimed at the galactic center~\cite{Thomas:2005te}.  Also included
    is the
    anticipated background flux from a 2.3~TeV Wino at such a detector for different values of
    detector energy resolution, ranging from $\Delta E/E = 50\%$ (top line) to
    $\Delta E/E = 5\%$ (bottom line).
    It should be noted that the spread in the signal is
    smaller than the width of the energy bin.  From this, it is apparent that
    for the NFW and Moore
    et al. profiles, the prospects for detection increase with better angular
    resolution (decreasing
    $\Delta\Omega$).  For a 1.1~TeV Higgsino LSP, the resulting curves are
    similar, but the signal is two orders of magnitude lower.  The
    background flux is also increased by a factor of $\sim10$ over the 2.3~TeV Wino
    case, owing to the lower energy of the signal photons\label{fig:fluCh}}
\end{figure}

\begin{figure}[ht!]
  \begin{center}
    \includegraphics[width=15cm]{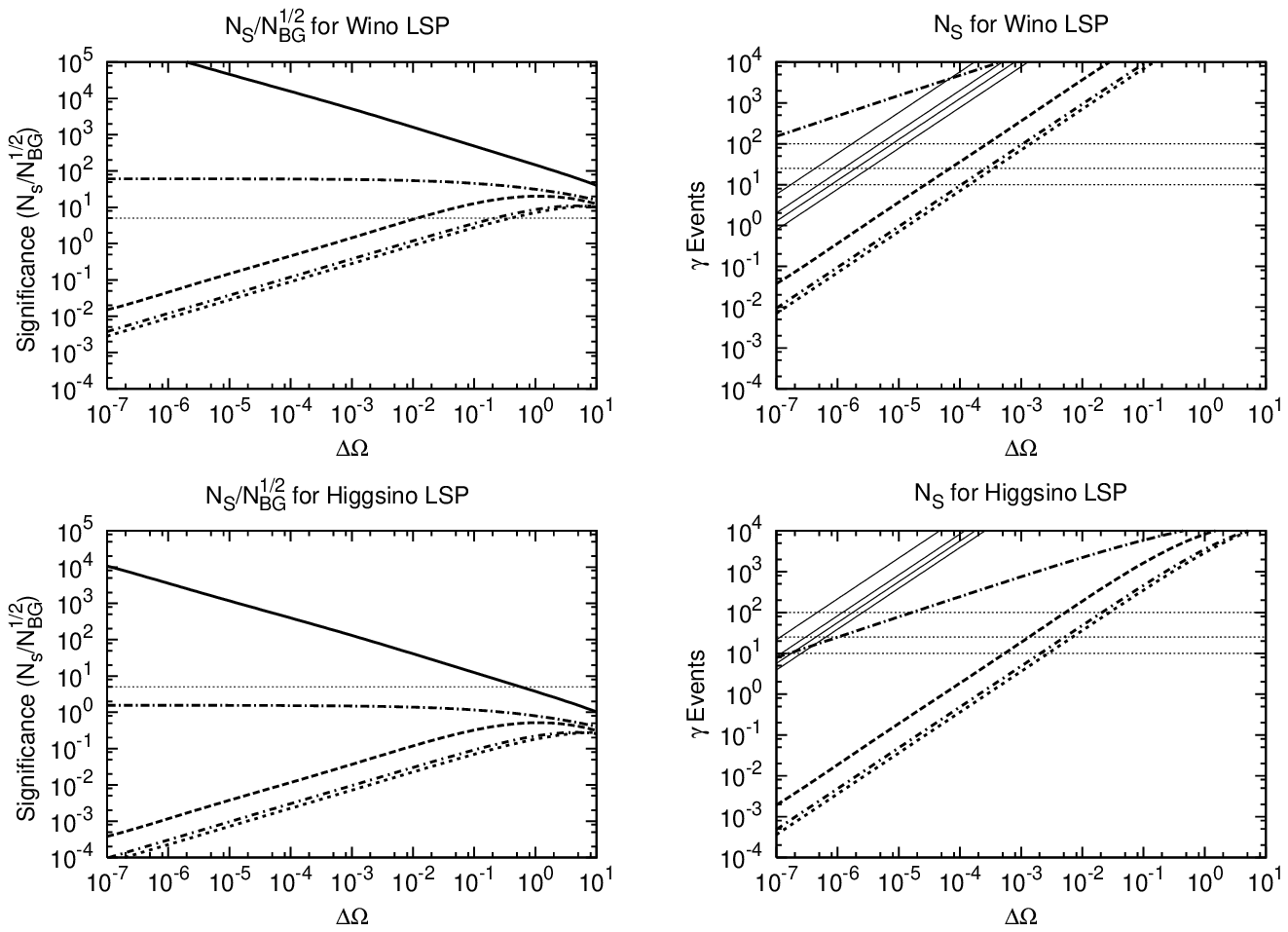}
  \end{center}
  \caption{The ratio of $N_{\mathrm{signal}}/\sqrt{N_{\mathrm{background}}}$
    (left panels) and total number of photons (right panels) collected by a generic
    Cherenkhov array with an effective area of
    $1.5\times 10^{9}~\mathrm{cm}^{2}$
    and an energy resolution of 10\%, over a range of $\Delta\Omega$, after
    $10^{7}$~s (about 1/3 of an active year) of viewing time, and for
    both a heavy Wino (top panels)
    and Higgsino LSP (bottom panels)~\cite{Thomas:2005te}.  See figure~\ref{fig:fluCh}
    caption for the halo model key.  The
    threshold for $5\sigma$ discovery has been
    included for reference in the significance graphs, and contours corresponding
    to 10, 25, and 100 events have been included in the event
    count graphs.  It can be seen here that there are real prospects for detection
    with such a Cherenkhov detector, provided that the galactic CDM halo density
    resembles the NFW or Moore et al.\ profiles.\label{fig:EventsSigCh}}
\end{figure}

\indent In figure~\ref{fig:fluCh}, we plot the $\gamma$-ray flux observed at an ACT a
function of angular acceptance for the different halo profiles surveyed along with
background flux contours corresponding to different values of \(\Delta E/E\).  The
background is larger than in the space telescope case (compare figure~\ref{fig:fluSp}) by
about a factor of ten.  In figure~\ref{fig:EventsSigCh}, we plot both the signal
significance and total number of signal photons recorded at an advanced, hypothetical
Cherenkhov array (the ``Next generation ACT'' from table~\ref{tab:detectors}) for a range
of \(\Delta\Omega\) and an exposure time of \(10^{7}\)~s, in both the Wino and Higgsino
LSP cases.  It is evident from this plot that despite the larger background flux and
poorer energy resolution, ACTs are far more likely to detect a statistically significant
$\gamma$-ray signal from heavy LSP annihilation than are satellite detectors, due to
their far greater effective area (on the order of \(10^{8} - 10^{9} \mbox{ cm}^{2}\)).
Since HESS and VERITAS have an angular resolution (and hence minimum \(\Delta \Omega\))
on the order of $10^{-7}$~steradians, detection would not be difficult if the dark matter
density conformed to one of the more cuspy profiles.  However, since the field of view
(and hence maximum \(\Delta \Omega\)) for an ACT is generally around
$10^{-3}$~steradians, the less sharply peaked dark matter distributions (the Burkert
profile and the two isothermal models) will still be out of reach.  It is an interesting
coincidence that the NFW profile nearly demarcates the line between detection and
non-detection for presently operational facilities: if the actual dark matter
distribution is cuspier than that given by the NFW profile, the \(\gamma\)-ray signature
of Wino dark matter in PeV-scale split supersymmetry should be detectable at the next
generation of ACTs; if the actual profile is much less sharply peaked, it is unlikely
that such a signal would ever be detectable at an ACT.

\indent As mentioned previously, the energy resolution for ACTs is limited by
considerations related to event reconstruction from the Cherenkhov
cascade~\cite{Weekes:2003}, and hence comparatively poor---around \(10-15\%\).  Due to
this intrinsic limitation, it is unlikely that future facilities will offer significant
(i.e.\ order of magnitude) improvements in \(\Delta E/E\) over present facilities.  Since
the splitting between the $\gamma\gamma$ and $Z\gamma$ lines~(\ref{eq:TwoLines}) requires
\(\Delta E/E=0.8\%\) when \(m_{\mathit{LSP}}=500\)~GeV, and even more precision for LSP
masses near the upper bound~(\ref{eq:mwUPb})\,-\,(\ref{eq:mhUPb}) from thermal
generation, the prospects for differentiating these lines at an ACT are effectively nil.
Still, even if these lines cannot be resolved, the detection of any effectively
monoenergetic photon signal in the TeV range would serve as provocative evidence for
heavy neutralino dark matter.

\begin{figure}[ht!]
  \centering
    \includegraphics[width=15cm]{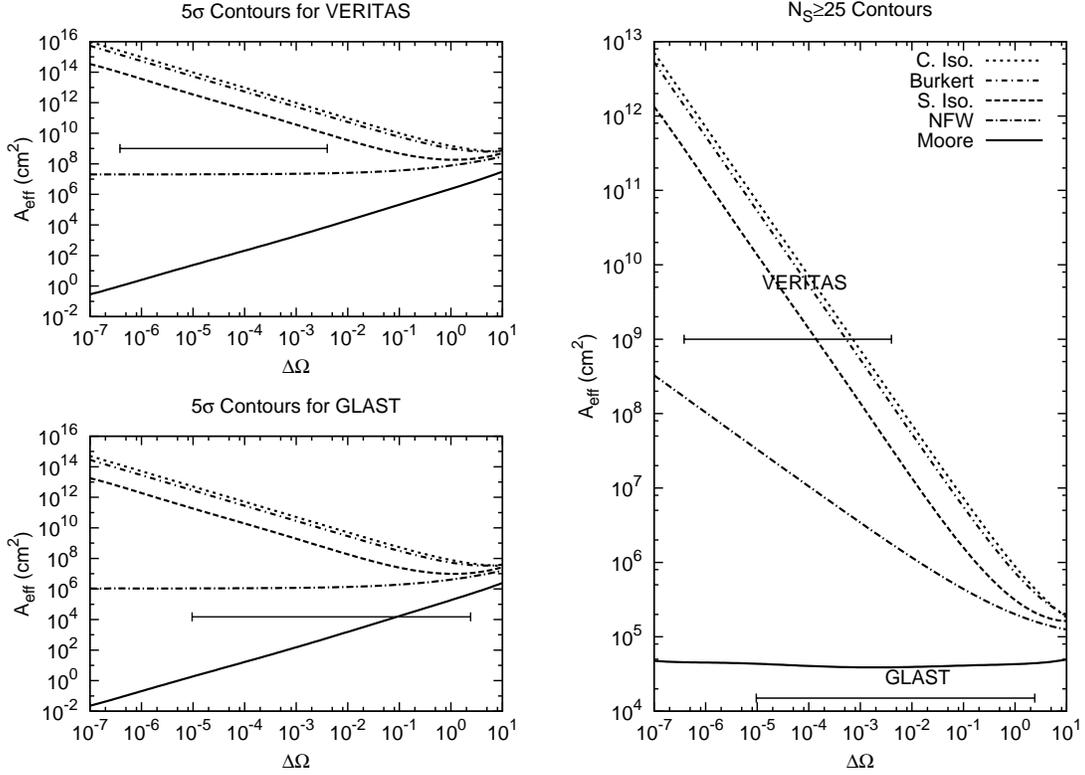}
  \caption{Detection boundary contours in $A_{\mathrm{eff}}\,-\,\Delta\Omega$
    parameter space for the $\gamma$-ray signature of a 2.3 TeV Wino,
    based on the $5\sigma$ significance requirement from
    equation~\ref{eq:fivesigma} for VERITAS (top left panel), with
    $A_{\mathrm{eff}}=1\times 10^{9}\mbox{ cm}^{2}$ and $\Delta E/E$, and
    GLAST (bottom left panel), with
    $A_{\mathrm{eff}}=1.5\times 10^{4}\mbox{ cm}^{2}$;
    and based on the $N_{S}\geq25$ event count requirement from
    equation~\ref{eq:eventcount} (right panel), for a variety of halo profiles.  Bars
    showing the range of angular acceptances that can be chosen at VERITAS and
    GLAST have
    also been included.  In order to register a discovery at either of these
    facilities for a given halo model, \textit{both} the $5\sigma$ and
    $N_{s}\geq25$
    contours for that model must lie below the bar corresponding to that facility~\cite{Thomas:2005te}.
    \label{fig:ContCh}}
\end{figure}

\indent Thus far, we have examined the potential for detecting energetic photons from
heavy neutralino annihilation at current and planned facilities (or hypothetical ones
only slightly more advanced than these), but it is also interesting to take a slightly
different point of view and ask exactly what attributes would a detector need in order to
conclusively register a discovery.  We have seen that the effective area
\(A_{\mathit{eff}}\) and angular acceptance \(\Delta \Omega\) (which may be tuned to any
value between the detector's angular resolution and its field of view) are the most
important of the detector attributes in the detection of the LSP annihilation signal, so
these variables will become our primary focus.  In figure~\ref{fig:ContCh}, we plot
contours from both the significance constraint~(\ref{eq:fivesigma}) and the event count
constraint~(\ref{eq:eventcount}) in \(A_{\mathrm{eff}}\)-\(\Delta\Omega\) space.  Since
the significance contours are slightly different for space telescopes and ACTs, which
``see'' different backgrounds, we present a separate plot for each case (an ACT in the
top left panel; a space telescope in the bottom left).  The results shown are for
\(10^{7}\mbox{ s}\) of viewing time, and the \(\Delta E/E\) values used in computing the
significance limits are those given in table~\ref{tab:detectors} for the GLAST telescope
and the VERITAS array.  The horizontal bars correspond to GLAST and VERITAS themselves,
and the endpoints thereof are given by each detector's respective angular resolution and
field of view.  The criterion for discovery at either of these facilities is that for a
given halo model, both the \(5\sigma\) and \(N_{s}\geq25\) contours for that model lie
below the corresponding bar.  From this figure, it is once again evident that GLAST,
primarily due to its small effective area, would be unable to detect CDM from PeV-scale
split supersymmetry at all, while VERITAS would have far better hopes for detection.

\indent Perhaps more importantly, however, it is also evident from
figure~\ref{fig:ContCh} what improvements in detector area, angular resolution, and field
of view would be necessary in order to detect a signal in even the least sharply peaked
of halo models surveyed.  A space telescope would need about a factor of 5 increase in
\(A_{\mathit{eff}}\) in order to register the requisite number of signal events to claim
a discovery even in the Moore et al.\ profile case, and this would require significant
feats of engineering.  The best way to increase the effective area of an ACT is to add
further telescopes to the detector array, but the increase in \(A_{\mathit{eff}}\) from
each such addition is merely additive.  We see from figure~\ref{fig:ContCh} that in order
to register a discovery when the dark matter profile is one of the those included in our
survey less sharply peaked than the NFW profile (the Burkert profile and the two
isothermals), a factor of ten increase in \(A_{\mathit{eff}}\) is required (one could
theoretically also widen the field of view by several orders of magnitude, but in
practice this is even more difficult). This is in principle possible, of course, if one
adds enough telescopes to a given array, but this is an expensive and somewhat
impractical proposition.


\indent In addition to the discrete lines arising from to
\(\tilde{N}_{1}\tilde{N}_{1}\rightarrow \gamma\gamma\) and
\(\tilde{N}_{1}\tilde{N}_{1}\rightarrow \gamma Z\), the \(\gamma\)-ray spectrum from
heavy neutralino annihilation has a continuum component which arises primarily from pion
decay.  The requirement that this continuum contribution not exceed the flux observed by
EGRET~\cite{Sreekumar:1997un} provides another phenomenological check on the model.  It
was shown in~\cite{Ullio:2001qk} that the continuum photon flux produced by the decay of
a heavy Wino or Higgsino LSP will be low enough that no conflict arises between the
predictions of the dark matter scenario model and EGRET data.  This is one advantage of
heavy Wino or Higgsino dark matter in AMSB scenarios: this constraint is not a trivial
one, and may be relevant in many models with a lighter LSP.

\indent So far we have considered only the contribution to the \(\gamma\)-ray flux
arising from annihilations in our own galactic halo, but it is also interesting to
consider the effects of dark matter annihilation in other galaxies to the total observed
photon background.  The \(\gamma\)-ray flux arising from extragalactic WIMP annihilation
has been investigated by several
authors~\cite{Ullio:2002pj,Taylor:2002zd,Ando:2005hr,Oda:2005nv}.  The results depend
significantly on the choice of halo profile and on other astrophysical inputs about which
there is substantial uncertainty, including the number density distribution (often also
called the cosmological mass function) \(\frac{dn}{dM}(M,z)\) of halos in the universe as
a function of halo mass \(M\) and redshift \(z\) and the correlation between halo mass
and dark matter concentration. The situation is further complicated by the possibility
that a nontrivial fraction of dark matter could be bound in small subhalos within larger,
virialized halos~\cite{Ghigna:1998vn,Klypin:1999uc}, for in this case the contribution to
the \(\gamma\)-ray flux from extragalactic dark matter annihilation can be substantially
increased.  Owing to these uncertainties, the best way to proceed is to choose a
reasonable astrophysical model that is not too conservative (in terms of the resulting
\(\gamma\)-ray flux) and ascertain whether the continuum spectrum from LSP annihilation
is consistent with EGRET data.  Results corresponding to the choice of a
Press-Schechter~\cite{Press:1973iz} cosmological mass function, an NFW halo profile
without subhalos, and a Wino or Higgsino LSP with a mass of 180~GeV were derived
in~\cite{Ullio:2002pj}.  It was shown that for the relatively cuspy NFW profile, the
continuum component of this flux was not large enough to conflict with EGRET data, and
one would expect a heavier LSP of the same sort would also be compatible with the
observed \(\gamma\)-ray spectrum at energies \(E_{\gamma}\lesssim 100\)~GeV.  For a
cuspier distribution, such as the Moore et.\ al profile, or a model where a substantial
fraction of the dark matter is concentrated in subhalos, however, the continuum flux
could potentially come into conflict with these measurements.

\indent In addition to the continuum flux, the extragalactic photon spectrum from LSP
annihilation also has a ``discrete'' component---the
\(\tilde{N}_{1}\tilde{N}_{1}\rightarrow \gamma\gamma\) and
\(\tilde{N}_{1}\tilde{N}_{1}\rightarrow \gamma Z\) lines, which will be smeared out
somewhat due to absorption and redshifting effects.  If the diffuse \(\gamma\)-ray
background in the TeV range indeed obeys the same power-law
relation~(\ref{eq:alphaSpace}) that it does at lower energies, the \(\gamma\gamma\) and
\(Z\gamma\) peaks will likely be difficult to detect in particle physics models of this
sort.  However, if the contribution to the \(\gamma\)-ray spectrum from all other sources
drops significantly when \(E_{\gamma}\gtrsim 100\)~GeV, as is the case in certain models
where blazar emissions dominate this spectrum at high energies~\cite{Ullio:2002pj}, these
peaks could dominate over the rest of the background by several orders of magnitude. In
practice, however, it is unlikely that the signal from the extragalactic annihilation of
a TeV-scale Wino or Higgsino LSP offers much in the way of additional prospects for
indirect detection.  It was illustrated in figure~\ref{fig:ContCh} that GLAST, like all
space telescopes, is hindered in its ability to meet the event count requirement
(\ref{eq:eventcount}) necessary for signal detection by its small effective area.  Even
when \(\Delta\Omega\) is adjusted to near the telescope's field of view, the
extragalactic photon flux from LSP annihilation will likely be insufficient to meet this
requirement, and hence there is good reason to think that GLAST will fare little better
in detecting extragalactic LSP annihilation than it will in detecting LSP annihilation at
the center of our own galaxy, no matter what the shape of the diffuse photon background
is at high energies.  ACTs are also unlikely to detect this signal, since the background
they see~(\ref{eq:alphaACThad})\,-\,(\ref{eq:alphaACTlep}) is an order of magnitude
larger than that seen by space telescopes and does not depend on the shape of the diffuse
\(\gamma\)-ray background.

\indent The message here is that while the \(\gamma\)-ray spectrum from extragalactic LSP
annihilation is interesting both because it provides a consistency check on a given
particle physics model (that the continuum spectrum does not conflict with EGRET data)
and because it contains distinctive structures (the ``monoenergetic'' peaks) at energies
as yet unprobed, the former consideration is not damning for the case of heavy Wino or
Higgsino dark matter, and the latter is not likely to yield any new possibilities for
indirect detection.  As with the case with the signal from LSP annihilations in our own
galactic halo, the difficulty in detecting the signal from extragalactic LSP annihilation
stems from the limitations of satellite detectors due to their small collection areas.


\indent To summarize the results of this chapter, although energy resolution
considerations rule out discrimination between the $\gamma\gamma$ and $Z\gamma$ lines
from heavy LSP decay, the detection of a single monoenergetic photon ``line'' is still a
possibility at atmospheric Cherenkhov detectors like HESS and VERITAS, which have a large
enough effective collection area to register a statistically significant discovery. Space
telescopes like GLAST, on the other hand, have far smaller effective areas and hence will
be unlikely ever to observe the signal in question.  Detection at Cherenkhov facilities
is far from guaranteed, however, and prospects for it depend significantly on the dark
matter distribution in our galaxy.  Continuum photons are also produced during LSP
annihilations, both in our galactic halo and in the halos of other galaxies, but these
are not likely to run afoul of current experimental constraints on the \(\gamma\)-ray
background unless the universal dark matter profile is extremely cuspy or involves
substantial substructure.

\chapter{CONCLUSIONS\label{ch:Conclusions}}

\indent

Leptogenesis is an attractive model, and the prospect that its ability to relate the
smallness of neutrino masses to the mechanism responsible for baryogenesis could be
realized in a scenario where neutrinos are purely Dirac is an interesting one for many
reasons. We have shown that thermal Dirac leptogenesis is not only an intriguing
theoretical curiosity, but also a genuinely viable phenomenological model. Not only is it
capable of producing a baryon-to-photon ratio \(\eta\) for the universe that matches that
observed by WMAP, but it can also satisfy all relevant constraints from cosmology, flavor
physics, etc.\ while reproducing the observed neutrino spectrum.

\indent Specifically, from the research presented in this work and that originally
presented in the publications~\cite{Thomas:2005rs,Thomas:2006gr,Thomas:2005te} and
reviewed herein, the following results and conclusions were obtained:

\begin{itemize}

\item {\bf There is a particularly simple, theoretically-motivated realization of Dirac
leptogenesis, constrained hierarchical Dirac leptogenesis (CHDL), which reproduces the
observed neutrino spectrum.}

In its most general form, the Dirac leptogenesis
superpotential~(\ref{eq:DiracLepSuperpotential}) contains a large number of new free
parameters, including the heavy particle masses \(M_{\Phi_{i}}\), the scalar VEV \(\chi\)
and the elements of the general, complex trilinear coupling matrices \(\lambda\) and
\(h\).  It is thus not terribly surprising that some configuration of parameters is
capable of satisfying all experimental bounds on neutrino masses and mixings, but it is
of interest that this set of constraints can be satisfied in a simple,
theoretically-motivated model that contains only five free parameters (four, if one does
not include the highly constrained coupling matrix entry \(b_{3}\)).
In~\cite{Thomas:2005rs}, it was shown that a particularly simple model of Dirac
leptogenesis, which was dubbed constrained hierarchical Dirac leptogenesis (CHDL), in
which the trilinear coupling matrices \(\lambda\) and \(h\) are antisymmetric, the mass
matrix \(M_{\Phi}\) is real and diagonal, and the small ratio
\(\delta=M_{\Phi_{1}}/M_{\Phi_{1}}\) dictates the textures of the neutrino mass matrix,
the experimental constraints are readily satisfied.  The model is simple, robust and
well-motivated by flavon physics (see section~\ref{sec:FlavonExtension}).  Furthermore,
as was shown in~\cite{Thomas:2005rs} and reviewed in chapter~\ref{sec:NumSolBigF}, this
model is compatible with the additional constraints arising from baryogenesis,
astrophysics, and cosmology.

\item {\bf Dirac leptogenesis strongly prefers a heavy gravitino with $m_{3/2}\gtrsim 10^{5}$~GeV and
is compatible with split supersymmetry.}

Dirac leptogenesis must respect a  battery of astrophysical constraints in order to be
considered a phenomenologically viable theory, and the most stringent of these are
related to gravitino cosmology.  The connection between this subject and leptogenesis
occurs through the reheating temperature \(T_{R}\).  On the one hand, \(T_{R}\lesssim
M_{\Phi_1}\) is required for thermal leptogenesis; on the other, for certain values of
the gravitino mass \(m_{3/2}\), considerations related to BBN and nonthermal LSP
production place stringent limits on \(T_{R}\).  In~\cite{Thomas:2005rs}, it was shown
that baryogenesis considerations require $M_{\Phi_1}\gtrsim 10^{10}$~GeV, unless some
additional mechanism---such as resonant leptogenesis---is invoked.  Thus, when
\(m_{3/2}\lesssim 10^{5}\)~GeV, nucleosynthesis bounds on \(T_{R}\) from late gravitino
decay render Dirac leptogenesis in its simplest form essentially unworkable; when
\(10^5\)~GeV~\(\lesssim m_{3/2}\lesssim 10^8\)~GeV things are better, but constraints
related to nonthermal LSP creation also cause problems when the splitting between
\(m_{3/2}\) and \(m_{\mathit{LSP}}\) is small. These results are summarized in
table~\ref{tab:M32Regimes}.  For this reason, Dirac leptogenesis strongly prefers a heavy
gravitino with \(m_{3/2}\gtrsim 10^{5}\)~GeV.  Since this is a natural consequence of
split supersymmetry, this scenario becomes one possible context for Dirac leptogenesis.
This scenario has the added advantage of keeping squark and slepton masses heavy to
alleviate flavor-violation concerns, while keeping gauginos light to provide a
(characteristic) dark matter candidate.

\item {\bf Dirac leptogenesis can succeed in models with light ($m_{s}=200$~GeV) scalar masses.}

In any model where supersymmetric scalar particles are light and where the mass
eigenstates of these particles are not diagonal in the flavor basis of the Standard Model
quarks and leptons, loop diagrams involving such particles can potentially give rise to
unacceptable levels of flavor violation.  In Dirac leptogenesis, the renormalization
group running of slepton soft masses will tend to generate off-diagonal terms in the
sneutrino and charged slepton mass matrices. It was shown in~\cite{Thomas:2006gr} that
flavor violation constraints do not rule out CHDL: the model is successful for a
universal scalar mass \(m_{s}\) as low as \(200\)~GeV, respecting all
lepton-flavor-violation constraints from processes such as $\mu\to e\gamma$ and $\tau\to
\mu\gamma$.  Dirac leptogenesis therefore does not require split supersymmetry and will
be phenomenologically viable in a wide range of contexts.  More general
AMSB~\cite{Randall:1998uk,Giudice:1998xp} models, for example, can yield a splitting
between \(m_{3/2}\) and \(m_{LSP}\) large enough to permit a gravitino with a mass
slightly above the BBN bound (\(m_{3/2}\lesssim 10^{5}\)~GeV) and sparticle masses of
\(\mathcal{O}(200\mbox{ TeV})\) without running afoul of any of the late gravitino decay
constraints.

\item {\bf Obtaining the appropriate superpotential for Dirac leptogenesis can be arranged in
a consistent, anomaly-free manner through the introduction of a hidden sector
\(U(1)_{N}\) symmetry.}

The Dirac leptogenesis superpotential requires some additional symmetry to forbid
tree-level Dirac and Majorana masses for right-handed neutrinos.  It was shown in
section~\ref{sec:sneutrinoCDM} that this can be accomplished in a simple, consistent,
anomaly-free manner by enlarging the hidden sector of the model.  It was shown that the
necessary VEVs are obtained to yield small but nonzero masses and to ensure that all
\(R\)-parity odd fields in the hidden sector become massive after supersymmetry is
broken.

\item {\bf Right-handed sneutrino dark matter is a viable possibility in Dirac
leptogenesis.}

Unlike in Majorana leptogenesis, in Dirac leptogenesis the lightest right-handed
sneutrino \(\tilde{\nu}_{R_1}\) can be light, and as it is neutral under the Standard
Model gauge group, it is a potential dark matter candidate if it is the LSP. Since the
interactions of this particle are extremely weak (as they are suppressed by the powers of
the effective neutrino Yukawa coupling), its relic abundance will in general be far
larger than the upper bound~(\ref{eq:WMAPBounds}) on \(\Omega_{\mathit{CDM}}\) obtained
from WMAP data.  However, Dirac leptogenesis comes equipped with a mechanism for
alleviating this problem: the theory requires an additional symmetry (to forbid Majorana
masses for neutrinos) under which \(\tilde{\nu}_{R_1}\) is necessarily charged.  In
section~\ref{sec:sneutrinoCDM}, it was shown that if this symmetry is a gauged \(U(1)\),
annihilations through the gauge bosons and gauginos associated with this additional
symmetry can reduce the right-handed sneutrino abundance to an appropriate level.

\item {\bf Linking \(\Omega_{CDM}\) to \(\Omega_{b}\) via an asymmetry in some conserved quantum number
is generally quite difficult, at least in supersymmetric theories, and requires a
hierarchy among annihilation rates.}

The possibility that the cold dark matter abundance \(\Omega_{\mathit{CDM}}\) and the
relic density \(\Omega_{b}\) of baryonic matter in the universe, which are of roughly the
same order, might be linked is an intriguing one.  One way of arranging this is the
method of~\cite{Hooper:2004dc}, in which the association is forged by charging a
hidden-sector dark matter candidate under a globally conserved quantum number.  Then dark
matter particles annihilate with their antiparticles until only the excess given by the
asymmetry is left, and this excess becomes the dark matter abundance.  This cannot work
in the usual neutralino dark matter scenario, since Majorana neutralinos cannot develop a
charge asymmetry, but a right-handed sneutrino LSP in Dirac leptogenesis, which carries
lepton number charge, has the right properties to allow this mechanism to work.
Unfortunately, linking the right-handed sneutrino abundance to the lepton number
\(L_{\nu_{R}}\) stored in \(\tilde{\nu}_{R_1}\) in this manner proves difficult in
practice.  It was shown in section~\ref{sec:sneutrinoCDM} that this required the
\(L_{\nu_{R}}\)--violating rate \(\Gamma_{\tilde{\nu}\tilde{\nu}}\) to be small, so as to
preserve the lepton number asymmetry at the appropriate level, whereas it required the
\(L_{\nu_{R}}\)--conserving rate \(\Gamma_{\tilde{\nu}\tilde{\nu}^{\ast}}\) to be several
orders of magnitude larger, so that right-handed sneutrinos could annihilate efficiently
and dark matter would not consequently be overproduced.  However, in a supersymmetric
model, \(s\)-channel diagrams in which a dark matter particle and its antiparticle
annihilate through some virtual \(Z'\) are accompanied by \(t\)-channel diagrams (with
Majorana gaugino intermediaries) that violate \(L_{\nu_{R}}\).  It is worth restating
that the evolution of $L_{\nu_R}$ during the CDM freeze-out epoch will not affect the
baryon number $B$ of the universe whatsoever (the effect of
\(\Gamma_{\tilde{\nu}\tilde{\nu}}\) is simply to shuffle lepton number from
$\tilde{\nu}_{R_1}$ to \(\nu_{R}\) and other light hidden sector fields, and not to
change $\Lhid$), and so right-handed sneutrino dark matter is perfectly compatible with
baryogenesis in Dirac leptogenesis scenarios.  However, it was shown in
section~\ref{sec:sneutrinoCDM} that while it is possible to obtain the correct
\(\Omega_{\mathit{CDM}}\) with a right-handed sneutrino, the inherent tension between
\(\Gamma_{\tilde{\nu}\tilde{\nu}}\) and \(\Gamma_{\tilde{\nu}\tilde{\nu}}^{\ast}\) makes
it very difficult simultaneously to prevent \(L_{\nu_{R}}\) from being washed out to the
extent where it will have no bearing on the dark matter abundance.  It was shown that
there are ways of overcoming this hurdle, such as adding a large number of light scalars
to the theory, but these methods are somewhat contrived.  The prospects may be brighter
for linking \(\Omega_{CDM}\) and \(\Omega_{b}\) in other situations where a dark matter
candidate is charged under some conserved quantum number, but since it is the
supersymmetrization of annihilation processes that lies at the root of the problem, the
issue is a general one---especially since supersymmetric models naturally provide a
compelling solution to the dark matter problem, in the form of the LSP, when \(R\)-parity
is conserved.

\item {\bf The prospects for detecting the monoenergetic photon signal of heavy Wino or Higgsino dark
matter reasonably good at atmospheric Chernenkhov telescopes (ACTs) like HESS and
VERITAS, but slim at space-based facilities like GLAST.}

Since split supersymmetry emerges as one promising context for CHDL, it is relevant to
the study of Dirac leptogenesis, and of more general interest as well, to examine the
discovery potential for the dark matter candidate that naturally emerges from this
scenario: a predominately Wino or Higgsino LSP with a mass around 1~TeV.  The direct
detection prospects for such a particle are somewhat dim, and the easiest way of
discovering such a particle is through the observation of the effectively monoenergetic
photon signal produced by its annihilations at the galactic center.
In~\cite{Thomas:2005te}, we show that space telescopes such as GLAST are highly unlikely
to record the necessary number of signal events to claim a statistically significant
discovery, due to their small effective collection area \(A_{\mathit{Eff}}\).
Ground-based Cherenkhov arrays, however, have far larger effective areas and hence far
better prospects for the discovery of a heavy Wino or Higgsino LSP.

\end{itemize}

\begin{table}[ht!]
\begin{center}
\small{\begin{tabular}{|c|c|c|l|} \hline
\parbox{2.5cm}{\begin{center}Gravitino Mass Range (GeV)\end{center}} &
\parbox{2.0cm}{\begin{center}Maximum \(M_{\Phi_{1}}\) (GeV)\end{center}} & \parbox{2.0cm}{\begin{center}Workability of CHDL?\end{center}} & Comments \\
\hline \tiny \(m_{3/2}\lesssim 10^{5}\)&  \tiny \(10^{6} - 10^{8}\) & \tiny Very Low &
\parbox{4.5cm}{\begin{flushleft}\tiny\(\tilde{G}\) decay during or after BBN.  Insufficient \(\eta\)
generated.\end{flushleft}}\\
\hline \tiny \(10^{5}\lesssim m_{3/2}\lesssim 10^{8}\) & \tiny
\parbox{1.5cm}{\begin{center}\tiny \(10^{9} - 10^{10}\) or higher \end{center}} &
\parbox{2.0cm}{\begin{center}\tiny Depends on \(m_{\mathit{LSP}}\) and the ratio \(m_{\mathit{LSP}}/m_{3/2}\)\end{center}} & \parbox{4.5cm}{\begin{flushleft}\tiny LSP
annihilations ineffective unless \(m_{\mathit{LSP}}/m_{3/2}\) is small. \(T_{R}\)
constrained by nonthermal LSP abundance from \(\tilde{G}\) decay.
Loop-split SUSY works only for \(m_{3/2}\sim10^{5}\)~GeV.  More general split SUSY theories can be successful. \end{flushleft}}\\
\hline \tiny \(10^{8}\lesssim m_{3/2}\lesssim 10^{10}\) & \tiny None & \tiny Excellent &
\parbox{4.5cm}{\begin{flushleft}\tiny \(\Omega_{LSP}\) is thermal, since \(\tilde{G}\) decays
before LSP freeze-out. \(M_{\Phi_{1}}>10^{11}\)~GeV
allowed. \(\nu\) spectrum requirements compatible with CHDL.\end{flushleft}}\\
\hline \tiny \(10^{10}\lesssim m_{3/2}\lesssim 10^{12}\) & \tiny None &
\parbox{2.0cm}{\begin{center}\tiny Questionable (depends on gluino properties)\end{center}} &
\parbox{4.5cm}{\begin{flushleft}\tiny \(\nu\) sector and baryogenesis
okay, but model may have a cosmological gluino problem.\end{flushleft}}\\
\hline \tiny \(  10^{12}\lesssim m_{3/2}\) & \tiny None & \tiny Very Low &
\parbox{4.5cm}{\begin{flushleft}\tiny Potential gluino problem becomes a serious concern.\end{flushleft}} \\ \hline
\end{tabular}}
\end{center}
\caption{The various gravitino mass regimes for split supersymmetry models and the
viability
  of thermal Constrained Hierarchical Dirac Leptogenesis (CHDL) in each case.\label{tab:M32Regimes}}
\end{table}

\indent It should be mentioned that there are a number of experimental checks on the
viability of Dirac leptogenesis in general and on the specific model we have dubbed CHDL.
The major prediction of Dirac leptogenesis is that neutrinoless double-beta decay will
not be observed to any degree, for this process relies on the existence of a Majorana
mass term for right-handed neutrinos.  The discovery of such a process experimentally
would rule the theory out.  In addition, forthcoming results from
MiniBooNE~\cite{Ray:2007yd} should either confirm or deny the LSND result, which will
reveal whether or not the neutrino spectrum produced by CHDL is in fact the one present
in nature (if MiniBooNE confirms the LSND result, Dirac leptogenesis should still be
workable, albeit in a more complicated manifestation than CHDL). Finally, CHDL places
constraints on the neutrino mixing parameter \(\sin\theta_{13}\) (see
figure~\ref{fig:NeutrinoMassPlots}), the value of which will be measured in future
experiments.

\startappendices                     
\chapter{Effective Couplings for Flavor-Violating Calculations\label{app:Looplitudes}}

\indent

For completeness, we list here the results used in our analysis for lepton flavor
violating processes. The amplitudes $A^{L}$ and $A^{R}$ in
equation~(\ref{eq:ViolationRate}) were computed in \cite{Hisano:1995cp} and, with a
trivial extension to include three right-handed sneutrinos, are given by
\begin{equation}
  A^{L}=A^{(c),L}+A^{(n),L}\hspace{0.75cm}\mathrm{and}\hspace{0.75cm}A^{R}=A^{(c),R}+A^{(n),R},
\end{equation}
where the individual amplitudes $A^{(c),L}$, $A^{(n),L}$, $A^{(c),R}$, and $A^{(n),R}$
are
\begin{eqnarray}
A^{(n)L}&=&\frac{1}{32 \pi^2}\sum_{A=1}^{4}\sum_{X=1}^{6}\frac{1}{m^2_{\tilde{\ell}_X}}
\left[
 N_{iAX}^{L} N_{jAX}^{L*}
\frac{1}{6 (1-x_{AX})^4} \right.
\nonumber \\
&&\times (1-6x_{AX}+3x_{AX}^2+2x_{AX}^3-6x_{AX}^2\ln x_{AX})
\nonumber \\
&&\left. +N_{iAX}^{L} N_{jAX}^{R*} \frac{M_{\tilde{\chi}_A^0}}{m_{l_j}}
\frac{1}{(1-x_{AX})^3} (1-x_{AX}^2+2x_{AX} \ln x_{AX}) \right], \label{eq:ANL}
\\
A^{(c)L}&=&-\frac{1}{32 \pi^2} \sum_{A=1}^{2}\sum_{X=1}^{6}\frac{1}{m^2_{\tilde{\nu}_X}}
\left[
 C_{iAX}^{L} C_{jAX}^{L*}
\frac{1}{6 (1-x_{AX})^4} \right.
\nonumber \\
&&\times (2+3x_{AX}-6x_{AX}^2+x_{AX}^3+6x_{AX} \ln x_{AX})
\nonumber \\
&&\left. +C_{iAX}^{L} C_{jAX}^{R*} \frac{M_{\tilde{\chi}_A^-}}{m_{l_j}}
\frac{1}{(1-x_{AX})^3}
(-3+4x_{AX}-x_{AX}^2-2 \ln x_{AX}) \right],\label{eq:ACL}\\
A^{(n,c)R}&=&A^{(n,c)L}|_{L \leftrightarrow R}\label{eq:ACRANR}.
\end{eqnarray}
Here, the indices \(A\) and \(X\) respectively label the gaugino (chargino or neutralino)
and slepton (sneutrino or charged slepton) mass eigenstates, $x_{AX}\equiv
m^{2}_{\chi_{A}}/m^{2}_{\phi_{X}}$, and $C_{iAX}^{L,R}$ ($N_{iAX}^{L,R}$) denote the
effective couplings of charged lepton \(i\) to chargino (neutralino) \(A\) and sneutrino
(charged slepton) \(X\). The flavor mixing terms in~(\ref{eq:MLL})\,-\,(\ref{eq:MLR})
enter into the overall rate~(\ref{eq:ViolationRate}) through $C_{iAX}^{L,R}$ and
$N_{iAX}^{L,R}$, which contain elements of the matrices $U_{\nu}$ and \(U_{\ell}\) that
diagonalize the mass-squared matrices for sneutrinos and charged sleptons, respectively.
The slepton masses also enter into the partial amplitudes~(\ref{eq:ANL}-\ref{eq:ACRANR}).

The effective couplings $N^{L,R}_{iAX}$ and $C^{L,R}_{iAX}$ are
\begin{eqnarray}
  N^{R}_{iAX}&=& -\frac{g_2}{\sqrt{2}} \left(
       [-(U_N)_{A2} -(U_N)_{A1} \tan \theta_W] U^{\ell}_{X,i}
        + \frac{m_{l_i}}{m_W\cos\beta} (U_N)_{A3} U^{\ell}_{X,i+3}
        \right),
\nonumber \\
  N^{L}_{iAX} &=& -\frac{g_2}{\sqrt{2}} \left(
           \frac{m_{l_i}}{m_W\cos\beta} (U_N)_{A3}
           U^{\ell}_{x,i}
           +2 (U_N)_{A1} \tan \theta_W U^{\ell} _{X,i+3} \right),
\nonumber\\
 C^{R}_{iAX}& =& -g_2(O_R)_{A1} U^{\nu}_{X,i},~~~\mathrm{and}
\nonumber \\
 C^{L}_{iAX}& = & g_2\frac{m_{l_i}}{\sqrt{2}m_W\cos\beta}(O_L)_{A2}
                    U^{\nu}_{X,i}
\end{eqnarray}
in terms of the chargino mixing matrices $(O_R)_{A,i}$ and $(O_L)_{A,i}$ the neutralino
mixing matrix $U^{N}_{X,i}$, and the sneutrino and charged slepton mixing matrices
$U^{\nu}_{X,i}$ and $U^{\ell}_{X,i}$.  The chargino mixings matrices are defined by the
relation
\begin{equation}
 M_{c}^{\mathit{diag}}=(O_R)M_{c}(O_L)^{T},
\end{equation}
where
\begin{equation}
  M_{c}=\left(\begin{array}{cc} 0 & X \\ X^{T} & 0
  \end{array}\right),~~~\mathrm{where}~~~ X=\left(\begin{array}{cc} M_{2} & \sqrt{2}M_W\cos\beta \\
  \sqrt{2}M_{W}\sin\beta & \mu \end{array}\right)
\end{equation}
and $M_{c}^{\mathit{diag}}$ is diagonal.  The sneutrino mixing matrix $U^{\nu}_{X,i}$ and
the charged slepton mixing matrix $U^{\ell}_{X,i}$ are defined by the relations
\begin{equation}
  (m_{\tilde{\ell}^{\pm}}^{2})^{\mathit{diag}}=U^{\ell}m_{\tilde{\ell}^{\pm}}^{2}U_{\ell}^{\dagger}~~~,~~~
  (m_{\tilde{\nu}^{\pm}}^{2})^{\mathit{diag}}=U^{\ell}m_{\tilde{\nu}}^{2}U_{\ell}^{\dagger},
\end{equation}
where the matrices $m_{\tilde{\ell}^{\pm}}^{2}$ and $m_{\tilde{\nu}}^{2}$ are given by
the sum of the MSSM contribution
 and the respective Dirac leptogenesis
contributions in~(\ref{eq:MassMatricesDiracLep}).  The neutralino mixing matrix $U_{N}$
is defined by the relation
\begin{equation}
  (m_{\tilde{N}})^{\mathit{diag}}=U_{N}m_{\tilde{N}}U_{N}^{\dagger},
\end{equation}
where
\begin{equation}
  m_{\tilde{N}}=\left(\begin{array}{cccc}
  M_1 & 0 & -M_Z\sin\theta_w\cos\beta & M_Z\sin\theta_w\sin\beta \\
  0 & M_2 & M_Z\cos\theta_w\cos\beta & -M_Z\cos\theta_w\cos\beta \\
  -M_Z\sin\theta_w\cos\beta & M_Z\cos\theta_w\cos\beta & 0 & -\mu \\
  M_Z\sin\theta_w\sin\beta & -M_Z\cos\theta_w\sin\beta & -\mu &
  0 \end{array}\right).
\end{equation}

\chapter{Derivation of the Boltzmann Equations for Dirac Leptogenesis\label{app:BoltzDiracLep}}


\indent

In this appendix, we derive the Boltzmann
equations~(\ref{eq:BoltzB})\,-\,(\ref{eq:BoltzLnet}) for the evolution of baryon and
lepton number in Dirac leptogenesis, following the methods and notation
of~\cite{Kolb:1979qa}.  We begin by observing that the Boltzmann equation for any
particle species \(a\) in the early universe can be written in terms of the number
density \(n_a\) of \(a\)  as
\begin{eqnarray}
  \frac{dn}{dt}_a+3Hn_a&=& \int {d^{3}p_{a}\over (2\pi)^{3}} {d^{3}p_{i}\over
(2\pi)^{3}} {d^{3}p_{j}\over (2\pi)^{3}}\ldots{d^{3}p_{k}\over (2\pi)^{3}}
(2\pi)^{4}\delta(\sum_{n=a,i,j,...k}p_{n})\non \\
& & \left[\sum_{\mathit{int.}}|\mathcal{M}(a...i \to
j...k)|^{2}(f_{a}...f_{i})\right.\non\\& &\left.-\sum_{\mathit{int.}}|\mathcal{M}(j...k
\to a...i)|^{2}(f_{j}...f_{k})\right],\label{eq:DefBoltzindndt}
\end{eqnarray}
where \(f_{i}\) is the phase-space distribution function of particle \(i\),
\(|\mathcal{M}(a...i\rightarrow j...k)|^{2}\) are the squared matrix elements for
particle-number-changing interactions involving \(a\), and the sums are over all
interaction processes which create or destroy \(a\).  The term proportional to the Hubble
parameter \(H\) can be absorbed by rewriting equation~(\ref{eq:DefBoltzindndt}) in terms
of the ratio \(Y_{a}\equiv n_{a}/s\), where \(s\) is the entropy density of the
unverse~\cite{Kolb:1990vq}:
\begin{eqnarray}
\frac{dY_{A}}{dt} &=&{1\over s} \int {d^{3}p_{a}\over (2\pi)^{3}} {d^{3}p_{i}\over
(2\pi)^{3}} {d^{3}p_{j}\over (2\pi)^{3}}\ldots{d^{3}p_{k}\over (2\pi)^{3}}
(2\pi)^{4}\delta(\sum_{n=a,i,j,...k}p_{n})\non \\
& & \left[\sum_{\mathit{int.}}|\mathcal{M}(a...i \to
j...k)|^{2}(f_{a}...f_{i})-\sum_{\mathit{int.}}|\mathcal{M}(j...k \to
a...i)|^{2}(f_{j}...f_{k})\right], \label{eq:DefBoltz}
\end{eqnarray}
In the high temperature limit \(T\gg m_a\), the ratio \(Y_{a}\) is well-approximated by
\begin{equation}
  Y_{a}= n_a/s \simeq{1\over2} \frac{g_a}{g_{*s}(T)} e^{\mu_a/T},
\end{equation}
where \(g_{a}\) and \(\mu_a\) respectively denote the the number of degrees of freedom
and chemical potential of species \(a\) and $g_{*s}(T)$ represents the total number of
relativistic interacting degrees of freedom at temperature \(T\).\footnote{During the
epoch of interest, $g_{*s}(T)$ remains constant, and thus we will henceforth drop the
\(T\)-dependence in our notation.}  If an asymmetry \(A_{a}=Y_{a}-Y_{a^{c}}\) develops
between \(a\) and its antiparticle \(a^{c}\), the chemical potential \(\mu_a\) can be
expressed in terms of this asymmetry (assuming \(\mu_{a},\,A_{a}\ll 1\)) using the
relation
\begin{equation}
  {A_a}\simeq {1\over2}\frac{g_a}{g_{*s}} e^{-\mu_a/T} (e^{2\mu_a/T}-1)
  \simeq {1\over2}\frac{g_a}{g_{*s}} (e^{2\mu_a/T}-1),
\end{equation}
which implies that
\begin{equation}
  e^{\mu_a/T}\simeq 1+{A_a}\frac{g_{*s}}{g_a}.\label{eq:MuAsymmetryRel}
\end{equation}

\indent In deriving Boltzmann equations for the heavy fields, we will begin with those
for the scalars \(\phi\) and \(\phi^{c}\).  To first order in these fields, the leading
processes include only decays and inverse decays, the Boltzmann equations for the
abundances \(Y_{\phi}\) and \(Y_{\overline{\phi}}\) are given by
\begin{eqnarray}
{dY_{\phiphi} \over dt} &=&-{1\over s} \Lambda^{{}^{\phiphi}}_{12}\left[ f_{\phiphi}\
|{\cal M}(\phiphi\to\tilde{\ell}\phichi)|^2
+f_{\phiphi}\ |{\cal M}(\phiphi\to\nu_R^c \widetilde{H}_u^c)|^2 \right.\non\\
&&-\left. f_{\phichi}f_{\widetilde{\ell}}\ |{\cal
M}(\widetilde{\ell}\phichi\to\phiphi)|^2 -f_{\nu_R^c}f_{ \widetilde{H}^c_u}\ |{\cal
M}(\nu_R^c \widetilde{H}^c_u\to\phiphi)|^2 \right] \label{eq:YPhiBasic}
\end{eqnarray}
and
\begin{eqnarray}
{dY_{\phiphi^c} \over dt} &=&-{1\over s} \Lambda^{{}^{\phiphi^c}}_{12}\left[
f_{\phiphi^c}\ |{\cal M}(\phiphi^c\to\widetilde{\ell}^c\phichi^c)|^2
+f_{\phiphi^c}\ |{\cal M}(\phiphi^c\to\nu_R \widetilde{H}_u)|^2 \right.\non\\
&&\left. -f_{\phichi^c}f_{\widetilde{\ell}^c}\ |{\cal
M}(\widetilde{\ell}^c\phichi^c\to\phiphi^c)|^2 -f_{\nu_R}f_{ \widetilde{H}_u}\ |{\cal
M}(\nu_R \widetilde{H}_u\to\phiphi^c)|^2 \right],\label{eq:YPhibarBasic}
\end{eqnarray}
where \(\Lambda^{a...i}_{j...k}\) has been used as a shorthand to denote the appropriate
phase-space integral and we use the notation $\ell$ and $H_u$ to denote the usual Lepton
and Higgs-up doublets of $SU(2)$.  These equations can be simplified by noting that
energy conservation implies that the inverse decay processes of the form
\((a+i\rightarrow b)\) appearing in equations~(\ref{eq:YPhiBasic})
and~(\ref{eq:YPhibarBasic}), in which particle \(i\) is in chemical equilibrium with
other particles in the thermal bath (i.e. \(\mu_{i}=0\)), the previous relation allows us
to write
\begin{equation}
    f_{a}\ f_{i}\simeq f_{b}^{\mathit{eq}}(1+{s\over n_\gamma}{A_a\over g_{a}}),
    \label{eq:feqsimplify}
\end{equation}
where \(f_{a}^{eq}\) is the equilibrium distribution of $a$. The assumption of
\(CPT\)-invariance also simplifies these expressions by enforcing the relation
\begin{equation}
 |{\cal M}(a\to ij)|^2=  |{\cal M}(i^{c}j^{c}\to a^c)|^2.\label{eq:CPTMatrixElRel}
\end{equation}

\indent The particle asymmetries with which we will be concerned here are the lepton
number abundances \(L_{i}=Q_{L_i}(n_{i}-n^{c}_{i})/s\) carried by each field \(i\) in the
theory with lepton number charge \(Q_{L_i}\).  From the charge assignments given in
table~\ref{tab:U1Charges} the nonzero \(L_{i}\) are given by
\begin{eqnarray}
    L_\ell & = & Y_{\ell} - Y_{\ell^{c}} \\
    L_{\nu_R} & = & -( Y_{\nu_{R}} -  Y_{\nu_{R}^{c}}) \\
    L_{\widetilde{\ell}} & = &  Y_{\widetilde{\ell}} -  Y_{\widetilde{\ell}^{c}}  \\
    L_{\widetilde{\nu}_R} & = &-(  Y_{\widetilde{\nu}_{R}} - Y_{\widetilde{\nu}_{R}^{c}})\\
    L_{\phiphi} & = & Y_{\phiphi}-Y_{\phiphi^c}\\
    L_{\phiphibar} & = & -(Y_{\phiphibar}-Y_{\phiphibar^c})
\end{eqnarray}
Using these definitions and relations~(\ref{eq:feqsimplify})
and~(\ref{eq:CPTMatrixElRel}), the Boltzmann equations for \(\phi\) and \(\phi^{c}\)
become
\begin{eqnarray} {dY_{\phiphi} \over dt}
&=&-{1\over s}  \int{d^3p_{\phiphi}\over(2\pi)^3}\left[ (f_{\phiphi} - f_{\phiphi}^{eq})
\Gamma_{D} - {s\over2n_\gamma} f_{\phiphi}^{eq} \left( {L_{\widetilde{\ell}}\over2}\
\Gamma^c_L+
L_{\nu_R}\ \Gamma^c_R \right)\right] \label{eq:nphieqnew} \\
{dY_{\phiphi^c} \over dt} &=&-{1\over s} \int{d^3p_{\phiphi}\over(2\pi)^3}\left[
(f_{\phiphi^c} - f_{\phiphi}^{eq}) \Gamma_{D} + {s\over2n_\gamma} f_{\phiphi}^{eq} \left(
{L_{\widetilde{\ell}}\over2}\ \Gamma_L+ L_{\nu_R}\ \Gamma_R\right) \right],
\label{eq:nphiceqnew}
\end{eqnarray} where the interaction rates
\(\Gamma_{L}=\Gamma(\phi\to \widetilde{\ell}+\chi)\) and \(\Gamma_{R}=\Gamma(\phi\to
\nu^c_R+\tilde{H}^c_u)\) are defined by the relations \begin{eqnarray}
    \Gamma_{L}&=&\int \frac{d^{3}p_{i}}{(2\pi)^{3}}
    \frac{d^{3}p_{i}}{(2\pi)^{3}} |\mathcal{M}(\phi\to \ell\chi)|^{2}\\
 \Gamma_{R}&=&\int \frac{d^{3}p_{i}}{(2\pi)^{3}}
    \frac{d^{3}p_{i}}{(2\pi)^{3}} |\mathcal{M}(\phi\to \nu_R^c\tilde{H}^c_u)|^{2},
\end{eqnarray}
with $\Gamma^c_L,R$ being the rates of the conjugate processes and
\(\Gamma_{D}=\Gamma_L+\Gamma_R\) is the total decay rate for \(\phi\),
\(\overline{\phi}\), etc.\ given in~(\ref{eq:GammaD}). Because of supersymmetry, the
total decay rate of the fermion components of the heavy supermultiplets $\Phi$ and
$\overline{\Phi}$ will also be \(\Gamma_{D}\), with the same $\Gamma_L$ and $\Gamma_R$ as
their partial rates.

\indent It will be more convenient for our purposes to express the Boltzmann equations in
terms of \(Y_{\phiphi^c}\) and $L_{\phiphi}$.  Subtracting~(\ref{eq:nphiceqnew})
from~(\ref{eq:nphieqnew}) yields
\begin{eqnarray}
{dL_{\phiphi} \over dt} &=&-{1\over s}  \int{d^3p_{\phiphi}\over(2\pi)^3}\left[
(f_{\phiphi} - f_{\phiphi}^{c}) \Gamma_{D}\rule{0pt}{18pt} \right.\nonumber
\\&&\left.- {s\over2n_\gamma} f_{\phiphi}^{eq} \left( {L_{\widetilde{\ell}}\over2}\
(\Gamma_L+\Gamma^c_L)+ L_{\nu_R}\ (\Gamma_R+\Gamma^c_R) \right)\right]
\label{eq:Lphieqnew}
\end{eqnarray}
After integrating equations~(\ref{eq:nphiceqnew}) and~(\ref{eq:Lphieqnew}) over the
incoming momentum $\vec{p}_{\phi}$, using
relations~(\ref{eq:1steps})\,-\,(\ref{eq:4theps}) to express the result in terms of the
decay asymmetry \(\epsilon\), and averaging over time-dilation factors~\cite{Kolb:1979qa}
we obtain
\begin{eqnarray}
{dL_{\phiphi} \over dz} &=& -\langle \Gamma_{D}\rangle \left[L_{\phiphi}-
{s\over2n_\gamma} Y_{\phiphi}^{eq}\epsilon
\left({L_{\widetilde{\ell}}\over2}-L_{\nu_R}\right)\right]\nonumber\\&&+
Y_{\phiphi}^{eq}{s\over n_\gamma}\left[ {L_{\widetilde{\ell}}\over2}\
\langle \Gamma_L\rangle +L_{\nu_R} \langle \Gamma_R\rangle \right]\label{eq:Lphidiffeq}\\
{dY_{\phiphi^c} \over dt} &=&-\left[ (Y_{\phiphi^c} - Y_{\phiphi}^{eq})
\langle\Gamma_{D}\rangle + {s\over2n_\gamma} Y_{\phiphi}^{eq} \left(
{L_{\widetilde{\ell}}\over2}\ \langle\Gamma_L\rangle+ L_{\nu_R}\
\langle\Gamma_R\rangle\right) \right], \label{eq:Yphicdiffeq},
\end{eqnarray}
were, the time-dilation-averaged rates \(\langle \Gamma_i\rangle\) are given in terms of
the naive rates $\Gamma_i$
 factors \cite{Kolb:1979qa}:
\begin{equation}
\langle\Gamma_i\rangle={K_1(M_\Phi/T)\over K_2(M_\Phi/T)}\ \Gamma_i
\label{eq:timedilation}
\end{equation}
where $K_1(x)$ and $K_2(x)$ are modified Bessel functions.

\indent The Boltzmann equations for the scalar component of the \(\overline{\Phi}\)
superfield (and its conjugate), as well as the ones for the fermion components of $\Phi$
and $\overline{\Phi}$, are obtained in a similar manner.  They turn out to be
\begin{eqnarray}
{dL_{\phiphibar} \over dt} &=& -\langle \Gamma_{D}\rangle \left[L_{\phiphibar}+ {s\over2
n\gamma}\epsilon\ Y_{\phiphi}^{eq}\left({L_{\ell}\over 2}-
{L_{\tilde{\nu}_R}}\right)\right]\nonumber\\&& + Y_{\phiphi}^{eq}\left[{ L_{\ell}\over
2}\langle \Gamma_{L}^c\rangle+ L_{\tilde{\nu}_R} \langle
\Gamma_{R}^c\rangle \right)]\label{eq:Lphibardiffeq}\\
{dY_{\phiphibar} \over dt} &=&-\left[ (Y_{\phiphibar} - Y_{\phiphi}^{eq})
\langle\Gamma_{D}\rangle - {s\over2n_\gamma} Y_{\phiphi}^{eq} \left( {L_{\ell}\over2}\
\langle\Gamma_L\rangle- L_{\tilde{\nu}_R}\ \langle\Gamma_R\rangle\right) \right.\non\\
&&+ \left.{s\over2n_\gamma}\epsilon\langle\Gamma_{D}\rangle
Y_{\phiphi}^{eq}\left(\frac{L_{\ell}}{2}-L_{\tilde{\nu}_R}\right)\right]
 \label{eq:Yphibardiffeqthing},
\end{eqnarray}
where we have used the fact that \(Y_{\phiphi}^{eq}=Y_{\phiphibar}^{eq}\).  Dropping
negligibly small terms proportional to \(\epsilon L_{i}\), we notice that Boltzmann
equations for the combinations $Y_{\phiphi}^{+}\equiv=Y_{\phiphi}+Y_{\phiphi^c}$ and
$Y_{\phiphibar}^{+}\equiv=Y_{\phiphibar}+Y_{\phiphibar^c}$
\begin{eqnarray}
  \frac{d Y_{\phiphi}^{+}}{dt}&=&-\langle\Gamma_{D}\rangle
    \left(Y_{\phiphi}^{+}-2Y_{\phiphi}^{eq}\right)\\
  \frac{d Y_{\phiphibar}^{+}}{dt}&=&-\langle\Gamma_{D}\rangle
    \left(Y_{\phiphibar}^{+}-2Y_{\phiphi}^{eq}\right)\\
\end{eqnarray}
are redundant.  We thus only require three equations to describe the dynamics of the
heavy field sector.

\indent  We now turn to address the evolution of the Lepton number abundance of the
particle species \(\ell\), $\nu_{R}$, $\tilde{\ell}$, and $\tilde{\nu_{R}}$, in which we
must take into account the effect of \(2\leftrightarrow2\) processes which transfer
lepton number between \(L_\ell\), $L_{\nu_R}$, $L_{\widetilde{\ell}}$, and
$L_{\widetilde{\nu}_R}$.  For the moment, we will not concern ourselves with the exact
form these interaction terms will take, but will will make one important observation: the
rates for interactions which shuffle lepton number between \(\ell\), \(\tilde{\ell}\),
and the right-handed charged lepton and sleptons fields \(e_{R}\) and \(\tilde{e}_{R}\)
will be much larger than those for the interactions which shuffle lepton number between
\(\nu_{R}\) and any of these other fields.  This is because the \(\nu_{R}\) interact only
via processes pictured in figure~\ref{fig:2to2SleptonProc}, which involve a virtual
\(\phiphi\), \(\phiphibar\), etc.\; while all of the other fields either take part in
\(SU(2)\) and/or \(U(1)_{Y}\) gauge interactions.  We will represent the effects of these
rapid equilibration processes by including terms \(\Sigma_{A}\) (where \(A\) is the
relevant particle asymmetry) to represent them in the Boltzmann equations. The slower
\(2\leftrightarrow2\) processes through which right handed neutrinos \(\nu_R\) interact
with lepton doublets $\ell$ and \(\widetilde{\ell}\) and with right handed sneutrino
\(\tilde{\nu}_R\) will be included separately as \(C_{\nu_R\leftrightarrow \ell
\vphantom{\tilde{A}}}\), \(C_{\nu_R\leftrightarrow\widetilde{\ell}}\) and
\(C_{\nu_R\leftrightarrow\widetilde{\nu}_R}\).  As discussed in
chapter~\ref{ch:Boltzmann} and elsewhere, \(\tilde{\nu}_{R}\) may or may not also take
part in rapid equilibration processes, depending on the value of \(\langle
F_{\chi}\rangle\).

To simplify further the notation we will define the terms \(F_{A}\) to account for the
collective contribution from decays (and inverse decays) of the fermionic components of
\(\Phi\) and \(\overline{\Phi}\) (which we will not write explicitly, being of similar
form to the contribution from the scalar components).


\indent Expressed in the above notation, the equations for the left-handed lepton field
\(\ell\) and its conjugate \(\ell^{c}\) are
\begin{eqnarray}
{dY_{\ell} \over dt} &=&{1\over n_\gamma} \Lambda^{{}^{\phiphibar^c}}_{12}
\left[f_{\phiphibar^c}\ |{\cal M}(\phiphibar^c\to \ell\psichi)|^2 - f_{\phichi}f_{\ell}\
|{\cal M}(\ell\psichi\to\phiphibar^c)|^2
\right]\non\\&&+F_L(\psiphi^c) + \Sigma_L. \\
{dY_{\ell^c} \over dt} &=&{1\over n_\gamma} \Lambda^{{}^{\phiphibar}}_{12}
\left[f_{\phiphibar}\ |{\cal M}(\phiphibar\to \ell^c\psichi^c)|^2 -
f_{\psichi^c}f_{\ell^c}\ |{\cal M}(\ell^c\psichi^c\to\phiphibar)|^2
\right]\non\\
&&+ F^c_L(\psiphi) +\  \Sigma_L^c
\end{eqnarray}
The lepton number abundance \(L_\ell\) is obtained by subtracting the second of these
equations from the first, and the result is
\begin{eqnarray}
{dL_\ell\over dt} &=& - {1\over n_\gamma} \int d^3p_{\phiphi}\left[ f_{\phiphibar}\
\Gamma^{\overline{\phi}}_{\ell^c\tilde{\chi}^c} - f_{\phiphibar}^{eq}
(1-L/4)\Gamma^{\overline{\phi}^c}_{\ell\tilde{\chi}} -f_{\phiphibar^c}\
\Gamma^{\overline{\phi}^{c}}_{\ell\tilde{\chi}}\right. \non
\\ & & + \left. f_{\phiphi}^{eq}
(1+L/4)\Gamma^{\overline{\phi}}_{\ell^c\tilde{\chi}^c}
\right] + (F_L-F^c_L) +(\Sigma_{L}-\Sigma^{c}_{L})\non\\
&=&-\epsilon\ \langle\Gamma_{D}\rangle \Big(Y_{\phiphibar^c}+Y^{eq}_{\phiphi}\Big)+
L_{\phiphibar} \langle\Gamma_L\rangle-Y^{eq}_{\phiphi}{s\over n_\gamma} L_\ell\
\Big(\langle\Gamma_L\rangle-{1\over2}\epsilon
\langle\Gamma_{D}\rangle \Big)\non\\
&&  +F_\ell+\Sigma_\ell +{C}_{\nu_R\leftrightarrow \ell\vphantom{\widetilde{A}}}
\label{eq:lepnoLL}
\end{eqnarray}
The equations for \(L_{\nu_R}\), \(L_{\widetilde{\nu}_R}\), and \(L_{\widetilde{\ell}}\)
are determined in a similar manner:
\begin{eqnarray}
{dL_{\nu_R}\over dt}&=& \epsilon\ \langle \Gamma_{D}\rangle
\Big(Y^c_{\phiphi}+Y^{eq}_{\phiphi}\Big) -L_{\phiphi} \langle\Gamma_{R}\rangle +
Y^{eq}_{\phiphi} {s\over n_\gamma} L_{\nu_R}
\Big(\langle\Gamma_R\rangle-{1\over2}\epsilon
\Gamma_{D})\Big)\non\\
&&  + F_{\nu_R}
-C_{\nu_R\leftrightarrow\widetilde{\ell}}-{C}_{\nu_R\leftrightarrow\widetilde{\nu}_R}-{C}_{\nu_R\leftrightarrow
\ell\vphantom{\widetilde{A}}}
\label{eq:Rdiffeq}\\
{dL_{\widetilde{\ell}}\over dt}&=& - \epsilon\ \langle \Gamma_{D}\rangle
 (Y_{\phiphi^c}+Y^{eq}_{\phiphi})
+ L_{\phiphi}\langle\Gamma_L\rangle - Y^{eq}_{\phiphi} {s\over n_\gamma}
L_{\widetilde{\ell}} \Big({1\over2}\langle\Gamma_L\rangle+{1\over2}\epsilon
\langle\Gamma_{D}\rangle\Big)\non\\
&&+F_{\widetilde{\ell}}+\Sigma_{\tilde{\ell}}+\ C_{\nu_R\to\widetilde{\ell}} \\
{dL_{\widetilde{\nu}_R}\over dt}&=&  \epsilon\ \langle \Gamma_{D}\rangle
(Y_{\phiphibar^c}+Y^{eq}_{\phiphibar}) - L_{\phiphibar}\langle\Gamma_R\rangle
-Y^{eq}_{\phiphibar}{s\over n_\gamma} L_{\widetilde{\nu}_R}
\Big(\langle\Gamma_R\rangle+{1\over2}\epsilon
\Gamma_{D}\Big)\non\\
&& +F_{\widetilde{\nu}_R}+\Sigma_{\widetilde{\nu}_R}+ C_{\nu_R\to\widetilde{\nu}_R}.
\label{eq:lastofindividualLepnos}
\end{eqnarray}

Before going any further we still need to compute the terms
$C_{\nu_R\leftrightarrow\widetilde{\ell}}$, ${C}_{\nu_R\leftrightarrow\widetilde{\nu}_R}$
and ${C}_{\nu_R\leftrightarrow \ell\vphantom{\widetilde{A}}}$ corresponding to the
$2\leftrightarrow 2$ processes mediated by heavy fields.  We begin by calculating
\(C_{\nu_R\leftrightarrow\tilde{L}}\), which is given by
\begin{eqnarray}
C_{\nu_R\leftrightarrow\widetilde{\ell}} &=&{2\over
n_\gamma}\Lambda^{34}_{12}e^{-(E_1+E_2)/T} \left[\Big(|{\cal M'}(\widetilde{H}_u^c
\nu_R^c\to\widetilde{\ell}\chi)|^2-|{\cal M'}(\widetilde{\ell}\chi\to\nu_R^c
\widetilde{H}^c_u)|^2\Big)
\vphantom{\int_\int}\right.\non\\
&& \left.+ {(L_{\nu_R}-{1\over2}L_{\widetilde{\ell}})\over2}\Big(|{\cal
M'}(\widetilde{H}_u^c \nu_R^c\to\widetilde{\ell}\chi)|^2 +|{\cal
M'}(\widetilde{\ell}\chi\to\nu_R^c \widetilde{H}^c_u)|^2\Big)\vphantom{\int_\int} \right]
\label{eq:CRL}
\end{eqnarray}
${\cal M'}$ here refers to the amplitude for the specified $2\leftrightarrow 2$ process
to which we have substracted the contribution from the resonant intermediate state (RIS)
in which a real field $\phiphi$ is produced and then decayed into the 2 particle final
state. The RIS contribution must be substracted since we have already counted
contributions from decays of real $\phiphi$ fields.

\indent The leading term in the difference between a $2\leftrightarrow 2$ process of the
form \(ab\to ij\) involving a heavy intermediary \(k\) and its conjugate process depends
on the contribution of the on-shell (resonant) intermediate state: \beq \!\!\!|{\cal
M'}(ab\to ij)|^2-|{\cal M'}(ij\to ab)|^2 =|{\cal M_{RIS}}(ij\to ab)|^2 -|{\cal
M_{RIS}}(ab\to ij)|^2 \eeq with \beq \hspace{-.5cm}|{\cal M_{RIS}}(ab\to ij)|^2 \simeq
{\pi\over m_\phi \Gamma_{D}} \delta(s-m_\phi^2) |{\cal M}(ab\to k)|^2\times|{\cal M}(k\to
ij)|^2
\eeq where \(k\) represents the intermediate-state particle, and $s$ is the usual
kinematic variable $s=(p^{in}_1+p^{in}_2)^2$.  In our case, making use of the equality
$\Gamma_L^c\Gamma_R-\Gamma_L\Gamma_R^c=\epsilon \Gamma_{D}$ we find \bea |{\cal
M_{RIS}}(\widetilde{\ell}\chi\to\nu_R^c \widetilde{H}^c_u)|^2 -|{\cal
M_{RIS}}(\widetilde{H}^c_u \nu_R^c\to\widetilde{\ell}\chi)|^2 \simeq \epsilon {\pi\over
m_\phi \Gamma_{D}}\delta(s-m_\phi^2) |{\cal M}^\phi_{tot}|^4 \eea and substituting this
result in eq.~(\ref{eq:CRL}) we finally obtain \bea
C_{\nu_R\leftrightarrow\widetilde{\ell}}&=& 2 \epsilon Y_{\phiphi}^{eq} \langle
\Gamma_{D}\rangle + (L_{\nu_R}-{1\over2}L_{\widetilde{\ell}})\ n_\gamma \langle
v\sigma_{\nu_R\to \widetilde{\ell}}+v\sigma_{\widetilde{\ell}\to \nu_R}\rangle. \eea For
the other sets of \(2\leftrightarrow2\) processes, \(C_{\nu_R\to\widetilde{\nu}_R}\) and
\(C_{\nu_R\to \ell}\), the procedure is essentially the same. The rates
\(\langle\Gamma_{\nu_R\leftrightarrow \widetilde{\ell}}\rangle\equiv n_\gamma \langle
v\sigma_{\nu_R\to \widetilde{\ell}}+v\sigma_{\tilde{\ell}\to \nu_R}\rangle\),
\(\langle\Gamma_{\nu_R\leftrightarrow \ell}\rangle\equiv n_\gamma \langle
v\sigma_{\nu_R\to \ell}+v\sigma_{\ell\to \nu_R}\rangle\) and
\(\langle\Gamma_{\nu_R\leftrightarrow\widetilde{\nu}_R}\rangle\equiv n_\gamma \langle
v\sigma_{\nu_R\to \widetilde{\nu}_R}+v\sigma_{\widetilde{\nu}_R\to \nu_R}\rangle\),
associated with these interactions are calculated in section~\ref{sec:NumSolBigF}. We
will denote the total contribution from these processes as
\(\langle\Gamma_{2\leftrightarrow2}\rangle\).

\indent  At this point, the full Boltzmann system comprises fifteen individual
differential equations: four to represent the evolution of \(\phiphi\), \(\phiphibar\),
and their conjugates (in whatever basis we choose); an additional four for the fermionic
superpartners in the \(\Phi\) and \(\overline{\Phi}\) supermultiplets; six for the the
Lepton asymmetries $L_{\ell}$, $L_{\nu_R}$, $L_{\widetilde{\ell}}$,
\(L_{\widetilde{\nu}_R}\), \(L_{e_R}\equiv e^c_{R}-e_{R}\), and \(L_{\widetilde{e}_R}\)
stored in various individual lepton and slepton species; and one for the overall baryon
number \(B\) of the universe, which interacts with $L_{\ell}$ via sphaleron processes of
the form given in equation~\ref{eq:SphalOpDef}.\footnote{Keeping track of individual
baryon numbers is unnecessary, since QCD gauge interactions among the quarks are rapid.}
We have already noted that equations~(\ref{eq:Yphicdiffeq})
and~(\ref{eq:Yphibardiffeqthing}) are redundant up to terms of \(\mathcal{O}(\epsilon
L_{i})\), and the assumption of unbroken supersymmetry during the leptogenesis epoch
allows us to equate the abundances of the fermionic and bosonic components of
\(\Phi_{1}\) and \(\overline{\Phi}_{1}\), which reduces the number of equations in our
system to ten.  As discussed in section~\ref{sec:BoltzEqIntro}, we can also make use of
the fact that many of the light fields in our theory will be brought into chemical
equilibrium by rapid \(SU(2)\times U(1)_{Y}\) gauge interactions, Yukawa interactions,
\(A\)-terms, etc.\ whose collective rates we have denoted \(\Sigma_{A}\).  In this case,
any lepton number stored in \(\ell\), \(\tilde{\ell}\), \(e_{R}\), or \(\tilde{e}_{R}\)
should be rapidly distributed among all of these particles in proportion to the relative
number of degrees of freedom for each field.  Whether \(\tilde{\nu}_{R}\) is also in
equilibrium with these fields depends on the size of the effective
$A$-term~(\ref{eq:EffectiveAtermL}) induced by $\langle F_{\chi}\rangle$: if $\langle
F_{\chi}\rangle$ is large, the \(\tilde{\nu}_{R}\) fields will equilibrate chemically
with the left-handed leptons and sleptons; if $\langle F_{\chi}\rangle$ is small (or
zero), they will not, but can be seen as part of the hidden sector which contains the
right-handed neutrino fields.  In either case, we can define two aggregate lepton numbers
\(\Lvis\) and \(\Lhid\) for the fields in the visible and hidden sectors respectively and
reduce the number of equations in our Boltzmann system to six.  In the case where
$\langle F_{\chi}\rangle$ is large and \(\tilde{\nu}_{R}\) is in the visible sector, we
have
\begin{eqnarray}
  \Lvis&=&L_\ell+L_{\tilde{\nu}_R}+L_{\tilde{\ell}}+L_{e_R}+L_{\tilde{e}_R}\label{eq:LvisDefBigF}\\
  \Lhid&=&L_{\nu_{R}},\label{eq:LhidDefBigF}
\end{eqnarray}
When \(\tilde{\nu}_{R}\) is in the hidden sector, we have
\begin{eqnarray}
  \Lvis&=&L_\ell+L_{\tilde{\nu}_R}+L_{\tilde{\ell}}+L_{e_R}\label{eq:LvisDefSmallF}\\
  \Lhid&=&L_{\nu_{R}}+L_{\tilde{e}_R}.\label{eq:LhidDefSmallF}
\end{eqnarray}
In either case, chemical equilibrium enforces that the lepton number stored in each
individual particle species be distributed evenly among them in proportion to the number
of degrees of freedom of each, or in other words
\begin{equation}
  \frac{7}{2}L_\ell= \frac{7}{2}L_{\tilde{\ell}}=
  7L_{\tilde{\nu}_R}=7L_{e_R}=7L_{\tilde{e}_R}=\Lvis\label{eq:DefofLeqBigF}
\end{equation}
in the large-$\langle F_{\chi}\rangle$ scenario, and
\begin{equation}
  3L_\ell=3L_{\tilde{\ell}}=6L_{e_R}=L_{\tilde{e}_R}=\Lvis\label{eq:DefofLeqLittleF}
\end{equation}
in the small-$\langle F_{\chi}\rangle$ scenario.  In the latter case, if chemical
equilibrium is established among the fields in the hidden sector (for example, via the
gauge interactions of the gauged \(U(1)_{Y}\) discussed in
section~\ref{sec:sneutrinoCDM}), one obtains the similar relation
\begin{equation}
  \frac{1}{2}L_{\nu_{R}}=\frac{1}{2}L_{\tilde{e}_R}=\Lhid.
\end{equation}

\indent In the large-$\langle F_{\chi}\rangle$ scenario, the Boltzmann equation that
describes the evolution of \(\Lvis\), which is obtained by summing
equations~(\ref{eq:lepnoLL})\,-\,(\ref{eq:lastofindividualLepnos}) with the appropriate
numerical prefactor from~(\ref{eq:DefofLeqBigF}), is
\begin{eqnarray}
{d \Lvis\over dt} &=& -\epsilon\langle \Gamma_{D}\rangle
\left(Y_{\phiphi}^c-Y_{\phiphi}^{eq}\right)+\langle \Gamma_L\rangle \left(L_{\phiphi} +
L_{\phiphibar}\right) + \langle\Gamma_R\rangle
L_{\phiphibar}\non\\
&& - Y_{\phiphi}^{eq} {s\over 2 n_\gamma} \Lvis \Big(\langle\Gamma_D\rangle
+\langle\Gamma_L\rangle \Big)+ (\Lhid-{1\over7}\Lvis)\
\langle\Gamma_{2\leftrightarrow2}\rangle,\label{eq:Lnetdiffeq}
\end{eqnarray}
where small terms proportional to $\epsilon$ times $L_\ell$, $L_{\nu_R}$,
$L_{\widetilde{\ell}}$ or $L_{\tilde{\nu}_R}$ have been dropped. Here, we have used the
fact that the Boltzmann equations for \(L_{e_R}\) and \(L_{\tilde{e}_R}\) are trivial,
consisting of only \(\Sigma_{A}\) terms, which all cancel after taking the sum of all
Lepton abundances.

\indent In the event that annihilation processes of the form
\(\phiphi\phiphi^c\rightarrow X\), where \(X\) represents some final state comprising
light fields, are effective in reducing the abundance of \(\phiphi\) and its conjugate,
equation~(\ref{eq:Yphicdiffeq}) must be modified slightly to account for the presence of
source and sink terms
\begin{equation}
{dY_{\phiphi^c} \over dt}=\ldots\ -{1\over s} \Lambda^{{}^{34}}_{12}\left[
f_{\phiphi}f_{\phiphi^c}\ |{\cal M}(\phiphi\phiphi^c\to ij)|^2 -f_{i}f_{j}\ |{\cal
M}(ij\to\phiphi\phiphi^c)|^2\right].
\end{equation}
Here, \(i\) and \(j\) represent the unspecified final states products of
\(\phiphi\,-\,\phiphi^s\) annihilation (in which $2\leftrightarrow 2$ processes will
again dominate) and \(\Lambda^{{}^{34}}_{12}\) is the appropriate four-particle
phase-space integral.  Ignoring negligible $CP$-violation effects and invoking
conservation of momentum, we obtain the relation $f_{i}f_{j}=(f_{\phiphi}^{eq})^2$, which
leads to the result
\begin{equation}
  {dY_{\phiphi^c} \over dt}=\ldots\ - \langle \sigma(\phiphi\phiphi^c\to X)|v|\rangle
  \left[Y_{\phiphi^c}^2-(Y_{\phiphi}^{eq})^2\right].
\end{equation}
Using equation~(\ref{eq:RelBetweenGammasandSigmas}), this can be written in terms of the
effective annihilation rate \(\Gamma_{A}\).

\indent The only task left is to couple the Bolztmann equations for \(B\) and \(\Lvis\)
via electroweak sphaleron interactions.  The high-temperature rate for these
interactions, \(\Gamma_{\mathit{sph}}\), is given in equation~(\ref{eq:HighTempSphRate}),
and the proportionality constant between \(B\) and \(\Lvis\) is given, in the
supersymmetric case, by equation~(\ref{eq:BLrelSUSY}), and since \(B\) is otherwise
conserved, the equation for its evolution is
\begin{equation}
  {d B \over dt}
  =-\langle\Gamma_{\mathit{sph}}\rangle\Big(B+\frac{8}{15}\Lvis\Big)\label{eq:TheBequation}
\end{equation}
We are now ready to write down the complete Boltzmann system for the evolution of baryon
and lepton number in Dirac leptogenesis.  Collecting equations~
(\ref{eq:Lphidiffeq})\,-\,(\ref{eq:Yphicdiffeq}), (\ref{eq:Lphibardiffeq}),
(\ref{eq:Rdiffeq}), (\ref{eq:Lnetdiffeq}), and~(\ref{eq:TheBequation}) and defining the
shorthand expressions
\begin{equation}
\langle\Gamma_D\rangle_{{}_{ID}}={1\over7}{n_{\phiphi}^{eq}\over
n_\gamma}\langle\Gamma_D\rangle~~\hspace{.5cm}~~
\langle\Gamma_L\rangle_{{}_{ID}}={1\over7}{n_{\phiphi}^{eq}\over
n_\gamma}\langle\Gamma_L\rangle~~\hspace{.5cm}~~
\langle\Gamma_R\rangle_{{}_{ID}}={n_{\phiphi}^{eq}\over n_\gamma}\langle\Gamma_D\rangle
\end{equation}
for the inverse decay rates, we obtain the following result:
\begin{eqnarray}
{d B \over dt}
  &=&-\langle\Gamma_{\mathit{sph}}\rangle\Big(B+\frac{8}{15}\Lvis\Big)\\
{d \Lvis\over dt}&=& 
-2\epsilon\langle \Gamma_{D}\rangle(Y_{\phiphi^c}-Y_{\phiphi}^{{eq}})+\langle
\Gamma_L\rangle (L_{\phiphi} + L_{\phiphibar}) + \langle\Gamma_R\rangle
L_{\phiphibar}\non\\ && - 2 \Lvis\Big(\langle \Gamma_{D}\rangle_{{}_{ID}}+\langle
\Gamma_L\rangle_{{}_{ID}}\Big) + (\Lhid -{1\over7}\Lvis) \langle
\Gamma_{2\leftrightarrow2}\rangle\non\\& & -\langle\Gamma_{\mathit{sph}}\rangle
\Big(B+\frac{8}{15}\Lvis\Big)\\ 
{d\Lhid\over dt}&=& 2\epsilon \langle\Gamma_{D}\rangle(Y_{\phiphi^c}-Y_{\phiphi}^{eq}) +
L_{\phiphi} \langle\Gamma_R\rangle - 2 \Lhid \langle \Gamma_R\rangle_{{}_{ID}}\non \\
& & -(\Lhid
-{1\over7}\Lvis) \langle \Gamma_{2\leftrightarrow2}\rangle\\
{dY_{\phiphi^c} \over dt} &=&-\langle\Gamma_{D}\rangle(Y_{\phiphi^c} - Y_{\phiphi}^{eq})
+{1\over2} \Lvis
\langle\Gamma_L\rangle_{{}_{ID}}+{1\over2} \Lhid\ \langle\Gamma_R\rangle_{{}_{ID}}\non\\
& & -\langle\Gamma_A\rangle \left[\left(Y_{\phiphi^c}/Y_{\phiphi}^{eq}\right)^2 - 1 \right]\\
{dL_{\phiphi} \over dt} &=& -\langle \Gamma_{D}\rangle L_{\phiphi} +2 \Lvis\langle
\Gamma_L\rangle_{{}_{ID}}+2 \Lhid
\langle \Gamma_R\rangle_{{}_{ID}}\\
{dL_{\phiphibar} \over dt} &=& -\langle \Gamma_{D}\rangle
 L_{\phiphibar} +2 \Lvis\langle
 \Gamma_{D}\rangle_{{}_{ID}},
\end{eqnarray}
where once again, negligibly small terms proportional to \(\epsilon \Lvis\) or \(\epsilon
\Lhid\) have been dropped.  Changing variables from \(t\) to \(z=M_{\Phi_{i}}/T\), one
obtains equations~(\ref{eq:BoltzB})\,-\,(\ref{eq:BoltzLPhibar}).

\indent In the ``drift and decay'' limit, in which all the rates in
$\langle\Gamma_A\rangle$ and $\langle\Gamma_{2\leftrightarrow 2}\rangle$  are assumed to
be much smaller than the rate of expansion of the universe \(H\), we note that this
system simplifies considerably.  In this case
\begin{eqnarray}
{d \Lvis\over dt}&=& 
-2\epsilon\langle
\Gamma_{D}\rangle(Y_{\phiphi^c}-Y_{\phiphi}^{{eq}})\\
{d\Lhid\over dt}&=& 2\epsilon
\langle\Gamma_{D}\rangle(Y_{\phiphi^c}-Y_{\phiphi}^{eq}) \\
{dY_{\phiphi}^c \over dt} &=&-\langle\Gamma_{D}\rangle(Y_{\phiphi^c} - Y_{\phiphi}^{eq}),
\end{eqnarray}
and since the right sides of these expressions are proportional to one another, the
Boltzmann system reduced to a single differential equation, whose solution is
\begin{eqnarray}
\Lhid(t\to\infty)=-\Lvis(t\to\infty)=2\epsilon Y_{\phiphi}^{eq}(t=t_0),
\end{eqnarray}
given the boundary conditions \(\Lvis(t=0)=\Lhid(t=0)=0\) and
\(Y_{\phiphi}^{eq}(t\to\infty)=0\) and approximating $Y_{\phiphi^c}$ by its equilibrium
abundance at time \(t_0\), defined as the time at which \(T=M_{\Phi_{1}}\).

\chapter{Derivation of the Boltzmann Equations for Hidden Sector Dark Matter\label{app:BoltzRHSneutDM}}


\indent

Here we present a derivation of the Boltzmann equations for right-handed sneutrino dark
matter, using the methods of~\cite{Kolb:1979qa}.  We begin by writing down the equations
in terms of the abundances:
 $Y_{\RHSn}=n_{\RHSn}/s$ and $Y_{\RHSn^c}=n_{\RHSn^c}/s$:
 \begin{eqnarray}
{dY_{\RHSn} \over dt} &=& {1\over s}\Lambda^{34}_{12} \left[2 f_{\nu_R}f_{\nu_R}\ |{\cal
M}(\nu_R \nu_R\to\RHSn\RHSn)|^2
-2 f_{\RHSn}f_{\RHSn}\ |{\cal M}(\RHSn\RHSn\to\nu_R \nu_R)|^2\right.\non\\
&&\hspace{.5cm} \left.+f_{\nu_R^c}f_{\nu_R}\ |{\cal M}(\nu_R^c \nu_R\to\RHSn \RHSn^c)|^2
-f_{\RHSn^c}f_{ \RHSn}\ |{\cal M}(\RHSn^c \RHSn\to\nu_R^c
\nu_R)|^2\right]\label{eq:sneutBoltzBasicA} \\
{dY_{\RHSn^c} \over dt} &=& {1\over s}\Lambda^{34}_{12} \left[2 f_{\nu_R^c}f_{\nu_R^c}\
|{\cal M}(\nu_R^c \nu_R^c\to\RHSn^c\RHSn^c)|^2
-2 f_{\RHSn^c}f_{\RHSn^c}\ |{\cal M}(\RHSn^c\RHSn^c\to\nu_R^c \nu_R^c)|^2\right.\non\\
&&\hspace{.5cm} \left.+f_{\nu_R^c}f_{\nu_R}\ |{\cal M}(\nu_R^c \nu_R\to\RHSn\RHSn^c)|^2
-f_{\RHSn^c}f_{ \RHSn}\ |{\cal M}(\RHSn^c \RHSn\to\nu_R^c
\nu_R)|^2\right].\label{eq:sneutBoltzBasicB}
\end{eqnarray}
Here the phase space integral \(\Lambda_{12}^{34}\) is defined as in
equation~(\ref{eq:YPhiBasic}).  The energy-conservation condition~(\ref{eq:feqsimplify})
implies that
\begin{eqnarray}
  f_{\nu_R}f_{\nu_R} &=& e^{-(E_{{N_R}_1}+E_{{N_R}_2})/T} (1+g_{*s} L_{\nu_R})\\
  f_{\nu_R^c}f_{\nu_R^c} &=&  e^{-(E_{{N_R}_1}+E_{{N_R}_2})/T} (1-g_{*s} L_{\nu_R})\\
  f_{\nu_R}f_{ \nu_R^c} &=&   e^{-(E_{{N_R}_1}+E_{{N_R}_2})/T},
\end{eqnarray}
and we also will neglect \(CP\) violation (a good assumption at scales \(T\ll
M_{\Phi_1}\), and define
\begin{eqnarray}
  |{\cal M}(\nu_R^c \nu_R^c\to\RHSn^c\RHSn^c)|^2&=&|{\cal M}(\nu_R
  \nu_R\to\RHSn\RHSn)|^2=|{\cal M}_{RR}|^2\\
  |{\cal M}(\nu_R \nu_R^c\to\RHSn\RHSn^c)|^2&=&|{\cal M}_{RR^c}|^2.
\end{eqnarray}
With these simplifications, the Boltzmann equations become
\begin{eqnarray}
  {dY_{\RHSn}
  \over dt} &=& {1\over s}\Lambda^{34}_{12}\ e^{-\left(E_{{N_R}_1}+E_{{N_R}_2}\right)/T}
  \left[2\ |{\cal M}_{RR}|^2 \left(1+g_{*s} L_{\nu_R}-
  \left({n_{\RHSn}\over n^{MB}_{\RHSn}}\right)^2\right)\ \right.\non\\
  &&\hspace{5cm} \left.+\  |{\cal M}_{RR^c}|^2  \left( 1-
  {n_{\RHSn} n_{\RHSn^c}\over n^{MB}_{\RHSn^c} n^{MB}_{\RHSn}}\right)\ \right]\\
  {dY_{\RHSn^c} \over dt} &=& {1\over s}\Lambda^{34}_{12}\
  e^{-\left(E_{{N_R}_1}+E_{{N_R}_2}\right)/T}
  \left[2\ |{\cal M}_{RR}|^2 \left(1-g_{*s} L_{\nu_R}-
  \left({n_{\RHSn^c}\over n^{MB}_{\RHSn^c}}\right)^2\right)\ \right.\non\\
  &&\hspace{5cm} \left.+\  |{\cal M}_{RR^c}|^2  \left( 1-
  {n_{\RHSn} n_{\RHSn^c}\over n^{MB}_{\RHSn^c} n^{MB}_{\RHSn}}\right)\ \right]
\end{eqnarray}
It will be more convenient for our purposes (to relate the dark matter abundance to the
hidden-sector lepton asymmetry $\Lhid$) to write these equations in terms of
$L_{\RHSn}=Y_{\RHSn}-Y_{\RHSn^c}$ and $Y_{DM}= Y_{\RHSn} + Y_{\RHSn^c}$.  We also define
the effective rates
\begin{eqnarray}
  \Gamma_{\tilde{\nu}\tilde{\nu}}&=& \frac{1}{(Y^{eq}_{\RHSn})^2}
  \Lambda^{34}_{12}\ e^{-\left(E_{{N_R}_1}+E_{{N_R}_2}\right)/T}
  |{\cal M}_{RR}|^2 = \frac{\gamma _{\RHSn\RHSn}}{(Y^{eq}_{\RHSn})^2} \\
  \Gamma_{\tilde{\nu}\tilde{\nu}}^{\ast}&=& \frac{1}{(Y^{eq}_{\RHSn})^2}
  \Lambda^{34}_{12}\ e^{-\left(E_{{N_R}_1}+E_{{N_R}_2}\right)/T}
  |{\cal M}_{RR^c}|^2 = \frac{\gamma _{\RHSn\RHSn^c}}{(Y^{eq}_{\RHSn})^2},
\end{eqnarray}
where $Y^{eq}_{\RHSn}\equiv Y^{MB}_{\RHSn}(\mu=0)$, in terms of which the Boltzmann
equations become
\begin{eqnarray}
  {dL_{\RHSn} \over
  dt} &=&{2\over s}\Lambda^{34}_{12}\ e^{-\left(E_{{N_R}_1}+E_{{N_R}_2}\right)/T} |{\cal
  M}_{RR}|^2 \left(2g_{*s} L_{\nu_R}+ \left({n_{\RHSn^c}\over n^{MB}_{\RHSn^c}}\right)^2
  - \left({n_{\RHSn}\over n^{MB}_{\RHSn}}\right)^2\right)\non\\
  &=&-{2\over s } \Gamma_{\RHSn\RHSn}
  \left(
  L_{\RHSn} Y_{DM} - 2g_{*s} L_{\nu_R}(Y^{eq}_{\RHSn})^2 \right)\\
  {dY_{DM} \over dt} &=&{2\over
  s}\Lambda^{34}_{12}\ e^{-\left(E_{{N_R}_1}+E_{{N_R}_2}\right)/T} \left[ |{\cal M}_{RR}|^2
  \left(2-\left({n_{\RHSn}\over n^{MB}_{\RHSn}}\right)^2-\left({n_{\RHSn^c}\over
  n^{MB}_{\RHSn^c}}\right)^2\right)
  \ \right.\non\\
  &&\hspace{5cm} \left.+\  |{\cal M}_{RR^c}|^2 \left( 1- {n_{\RHSn}
  n_{\RHSn^c}\over n^{MB}_{\RHSn^c} n^{MB}_{\RHSn}}\right)\ \right]\non\\
  &=&{2\over s}
  \left[\left(\Gamma_{\RHSn\RHSn^{\ast}}+2\Gamma_{\RHSn\RHSn}\right)\left({Y^{eq}_{\RHSn}}^2-{Y^2_{DM}\over4}\right)
  +\left(\Gamma_{\RHSn\RHSn^{\ast}}-2\Gamma_{\RHSn\RHSn}\right) {L_{\RHSn}^2\over4} \right]
\end{eqnarray}
Defining the quantities $\Gamma^{(\pm)}=\Gamma_{\RHSn\RHSn^c}\pm2\Gamma_{\RHSn\RHSn}$ and
$Y^{eq}_{DM}= 2 Y^{eq}_{\RHSn}$, these two equations take the far more manageable form
\begin{eqnarray}
  {dL_{\RHSn}
  \over dt} &=&-{2\over s } \Gamma_{\RHSn\RHSn}\ \left(L_{\RHSn} Y_{DM} -
  2g_{*s} (L^{tot}_R - L_{\RHSn}) {Y^{eq^2}_{\RHSn}} \right)\label{Leq}\\
  {dY_{DM} \over dt} &=&{1\over2 s} \left[\Gamma^{(+)}\
  \left({Y^{eq}_{DM}}^2-{Y^2_{DM}}\right) +\Gamma^{(-)}\ {L_{\RHSn}^2} \right] \label{Yeq}
\end{eqnarray}
given in~(\ref{eq:LeqSnoot}\,-\,\ref{eq:YeqSnoot}).

\chapter{Thermally Averaged Cross Sections\label{app:SneutInts}}


\begin{figure}
\begin{center}
\begin{fmffile}{NNann8}
  \fmfframe(20,20)(20,20){\begin{fmfchar*}(100,80)
    \fmfleft{i1,i2}
    \fmfright{o1,o2}
    \fmflabel{$\tilde{\nu}_{R_1}$}{i1}
    \fmflabel{$\tilde{\nu}_{R_1}$}{i2}
    \fmflabel{$\nu_R$}{o2}
    \fmflabel{$\nu_R$}{o1}
    \fmf{scalar,label=$p_1$}{i1,v1}
    \fmf{scalar,label=$p_2$}{i2,v2}
    \fmf{fermion,label=$p_4$}{v2,o2}
    \fmf{fermion,label=$p_3$}{v1,o1}
    \fmf{vanilla,label=$\chi^0_i$}{v1,v2}
  \end{fmfchar*}}
\end{fmffile}
\end{center}
  \caption{Processes contributing to $\hat{\sigma}(\tilde{\nu}\tilde{\nu})$ in the
  right-handed sneutrino dark matter scenario.\label{fig:ViolateLsneutdiags}}
\end{figure}
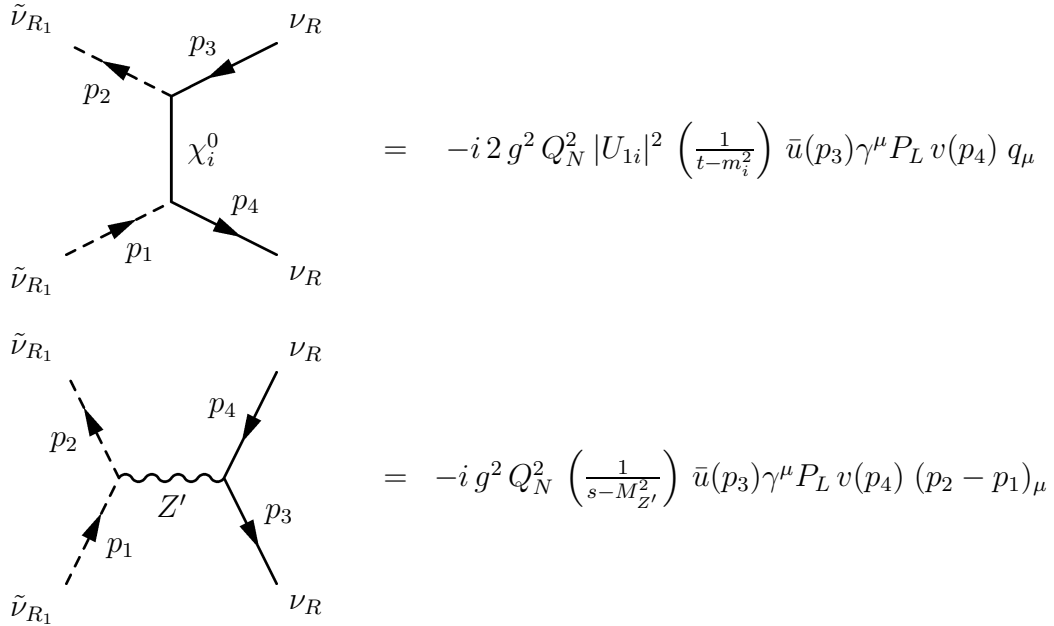

\indent Here we provide a computation of \(\hat{\sigma}(\tilde{\nu}\tilde{\nu})\) and
\(\hat{\sigma}(\tilde{\nu}\tilde{\nu}^{\ast})\).  Let us begin with
\(\hat{\sigma}(\tilde{\nu}\tilde{\nu})\).  For a right-handed sneutrino LSP, the only
relevant process is annihilation into neutrinos via $t$-channel exchange of hidden sector
neutralinos \(\chi^{0}_{j}\) shown in figure~\ref{fig:ViolateLsneutdiags}.  It it the
\(\tilde{Z}'\) component of each neutralino that mediates this exchange, hence neutralino
mass mixing will introduce factors of $U_{1j}$, where $U_{ij}$ is unitary matrix that
diagonalizes~(\ref{eq:HSneutralinoMassMat}).  The squared amplitude for this process is
\begin{equation}
\sum_{s,s'}|\mathcal{M}|^2 = 4g^4\sum_{i,j}{m_{\chi_i}^2}\,{s}\,
\left(\frac{1}{t-m_{\chi_i}^2} +
\frac{1}{u-{\chi_i}^2}\right)\left(\frac{1}{t-m_{\chi_j}^2} +
\frac{1}{u-{\chi_j}^2}\right).
\end{equation}
in terms of the Mandelstam variables \(s\), $t$, and $u$; the \(U(1)_{N}\) gauge coupling
\(g\) and the neutralino masses \(m_{\chi_i}\).  Let us assume that \(m_{\chi_1}\ll
m_{\chi_{i}}\) for all $i>1$.  Then if we define the the quantities
$x_{\chi_1}=s/m^2_{\chi_1}$, $x_n=s/m_{\tilde{\nu}_{R}}^2$,
$A=(2x_n^{-1}-2x_\chi^{-1}-1)$, this expression simplifies to
\begin{eqnarray}
  \sum_{s,s'}|\mathcal{M}|^2 &=& 4g^4{m_{\chi_i}^2}\,{s}\, \left(\frac{1}{t-m_{\chi_i}^2} +
  \frac{1}{u-{\chi_i}^2}\right)^2\nonumber\\&=&
  64g^4{m_{\chi_i}^2\over s}\left({A\over A^2-Y^2}\right)^2.
\end{eqnarray}
Using equation~(\ref{eq:SigmaWithHatOnIt}) and defining
$r\equiv\sqrt{1-4m_{\tilde{\nu}_R}^2/s}$, we find
\begin{equation}
  \hat{\sigma}(s)={g^4\over \pi}{1\over x_\chi}\left({1\over
  A}\ln{\left(A+r\over A-r\right)^2}-4{r\over r^2-A^2}\right)
\end{equation}
where the \(s\)-dependence comes in through both $A$ and $r$.

\begin{figure}[h!]
\begin{center}
\begin{displaymath}
\begin{array}{ccc}
\begin{fmffile}{NNstarann8}
  \fmfframe(20,20)(20,20){\begin{fmfchar*}(100,80)
    \fmfleft{i1,i2}
    \fmfright{o1,o2}
    \fmflabel{$\tilde{\nu}_{R_1}$}{i1}
    \fmflabel{$\tilde{\nu}_{R_1}$}{i2}
    \fmflabel{$\nu_{R}$}{o2}
    \fmflabel{$\nu_R$}{o1}
    \fmf{scalar,label=$p_1$}{i1,v1}
    \fmf{scalar,label=$p_2$}{v2,i2}
    \fmf{fermion,label=$p_4$}{v1,o1}
    \fmf{fermion,label=$p_3$}{o2,v2}
    \fmf{vanilla,label=$\chi^0_i$}{v1,v2}
  \end{fmfchar*}}
\end{fmffile}
&\raisebox{2cm}{$=$}&\raisebox{2cm}{$-i\,2\,g^2\,Q_N^2\,|U_{1i}|^2\,
  \left(\frac{1}{t-m_i^2}\right)\,\bar{u}(p_3)\gamma^{\mu}P_L\,v(p_4)\;q_{\mu}$}\\
\begin{fmffile}{NNstarannZ8}
  \fmfframe(20,20)(20,20){\begin{fmfchar*}(100,80)
    \fmfleft{i1,i2}
    \fmfright{o1,o2}
    \fmflabel{$\tilde{\nu}_{R_1}$}{i1}
    \fmflabel{$\tilde{\nu}_{R_1}$}{i2}
    \fmflabel{$\nu_R$}{o2}
    \fmflabel{$\nu_R$}{o1}
    \fmf{scalar,label=$p_1$}{i1,v1}
    \fmf{scalar,label=$p_2$}{v1,i2}
    \fmf{fermion,label=$p_4$}{o2,v2}
    \fmf{fermion,label=$p_3$}{v2,o1}
    \fmf{photon,label=$Z'$}{v1,v2}
  \end{fmfchar*}}
\end{fmffile}&\raisebox{2cm}{$=$}&\raisebox{2cm}{$-i\,g^2\,Q_N^2\,\left(\frac{1}{s-M_{Z'}^2}\right)\,
  \bar{u}(p_3)\gamma^{\mu}P_L\,v(p_4)\;(p_2-p_1)_{\mu}$}
\end{array}
\end{displaymath}
\end{center}
  \caption{Processes contributing to $\hat{\sigma}(\tilde{\nu}\tilde{\nu}^{\ast})$ in the
  right-handed sneutrino dark matter scenario.\label{fig:ConserveLsneutdiags}}
\end{figure}

\indent The reduced cross-section \(\hat{\sigma}(\tilde{\nu}\tilde{\nu}^{\ast})\) for
\(L_{\tilde{\nu}}\)-conserving processes is slightly more complicated, as we have not
only a \(t\)-channel contribution from neutralino exchange, but also an \(s\)-channel
contribution mediated by the $Z'$ gauge boson of \(U(1)_{N}\), as shown in
figure~\ref{fig:ConserveLsneutdiags}.  The squared amplitude for each of these processes
is
\begin{eqnarray}
  \sum_{s,s'}|\mathcal{M}_{Z'Z'}|^2 &=&g^4\,\left(\frac{1}{s-M_Z^2}\right)^2 s^2
  (r^2-Y^2)\\
  \sum_{s,s'}|\mathcal{M}_{\chi_1\chi_1}|^2
   &=& g^4\,\left(\frac{1}{u-M_\chi^2}\right)^2 s^2 (r^2-Y^2)\nonumber\\
   &=&4 g^4\,\frac{r^2-Y^2}{(A+Y)^2},
\end{eqnarray}
and the interference term between them is
\begin{eqnarray}
  \sum_{s,s'}|\mathcal{M}_{Z'\chi_1}|^2&=& g^4\,\frac{1}{u-M_\chi^2}\frac{1}{s-M_Z^2} s^2 (r^2-Y^2)\non\\
  &=&2 g^4\,\frac{s}{s-M_Z^2} \frac{r^2-Y^2}{A+Y}.
\end{eqnarray}
Straightforward calculation yields the reduced cross-section contributions
\begin{eqnarray}
  \hat{\sigma}_{Z'Z'}&=&{g^4\over 12\pi}\left(\frac{s}{s-M_Z^2}\right)^2 r^3\\
  \hat{\sigma}_{\chi\chi}&=&{g^4\over4\pi}\left(-4r+A \ln{\left(A+r\over
  A-r\right)^2}\right)\\
  \hat{\sigma}_{Z'\chi_1}
  &=&{g^4\over8\pi}\left(\frac{s}{s-M_Z^2}\right)\left(2Ar+
  {(r^2-A^2)\over2} \ln{\left(A+r\over A-r\right)^2}\right),
\end{eqnarray}
the sum of which appears in
equations~(\ref{eq:SigmaHatsofSneutConserve})\,-\,(\ref{eq:SigmaHatsofSneutViolate}).


\newpage

\end{document}